\documentclass[preprint,11pt]{elsarticle}

\usepackage{lineno}
\modulolinenumbers[5]

\usepackage{charter}
\usepackage{wrapfig}
\usepackage{arydshln}

\usepackage[colorlinks,citecolor=blue,linktoc=all,linkcolor=cyan]{hyperref}
\usepackage{graphicx}
\usepackage[usenames]{color}

\usepackage[T1]{fontenc}
\usepackage{dsfont}               
\usepackage{mathrsfs}             
\usepackage{slashed}              
\usepackage{amsmath}
\usepackage{amssymb}
\usepackage{amsbsy}
\usepackage{amsfonts}
\usepackage{booktabs}
\usepackage[applemac]{inputenc} 

\usepackage{tabularx}
\newcolumntype{R}{>{\raggedleft\arraybackslash}X} 
\newcolumntype{L}{>{\raggedright\arraybackslash}X}
\newcolumntype{C}{>{\centering\arraybackslash}X}  

\usepackage{nicefrac}
\usepackage{upgreek}
\usepackage{esint}
\usepackage{multirow}
\usepackage{caption}
\usepackage{floatrow}
\usepackage{subcaption}
\usepackage{bbm}
\usepackage{import}
\usepackage{empheq}

\numberwithin{equation}{section}
\numberwithin{table}{section}
\numberwithin{figure}{section}

\biboptions{numbers,sort&compress}

    \usepackage{sectsty}
    \usepackage{calc}

\usepackage[usenames]{color}
\usepackage{colortbl}

\definecolor{violet}{RGB}{111,0,255}
\definecolor{webgreen}{rgb}{0,0.75,0}
\definecolor{webred}{rgb}{0.75,0,0}
\definecolor{webblue}{rgb}{0,0,0.75}
\definecolor{darkblue}{rgb}{0,0,0.6}
\definecolor{darkgreen}{rgb}{0,0.5,0.5}
\definecolor{darkpurple}{rgb}{0.5,0,0.5}
\definecolor{darkorange}{rgb}{1,0.5,0}
\definecolor{darkgrey}{rgb}{0.4,0.4,0.4}
\definecolor{lgray}{rgb}{0.95,0.95,0.95}
\definecolor{lgreen}{rgb}{0.95,1.00,0.90}
\definecolor{lred}{rgb}{1.00,0.90,0.80}
\definecolor{lblue}{rgb}{0.2,0.35,1.00}
\definecolor{shadecolor}{rgb}{1.00,0.92,0.82}

\usepackage[vcentermath]{youngtab}

\usepackage{enumitem}
\setitemize{noitemsep,topsep=0ex,parsep=0.25ex,partopsep=0pt,leftmargin=20pt}
\setenumerate{topsep=0pt,parsep=0pt,partopsep=0pt} 

\usepackage{bm}
\newcommand{\vect}[1]{\bm{#1}}

\newcommand{\Tr}{\mathrm{Tr}}

\newcommand{\vp}{\vect{p}}

\newcommand{\op}{\omega_{p}}
\newcommand{\oq}{\omega_{q}}
\newcommand{\ok}{\omega_{k}}

\newcommand{\beq}{\begin{equation}}
\newcommand{\eeq}{\end{equation}}
\newcommand{\beqa}{\begin{eqnarray}}
\newcommand{\eeqa}{\end{eqnarray}}
\newcommand{\be}{\begin{equation}}
\newcommand{\ee}{\end{equation}}

\usepackage{titlesec}
\usepackage{sectsty}
\titleformat{\section}{\normalfont\Large\bfseries}{\thesection}{1em}{}
\titleformat{\subsection}{\normalfont\large\bfseries}{\thesubsection}{1em}{}
\titleformat{\subsubsection}{\normalfont\normalsize\bfseries}{\thesubsubsection}{1em}{}

\def\Slash#1{\setbox0=\hbox{$#1$} 
\dimen0=\wd0 
\setbox1=\hbox{/} \dimen1=\wd1 
\ifdim\dimen0>\dimen1 
\rlap{\hbox to \dimen0{\hfil/\hfil}} 
#1 
\else 
\rlap{\hbox to \dimen1{\hfil$#1$\hfil}} 
/ 
\fi}

\newcommand{\Pslash}{\Slash{P}}
\newcommand{\Dslash}{\Slash{D}}
\newcommand{\pslash}{\Slash{p}}
\newcommand{\qslash}{\Slash{q}}

\newcommand{\vpslash}{\Slash{\vect{p}}}

\newfloatcommand{capbtabbox}{table}[][\FBwidth]

\begin{document}

\journal{Progress in Particle and Nuclear Physics}

\begin{frontmatter}

\title{QCD at finite temperature and chemical potential from Dyson-Schwinger equations}

\author[L1,L2]{Christian~S.~Fischer~\fnref{myfootnote3}}

\address[L1]{Institut f\"ur Theoretische Physik, Justus-Liebig--Universit\"at Giessen, 35392 Giessen, Germany}
\address[L2]{HIC for FAIR Giessen, 35392 Giessen, Germany}

\fntext[myfootnote1]{christian.fischer@physik.uni-giessen.de}

\begin{abstract}
We review results for the phase diagram of QCD, the properties of quarks and gluons 
and the resulting properties of strongly interacting matter at finite temperature and 
chemical potential. The interplay of two different but related transitions in QCD, 
chiral symmetry restoration and deconfinement, leads to a rich phenomenology when 
external parameters such as quark masses, volume, temperature and chemical potential
are varied. We discuss the progress in this field from a theoretical perspective,
focusing on non-perturbative QCD as encoded in the functional approach via Dyson-Schwinger 
and Bethe-Salpeter equations. We aim at a pedagogical overview on the physics associated 
with the structure of this framework and explain connections to other approaches, in particular 
with the functional renormalization group and lattice QCD. We discuss various aspects associated
with the variation of the quark masses, assess recent results for the QCD phase diagram including 
the location of a putative critical end-point for $N_f=2+1$ and $N_f=2+1+1$, discuss results 
for quark spectral functions and summarise aspects of QCD thermodynamics and fluctuations.  
\end{abstract}


\end{frontmatter}

\newpage

\thispagestyle{empty}
\tableofcontents

\newpage

\section{Introduction}\label{intro}

Exploring the structure of the phase diagram of QCD, unravelling the potential 
existence of a critical end point (CEP) and mapping out the region of a first order transition at large 
chemical potential are major goals of current and future experimental programs at the 
Relativistic Heavy Ion Collider (RHIC) at the Brookhaven National Laboratory (BNL) and future experiments 
at the FAIR facility in Darmstadt and NICA in Dubna. These experiments seek to probe 
the chiral as well as the deconfinement transition from the hadronic state of matter to the 
quark-gluon plasma phase. The experimental verification of a CEP would be a major step in our 
understanding of QCD and an important cornerstone in the further exploration of the QCD phase diagram.

Non-perturbative methods are mandatory in this endeavour. Lattice QCD 
has firmly established the notion of an analytic crossover at zero chemical potential     
\cite{Aoki:2006we,Aoki:2009sc,Borsanyi:2010bp,Bazavov:2011nk,Bhattacharya:2014ara,Bazavov:2014pvz}.
However, the situation is much less clear at (real) chemical potential, where lattice calculations 
are hampered by the notorious fermion sign problem. Although much progress has been made in the past 
years it seems fair to say that a satisfactory solution of this problem is yet to be found. Various
extrapolation methods from zero or imaginary chemical potential into the real chemical potential region
have been explored and agree with each other for chemical potentials $\mu_B/T < 2$. Then errors
accumulate rapidly and solid predictions are not yet possible.

Larger chemical potentials are accessible by continuum methods, i.e. effective models and the 
functional approach. The Polyakov-loop enhanced effective models such as the Polyakov-loop 
Nambu--Jona-Lasinio model (PNJL) \cite{Fukushima:2003fw,Megias:2004hj,Ratti:2005jh} and the 
Polyakov-loop quark-meson model (PQM) \cite{Schaefer:2007pw,Skokov:2010wb,Herbst:2010rf} are 
excellent tools to study a range of fundamental and phenomenological questions related to the 
QCD phase diagram, see e.g. \cite{Drews:2016wpi,Fukushima:2017csk} for recent review articles and comprehensive
guides to the literature. These models rely on a chiral effective action augmented by the Polyakov loop
potential which serves as a background that couples the physics of the Yang-Mills theory (in particular
its confining aspects) to the chiral dynamics. However, gluons are no active degrees of freedom and
their reaction to the medium can neither be studied nor directly taken into account.   

This is possible within functional approaches to QCD. Dyson-Schwinger equations,
the functional renormalisation group, the Hamilton variational approach and the Gribov-Zwanziger formalism
work with the quark and gluon degrees of freedom and
determine the phase structure of QCD from order parameters extracted from Green's functions. 
In general, the functional approach is restricted by the need to truncate
an infinite system of equations to a level which can be dealt with numerically. The truncation 
assumptions, however, are not arbitrary and can be systematically assessed. 
Consequently, in the past decade functional methods have contributed
substantially to our understanding of the phase diagram, the properties of quarks and gluons 
and observable consequences related to thermodynamics, transport and fluctuations. 

Certainly, a review of this size cannot be complete and we selected the material reflecting our personal 
interest. We focus a choice of topics that has been addressed via the framework of Dyson-Schwinger (DSE)
and Bethe-Salpeter (BSE) equations and reflect the progress that has been made in the almost twenty years
since the renowned review of Roberts and Schmidt \cite{Roberts:2000aa}. Whenever appropriate, we will also 
make contact with results obtained using other functional approaches and provide links to results 
from lattice gauge theory. Nevertheless, it is clear that many equally interesting topics cannot be properly 
done justice to the given amount of space and time.  

We start the discussion in section \ref{gen} with general remarks on the QCD phase diagram, possible 
phases in the temperature and real chemical potential plane and comments on interesting extensions into 
several directions such as imaginary chemical potential, isospin chemical potential or non-zero magnetic 
field. We focus in some detail on the physics of the Columbia plot of non-physical quark masses and
highlight the interesting interplay of chiral and deconfinement transitions in various limits. 
In section \ref{DSE} we summarise the DSE approach and discuss truncation strategies. A brief overview
on selected results in the vacuum serves as a basis for the subsequent presentation of results
at finite temperature and chemical potential in section \ref{results}. We
walk through the Columbia plot starting with the pure gauge corner and the region of first and second order 
deconfinement transitions in section \ref{results:heavy}, spend some time in the upper left corner
of a potential second order chiral transition in section \ref{results:chiral} and then concentrate
on results for physical quark masses in section \ref{results:CEP}. In section \ref{results:thermo}
we summarise results on observable quantities related to thermodynamics and fluctuations and
discuss the issue of quark spectral functions and positivity in section \ref{results:spectral}. We then 
conclude the section with a brief overview on DSE-results for the color superconducting region of the
QCD phase diagram \ref{results:colorSC}. A general outlook is given in section \ref{sec:sum}. 
Some technical details are relegated to an appendix. 

Readers mostly interested in the context of the research field and the results and solutions offered 
by the functional approach are encouraged to skip section \ref{DSE} in a first reading and directly 
progress from section \ref{gen} to section \ref{results}. In a second reading the technical details 
and general considerations on DSEs offered in section \ref{DSE} may then be beneficial.
\newpage
\newpage

\section{Generalities}\label{gen}
\subsection{QCD phase diagram}\label{gen:phase}

\subsubsection{A sketch}\label{gen:sketch}

\begin{figure}[t]
    \centering
    \includegraphics[scale=0.50]{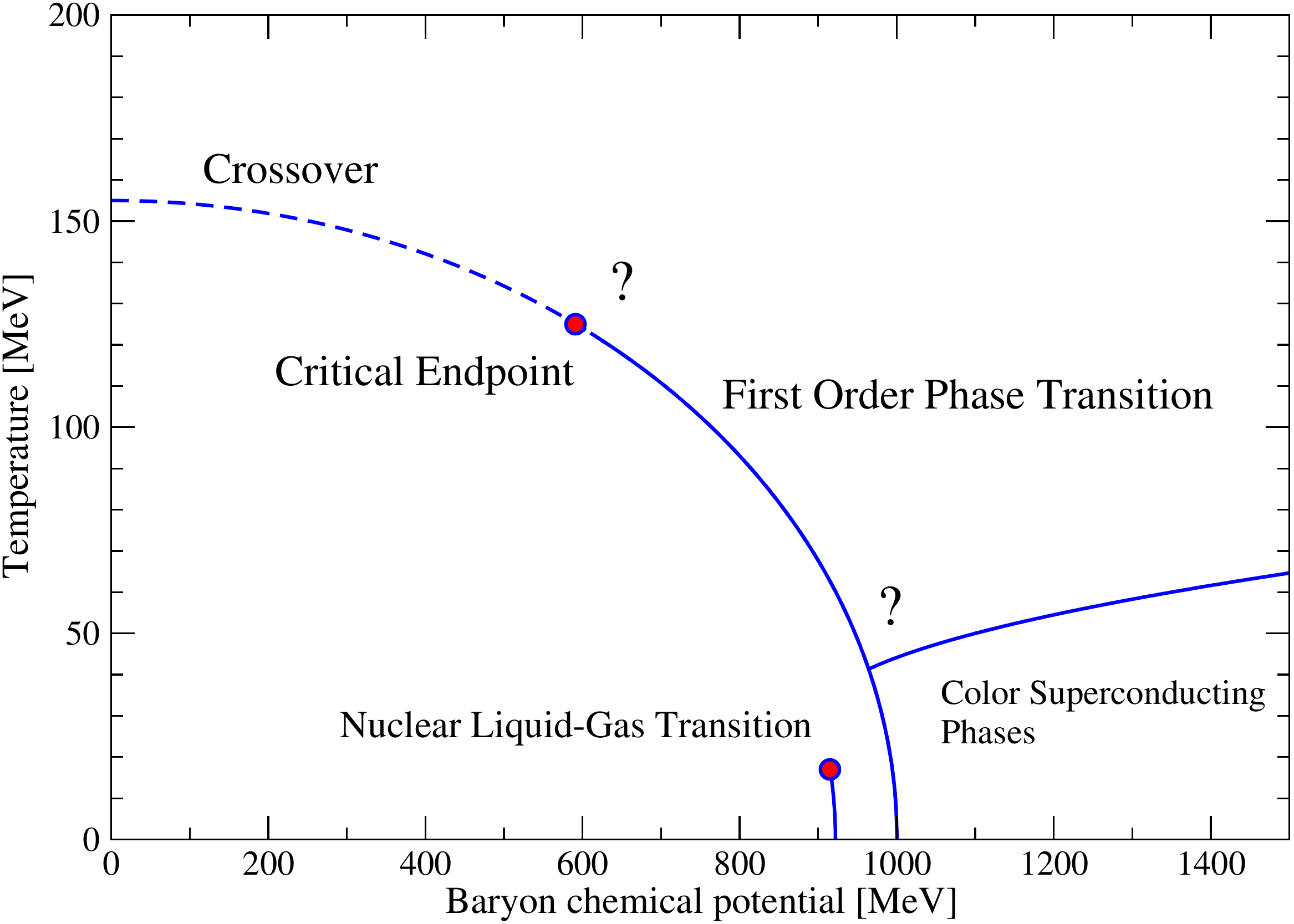}    
    \caption{Sketch of the QCD phase diagram in the temperature and baryon chemical potential plane. \label{fig:phase}}
\end{figure}

Even after decades of theoretical and experimental exploration, the sketch of the QCD phase diagram 
given in Fig.~\ref{fig:phase} is largely driven by (more or less well-grounded) speculation.
There is widespread agreement that results from lattice QCD  \cite{Aoki:2006we,Aoki:2009sc,Borsanyi:2010bp,Bazavov:2011nk,Bhattacharya:2014ara,Bazavov:2014pvz}
demonstrate an analytic cross-over at zero chemical potential from a low-temperature phase characterised 
by confinement and chiral symmetry breaking to a high-temperature deconfined and (partially) chirally 
restored phase where the quark-gluon plasma (QGP) is realized. 
The corresponding pseudo-critical temperature for the chiral transition has been localized
at $T_c \approx 155$ MeV with an error margin below ten MeV \cite{Borsanyi:2010bp,Bazavov:2011nk}.
Furthermore, the thermodynamic properties of 
the hot matter in a broad temperature range around $T_c$ have been pinned down with great accuracy
\cite{Borsanyi:2010cj,Borsanyi:2013bia,Bazavov:2014pvz,Ding:2015ona,Bazavov:2017dus} and serve as
input and benchmark for a large number of phenomenological applications. Thus the emerging standard 
picture of the situation at zero baryon chemical potential $\mu_B$ and physical quark masses is that 
of a continuous 
cross-over characterized by narrow but finite peaks in various susceptibilities. 
 
Many model calculations suggest, that this continuous cross-over becomes steeper with increasing 
chemical potential and finally merges into a second-order order critical end-point (CEP) followed
by a region of first-order phase transition at large chemical potential 
\cite{Asakawa:1989bq,Stephanov:1998dy,Stephanov:1999zu}. In Ref.~\cite{Stephanov:2004wx} 
one of the main arguments for the existence of such a critical end-point has been formulated accordingly:
(i) we know that there is an analytic cross-over at finite temperature and zero chemical potential;
(ii) we believe (from model studies) that the chiral phase-transition along the zero temperature and finite
chemical potential axis is first-order. It is then highly suggestive (if not thermodynamically unavoidable)  
that there has to be a critical end-point somewhere in the QCD phase diagram. Another argument along similar
lines even suggests the universality class of the CEP \cite{Halasz:1998qr,Berges:1998rc,Stephanov:1998dy}: 
From symmetry arguments (cf. the discussion in section \ref{gen:columbia} below) one is led to believe that 
the $T$-$\mu_B$-phase diagram of the two-flavour theory in the chiral limit features a line of second order phase 
transition points that merges into a tricritical point followed by a region of first order transition 
at large chemical potential - similar to the sketch of Fig.~\ref{fig:phase} but with cross-over replaced by
second order and critical endpoint replaced by tricritical point. This second order line is expected
to be in the universality class of O(4) spin models in three dimensions \cite{Pisarski:1983ms} with
three pseudoscalar pion fields and one scalar sigma field as massless degrees of freedom. In the theory with
massive quarks (Fig.~\ref{fig:phase}) the second order line collapses into a single second order point (the CEP), 
and the pion fields are no longer Goldstone bosons but become massive.
The remaining massless sigma field at the CEP places the theory then in the Z(2) universality class of the Ising
model in three dimensions\footnote{This picture including the O(4) and Z(2) scaling behaviour has been 
confirmed in renormalization group studies of the PQM model \cite{Schaefer:2006ds}.}. 

In principle, there are at least two obvious strategies to make these suggestions more rigorous: first,
one could follow the cross-over line into the phase diagram until one hits the critical end-point and second,
one could aim to nail down the first-order nature of the phase-transition at zero temperature and 
large chemical potential. The first strategy has been followed by lattice QCD and functional methods.
In general, lattice calculations suffer from the notorious sign-problem at finite chemical potential 
and therefore need to involve extrapolations from zero to positive real chemical potential. 
Methods like Taylor expansion, re-weighting schemes or extrapolation from imaginary chemical potential
(where the sign problem is not present) thus allow for indirect access to quantities at moderate chemical 
potential \cite{Karsch:2003jg,deForcrand:2006pv,Kaczmarek:2011zz,Endrodi:2011gv}. These methods have 
been refined over the years and work well up to the region of $\mu_B/T \lesssim 2$		
\cite{Bellwied:2015rza,Bazavov:2017dus}, after which errors accumulate rapidly. In contrast to early
lattice studies \cite{Fodor:2001pe,Fodor:2004nz,Datta:2012pj} which indicated the presence of a critical 
end-point at rather small chemical potentials, there seems to be agreement from recent studies that 
a potential CEP may only be located in the region $\mu_B/T > 2$ \cite{Bellwied:2015rza,Bazavov:2017dus}. 
As we will see in the course of this review, this finding is in agreement with the ones from 
Dyson-Schwinger studies \cite{Fischer:2014ata,Eichmann:2015kfa}. 

The second strategy, aiming at clarifying the situation at zero (or small) temperature and large chemical 
potential is hampered by the potentially rich physics in this region of the QCD phase diagram. It is 
probably fair to say, that there is currently no approach that captures all features of this physics and
therefore to quite some degree this is the realm of speculation, based on more or less rigorous studies. 
For zero temperature and small chemical potential, a well-founded expectation is the silver-blaze
property of QCD: unless the baryon chemical potential is larger than the lowest baryon mass in medium
(i.e. roughly the mass of the nucleon minus 16 MeV binding energy), the system must stay in the vacuum 
ground state and all observable quantities are similar to the vacuum. In the QCD path integral formulation, 
this property can be shown analytically for the case of finite isospin chemical potential, but on physical 
grounds it is at least extremely plausible also for the case of finite baryon chemical potential 
\cite{Cohen:1991nk,Cohen:2004qp}. In a lattice calculation with heavy but dynamical quarks the silver blaze 
property of QCD has been demonstrated in Ref.~\cite{Fromm:2012eb}. In the Dyson-Schwinger framework suitable
truncations that respect the silver blaze problem are discussed in \cite{Muller:2016fdr}. 

As soon as the chemical potential reaches values beyond the silver blaze region the system is able to produce 
baryonic matter. In model calculations (see e.g. \cite{Fukushima:2013rx,Drews:2016wpi} for overviews)
this is associated with a first-order phase transition which can be identified with the liquid-gas transition 
of infinite nuclear matter. The associated order parameter is the baryon density, which jumps from zero to 
$\rho_0 = 0.17/\mbox{fm}^3$. Signals for this transition have again been observed in an effective lattice theory 
\cite{Langelage:2014vpa} and are also visible in contemporary chiral mirror meson-baryon models 
\cite{Weyrich:2015hha}. 
At even larger chemical potential, model studies find the above mentioned first-order chiral transition, which 
in sufficiently rich models progresses directly into a (number of) color superconducting phase(s), see e.g.
\cite{Alford:1997zt,Rapp:1997zu,Alford:1998mk,Berges:1998rc,Alford:1999pa,Rajagopal:2000wf,Rischke:2003mt,
Buballa:2003qv,Alford:2007xm}. 
In the Dyson-Schwinger approach, superconducting phases have been studied in 
\cite{Nickel:2006vf,Nickel:2006kc,Marhauser:2006hy,Nickel:2008ef,Muller:2013pya,Muller:2016fdr}
with the aim to clarify the interplay between the superconducting 2SC phase (only up- and down-quarks 
form Cooper pairs) and the color-flavor-locked (CFL) phase (up- down- and strange-quarks are paired symmetrically).  
We will discuss this topic in more detail in section \ref{results:colorSC}. It is probably worth mentioning, that
in principle there could be a gap between the phases with broken chiral symmetry and the one with 
colour superconductivity. There has been some debate about this possibility in the context of QED$_3$ 
as an effective model for high temperature superconductors, see e.g.
Refs.~\cite{Franz:2001zz,Herbut:2002wd,Franz:2002qy,Fischer:2004nq}. In this context the gap is generated 
by a region with chirally restored but non-superconducting matter. There are, however, other possibilities 
as will be summarised in the next subsection.

\subsubsection{More structure and additional axis'}

So far we discussed the sketch in Fig.~\ref{fig:phase}, but there is much more.
An important possibility, well-known in solid state physics, has been 
suggested for QCD in \cite{Deryagin:1992rw,Shuster:1999tn,Park:1999bz,Rapp:2000zd}
and reviewed in \cite{Buballa:2014tba}. It is the appearance
of an inhomogeneous phase, where the quark condensate is spatially modulated at 
moderate temperatures and high densities. 
In the sketch of \ref{fig:phase} this corresponds to the region where the putative first order chiral
transition between the hadronic and the quark-gluon plasma or color superconducting phases
takes place. Investigations using Ginzburg-Landau theory, effective models (NJL, QM, PQM), 
large $N_c$ expansions or Dyson-Schwinger techniques indicate, that this possibility
has to be taken seriously. However, most of these studies are only performed on the mean field level.
Since it is known that fluctuations may have a significant influence on the phase structure of 
a theory, no firm conclusions can be made so far. A beyond mean field treatment has been performed 
within the framework of Dyson-Schwinger equations in Ref.~\cite{Muller:2013tya}, with a chiral-density-wave
like modulation of the condensate taken into account in the quark-sector of QCD (but not in closed quark loops).
One of the striking results of this calculation (in agreement with effective model approaches; see however
\cite{Carignano:2014jla} for counterexamples) is the
appearance of the inhomogeneous phase precisely at and around the first order transition with a 
so called Lifshitz point\footnote{I.e. the point where the inhomogeneous phase, the chirally symmetric 
and the chirally broken phase meet, c.f. \cite{Buballa:2014tba} p.20 for a discussion of the precise terminology.}
at the location of the critical end point. We will briefly come 
back to this result in section \ref{results:CEP}. It is an important task for the future to corroborate
(or reject) the existence of such inhomogeneous phases in more stringent approaches, in particular
with respect to their potential significance for signals in heavy ion experiments. 

Another possibility, introduced in Ref.~\cite{McLerran:2007qj} and reviewed in Ref.~\cite{Fukushima:2013rx} 
is the appearance of a phase of 'quarkyonic matter' in the same region of the phase diagram as for the 
inhomogeneous phase(s). Based on large-$N_c$ considerations, quarkyonic matter has been described as a state
of dense, strongly interacting baryons that has similar thermodynamic properties as quark matter. Its 
chiral properties are often associated with inhomogeneous condensates. Within this review, 
however, we will not cover this topic in any detail and we refer the interested reader to
the literature, see e.g. \cite{McLerran:2007qj,Kojo:2009ha,Torrieri:2010gz,Fukushima:2013rx} and 
references therein.    

There are also a number of possibilities to add more axis' to the sketch in Fig.~\ref{fig:phase}. A huge amount 
of literature is available that deals with the effects of non-zero magnetic field onto the various transitions
discussed above. Magnetic fields are interesting in connection with important physics applications: (i) in heavy 
ion-collisions huge (but short-lived) magnetic fields are created by the nuclei moving rapidly in opposite directions;
(ii) some compact stars, so called magnetars, are characterized by extremely high magnetic fields that may have 
a profound impact on the equation of state; (iii) magnetic fields may have played an important  role in the electroweak 
phase transition of the early universe. Comprehensive reviews on the effects of magnetic field are e.g.
Refs.~\cite{Andersen:2014xxa,Miransky:2015ava}. From a fundamental point of view, one of the most interesting 
effects of a non-vanishing magnetic field is the generation of magnetic catalysis, i.e. the fact that an external 
magnetic field initiates (or enhances) the dynamical generation of fermion masses even in theories with only weak 
interaction \cite{Klevansky:1989vi,Suganuma:1990nn}. This effect has been studied in detail over the years and
similarities and discrepancies with Cooper pairing effects have been worked out, see \cite{Miransky:2015ava} for
an overview. Perhaps surprisingly, results from lattice QCD also demonstrate that the opposite effect happens
for large enough temperatures: the dynamical mass generation is reduced and consequently the transition temperatures
for the crossover to the chirally restored phase is decreased
\cite{Bali:2011qj,Bali:2012zg,Bruckmann:2013oba,Bali:2013esa,Ilgenfritz:2013ara}. This effect, not present in 
simple models, can be traced back on the back-reaction of the quarks onto the Yang-Mills sector, and has been seen 
also in the Dyson-Schwinger framework \cite{Mueller:2015fka}. 
A further interesting consequence of non-vanishing magnetic fields, discussed extensively in the literature,
is the chiral magnetic effect of generated electric currents in heavy ion collisions, see e.g. \cite{Skokov:2016yrj} 
for a recent review. In this context the introduction of finite chiral chemical potential $\mu_5$ has been 
considered within lattice QCD \cite{Yamamoto:2011gk,Braguta:2010ej,Braguta:2015owi,Braguta:2016aov}, 
effective models \cite{Ruggieri:2011xc,Yu:2015hym,Ruggieri:2016xww,Ruggieri:2016ejz} and the 
Dyson-Schwinger framework \cite{Wang:2015tia,Xu:2015vna,Cui:2016zqp}.    

Another axis that can be added to the sketch in Fig.~\ref{fig:phase} is one of finite isospin chemical 
potential $\mu_I$, which
creates an imbalance between up- and down-quarks \cite{Son:2000xc}. This axis is by no means academic, since this imbalance
is present in heavy ion collisions (due to different numbers of protons and neutron in the colliding nuclei), as well
as in compact stars due to $\beta$-equilibrium and charge neutrality. The study of the QCD phase diagram with zero baryon chemical potential 
but non-zero $\mu_I$ is furthermore of systematic interest since it allows comparisons of the results of lattice QCD 
(which can be simulated at finite $\mu_I$ \cite{Kogut:2004zg,deForcrand:2007uz,Detmold:2012wc,Brandt:2017oyy}) with studies
in other approaches, see e.g. \cite{Splittorff:2000mm,Kamikado:2012bt,Loewe:2002tw,Klein:2003fy,Andersen:2015eoa, Stiele:2013pma} 
and references therein. The physics of finite isospin chemical potential is related to several phase transitions. For 
$\mu_I < \mu_\pi/2$ the silver-blaze phenomenon takes place, similar to the situation at finite baryon chemical potential.
One then finds a second order transition to a phase with pion condensation, which may even play a role in compact
stars, see e.g. \cite{Brandt:2018bwq} and references therein. The pion condensation phase extends to finite temperature and
may show another first order transition accompanied with a CEP inside \cite{Son:2000xc,Kamikado:2012bt}. More 
interesting phenomena (like the appearance of a Fulde-Ferrell-Larkin-Ovchinnikov (FFLO) 
phase \cite{Fulde:1964zz,larkin:1964zz,Son:2000xc}) occur when both the baryon chemical potential and the isospin 
chemical potential are considered.

\subsubsection{Curvature of the phase boundary}\label{sec:curvature}

\begin{figure}[t]
\begin{floatrow}
\ffigbox{\includegraphics[scale=0.35]{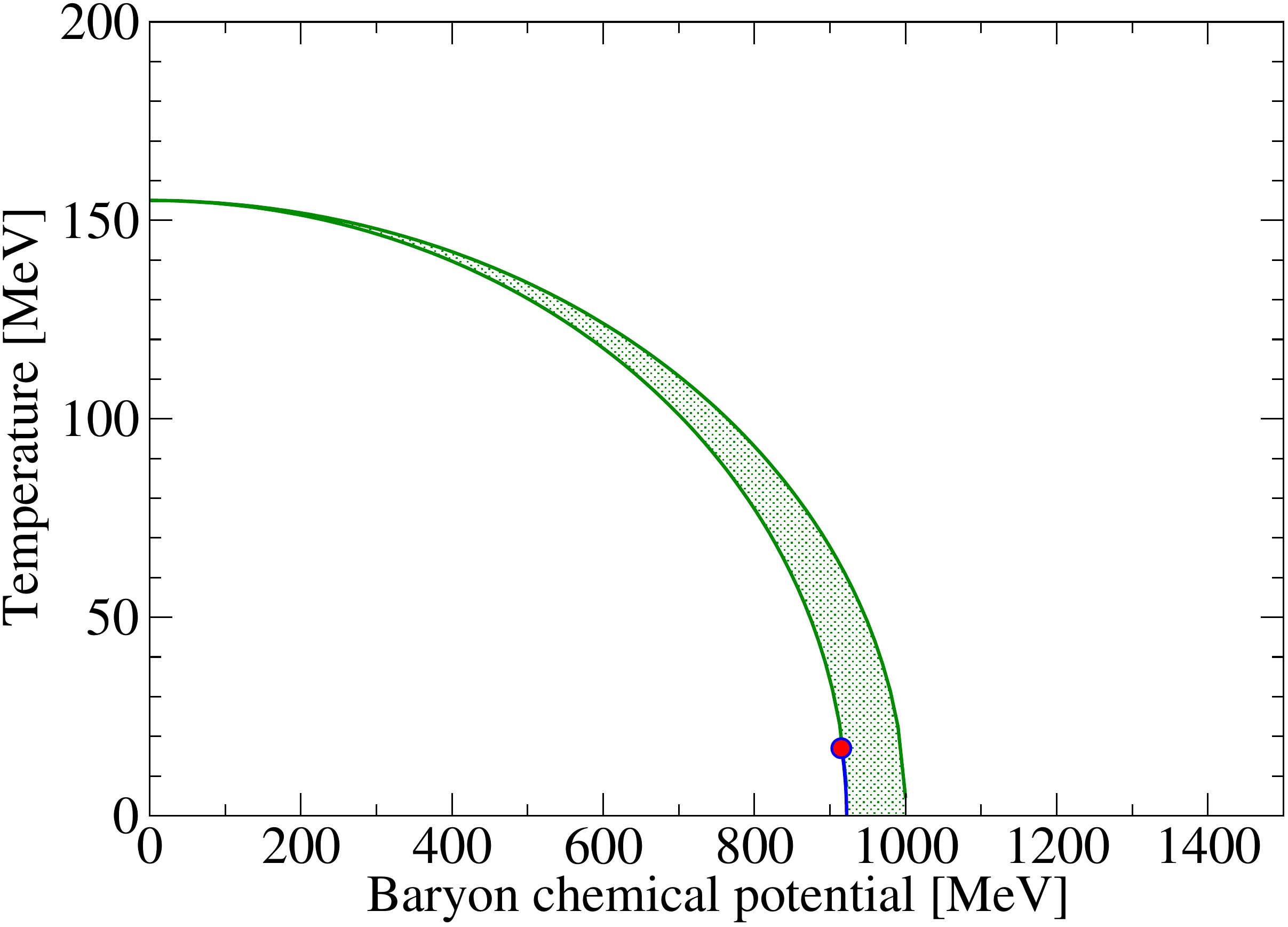}}
{\caption{Sketch of the QCD phase diagram - visualisation of the extrapolated curvature range 
$0.0120 < \kappa < 0.0141$ explained in the main text.}\label{fig:curvature}}\hfill
\capbtabbox{
\begin{tabular}{|c|c|}\hline\hline
								& $\kappa$ 		\\\hline\hline
Lattice \cite{Cea:2014xva}		& 0.0180 (40)	\\
Lattice \cite{Bonati:2014kpa}	& 0.0130 (30)	\\
Lattice \cite{Bonati:2015bha}	& 0.0135 (20)	\\
Lattice \cite{Bellwied:2015rza}	& 0.0149 (21)	\\
Lattice \cite{Bonati:2018nut}	& 0.0145 (25)	\\\hline
QM-model ($N_f=2$) \cite{Braun:2011iz}					& 0.0155 (7) 	\\
QM-model ($N_f=2$) \cite{Pawlowski:2014zaa}				& 0.0157 (1)	\\
														& 0.0160 (1)	\\
														& 0.0089 (1)	\\\hline
DSE \cite{Fischer:2014ata}								& 0.0238 (100)	\\\hline											
\end{tabular}
}
{\caption{Results for the curvature $\kappa$ of the transition line from different approaches.
The error budget of the QM-model contains errors from the fitting procedure only; the error of the DSE-result
is based on an estimate according to the systematic corrections introduced in \cite{Eichmann:2015kfa}. 
If not indicated otherwise, all calculations are performed with $N_f=2+1$ quark flavours.}\label{tab:curvature}}
\end{floatrow}
\end{figure}   

Let us now come back to the sketch of Fig.~\ref{fig:phase} and consider again the boundary of the chiral transition
in the region close to vanishing chemical potential. As has been mentioned in the introduction, lattice gauge theory
at non-zero chemical potential is hindered by the sign problem. However, various methods like Taylor expansion, 
re-weighting schemes or extrapolation from imaginary chemical potential have been developed and refined over the years,
such that reliable results in the region $\mu_B/T \lesssim 2$ are available. One of the most interesting issues from an
experimental point of view may be the determination of the pseudo-critical line separating the 
low-temperature phase from the high-temperature one. At small chemical potential, this line can be parametrised
by an expansion quadratic in the dimensionless ratio of chemical potential to temperature:
\begin{equation}\label{eq:kappa}
\frac{T_c(\mu_B)}{T_c} = 1-\kappa\left(\frac{\mu_B}{T_c}\right)^2 - \lambda \left(\frac{\mu_B}{T_c}\right)^4 \cdots\,,
\end{equation}   
with baryon chemical potential $\mu_B$, pseudo-critical temperature $T_c(\mu_B)$ and $T_c=T_c(0)$\footnote{A word of 
caution is
in order here: this expansion has been used in the literature in different forms, sometimes it is formulated not
in baryon but in quark chemical potential and sometimes factors of $\pi^2$ are included, resulting in trivial 
changes of the values of the expansion coefficients $\kappa$ and $\lambda$. Sometimes on the right hand side 
$T_c(\mu_B)$ is used instead of $T_c(0)$. The latter change
is immaterial for small chemical potential but has some impact on extrapolations at larger $\mu_B$. In this review
we will stick to the formulation Eq.~(\ref{eq:kappa}).}. The expansion is quadratic, since the grand canonical 
QCD partition function $Z$ is symmetric with respect to a change of sign in $\mu_B/T$ \cite{deForcrand:2002hgr}
\begin{equation}\label{eq:Z}
Z\left(\frac{\mu_B}{T}\right) = Z\left(-\frac{\mu_B}{T}\right)\,,
\end{equation}   
and therefore all odd powers of $\mu_B/T$ in the expansion have to vanish.

Which values for the curvature $\kappa$ of the critical transition line can we expect in Eq.~(\ref{eq:kappa})?
Suppose for the sake of the argument that the sketch of Fig.~\ref{fig:phase} represents the qualitative aspects 
of the QCD phase diagram accurately and the chiral
transition line is not affected by additional physics in the low-temperature and high chemical potential region
such as inhomogeneous condensates or other phenomena. 
A naive estimate of $\kappa$ can then be based on the following consideration: Because 
of Eq.~(\ref{eq:Z}) and analyticity at $\mu_B=0$ and $T=0$, the transition
lines cross the temperature and chemical potential axis' perpendicularly.\footnote{At zero temperature and finite 
chemical potential this is also guaranteed thermodynamically by the Clausius-Clapeyron relation, provided the transition
is first order.} 
A suitable functional form for the
entire transition line including the cross-over and first-order sections that reproduces both, the quadratic expansion
(\ref{eq:kappa}) around zero chemical potential but also its counterpart for the first order line around zero temperature
is elliptic, i.e.
\begin{equation}\label{eq:ellipse}
\left(\frac{T_c(\mu_B)}{T_c}\right)^2 = 1 - 2\kappa\left(\frac{\mu_B}{T_c}\right)^2\,.
\end{equation}
The two free parameters are the transition temperature at zero chemical potential $T_c\equiv T_c(\mu_B=0)$ and the
curvature $\kappa$ of the transition line. The former is well determined from lattice QCD, 
$T_c \approx 155$ MeV \cite{Borsanyi:2010bp,Bazavov:2011nk}. Using this value and the safe assumption that the chiral 
transition at $T=0$ cannot happen for chemical potentials smaller than the liquid-gas transition 
at $\mu_B^{lg} \approx 922$ MeV we immediately obtain an estimate for the lower bound of the curvature  
\begin{equation}\label{eq:kappaguess}
\kappa \le 0.0141\,. 
\end{equation}
It is interesting to compare this naive estimate with the results from lattice calculations and continuum approaches
displayed in table \ref{tab:curvature}. While earlier lattice studies obtained lower values for the curvature
\cite{Philipsen:2008gf,Kaczmarek:2011zz,Endrodi:2011gv}, the recent continuum extrapolated results now indicate 
convergence between different methods (Taylor expansion techniques and analytical continuation from
imaginary chemical potential) \cite{Cea:2014xva,Bonati:2014kpa,Bonati:2015bha,Bellwied:2015rza,Bonati:2018nut}. 
The naive estimate of Eq.~(\ref{eq:kappaguess}) agrees very well with the lattice results. It is also not 
too far from the results for the quark-meson model (although this comparison is not rigorous due to differences
in $N_f$). The DSE-result of \cite{Fischer:2014ata} is somewhat larger,
although the results of \cite{Eichmann:2015kfa} indicate that there is a systematic error that can account for 
this discrepancy; this is discussed in detail in sections \ref{results:2p1} and \ref{results:baryons}. 

Thus it seems as if our naive estimate is not so bad at all. Indeed, expanding Eq.~(\ref{eq:ellipse}) and comparing
with Eq.~(\ref{eq:kappa}) we can also predict the size of the coefficient $\lambda$ in Eq.~(\ref{eq:kappa}) to be 
given by 
\begin{equation}\label{eq:lambdaguess}
\lambda = \frac{1}{2} \kappa^2 \le 0.0001\,. 
\end{equation}
This value is very small. Recent lattice estimates indeed confirm that $\lambda$ may be orders of magnitude 
smaller than $\kappa$ \cite{SteinbrecherI,Steinbrecher:2018phh}, although the error bars are still
large.

Taking the error band of the lattice result in Ref.~\cite{Bonati:2018nut} as a lower limit we arrive at the
spread of transition lines show in Fig.~\ref{fig:curvature} representing the range $0.0120 < \kappa < 0.0141$.
This construction establishes a connection between rigorous results at finite temperature and zero chemical 
potential and the zero temperature large chemical potential region, and predicts a chiral transition at 
$ 922 \lesssim \mu_B \lesssim 1000$ MeV. Of course, this naive construction may very well be modified by 
dynamical effects at large chemical potential including the potential appearance of additional phases such 
as the inhomogeneous one discussed in the previous section. Nevertheless, it is amusing to see that this 
simple procedure leads to intuitive and interesting predictions.

\subsubsection{Detecting the critical end point by observables: fluctuations}\label{sec:fluct}

As mentioned in the introduction, it is one of the main goals of contemporary 
(Beam Energy Scan program at RHIC) and future (CBM/FAIR and NICA) experimental
programs to study the existence and the location of the critical end point in the QCD phase diagram. To this end it 
is vital to identify observables that connect the theoretical properties of the CEP with experimental data. 
This endeavour is reviewed extensively in Ref.~\cite{Luo:2017faz} and we therefore give only a brief overview here.
Provided the chemical freeze-out in heavy ion collisions is sufficiently close to the chiral critical line and
the CEP, it has been suggested \cite{Stephanov:1998dy,Stephanov:1999zu,Asakawa:2000wh,Jeon:2000wg,Koch:2005vg,Ejiri:2005wq,Friman:2011pf}
that fluctuations of conserved charges provide important information on the location of the CEP. In the experiments 
these appear as event-by-event fluctuations of the net baryon number $B$, the electric charge $Q$ or the strangeness $S$
of the heavy ion system. In particular, ratios of susceptibilities are expected to provide clean signals. 

In order to analyse these quantities theoretically one starts from the dimensionless pressure $P/T^4$ extracted 
from the QCD partition function via
\beq
\frac{P}{T^4} = \frac{1}{V T^3} \ln[Z(V,T,\mu_B,\mu_Q,\mu_S)]\,,
\eeq    
with Lagrange multipliers for the baryon chemical potential $\mu_B$, the charge $\mu_Q$ and the strangeness chemical
potential $\mu_S$. The normalized generalised susceptibilities are defined via 
\beq
\chi^{BSQ}_{lmn} = \frac{\partial^{l+m+n}(p/T^4)}{\partial(\mu_B/T)^l \partial(\mu_S/T)^m \partial(\mu_Q/T)^n}\,.
\eeq  
Experimentally, ratios of cumulants 
\beq
C^{BSQ}_{lmn} = V T^3 \chi^{BSQ}_{lmn}
\eeq
are extracted that do not depend explicitly on the volume (though there may be implicit dependencies) 
and can be directly compared with ratios of theoretical susceptibilities, see \cite{Luo:2017faz} for details. 
They are related to statistical quantities via
\begin{align}
\textrm{mean:}\hspace*{10mm} M_{B} &= C_1^{B}\,,  \nonumber\\
\textrm{variance:}\hspace*{10mm} \sigma^2_{B} &= C_2^{B}\,, \nonumber\\
\textrm{skewness:}\hspace*{10mm} S_{B} &= C_3^{B}/(C_2^{B})^{3/2}\,, \nonumber\\
\textrm{kurtosis:}\hspace*{10mm} \kappa_{B} &= C_4^{B}/(C_2^{B})^2\,, 
\end{align}
for the example of baryon number and analogous expressions for charge and strangeness. 

In terms of quark degrees of freedom, the chemical potentials for baryon number, strangeness and charge can be
related to the chemical potentials of the up, down and strange quarks via 
\begin{align}
\mu_u &= \mu_B/3 + 2\mu_Q/3\,,         \nonumber\\
\mu_d &= \mu_B/3 -  \mu_Q/3\,,                  \\
\mu_s &= \mu_B/3 -  \mu_Q/3 - \mu_S\,. \nonumber
\end{align}
In order to take into account the situation of heavy-ion collisions, these need to be adjusted appropriately.
Strangeness conservation in the colliding nuclei implies that the mean density of strange quarks vanishes, i.e.  
$\langle n_s \rangle = \chi^S_1 = 0$. On the other hand, typical ratios of the number of baryons to protons
in Au-Au and Pb-Pb collisions imply that $\langle n_Q \rangle = Z/A \,\langle n_B \rangle$ with $Z/A \approx 0.4$.
Thus the dependence of $\mu_Q$ and $\mu_S$ on $\mu_B$ or alternatively $\mu_u$, $\mu_d$ and $\mu_s$ need to be 
defined such that these conditions are satisfied. This has been studied in lattice simulations at small chemical 
potentials \cite{Bazavov:2012vg,Borsanyi:2013hza}.
Within errors both groups agree that for temperatures around $T=150$ MeV the leading order result is 
$\mu_Q \approx -0.02 \,\mu_B $, while $\mu_S \approx 0.2 \,\mu_B$. Thus to a good approximation one can choose 
$\mu_u=\mu_d$ and $\mu_s=0$ to explore the QCD phase diagram. For the location of the CEP this has been checked 
in the DSE approach \cite{Welzbacher:2016}, cf. also section \ref{results:CEP}.          

Ratios of generalised susceptibilities have been used to extract the freeze-out points from 
experimental data from first principles lattice calculations \cite{Bazavov:2012vg,Borsanyi:2013hza} that
can be compared with those obtained from the Hadron Resonance Gas model \cite{Karsch:2010ck}. 
Furthermore various higher order fluctuations have been determined as functions of temperature and chemical potential
\cite{Cheng:2008zh,Borsanyi:2011sw,Bazavov:2012jq,Bellwied:2015lba,DElia:2016jqh,Bazavov:2017tot,Borsanyi:2018grb}. 
In model calculations,
it has been shown that these fluctuations are sensitive to the critical end-point, see e.g.
\cite{Schaefer:2006ds,Schaefer:2011ex,Fu:2016tey,Almasi:2017bhq} and references therein. 
Important results include the discovery of the smallness of the critical region around the CEP 
in calculations using the renormalization group approach \cite{Schaefer:2006ds} and the appearance 
of different critical exponents depending on the path across the CEP \cite{Schaefer:2011ex}. Detailed 
comparisons between the model results and experimental data have been performed e.g. in \cite{Fu:2016tey,Almasi:2017bhq}. 

\subsection{The Columbia plot}\label{gen:columbia}

\subsubsection{A sketch}
\begin{figure}[t]
    \centering
    \includegraphics[scale=0.33]{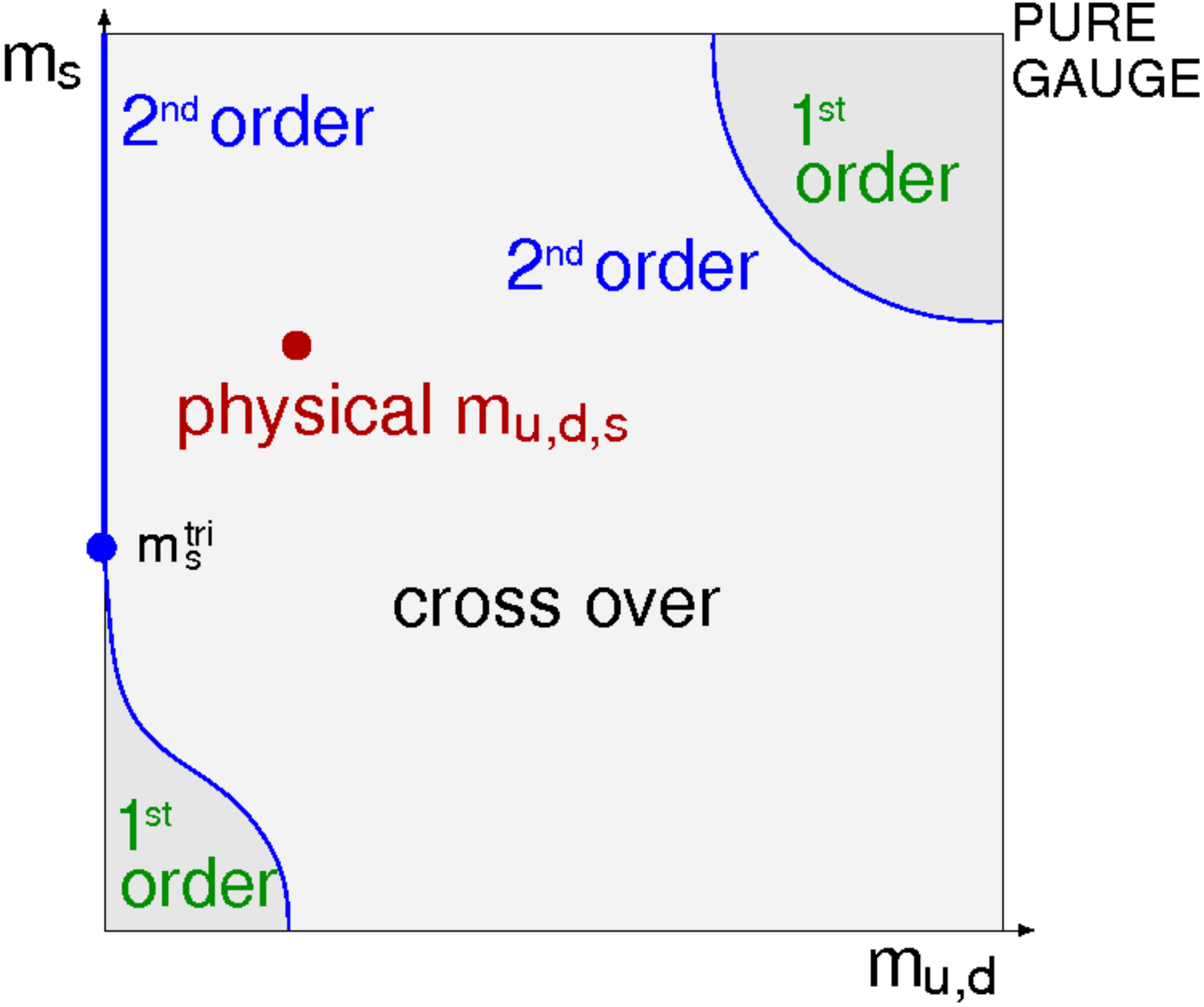}\hfill
    \includegraphics[scale=0.35]{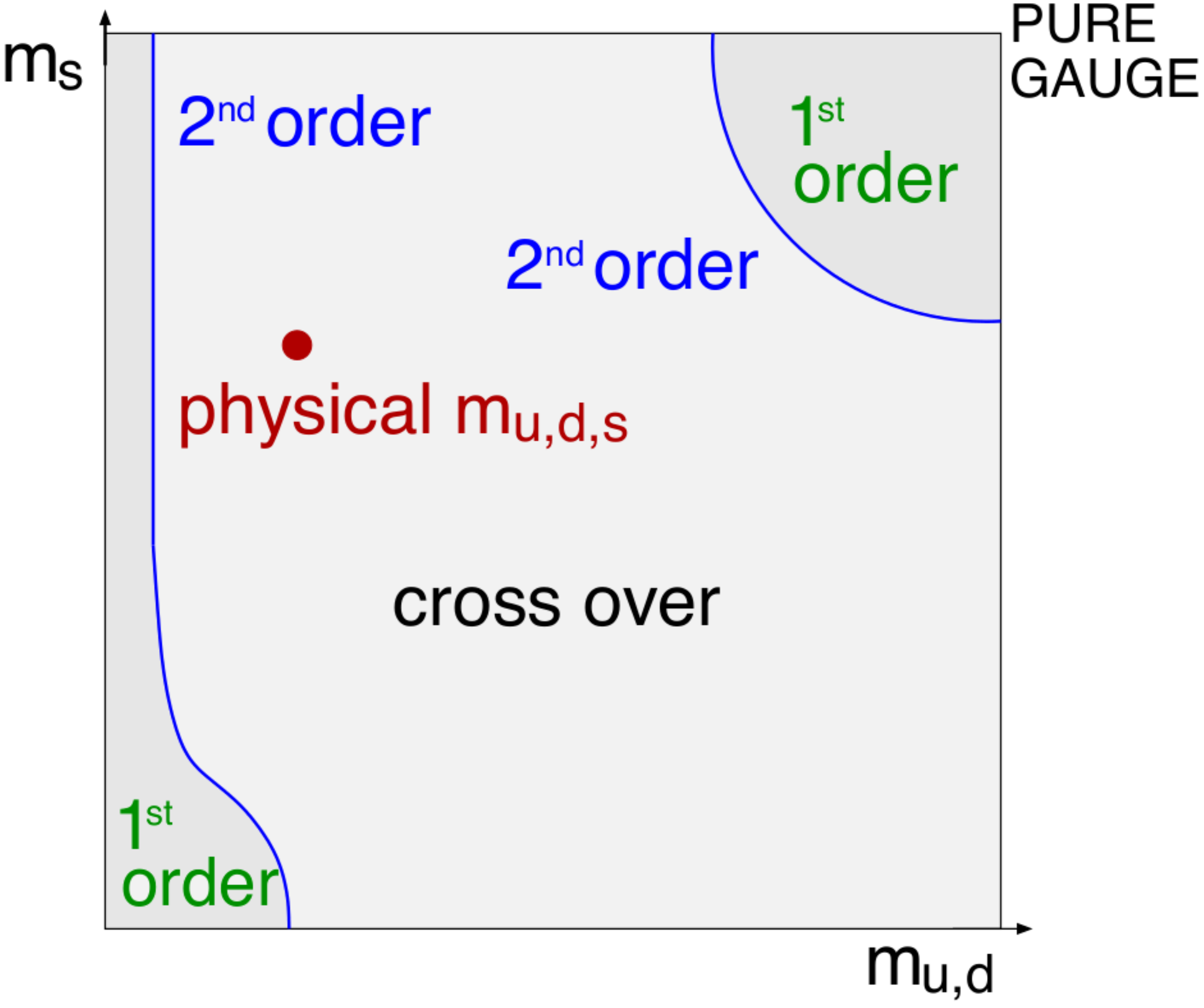}
    \caption{Left: Columbia plot of phase transition lines as functions of quark masses including $U_A(1)$ anomaly.\\
            Right: Same plot with $U_A(1)$ anomaly restored. 
            \label{fig:columbia}}
\end{figure}

The sketch of the QCD phase diagram discussed in the last section is, if at all, only valid for 'physical quark 
masses', i.e. quark masses that lead to a physical hadron spectrum in agreement with experiment. From a theoretical 
point of view it is also highly interesting to consider situations with unphysical values for the up-, down- and 
strange-quark masses. The variation of these reveals the intricate interplay of chiral and deconfinement transitions, 
sketched in Fig.(\ref{fig:columbia}), the 'Columbia plot' \cite{Brown:1990ev}. Each of these transitions is related to 
an underlying symmetry of QCD: chiral symmetry and center symmetry. Their explicit breaking due to non-vanishing
(chiral) or non-infinite (center) quark masses generates the pattern displayed in Fig.(\ref{fig:columbia}). 

Let us briefly discuss the various regions in the plot starting from the pure gauge theory limit of infinitely 
heavy masses in the upper right corner. In this limit the theory is symmetric with respect to center transformations
associated with the center of the gauge group, i.e. Z(3) for SU(3). Whereas center symmetry is maintained in the
low temperature phase, it is broken dynamically at large temperatures. The transition temperature of the associated
first-order deconfinement transition \cite{Yaffe:1982qf} is much larger than the pseudo-critical one at the physical 
point: for SU(3) it is around $T_c \approx 270$ MeV, see e.g.~\cite{Karsch:2003jg}, whereas other gauge groups result 
in different values \cite{Lucini:2005vg}. The transition can be traced by center sensitive order parameters such as the 
Polyakov-loop (cf. section \ref{DSE:order}). Finite-mass quarks in the 
fundamental representation of the gauge group furthermore break center symmetry explicitly and turn the first-order transition
into a cross-over at light enough masses. The second order separation line in the upper right corner of the 
Columbia plot is in the Z(2) universality class
\cite{deForcrand:2002hgr,deForcrand:2003vyj} and its location in the u/d-s-quark mass plane has been mapped 
out by lattice gauge theory \cite{deForcrand:2010he,Saito:2011fs,Fromm:2011qi}, 
effective models \cite{Kashiwa:2012wa,Lo:2014vba}, the Dyson-Schwinger approach \cite{Fischer:2014vxa} and 
background field techniques \cite{Reinosa:2015oua,Maelger:2017amh}. 
We come back to this issue in much more detail in section \ref{results:heavy}.

The low-mass corners of the Columbia plot are governed by the chiral transition. Massless QCD with $N_f$ quark 
flavours is invariant under a global flavour symmetry $U_V(1) \times U_A(1) \times SU_V(N_f) \times SU_A(N_f)$. 
Whereas $U_V(1)$ is conserved and related to the baryon number, $SU_V(N_f)$ is explicitly broken by differences 
in the finite quark masses of the QCD-Lagrangian. The most important properties of the chiral transitions of QCD 
are governed by the two axial symmetries $U_A(1) \times SU_A(N_f)$. Whereas the latter one is broken dynamically
at low temperatures (and always explicitly by finite quark masses), the former one is broken anomalously. The
corresponding current $J_\mu^5 = \bar{\Psi} \gamma_\mu \gamma_5 \Psi$ with quark fields $\Psi$ is not conserved,
\begin{equation}
\partial^\mu J_\mu^5 = \frac{g^2 N_f}{16 \pi^2} \textrm{tr}\left(\tilde{F}_{\mu \nu}F^{\mu \nu}\right)\,,
\end{equation}
due to the appearance of the topological charge density on the right hand side. Both, the dynamical and anomalous 
breaking can be restored at large temperatures, albeit the corresponding transition temperatures may very well 
differ from each other, since the underlying physics is different: any temperature effect that significantly 
reduces the interaction strength of QCD will lead to the restoration of $SU_A(N_f)$; in order to restore 
the chiral $U_A(1)$, however, one needs an effect that reduces the topological charge density. Although both
restoration mechanisms may be related, they are not necessarily so. 

The fate of the $U_A(1)$-symmetry is expected to affect the order of the chiral $SU_A(N_f)$ transition
as shown in the two versions of the Columbia plot given in Fig.~\ref{fig:columbia}. With an anomalously broken 
$U_A(1)$ at all temperatures it has been conjectured that the chiral transition for the two flavour theory
(upper left corner of the plot) is second order and in the universality class of the $O(4)$ theory, whereas
with restored $U_A(1)$ the transition may remain first order \cite{Pisarski:1983ms} (right diagram). In both scenarios the
chiral three-flavour theory (lower left corner in both diagrams) is expected to be first order\cite{Pisarski:1983ms}, 
since no three-dimensional $SU(N_f\le 3)$ second order universality class is known \cite{Butti:2003nu,deForcrand:2017cgb}.  
With restored $U_A(1)$ the
two first order corners are expected to be connected as shown in the right diagram of Fig.~\ref{fig:columbia}.
The corresponding scenario with broken $U_A(1)$ in the left diagram, however, features a tricritical strange quark
mass $m_s^{\textrm{tri}}$ where the first order region around the chiral three-flavour point merges into
the second order line connected to the chiral two-flavour point. 

It is currently an open question which of 
these scenarios is realised in QCD. For the theory with three degenerate flavours, lattice studies seem to 
support the existence of a first order region in the lower left corner of the Columbia plot
\cite{Karsch:2001nf,Karsch:2003va,deForcrand:2007rq,Ding:2011du,Jin:2014hea,Takeda:2016vfj,Bazavov:2017xul}.
However, the size of the first order region depends strongly on the formulation of the lattice action and 
the temporal extend of the lattice and has not yet been determined unambiguously. Even the possibility that 
the first order region vanishes in the lattice continuum limit is not yet excluded \cite{deForcrand:2017cgb}.
The situation in the upper left corner and, related, in the chiral limit of the $N_f=2+1$-theory with strange
quark mass fixed is also not clear and indications from lattice simulations vary between favouring either of 
the two scenarios of Fig.~\ref{fig:columbia}  \cite{Iwasaki:1996ya,DElia:2004uwa,DElia:2005nmv,Kogut:2006gt,Bonati:2014kpa,
Dick:2015twa,Philipsen:2016hkv,Cuteri:2017gci,Ding:2018auz}. 

Both scenarios of Fig.~\ref{fig:columbia} can be also realised in effective low energy QCD models such as the 
PQM or PNJL model, see e.g. \cite{Lenaghan:2000kr,Kovacs:2006ym,Fukushima:2008wg,Schaefer:2008hk,Mitter:2013fxa,Resch:2017vjs} 
and Refs. therein. In Ref.~\cite{Resch:2017vjs} it has been demonstrated that results on the Columbia plot 
from mean field approaches are substantially modified once fluctuations have been included using the 
functional renormalisation group (FRG). 
Depending on the strength of the t'Hooft interaction parametrizing the effects of the $U_A(1)$-anomaly one then 
ends up with a chiral phase structure precisely along the lines of Fig.~\ref{fig:columbia} with small first order
regions in both scenarios. First results on the chiral two-flavour theory are also available for functional 
FRG and DSE approaches to QCD: while the FRG-approach suggests a second order transition in the chiral limit
\cite{Braun:2009gm}, it has also been shown how $O(4)$-scaling might emerge in the DSE-approach \cite{Fischer:2011pk}. 
We will come back to this point in much more detail in section \ref{results:chiral}.

\subsubsection{Extension to real and imaginary chemical potential}\label{Columbia_ext}

\begin{figure}[t]
    \centering
    \begin{subfigure}[b]{0.48\textwidth}
    \includegraphics[scale=0.32]{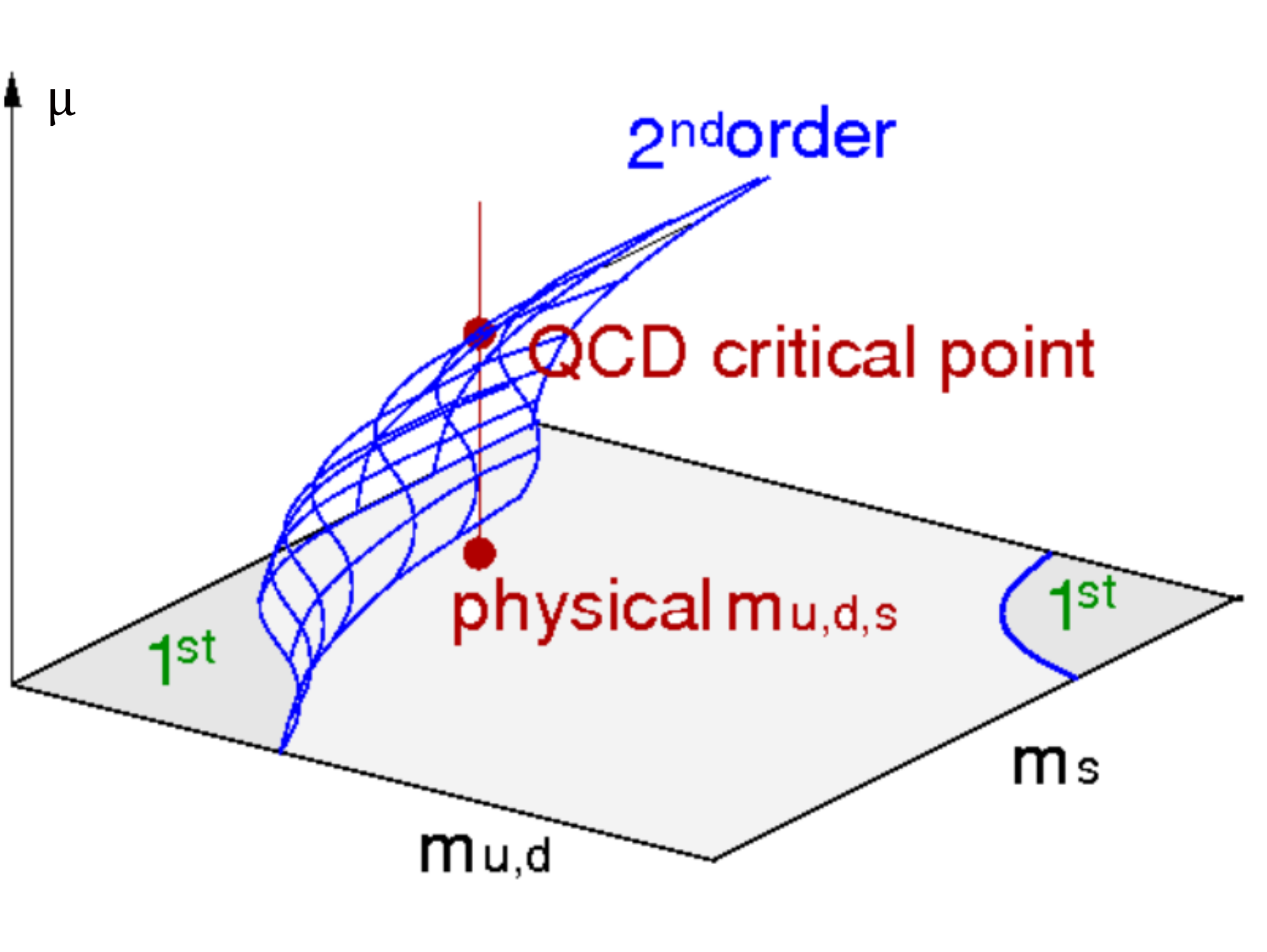}\hspace*{10mm}
    \caption{Real chemical potential (chiral transition only).        
            }\label{fig:columbia_ext_real}
    \end{subfigure}
    \begin{subfigure}[b]{0.48\textwidth}
    \includegraphics[scale=0.40]{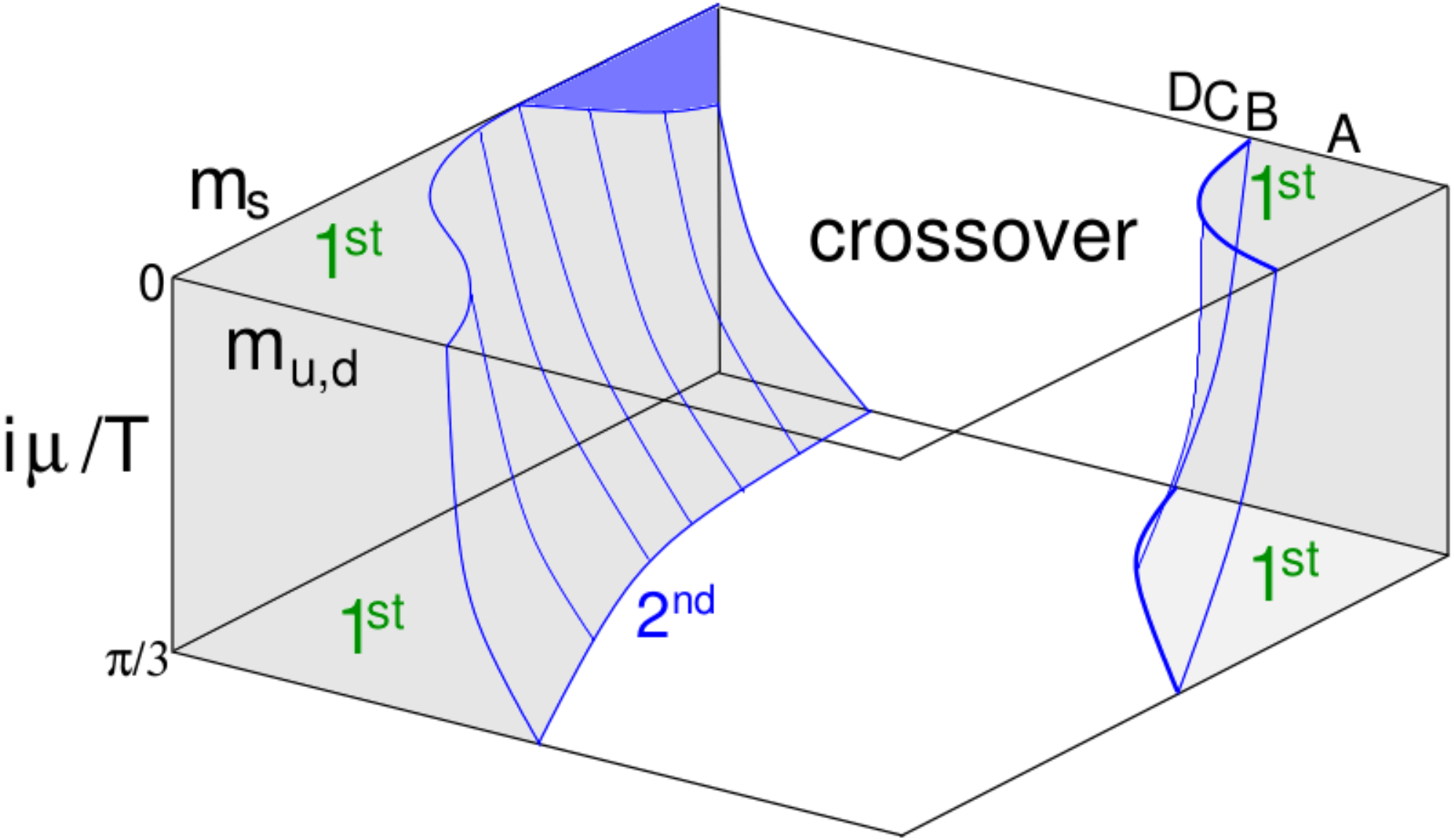}
    \caption{Imaginary chemical potential. 
            }\label{fig:columbia_ext_imag}
    \end{subfigure}\vspace*{5mm}
    \begin{subfigure}[b]{\textwidth}
    \includegraphics[scale=0.17]{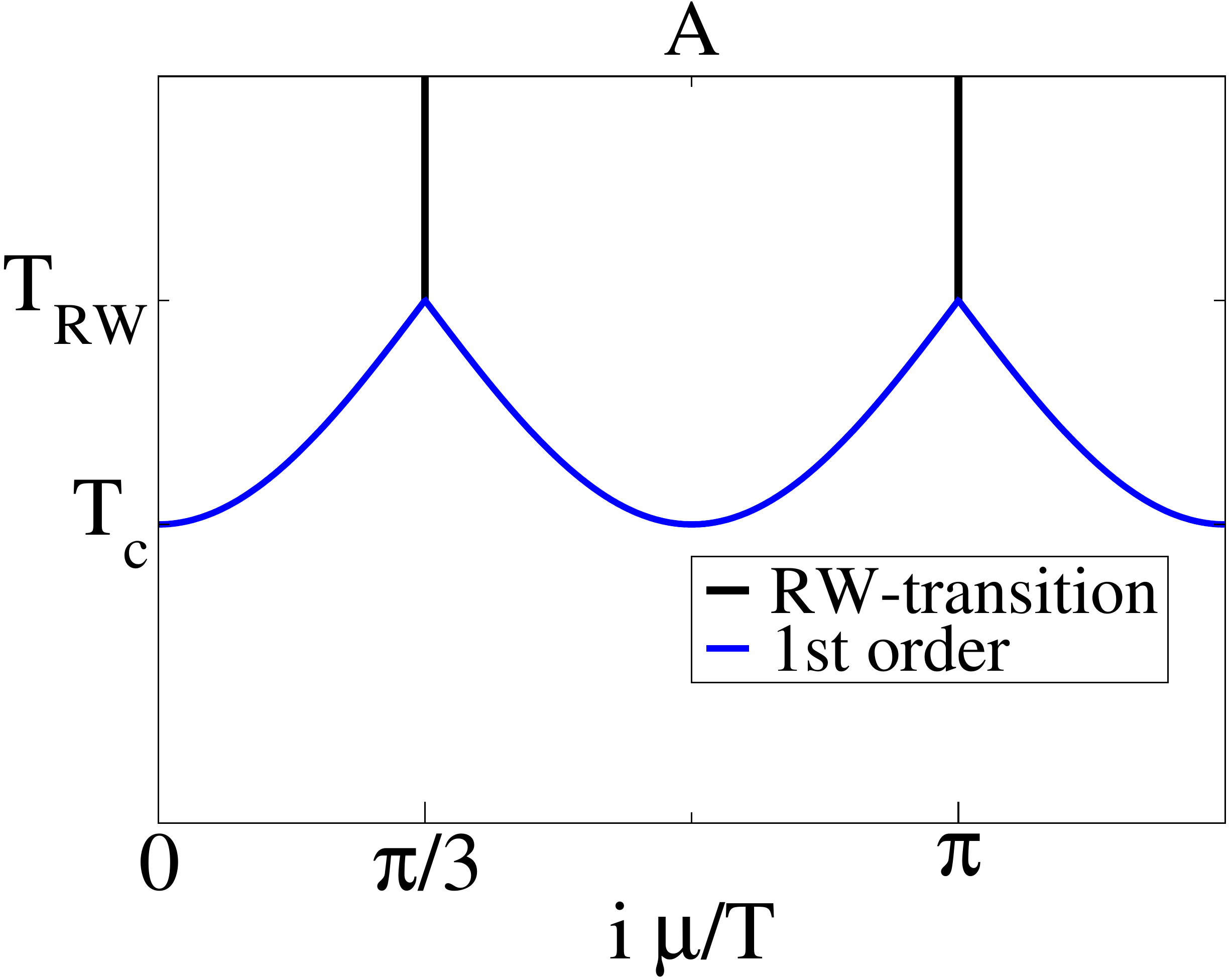}\hfill
    \includegraphics[scale=0.17]{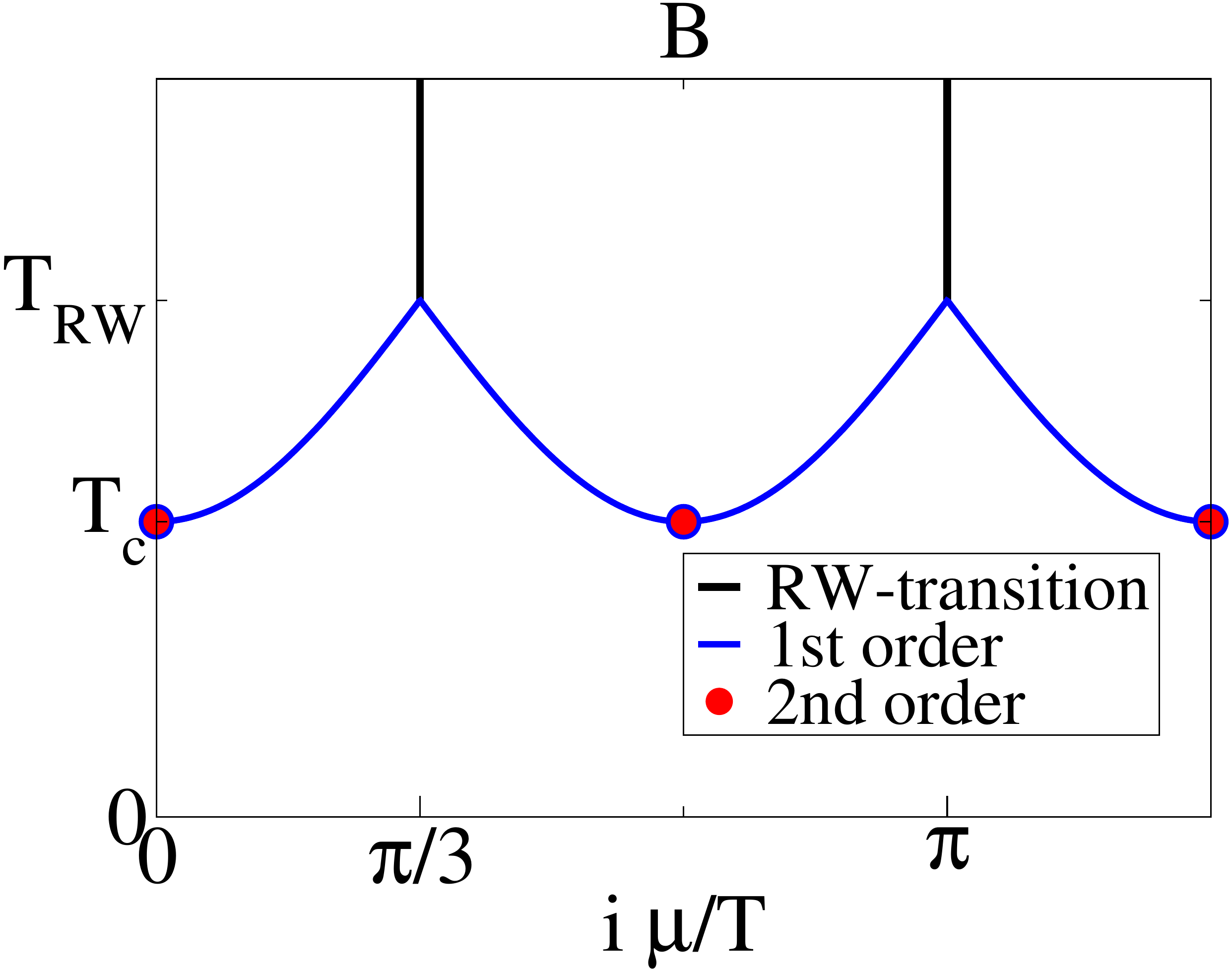}\hfill
    \includegraphics[scale=0.17]{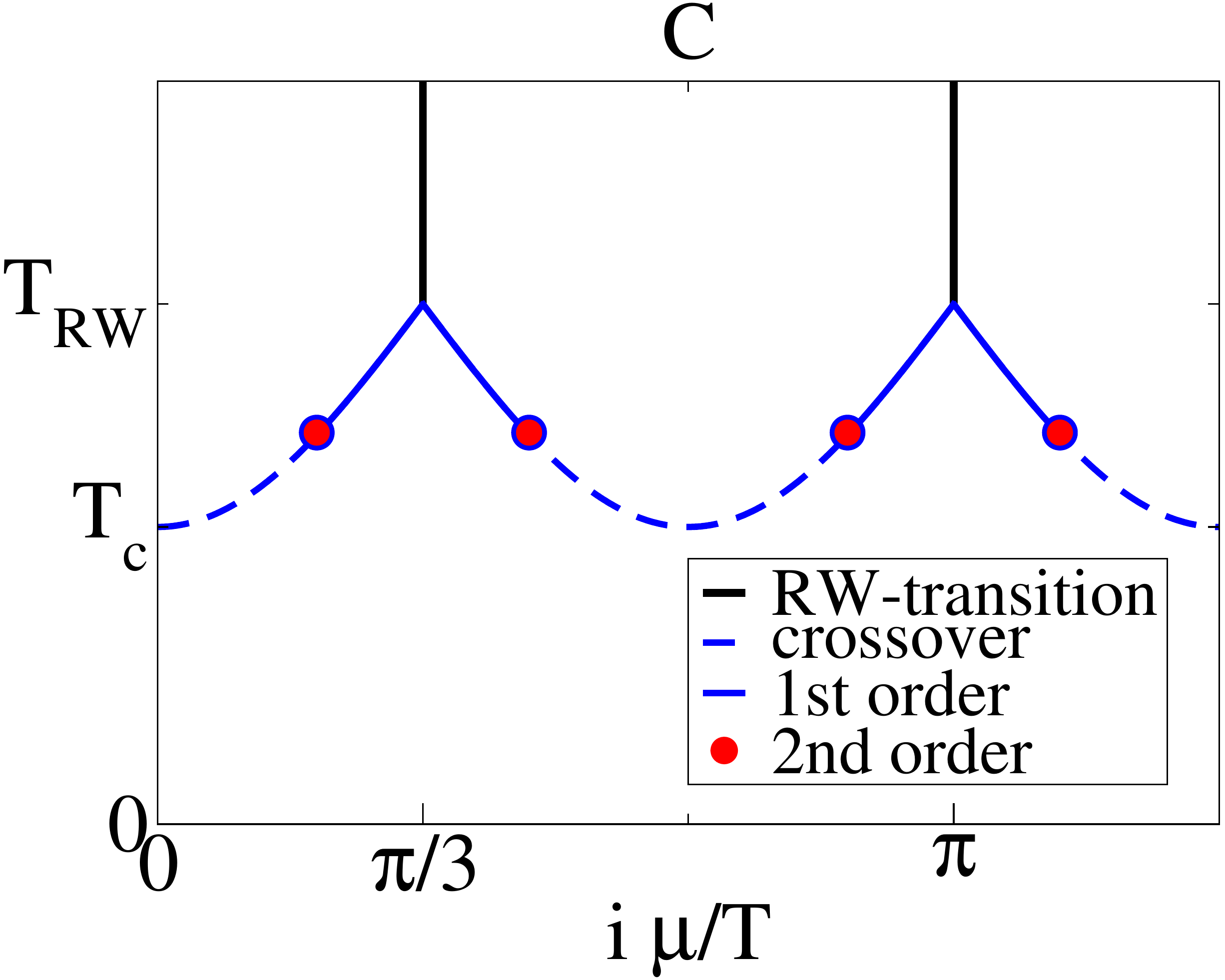}\hfill
    \includegraphics[scale=0.17]{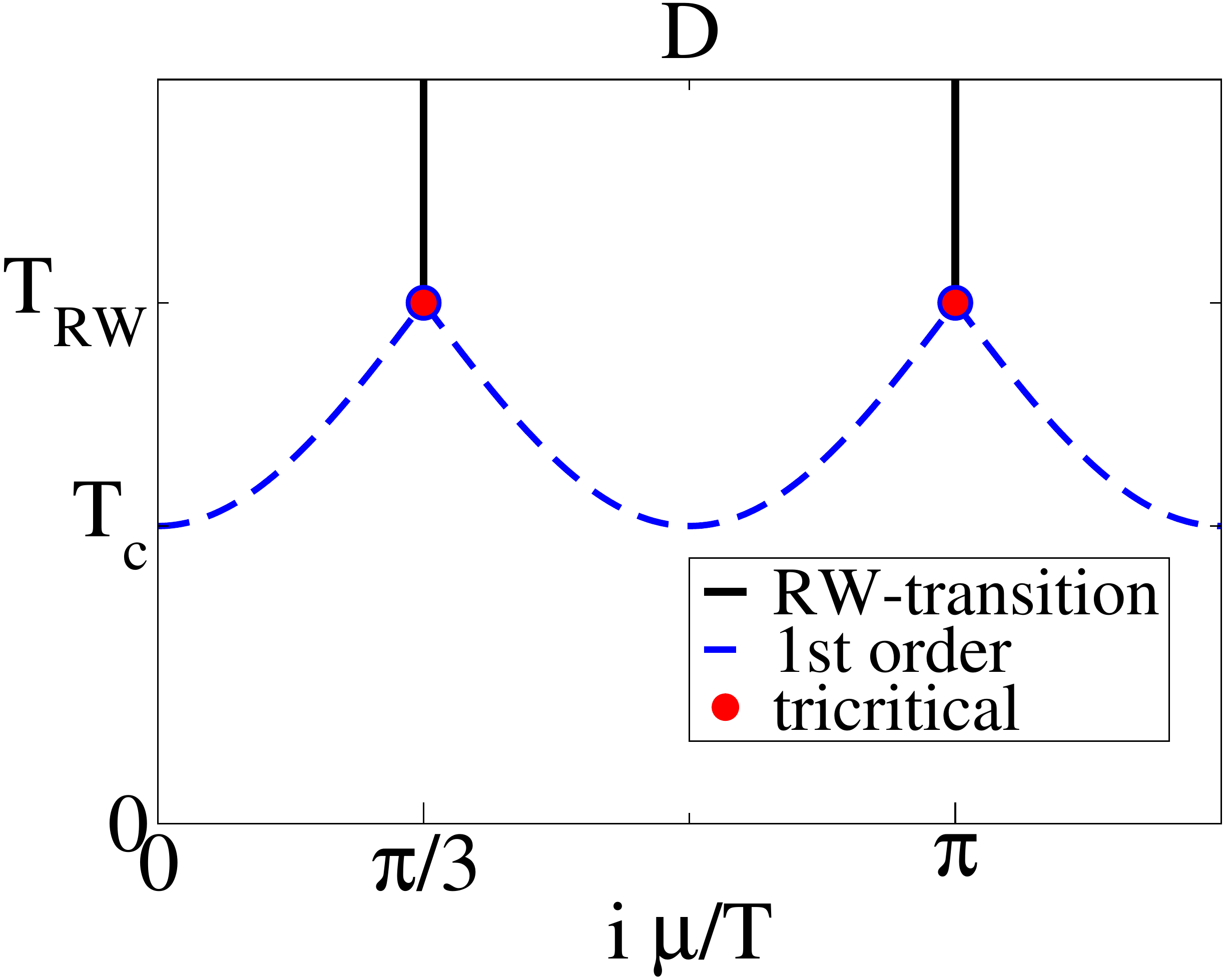}
    \caption{Left to right: phase diagrams in the temperature imaginary chemical potential plane at points A-D 
    of Fig.~\ref{fig:columbia_ext_imag} ($N_c=3$).}\label{fig:RW}
    \end{subfigure}\vspace*{-1mm}
    \caption{Extensions of Columbia plot to finite chemical potential and resulting phase diagrams.}
\end{figure}

The Columbia plot sketches the variation of the order of the QCD transitions at zero chemical potential. Possible
scenarios for the extension of the plot by a third axis denoting real or imaginary chemical potential are shown
in Figs.~\ref{fig:columbia_ext_real} and \ref{fig:columbia_ext_imag}. Let us first briefly discuss the extension
to real chemical potential. In Fig.~\ref{fig:columbia_ext_real} we show the 'default' scenario encountered in many
model calculations. The second order chiral critical line discussed above extends into the region of positive
real chemical potential and bends to the right, i.e. the first order region increases with chemical potential. 
At suitably large values of $\mu$ this surface then covers the point of physical
quark masses leading to the appearance of the QCD critical end point discussed above. This situation corresponds
to the one sketched in Fig.~\ref{fig:phase}. That this behaviour of the critical surface is by no means trivial
and actually might not be realised has been argued by de Forcrand and Phillipsen from the results of lattice 
simulations with three degenerate quark flavours \cite{deForcrand:2002hgr,deForcrand:2003vyj,deForcrand:2006pv,deForcrand:2010he}.
They used an extrapolation procedure from imaginary chemical potential to determine the curvature of the critical 
surface at $\mu =0$ and found that it bends towards smaller quark masses, i.e. in the other direction. This 
corresponds to a weakening of the chiral transition with increasing chemical potential instead of the strengthening 
shown in Fig.~\ref{fig:columbia_ext_real}. Whether this observation survives the continuum limit is an open question
that needs to be explored further. Furthermore, if the chiral first order region of the three-flavour theory is indeed 
small, as discussed above, then the corresponding behaviour of the critical surface at the three-flavour degenerate 
point might have nothing to do with the behaviour of the critical surface in the region of physical quark 
masses \cite{Bazavov:2017xul}. Again, this needs to be explored further. 

The extension of the Columbia plot to imaginary chemical potential, at least for heavy quark masses, stands on  
firmer ground. First, lattice calculations for imaginary chemical potential do not suffer from the sign problem.
Second, effective theories for heavy quarks are available that can be simulated with low CPU costs even at 
real chemical potential \cite{Langelage:2010yr,Fromm:2011qi}. Third, much can be learned from symmetry considerations
alone. Introducing the dimensionless variable $\theta_q \equiv \mbox{Im}(\mu_q)/T$ with quark chemical potential $\mu_q$ and 
taking center symmetry into account, Roberge and Weiss found that the physics of the theory is invariant under changes of 
$\theta \rightarrow \theta + 2\pi /N_c$ with $N_c$ the number of colours \cite{Roberge:1986mm}. This symmetry is smoothly
realised at low temperatures, but occurs via a first order phase transition at large temperatures beyond $T_{RW}$. The
situation in the temperature imaginary chemical potential plane is sketched in Fig.~\ref{fig:RW}. Above $T_{RW}$ the
regions $[0,\pi/N_c[$, $]\pi/N_c,3\pi/N_c[$, and $]3\pi/N_c,5\pi/N_c[$ are distinguished by the Polyakov loop 
$L = |L| e^{-i\phi}$, whose phase $\phi$ changes by $2\pi/N_c$ at every first order RW-transition line. The point where
the first order RW-transition lines end are touched by the confinement/chiral transition lines with temperature. 
Depending on the values of the quark masses, i.e. on the location in the Columbia plot, these transitions are
of different order, as visualised in Fig.~\ref{fig:RW} for the locations A (1st order region), B (second order critical
line), C (slightly in the cross-over region) and D at $\mu = 0$ displayed in Fig.~\ref{fig:columbia_ext_imag}.
The point D is defined to be on the line in the mass plane where the second order critical surface of the 
deconfinement transition intersects the plane with $i\mu_q/T = \pi/3$. For points further out in the crossover 
region the deconfinement transition is a crossover everywhere, regardless of the value of $i\mu_q/T$. It has been
shown \cite{Fromm:2011qi} that the tricritical point occurring at the intersection of the RW-transition with the
deconfinement transition for D strongly influences the whole second order critical surface in the sense that
it can be parametrised by tricritical scaling relations. We will came back to this issue in section \ref{results:heavy},
 where we discuss results from the DSE-framework for the critical surface. 

The critical surfaces for the chiral transition, also sketched in Fig.~\ref{fig:columbia_ext_imag} are much harder 
to evaluate on the lattice, since the corresponding quark masses are small and therefore simulations very cost intensive.
Nevertheless it is very interesting to map these out, since this may give vital clues on the issue of the order
of the transition in the two-flavour theory, discussed above. The situation sketched in Fig.~\ref{fig:columbia_ext_imag}
corresponds to the scenario with broken $U_A(1)$-anomaly, i.e. the left plot of Fig.~\ref{fig:columbia}. If it were established
that the critical surface intersects the left backplane of Fig.~\ref{fig:columbia_ext_imag} {\it above} the $\mu=0$ plane
for all quark masses, then the other scenario of Fig.~\ref{fig:columbia} is realised. Again, simulations have been performed
indicating that this may very well be the case \cite{Bonati:2014kpa,Philipsen:2016hkv,Cuteri:2017gci}, with the (important)
caveat that a continuum extrapolation has not yet been done.

\newpage
\newpage
\section{Non-perturbative quark and glue}\label{DSE}
In the previous section we discussed general aspects of the QCD phase diagram at finite temperature 
and baryon chemical potential, highlighted the merits of theoretical studies at unphysical external
parameters such as quark masses or imaginary chemical potential and summarised briefly the connection 
to heavy ion collision experiments via fluctuations of conserved charges. In this section we focus on
the technical aspects of the functional approach to QCD via Dyson-Schwinger equations (DSEs). We deal 
with the derivation of the DSEs, explain the extraction of order parameters from the correlation functions
of the theory and discuss strategies for devising truncations that can be systematically tested and improved.

Readers interested in the available results at finite $T$ and $\mu$ but not in the technical details of 
the framework are encouraged to skip this section in a first reading and proceed to section \ref{results}. 
In order to appreciate the details of the calculations and to understand the level of the rigorousness 
of the results, revisiting section \ref{DSE} afterwards, however, may be beneficial.   

\subsection{Functional equations}\label{DSE:generating functional}
We work with the Euclidean version of the QCD generating functional that describes
strongly interacting matter in thermodynamical equilibrium as a grand-canonical
ensemble. The gauge fixed partition function $Z[T,\mu]$ is given by\footnote{An introduction 
into path integral methods in quantum field theories is given {\it e.g.} in \cite{Rivers:1987hi}. 
In the following we adhere to the conventions of the review article \cite{Alkofer:2000wg}.}   
\beq
Z[T,\mu] = {\cal N} \int {\cal D} [A \bar{\Psi} \Psi c \bar{c}]  
\exp\left\{\phantom{\int}\hspace*{-2mm} - S_{QCD}[A,\Psi,\bar{\Psi}] - S_{gf}[A,c,\bar{c}]  \right\}, 
\label{genfunc} 
\eeq
with the QCD gauge invariant action
\beq
S_{QCD} = -\int_0^{1/T} dx_4 \int d^3x \left( \sum_{q=u,d,s,...} \bar{\Psi}_q \left( -\Dslash + m_q - \mu_q \gamma_4 \right) \Psi_q + 
\frac{1}{4} F_{\mu \nu}^a F_{\mu \nu}^a \right)\,,\label{action}  
\eeq
and the gauge fixing part 
\beq
S_{gf} =  \int_0^{1/T} dx_4 \int d^3x \left(\frac{\left(\partial_\mu A_\mu \right)^2}{2 \zeta} - i \partial_\mu \bar{c} D_\mu c\right) \,.
\label{gaugefix}
\eeq
Quarks with flavour $q$ and bare masses $m_q$ are represented by the Dirac fields $\Psi_q$ and $\bar{\Psi}_q$. 
Local gauge symmetry of the 
quark fields demands the introduction of a vector field $A_\mu^a$, which represents gluons. 
The gluon field strength $F_{\mu \nu}^a$ is given by    
\beq
F_{\mu \nu}^a = \partial_\mu A_\nu^a - \partial_\nu A_\mu^a -g f^{abc} A_\mu^b A_\nu^c \;,
\eeq
with the coupling constant $g$ and the structure constants $f^{abc}$ of the gauge group $SU(N_c)$,
where $N_c$ is the number of colours. 
The covariant derivative in the fundamental representation of the gauge group is given by
\beq
D_\mu = \partial_\mu + igA_\mu \;, \label{covder}
\eeq 
with $A_\mu = A_\mu^a t^a$ and the $t^a$ are the generators of the gauge group. We work with fixed gauge
using the Faddeev-Popov procedure \cite{Faddeev:1967fc} (see \cite{Pokorski:1987ed,Williams:2002dw} for 
pedagogical treatments of the subject) which introduces the Grassmann valued Faddeev-Popov ghost fields
$c$ and $\bar{c}$. The integral over the gauge group is absorbed in the normalisation ${\cal N}$. The gauge
parameter is denoted by $\zeta$. Below we always use Landau gauge, which is defined 
by the gauge condition $\partial_\mu A_\mu=0$ and gauge parameter $\zeta=0$. 
\footnote{Gauge fixing via the Faddeev-Popov
procedure is well-known not to be complete and leaves one to deal with the problem of Gribov-copies, i.e.
multivalued instances of $\partial_\mu A_\mu=0$ for gauge field configurations related by gauge transformations
(see the reviews \cite{Sobreiro:2005ec,Vandersickel:2012tz} for a detailed account of the problem). 
This problem has been studied intensively on the lattice
\cite{Cucchieri:1997dx,Silva:2004bv,Bogolubsky:2009dc,Maas:2009ph,Sternbeck:2012mf}
and found to be relevant for the behaviour of ghost and gluon propagators at very small momenta much below 
the temperature scales we are interested in. Thus for the topic of this review we can safely ignore this
problem.}    

Renormalization of the QCD action entails the introduction of suitable counterterms. The correspondence 
between the bare Lagrangian (\ref{action},\ref{gaugefix}) and its renormalised
version is given by the following rescaling transformations
\begin{eqnarray}
A_\mu^a &\rightarrow& \sqrt{Z_3}A_\mu^a, \hspace*{1cm} \bar{c}^a c^b \rightarrow \tilde{Z}_3 \bar{c}^a c^b,
\hspace*{1cm} \bar{\Psi}\Psi \rightarrow Z_2 \bar{\Psi}\Psi, \\
g &\rightarrow& Z_g g, \hspace*{2.1cm} 
\zeta \rightarrow Z_\zeta \zeta, 
\label{rescaling}
\end{eqnarray}
where five independent renormalisation constants $Z_3,\tilde{Z}_3,Z_2,Z_g$ and 
$Z_\zeta$ have been introduced. 
Furthermore five additional (vertex-) renormalisation constants are related to these via 
Slavnov--Taylor identities,
\beq
Z_1 = Z_g Z_3^{3/2}, \hspace{0.4cm} \tilde{Z}_1=Z_g \tilde{Z}_3 Z_3^{1/2}, \hspace{0.4cm} 
Z_{1F} = Z_g Z_3^{1/2} Z_2, \hspace{0.4cm}
 Z_4=Z_g^2 Z_3^2, \hspace{0.4cm} \tilde{Z}_4=Z_g^2 \tilde{Z}_3^2 . \hspace{0.4cm}
\label{Zsti}
\eeq
In general, these renormalisation constants depend on the renormalisation scheme, the renormalisation scale $\mu$ 
and the regularisation procedure. In the numerical treatment of Dyson-Schwinger equations, further detailed below, 
it is common to use either a hard cut-off or a Pauli-Villars type regulator, resulting in a generic dependence
of the renormalisation constants on a regularisation scale $\Lambda$, i.e. $Z_i = Z_i(\mu,\Lambda)$. Provided 
multiplicative renormalisability is not violated in the process of truncating the DSEs, all Green's functions 
extracted from the renormalised DSEs are independent of $\Lambda$ and therefore do not suffer from divergences
when $\Lambda$ is sent to infinity\footnote{Within quenched QED, the independence of the resulting Green's 
functions from the employed regularisation scheme has been studied and shown to hold in
Ref.\cite{Kizilersu:2000qd,Kizilersu:2001pd}}. 
This technical issue is well under control and further detailed in appendix \ref{renorm}.

The generating functional (\ref{genfunc}) depends on temperature via the restriction of the $x_4$-integration
from zero to $1/T := \beta$. In the imaginary time Matsubara formalism, which we adopt throughout this review, 
all fields obey (anti-)periodic boundary conditions such that $\phi(x_4)=\pm\phi(x_4+1/T)$ for a generic 
field $\phi$. Due to the Kubo-Martin-Schwinger condition bosons (i.e. gluons) need to have periodic boundary
conditions, while fermions (quarks) have anti-periodic boundary conditions. The Faddeev-Popov ghosts are the 
exception from this rule with periodic boundary conditions despite their Grassmann nature due to their origin 
from the gauge fixing procedure \cite{Bernard:1974bq}. In momentum space, this translates into momentum
vectors $p=(\vect{p},\omega_n)$ with Matsubara frequencies $\omega_n = \pi T (2n+1)$ for
fields with antiperiodic boundary conditions and $\omega_n = \pi T 2n$ for those
with periodic boundary conditions with integers $n$ running from minus to plus infinity.
 
The quark chemical potential $\mu_q$ is added to the QCD action via a Lagrange multiplier $-n_q \mu_q$ for the 
net quark density 
\beq
n_q = \int_0^{1/T} dx_4 \int d^3x \,\,\Psi^\dagger \Psi
\eeq
with $\Psi^\dagger = \bar{\Psi} \gamma_4$ and is subsequently absorbed into the quark part of the QCD Lagrangian,
cf. Eq~(\ref{action}). For non-zero chemical potential the Matsubara frequencies for fermions are modified into
$\tilde{\omega}_n = \omega_n - i\mu$.

Starting from the generating functional (\ref{genfunc}) one can derive the Dyson-Schwinger equations
for the Green's functions of the theory. In the following we outline the formalism in 
a very dense, symbolic notation. Readers interested in more details are referred to the textbooks
\cite{Itzykson:1980rh,Rivers:1987hi} or the reviews \cite{Roberts:1994dr,Alkofer:2000wg}. 
Dyson-Schwinger equations follow from the generating functional (\ref{genfunc}) and the fact that the 
integral of a total derivative vanishes, {\it i.e.}
\begin{align}
\hspace*{-1.8cm}
0 &= \int {\cal D} [A \bar{\Psi} \Psi c \bar{c}] \frac{\delta}{\delta \phi} 
\exp\left\{- S_{QCD} - S_{gf} + \int_0^{1/T} dx_4 \int d^3x\left(A J+\bar{\eta}\Psi+\bar{\Psi}\eta+\bar{\sigma}c+\bar{c}\sigma \right)\right\}
 \nonumber\\
\hspace*{-1.8cm}
&=
\left\langle -\frac{\delta (S_{QCD}+S_{gf})}{\delta \phi} 
+ j \right\rangle  \label{genDSE}
\end{align}
for any field $\phi \in \{A,\Psi,\bar{\Psi},c,\bar{c}\}$ and its corresponding source 
$j \in \{J,\eta,\bar{\eta},\bar{\sigma},\sigma\}$. 
Equation (\ref{genDSE}) is correct provided that the functional integral 
(\ref{genfunc}) is well-defined and the measure ${\cal D}[A \bar{\Psi} \Psi c \bar{c}]$ 
is translational invariant. Acting onto (\ref{genDSE}) with a suitable number of further 
functional derivatives and setting all sources to zero afterwards leads to the 
Dyson-Schwinger equation (DSE) for any desired full n-point function. A similar procedure 
applied to the generating functional $W=\ln(Z)$ or the effective action 
$\Gamma = W + \langle\phi\rangle j $ leads to the 
DSEs for connected Green's functions and the ones for one-particle irreducible 
Green's functions. The expression (\ref{genfunc}) and its functional derivatives constitute 
an infinite tower of coupled integral equations. Provided QCD is a local quantum field theory,
this infinite tower contains all physics of the original path integral, cf. the review article  
Ref.~\cite{Alkofer:2000wg}. Below we will analyse some of these equations in more detail.

Closely related to the formalism of Dyson-Schwinger equations is the one for the functional
renormalization group (FRG), see \cite{Wetterich:1992yh,Morris:1993qb} and the 
reviews \cite{Berges:2000ew,Pawlowski:2005xe,Gies:2006wv,Schaefer:2006sr}.
The basic idea of this approach is to endow the generating functional with an (infrared) cut-off 
scale $k$ which suppresses all quantum fluctuations with momenta $q^2 \stackrel{<}{\sim} k^2$ smaller 
than this scale. The resulting effective action $\Gamma_k$ depends on the cut-off scale and satisfies
a functional differential equation which controls this dependence. Lowering $k$ takes into account more 
and more quantum fluctuations characteristic for the scale $k$ in question. In this way one can explore
the relevant physics in a systematic way. Finally, sending $k \rightarrow 0$ recovers the full effective
action of the theory. By taking functional derivatives of the functional differential 
equation one obtains an infinite tower of coupled integro-differential equations for the Green's functions
of the theory, which are not too dissimilar in structure to the tower of Dyson-Schwinger equations. 
Thus, in principle it is possible to explore the physics content of the Green's functions of QCD from 
both frameworks, the DSE and the FRG tower of equations. Interestingly, the combination of both methods
sometimes leads to unique results, see \cite{Fischer:2006vf,Fischer:2009tn} for an example. In this review 
article we focus primarily on results for QCD at finite temperature and density obtained in the DSE approach,
but we make connections to corresponding or complementary results from the FRG-framework wherever
possible. 

A third functional framework that has been employed for QCD at finite temperature and chemical potential
is the variational Hamilton approach in various formulations
\cite{Reinhardt:2013iia,Quandt:2015aaa,Heffner:2015zna,Reinhardt:2016pfe,Reinhardt:2016xci,Quandt:2017poi,Quandt:2018bbu}. 
For all technical details we refer the interested reader to the original literature and the review article 
Ref.~\cite{Reinhardt:2017pyr}. Similar to other functional methods one ends up with a coupled set of 
integral equations for the Green's functions of the theory which have to be solved using numerical methods.
Again, in the following we do not review the corresponding results in any detail but make connections to the
DSE results discussed in this review wherever appropriate. 

Finally, there is growing interest in a fourth framework based on functional methods, the Gribov-Zwanziger
approach. This approach is based on studies of the effects of gauge fixing on the theory and the desire
to eliminate the effects of Gribov copies \cite{Gribov:1977wm,Zwanziger:1989mf}. As a result one arrives at an
effective action which contains a new scale, the Gribov parameter, which affects the gluon dispersion
relation and changes its infrared behaviour, see \cite{Vandersickel:2012tz} for a review on the technical details.
This approach was generalized to finite temperature in \cite{Zwanziger:2004np}. Somewhat related to the 
Gribov-Zwanziger approach is the one of \cite{Reinosa:2013twa,Reinosa:2014zta} based on the Curci-Ferrari model. 
Both, the Gribov-Zwanziger framework and the Curci-Ferrari model have been used at finite temperature and 
chemical potential, and heavy quark physics and transport properties of the quark-gluon plasma have been 
explored 
\cite{Fukushima:2013xsa,Su:2014rma,Florkowski:2015dmm,Reinosa:2015oua,Reinosa:2016iml,Maelger:2017amh,Maelger:2018vow}. 
Again, we will not discuss these results in detail but point the interested reader to the literature wherever appropriate.

\subsection{Quarks, gluons and order parameters}\label{DSE:order}
\begin{figure}[t]
    \centering
    \includegraphics[scale=0.85]{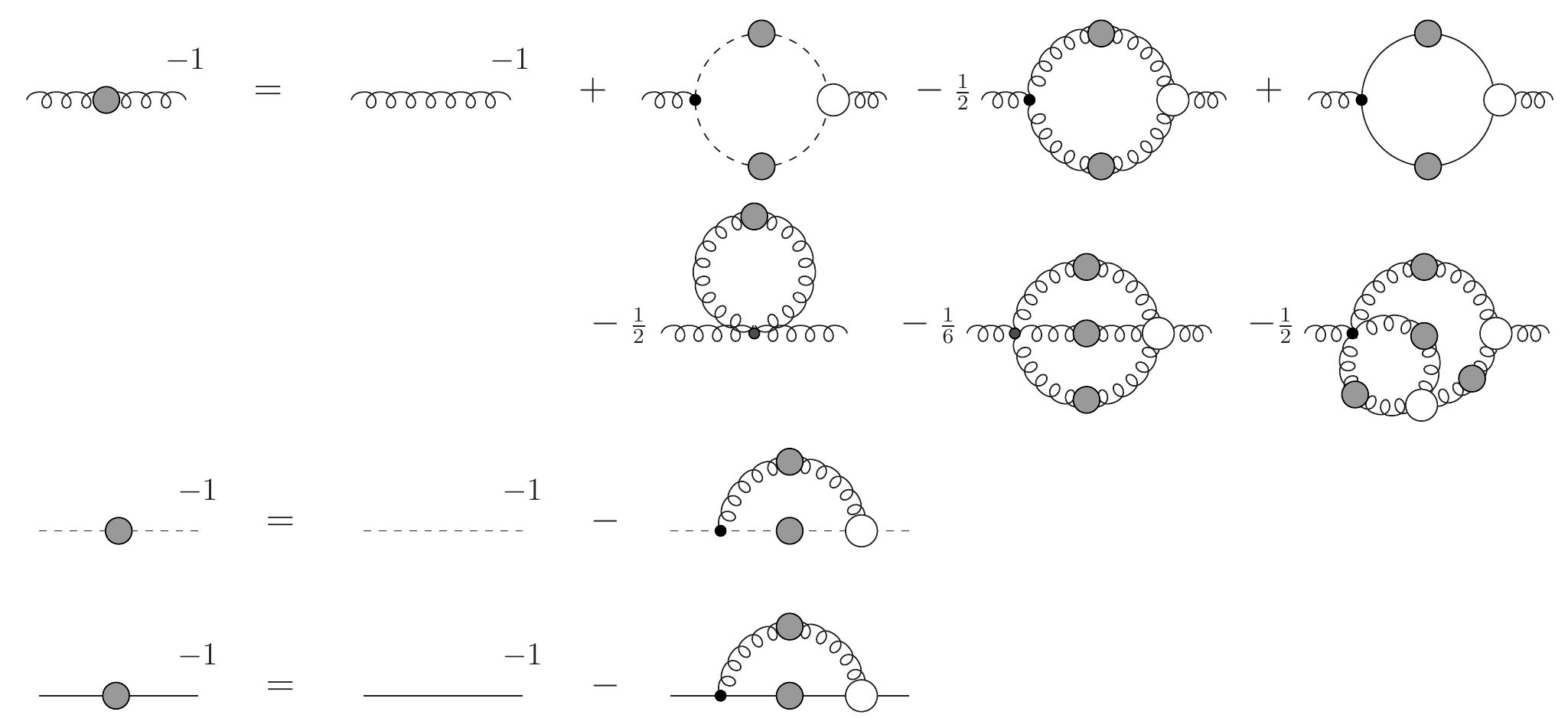}
    \caption{Dyson-Schwinger equations for the gluon (curly lines), the ghost (dashed lines) 
    and the quark (solid lines) propagator. 
            \label{fig:DSE-prop}}
\end{figure}
The DSEs for the ghost, gluon and quark propagators are given diagrammatically in 
figure \ref{fig:DSE-prop}. They form a coupled system of equations which demand  
dressed ghost-gluon, three-gluon, four-gluon and quark-gluon vertices as input.\footnote{
	The corresponding equations in the framework of the functional renormalization group display
	a number of interesting structural differences. First, they do not feature two-loop diagrams,
	as are present in the DSE for the gluon propagator. From the point of view of the practitioner
	aiming at numerical solutions for these equations, this is clearly an advantage. On the other 
	hand, however, the FRG equations contain more unknown four-point functions (such as a fully dressed
	ghost-gluon scattering kernel and others), which need to be specified in order to close the equations.
	Also, besides being integral equations, the FRGs contain derivatives with respect to the FRG-scale
	$k$. Thus, both formulations of functional equations present their own challenges which
	more or less balance out and it is often only a matter of personal taste and experience which ones to choose.}
These three- and four-point functions satisfy their own DSEs, which in turn contain four- and five-point
functions and so on. In order to solve the tower of DSEs one has to choose a truncation scheme, i.e.
a strategy to break down the infinite tower into a closed system of equations. We will come back to
ponder on strategies to make this procedure systematic in subsection 
\ref{DSE:truncation} below. Here we wish to focus first on the relation of the propagators of QCD
to suitable order parameters for the phase transitions at finite temperature and chemical potential.

\subsubsection{Chiral transition}\label{chiral}
Let us first discuss the chiral transition. The prime candidate for a suitable 
order parameter is the quark condensate $\langle\bar{\psi}\psi\rangle_f$. For a quark with flavor $f$ it
is given by the trace of the quark propagator $S^f(p)$ via
 \beq \label{eq:condensate}
\langle\bar{\psi}\psi\rangle_f = 
-Z_2 Z_m  N_c  T\sum_n\int\frac{d^3p}{(2\pi)^3}\mathrm{Tr}_D\left[S^f(p)\right],
\eeq
where $Z_2$ is the quark wave function renormalization constant, $Z_m$ the quark mass renormalization 
constant and $N_c=3$ the number of colours. The sum is over fermion Matsubara frequencies $\omega_n=\pi T(2n+1)$ 
and the momentum four-vector given by $p = (\vect{p},\omega_p)$. Logarithmic divergences of the integral when the
cut-off is sent to infinity are taken care of by the renormalisation factors $Z_2$ and $Z_m$. However, 
for all flavours with non-zero bare quark mass the condensate is also quadratically divergent and needs to be regularized. 
For dimensional reasons (the condensate has dimension three), the divergent part is also proportional to the bare 
quark mass (c.f. section \ref{sec:quark} for details) and therefore the difference 
\begin{equation}
\Delta_{l,h} = \langle\bar\psi\psi\rangle_l - \frac{m_l}{m_h}\langle\bar\psi\psi\rangle_h\,,
\label{eq:cond_renorm}
\end{equation}
fulfils this purpose: the divergent part of the light-quark condensate ($l \in \{u,d\}$) 
is cancelled by the divergent part of the heavy ($h$) quark condensate. 
In order to extract a clean signal for the chiral transition of the light quark flavour, the mass of the 
heavy quark needs to be sufficiently larger than the one of the light quark. For the extraction of the
light up/down-quark condensate a heavy quark mass of the order of the strange quark is already sufficient.

As discussed in section \ref{gen:columbia}, for physical quark masses and small chemical potential 
the chiral transition is a crossover which leads to ambiguities in the definition of a pseudo-critical 
temperature. Frequently used quantities to determine $T_c$ are the inflection point of the condensate, 
(i.e. the maximum of $\frac{\partial \langle\bar{\psi}\psi\rangle_l}{\partial T}$) or the maximum of 
the chiral susceptibility
\beq \label{eq:chisusz}
\chi_{\langle\bar{\psi}\psi\rangle} = \frac{\partial \langle\bar{\psi}\psi\rangle_l}{\partial m_{l}}\,.
\eeq

As an aside let us note, that it is also possible to directly use one of the Matsubara frequencies of 
the scalar dressing function of the quark propagator, $B(\vect{p},\omega_p)$, as indicator for the chiral 
transition. Formally, full chiral restoration requires all functions $B(\vect{p},\omega_p)$ to be zero (a condition
equivalent to a zero value of the quark condensate). In practise one finds, however, that the functions 
$B(\vect{p},\omega_p)$ for any Matsubara frequency $\omega_p$ react similarly to temperature, thus if one
of these goes to zero, all others do as well. Thus already the zeroth Matsubara frequency of $B$ alone
offers a reliable and quick way to extract $T_c$.

\subsubsection{Deconfinement transition} \label{sec:deconf}

As discussed in section \ref{gen:columbia}, there is a well-defined notion of the deconfinement phase transition
in the heavy quark limit of QCD associated with the breaking of center symmetry.\footnote{A concise and detailed 
discussion of many aspects of confinement can be found e.g. in
Refs.~\cite{Greensite:2003bk,Alkofer:2006fu,Greensite:2011zz}, see also \cite{Greensite:2017ajx}.} 
The extraction of a corresponding
order parameter from the propagators of the theory has been an unsolved problem for the DSE-community for many years.
In the past decade, however, this problem has been solved and we now have a range of good order parameters for
the deconfinement transition at our disposal. These are the dressed Polyakov loop \cite{Fischer:2009wc,Braun:2009gm},
the dual scalar quark dressing \cite{Fischer:2009gk}, the generalized $\varphi$-dependent quark 
condensate \cite{Mitter:2017iye} and the Polyakov loop potential \cite{Braun:2007bx,Fister:2013bh,Fischer:2013eca}.
Especially the latter is of considerable importance also in the context of Polyakov enhanced effective models
like the PNJL and the PQM model. As we will discuss further in section \ref{results:2p1}, the general capacity of the 
functional methods to perform calculations at arbitrary large chemical potential offers opportunities to not 
only study this quantity all across the QCD phase diagram, but also provide input for phenomenological model
calculations in regions not accessible to lattice QCD.

Let us start and briefly summarize the notion of the {\bf dressed Polyakov loop}, introduced first on the lattice
in \cite{Bilgici:2008qy} and adapted to functional methods in \cite{Fischer:2009wc}.
The method emerged from previous studies of spectral sums of the Dirac operator and their behaviour under 
center transformations explored in 
\cite{Gattringer:2006ci,Bruckmann:2006kx,Synatschke:2007bz,Bilgici:2008qy,Synatschke:2008yt}.
In order to appreciate the notion of the dressed Polyakov loop it is useful to first discuss the 'ordinary' 
Polyakov loop $L$ \cite{Polyakov:1978vu,Susskind:1979up}. It is defined by
\beq\label{PL}
L = \frac{1}{N_c} tr_F P\,, \hspace*{2cm}  P = {\cal P} \exp\left[ig \int_0^\beta dx_4 A_4(\vect{x},x_4) \right]\,.  
\eeq
where ${\cal P}$ denotes path ordering and the trace is in the fundamental representation of the gauge group $SU(N_c)$.
The Polyakov loop can be represented by a straight line winding once around the compactified time direction of 
the space-time manifold with $\beta=1/T$, see e.g. \cite{Fukushima:2011jc} for a graphical representation. The 
loop is closed by the temporal boundary conditions. The expectation value $\langle L \rangle$ of the Polyakov loop can be
interpreted as the partition function in the presence of one single static quark and is thus related to the free
energy $E$ of the static quark via
\beq
E = -T \ln \langle L \rangle \,.
\eeq
Thus the Polyakov loop presents a simple picture of quark confinement: in the confined phase the free energy 
is infinite and the expectation value $\langle L \rangle$ is zero, whereas in the deconfined phase $\langle L \rangle$ 
is finite and therefore also the free energy. This in turn means that single quarks can be generated. 
In a lattice formulation of QCD, the temporal Wilson line P is given by 
\beq
P = \Pi \,U_4(\vect{x},x_4)\,,
\eeq
with link variable $U_4$ in time direction and the product is over a line of temporal links between 
$x_4=0$ and $x_4=1/T$. Under a center transformation $U_4 \rightarrow z_k U_4$ of one element in the chain 
of links with respect to the center element $z_k = diag(e^{2\pi i k/N_c},\cdots,e^{2\pi i k/N_c})$ with
$k=0,1,\cdots,N_c-1$ the Polyakov loop transforms as
\beq
L \rightarrow z_k L\,.
\eeq
Hence center symmetry is only present in the confined phase with $\langle L \rangle = 0$ and the Polyakov loop 
is an order parameter for center symmetry breaking. 
The dressed Polyakov loop, or 'dual condensate' $\Sigma_1$ is defined via the Fourier-transform
\beq \label{dual}
\Sigma_n = -\int_0^{2\pi} \, \frac{d \varphi}{2\pi} \, e^{-i\varphi n}\,
\langle \overline{\psi} \psi \rangle_\varphi \,\,  \underset{|_{n=1}}\,,
\eeq
of the ordinary quark condensate $\langle \overline{\psi} \psi \rangle_\varphi$ evaluated using 
$U(1)$-valued boundary conditions with angle $\varphi$ in the temporal direction, i.e. 
$\Psi(1/T) = \Psi(0) e^{i\varphi}$. Thus instead
of the usual anti-periodic boundary conditions for fermions ($\varphi=\pi$) or periodic ones for
bosons ($\varphi=0$) here we vary $\varphi$ in the interval $[0,2\pi]$. This results in Matsubara 
modes $\op(n,\varphi) = (2\pi T)(n+\varphi/2\pi)$ in the $p_4$-direction. 
In order to explain why the quantity $\Sigma_1$ is of considerable interest, we again resort to the 
lattice formulation. It turns out,
that the $\varphi$-dependent quark condensate $\langle \overline{\psi} \psi \rangle_\varphi$ can be 
represented by a sum over all possible closed chains of link variables, i.e. closed loops $l$. 
One obtains
\beq \label{loop}
\langle \overline{\psi} \psi \rangle_\varphi =  
\sum_{l} \frac{e^{i\varphi n(l)}}{m^{|l|}} U(l) \,,
\eeq
where $U(l)$ denotes the closed chains of links including some sign and normalisation factors, 
see \cite{Bilgici:2008qy} for details. Each of these loops consists of $|l|$ links and is weighted 
by corresponding powers of the inverse quark mass $m$. Each time such a closed loop winds 
around the temporal direction of the lattice it picks up a factor $e^{\pm i\varphi}$ from the 
$U(1)$-valued boundary condition introduced above. Thus every loop is weighted by $e^{i\varphi n(l)}$, 
where $n(l)$ is the winding number of a given loop $l$. By a Fourier transform with respect to 
$e^{-i \varphi n}$ it is possible to project onto loops with $n(l)=1$ as done in Eq.~(\ref{dual}). 
The resulting quantity $\Sigma_1$ transforms under center transformation in the same way as the 
conventional Polyakov loop and is therefore an order parameter for 
the deconfinement transition. The numerical agreement between dressed and conventional Polyakov 
loops with respect to the location of the phase transition has been established in 
Ref.~\cite{Bruckmann:2008sy}.
Within functional methods this order parameter has been used in
\cite{Fischer:2009wc,Braun:2009gm,Fischer:2010fx,Fischer:2011mz,Fischer:2012vc}. 

Another order parameter for the deconfinement transition can be obtained by performing a similar transformation
as done in Eq.~(\ref{dual}) to the scalar quark dressing function, e.g. evaluated at lowest Matsubara frequency 
and zero momentum. At finite temperature and chemical potential the inverse quark propagator can be written as  
\beq \label{quark}
S^{-1}(\vp,\op) = i \gamma_4\, \tilde{\omega}_p C(\vp,\op) + i \gamma_i \, p_i A(\vp,\op) + B(\vp,\op) \,,
\eeq
with vector and scalar quark dressing functions $C,A,B$.\footnote{A further tensor component proportional 
to $\sigma_{\mu \nu}$ is possible in principle but can be omitted in all practical calculations, since its 
contribution to order parameters is extremely small \cite{Roberts:2000aa}.} 
The scalar quark dressing function $B(\vp,\op)$ 
evaluated at $\vp=0$ and $\op = \pi T$ is an order parameter for chiral symmetry breaking, 
similar to the quark condensate discussed above.
The '{\bf dual scalar quark dressing}' 
\beq 
\Sigma_B = \int_0^{2\pi} \, \frac{d \varphi}{2\pi} \, e^{-i\varphi}\, 
B(0,\op(0,\varphi))\,,
\eeq
is also sensitive to center transformation in a similar fashion as the dual quark condensate 
or other spectral sums \cite{Synatschke:2008yt}. This can be seen as follows \cite{Fischer:2009gk}:  
Using $B(0,\op(0,\varphi)) = 1/4 \,\mbox{tr} [S^{-1}(0,\op(0,\varphi))]$ we have
\beqa 
\Sigma_B = \int_0^{2\pi} \, \frac{d \varphi}{8\pi} \, e^{-i\varphi}\, 
\int d^3x \, \int_0^{1/T} dx_4 \,\, \mbox{tr} 
\langle \vect{x},x_4 | D_{\varphi}^{-1} | 0 \rangle^{-1}\,, \label{dualskalar}
\eeqa
where $D_{\varphi}$ is the (massive or chiral) Dirac operator evaluated
under presence of the $U(1)$-valued boundary conditions. A center 
transformation on $\Sigma_B$ introduces an additional phase factor
$z = e^{i 2 \pi k/N}$ with $k=0,\dots,N_c-1$ for $N_c$ colours, which adds
to the phase $e^{i\varphi}$ for our $U(1)$-valued boundary conditions.
We then obtain
\beqa 
^z\Sigma_B &=& \int_0^{2\pi}  \!\frac{d \varphi}{8\pi}  e^{-i\varphi} 
\!\!\int d^3x \!\!\int_0^{1/T} dx_4\,\, \mbox{tr} 
\langle \vect{x},x_4 | D_{\varphi+2\pi k/N}^{-1} | 0 \rangle^{-1}  \nonumber\\
           &=& z\, \int_0^{2\pi} \! \frac{d \varphi}{8\pi}  e^{-i\varphi}\, 
\!\!\int d^3x \!\!\int_0^{1/T} dx_4\,\, \mbox{tr} 
\langle \vect{x},x_4 | D_{\varphi}^{-1} | 0 \rangle^{-1}\,,
\eeqa
i.e. the dual scalar quark dressing $\Sigma_B$ transforms under center
transformations exactly like the conventional Polyakov loop and therefore
acts as order parameter for the deconfinement transition in the heavy quark 
limit. This has also been confirmed by numerical results discussed in \cite{Fischer:2009gk}.

Recently, the {\bf generalised $\varphi$-dependent quark condensate} has been suggested as 
yet another order parameter based on the idea of dual transform defined by \cite{Mitter:2017iye} 
\beqa
\Sigma^{(q)} &=& \int_0^{2\pi} \, \frac{d \varphi}{2\pi} \, e^{-i\varphi}\, 
\Sigma_\varphi^{(q)}\,,\nonumber\\
\Sigma_\varphi^{(q)} &=& T \sum_n \left[\frac{1}{4} \,\mbox{tr}\,  S(\vect{0},\omega_n(\varphi))\right]^2 \,. 
\eeqa
In contrast to the ordinary dual condensate which has to deal with the problem of divergences in 
the quark condensate away from the chiral limit, this quantity is always finite. First numerical 
results confirm its suitability especially for functional methods \cite{Mitter:2017iye}.

Finally, we discuss the {\bf Polyakov loop potential}. Its direct relation to the properties of the Yang-Mills 
sector of the theory (in contrast to the indirect relation offered by the dual quantities discussed above) 
makes it one of the most interesting tools to distinguish between the confined and deconfined phases.
The detailed properties of the Polyakov loop and its potential, its interpretative content and its
various applications in the context of Polyakov loop enhanced effective models have recently been reviewed
by Fukushima and Skokov \cite{Fukushima:2017csk}. They also give a comprehensive guide to the extensive 
literature on the subject. Therefore, here we restrict ourselves to some essentials which become
relevant later in the discussion of the results. In particular we focus on a related quantity that
is directly accessible by functional methods and therefore offers direct access to the confining properties
of the theory from its basic Green's functions. In the following we briefly summarise the derivation and the 
properties of this quantity, detailed discussion can be found in Refs.~\cite{Braun:2007bx,Marhauser:2008fz,Fister:2013bh}. 
The basic idea is to introduce a constant $A_4$ background field, which directly feeds into the Polyakov loop
and whose properties are therefore directly related to the deconfinement transition. To this end one splits 
the gauge field $A = \bar{A} + a$ in a background field $\bar{A}$ and fluctuations $a$ and implements the Landau-DeWitt  
gauge condition
\beq
D_\mu(\bar{A}) a_\mu = 0\,,
\eeq
with $D_\mu(A) = \partial_\mu - i g A_\mu$. With this gauge condition the effective action becomes dependent
on the background field, i.e. $\Gamma = \Gamma(a,\bar{A})$ and the effective potential $V$ is evaluated at 
$\Gamma(0,\bar{A})$. 
Above, we discussed the expectation value $\langle L[A_4] \rangle$ of the Polyakov loop evaluated for a gauge 
field $A_4$. For a constant background field in temporal direction, $\bar{A} = A_4$, it has been 
shown \cite{Braun:2007bx,Marhauser:2008fz} that also the quantity
\beq\label{PL2}
L[\langle A_4 \rangle] \ge \langle L[A_4] \rangle
\eeq
is an order parameter for confinement. The expectation value $\langle A_4 \rangle$ is determined from the 
minimum of the effective potential $V$ 
\beq
V[A_4] = \frac{1}{\beta \Omega} \Gamma[0,A_4]
\eeq
with three-dimensional spatial volume $\Omega$. In turn, the effective potential $V[A_4]$ can be computed 
in terms of dressed propagators and vertices using functional continuum methods. This can be exploited either
using the functional renormalisation group, see e.g.
\cite{Braun:2007bx,Marhauser:2008fz,Braun:2009gm,Braun:2010cy,Fister:2013bh,Herbst:2015ona}, 
the Dyson-Schwinger equations \cite{Fischer:2013eca,Fischer:2014vxa,Fischer:2014ata} or 
in the Hamilton approach \cite{Reinhardt:2012qe}. Furthermore, there are interesting applications in 
effective models, see e.g. \cite{Fukushima:2012qa,Kashiwa:2012td}. 

\begin{figure}[t]
\begin{center}
\includegraphics[width=0.85\textwidth]{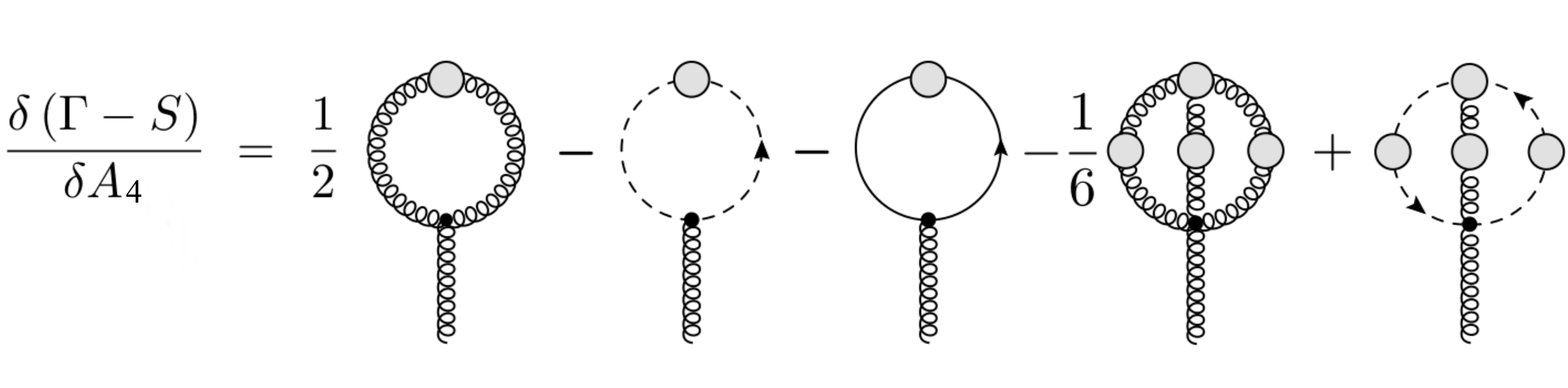}
\end{center}
\caption{The DSE for a background gluon one-point function. \label{fig:DSE-A}}
\end{figure}

The derivative of the effective potential with respect to the background field $A_4$ can be expressed in the 
FRG, the DSE and the 2PI-framework as shown in \cite{Fister:2013bh}. Written as a DSE it is given by 
\beq
\frac{\partial V[A_4]}{\partial A_4} = \frac{1}{\beta \Omega}\frac{\partial  \Gamma[A_4;0]}{\partial A_4} 
\eeq
and the important part of the right hand side is shown diagrammatically in Fig.~\ref{fig:DSE-A}. The one
and two-loop terms contain fully dressed correlators for fluctuating fields as well as mixed vertices with two or three 
fluctuation legs and one background leg associated with the cut external gluon line, see \cite{Fister:2013bh} 
for technical details. There it has also been argued that the two-loop terms can be neglected for many practical
purposes, leaving the one-loop terms with their dependence on the ghost (dashed line), gluon (curly line) and 
quark (solid line) propagators only. In principle, the DSEs for the corresponding propagators need to be evaluated
also in the background Landau-DeWitt gauge to take full advantage of the formalism. In practise, it has been argued
\cite{Braun:2007bx} that the ordinary Landau gauge propagators are already a good approximation for most purposes. 
Using solutions of the coupled system of DSEs for the gluon and quark propagators at finite temperature and chemical 
potential the one-loop equation Fig.~\ref{fig:DSE-A} has been applied to study the deconfinement of heavy quarks in 
Ref.~\cite{Fischer:2014vxa} and for QCD with $N_f=2+1$ quark flavours in Ref.~\cite{Fischer:2013eca}. The resulting 
deconfinement transition line in the QCD phase diagram agrees with the one from the dressed Polyakov loop. We will 
discuss this in more detail in section \ref{res:heavy} and \ref{results:CEP}.

\subsubsection{Positivity and spectral functions} \label{sec:positivity}

A physical particle that can be directly detected in experiments needs to have a propagator with a positive definite
spectral function. Conversely, if a certain degree of freedom has negative norm contributions in its propagator,
it cannot describe a physical asymptotic state, {\it i.e.\/} there is no K\"all\'en--Lehmann spectral representation 
for its propagator. The precise mathematical structure of this condition in the context of an Euclidean quantum 
field theory has been formulated by Osterwalder and Schrader in the so called axiom of {\it reflection positivity} 
\cite{Osterwalder:1973dx}. On the level of propagators this condition can be phrased as 
 \beqa 
  \Delta(t) &:=& \int d^3x \int \frac{d^4p}{(2\pi)^4}
  e^{i(t p_4+\vect{x}\cdot\vect{p})} \sigma(p^2)\,, \\ 
  &=& \frac{1}{\pi}\int_0^\infty
  dp_4 \cos(t p_4) \sigma(p^2_4) \;\ge 0 \,,
 \label{schwinger} 
 \eeqa
where $\sigma(p^2)$ is a scalar function extracted from the respective propagator. Eq.~(\ref{schwinger}) 
is a necessary condition, thus once $\Delta(t)$ develops negative values at some times $t$ the propagator 
cannot correspond to a stable asymptotic particle. In fact it can 
be shown \cite{Mandula:1987rh,Roberts:1994dr,Alkofer:2000wg} that $\Delta(t)$ even has to be convex, i.e.
\beq\label{convex}
\frac{d^2}{dt^2} \ln \Delta(t) \ge 0\,,
\eeq
in order to represent a particle with positive definite spectral function.

In the context of
the chiral transition of QCD it is especially interesting to monitor the behaviour of the Schwinger function
of the quark dressing functions across the phase transition. To this end it is instructive to introduce 
the (Euclidean) projectors on positive and negative energy solutions\footnote{Conventions are chosen such that 
Minkowsky and Euclidean gamma matrices are related by $\gamma_M^0 = -i \gamma_4$ and $\gamma_M^i = \gamma_i$.}
\beq
\Lambda^{\pm}_p = \frac{1}{2 \varepsilon_p} \left(\varepsilon_p \mp i \gamma_4 (\vpslash+m)\right) \,,
\eeq
with the limit $L_\pm = \frac{1}{2} \left(1 \mp i \gamma_4 \right)$ for $\vp =0$. In this limit we define
the quark propagator components $S_\pm$ by
\beq \label{0quark}
S(\vect{0},\op) = -i \left[ S_+(\op)L_+ + S_-(\op) L_- \right] \gamma_4 \,,
\eeq
and obtain
\beq \label{1quark}
S_\pm(\op) = \frac{-i \tilde{\omega}_p C(\vect{0},\op) \pm B(\vect{0},\op)}
                 {\tilde{\omega}_p^2 C^2(\vect{0},\op) + B^2(\vect{0},\op)}\,.
\eeq
The Fourier transform of these quantities
\beq \label{2quark}
S_\pm(\tau)  = T \sum_n e^{-i \omega_n \tau} S_\pm(\op) \,,
\eeq
with respect to time $\tau$ are Schwinger functions in the sense of 
Eq.~(\ref{schwinger}) and can be used to study the positivity properties
of the quark propagator at finite temperature and chemical potential. Corresponding results will be discussed in 
section \ref{results:spectral}. 

It is furthermore interesting to study potential spectral representations of the quark propagator.
Parametrizing the spectral function $\rho(\vp,\omega_p)$ by
\beq
\rho(\vp,\omega_p) = 2\pi\left(\rho_{4}(p,\omega_p)\gamma_4 +
                  \rho_{\rm v}(p,\omega_p) i \vpslash/p -\rho_{\rm s}(p,\omega_p) \right) \,, \label{eq:rhostruc} 
\eeq
with $p=|\vect{p}|$, the spectral representation is given by
\begin{align}\label{spectral}
S(\vp,\omega_p) &= \int_{-\infty}^\infty \!\frac{d\omega_p'}{2\pi}\, \frac{\rho(\vp,\omega_p')}{i\omega_p-\omega_p'} \,.
\end{align}
With a positive definite metric, which is not the case for gauge-fixed QCD, the components  
of the spectral function would furthermore obey the inequality
\begin{align}
\rho_{4}(p,\omega_p) &\geq \sqrt{\rho_{\rm v}(p,\omega_p)^2+\rho_{\rm s}(p,\omega_p)^2} \geq 0\,,
\label{eq:RhoInequality}
\end{align}
as well as the sum rules
\begin{align}
1&=Z_2\int_{-\infty}^\infty\!d\omega\, \rho_4(p,\omega_p)
\,,\label{eq:sum_rule1}\\
0&=\phantom{Z_2}\int_{-\infty}^\infty\!d\omega\, \rho_{\rm v}(p,\omega_p)
\,,\label{eq:sum_rule2}\\
0&=\phantom{Z_2}\int_{-\infty}^\infty\!d\omega\, \rho_{\rm s}(p,\omega_p)
\,,
\label{eq:RhoSumRules}
\end{align}
with wave function renormalization constant $Z_2$.

A useful component of the spectral function is $\rho_4$ with corresponding left hand side
of Eq.~\eqref{spectral} given by the $\gamma_4$ component of the quark propagator, i.e. 
\begin{align}
S_4(p,\omega_p)= \frac{-i \omega_p C(p,\omega_p)}{\omega_p^2 C^2(p,\omega_p)+ p^2 A^2(p,\omega_p)+B^2(p,\omega_p)}\,,
\end{align} 
which is particularly well suited for spectral reconstructions using e.g. Maximum Entropy methods (MEM), 
see \cite{Mueller:2010ah,Gao:2014rqa,Fischer:2017kbq} for technical details.  

Beyond the study of positivity, spectral functions are very interesting quantities to study. Thermal and 
transport properties of the QGP are encoded in the correlation functions of QCD and can be extracted from 
real time properties of the quark and gluon propagators
\cite{Nickel:2006mm,Harada:2007gg,Harada:2009zq,Karsch:2007wc,Karsch:2009tp,Mueller:2010ah,
Qin:2010pc,Qin:2013ufa,Gao:2014rqa,Christiansen:2014ypa,Ilgenfritz:2017kkp}. In heavy ion collisions,  
the dilepton production rate is directly related to the dispersion relation of quarks
\cite{Braaten:1990wp,Peshier:1999dt,Arnold:2002ja,Kim:2015poa}. Therefore, a detailed understanding 
of a potential quasi-particle spectrum in the QGP, in particular close to the chiral phase transition, is desirable.
We come back to this topic in the results section \ref{results:spectral}.

\subsection{Truncation strategies}\label{DSE:truncation}

In the previous section we discussed various order parameters for chiral symmetry breaking as well as the deconfinement
transition, which are all accessible by functional methods since they can be computed from gauge fixed Green's functions.
These Green's functions satisfy coupled sets of Dyson-Schwinger or functional renormalisation group equations that in
general need to be truncated in order to be solvable.\footnote{There are exceptions from this rule. One obvious exception
is the limit of large momenta, where asymptotic freedom allows the use of perturbation theory which effectively decouples
the tower of DSEs of FRGs and allows for an order by order approach. Another, highly non-trivial exception has been 
identified in \cite{Alkofer:2004it,Fischer:2006vf,Huber:2007kc,Fischer:2009tn}: for very small momenta, general
scaling laws for all one-particle irreducible Green's functions of the theory have been identified that solve
the complete tower of DSEs (and FRGs) self-consistently.}

In the Review article Ref.~\cite{Eichmann:2016yit} truncations of DSEs have been discussed in the context of vacuum
hadron physics\footnote{A corresponding discussion of truncations in the Yang-Mills sector of QCD can be found in 
Ref.~\cite{Huber:2018ned}.} starting from very simple truncation schemes that lead to NJL-type equations up to rainbow-ladder (RL) 
and beyond rainbow-ladder (BRL) constructions. There it has been argued, that rainbow-ladder type of truncations 
can be viewed as a water shed between phenomenological modelling and more systematic truncations of QCD. From a 
technical point of view this distinction is justified by two important properties of the RL and BRL schemes that 
are not present in simpler truncations: (i) the correct momentum running of the Green's functions in the perturbative,
asymptotically free region of QCD and in the transition region to non-perturbative physics; 
(ii) the preservation of multiplicative renormalizability. Thus while simpler truncations can be used to explore and 
model qualitative aspects of hadron physics, only RL and BRL have the merit that they can be improved systematically 
within QCD. We will se below that advanced schemes capture important physics that is generated from the Yang-Mills 
sector of the theory. Concerning applications at finite temperature and chemical potential this aspect turns out 
to be crucial in many respects. After all, this insight has led to the construction of the 'P'-models, i.e. the 
enhancement of the NJL and quark-meson (QM) models by supplementing the Polyakov-loop potential. 

In the following sections we therefore focus only on truncation schemes that fall into the RL and BRL class. Before 
we detail the properties of two of these scheme in section \ref{fulltrunc} and \ref{sec:RL}, however, we first
discuss in general which aspects of QCD we can reasonably hope to cover in feasible truncation schemes.

\subsubsection{General considerations}\label{general}

\begin{figure}[t]
    \centering
    \includegraphics[scale=0.80]{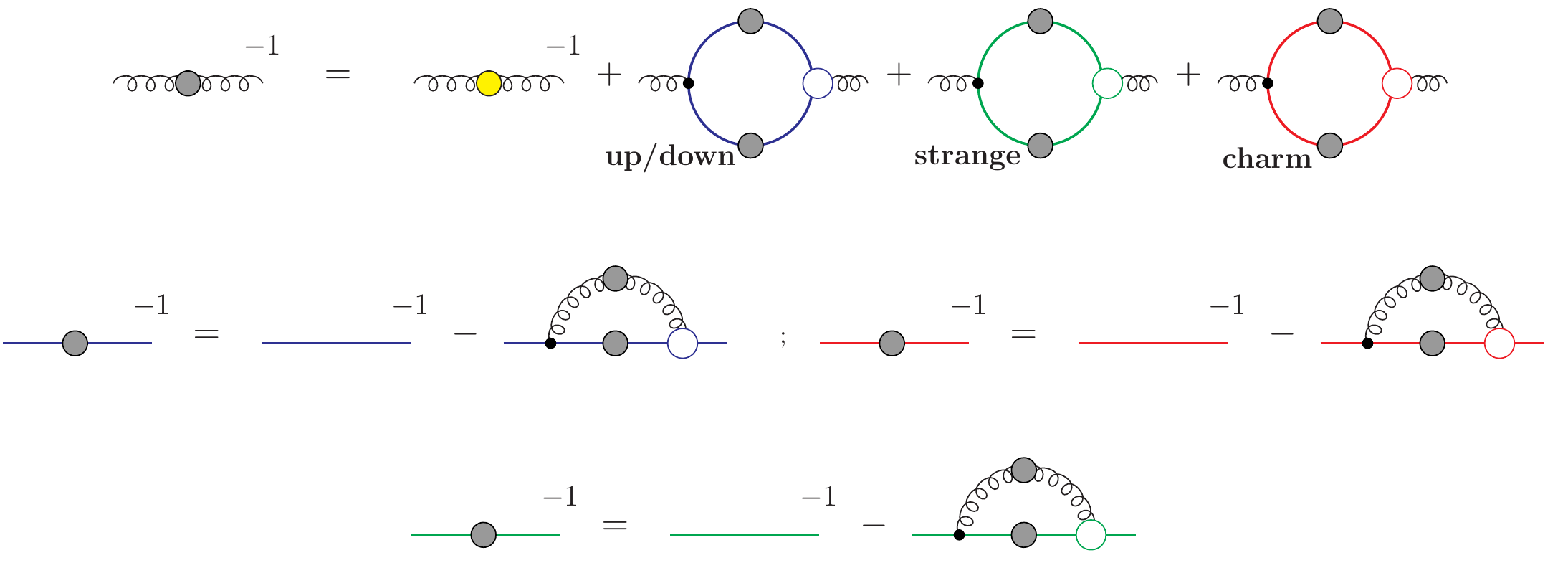}
    \caption{Dyson-Schwinger equations for the gluon (curly lines) and the quark propagator (solid lines), displaying
    explicitly the contributions from the up/down (blue), strange (green) and charm quarks (red). 
            \label{fig:DSE-prop2}}
\end{figure}

In the light of our discussion of the QCD phase diagram, the potential CEP and the Columbia plot in sections
\ref{gen:phase} and \ref{gen:columbia} one should wish for a truncation of the DSEs that includes the rich 
physics associated with the variation of quark masses and flavours and the resulting interplay of the chiral 
and the deconfinement phase transitions. To this end it is inevitable to explicitly take into account the
back-coupling of the quarks onto the Yang-Mills sector. To leading order in an expansion in $1/N_c$, this is accomplished
by the quark loop in the DSE for the gluon propagator, shown again in Fig.~\ref{fig:DSE-prop2}. Compared to 
Fig.~\ref{fig:DSE-prop} we have combined the inverse bare gluon propagators as well as all ghost and gluon diagrams into
one symbol, the inverse gluon propagator on the right hand side of the gluon-DSE 
with a yellow dot. For now, this is just an abbreviation.
Furthermore, we have made the contributions from the up/down, the strange and the charm quark loops explicit (with an implicit
factor of 2 in the up/down quark loop accounting for isospin degeneracy). Furthermore, we made explicit that the different
quark flavours each satisfy a separate DSE for their respective propagator. This serves to explain, how the physics of 
varying bare quark masses affects this system of equations: if e.g. one flavour becomes infinitely heavy, its quark loop
becomes zero and this flavour decouples from the equations. Thus for all quark masses going to infinity we arrive at the 
pure gauge limit of the Columbia plot, i.e the upper right corner. Keeping only the charm infinite and varying the
up/down and strange quark masses we are able to navigate through the Columbia plot and explore the associated 
changes of orders of the chiral and deconfinement transitions. 

What else can be expected ? In order to study the properties of the quark-gluon plasma, i.e. the state of matter reached
in high energy heavy ion collisions one would like to extract the properties of gluons under variation of temperature
and chemical potential. Of particular fundamental interest may be the question of the analytic structure of the
pure gauge gluon propagator below and above the deconfinement transition and the related property of positivity violation.
Furthermore interesting physics is expected to show up in the different components of the propagator: At finite 
temperature and chemical potential the gluon propagator splits into two different parts transverse and 
longitudinal to the heat bath.\footnote{For a discussion of hidden Lorenz invariance of this formulation 
see e.g. \cite{Maas:2011se}.} Both depend separately on the three momentum $\vect{p}$ and the energy $\omega$. 
In Landau gauge, the propagator is then given by 
\begin{align}\label{eq:qProp}
D_{\mu\nu}(p) &= P_{\mu\nu}^{T}(p)\frac{Z_{T}(p)}{p^2} + P_{\mu\nu}^{L}(p)\frac{Z_{L}(p)}{p^2} \\
              &= P_{\mu\nu}^{T}(p) D_{T}(p) + P_{\mu\nu}^{L}(p)D_{L}(p)\,, \nonumber
\end{align}
with momentum $p=(\vect{p},\omega_n)$. The Matsubara frequencies are $\omega_n=\pi T \, 2n$.
All dressing functions implicitly depend on temperature and chemical potential.
The projectors $P_{\mu\nu}^{{T},{L}}$ are transverse (${T}$) and longitudinal (${L}$) with respect
to the heat bath vector aligned in four-direction and given by
\begin{equation} \label{eq:projTL}
\begin{split}
P_{\mu\nu}^{T} &= \left(1-\delta_{\mu 4}\right)\left(1-\delta_{\nu 4}\right)\left(\delta_{\mu\nu}-\frac{p_\mu p_\nu}{\vect{p}^{\,2}}\right),   \\
P_{\mu\nu}^{L} &= P_{\mu\nu} - P_{\mu\nu}^{T} \,,
\end{split}
\end{equation}
where $P_{\mu\nu} = \delta_{\mu\nu} - p_\mu\, p_\nu/p^2$ is the covariant transverse projector.
As we will see later on, especially at or around the deconfinement phase transition, temperature effects act  
differently on the transverse (magnetic) and longitudinal (electric) part of the gluon propagator and lead to 
interesting effects such as the realisation of an electric screening mass. These effects can be studied from 
the gluon DSE. Explicit knowledge of the gluon is furthermore required in order to determine important 
observable quantities such as thermodynamics and transport coefficients, 
see e.g. \cite{Haas:2013hpa,Christiansen:2014ypa}.

While the Yang-Mills part of the tower of DSEs accounts for gluonic effects correlated with confinement, screening and
other important phenomena, the matter part matters when it comes to effects of chiral symmetry breaking. From the study
of highly simplified versions of the quark DSE (such as NJL-type of equations) it is known that chiral symmetry breaking 
is triggered by a certain strength of the interaction independent of its details. These simple
models also produce chiral restoration with temperature and even feature the appearance of a critical end point at
finite chemical potential. However, they are not predictive on a quantitative basis as can be seen e.g. by the variety 
of results from such simple models collected in Ref.~\cite{Stephanov:2004wx}. In order to quantitatively study important
questions like the location and even the very existence of the critical end point we need to take into account all we know
about the non-trivial structure of the gluon propagator and the quark-gluon vertex. As we will see later on in 
sections \ref{results:heavy} and \ref{results:2p1}, we
already have very detailed and accurate results for the temperature dependence (and presumably to a lesser extent also the
chemical potential dependence) of the gluon propagator at our disposal. From the quark-DSE in Fig.~\ref{fig:DSE-prop2} we
then identify the fully dressed quark-gluon vertex as the crucial quantity to determine the quark completely.

What is known about the quark-gluon vertex ? Unfortunately not as much as we would like. First of all, the vertex
is a complicated object. In Landau gauge, there are already 12 different Dirac tensor structures in the vacuum,
which inflate to 32 different structures at finite temperature and chemical potential due to the splitting in
directions longitudinal and transverse to the heat bath.\footnote{This can be seen as follows: Due to the external 
gluon and quark legs there is one Lorenz index and two Dirac indices. Furthermore momentum conservation
leaves two independent four-momenta. The building blocks for the vertex are therefore the three four-momenta 
$[p_\mu, q_\mu, \gamma_\mu]$ which can be multiplied each with one of the four possible different Lorenz contractions
$[\mathbbm{1},\pslash,\qslash,\qslash \pslash]$. Thus we obtain twelve different and linearly independent Dirac structures.
At finite temperature all four-vectors split up into a part longitudinal and a part transverse to the direction
of the heat bath. It can be shown that multiplying $[\vect{q},\vect{p},\vect{\gamma},\gamma_4] 
\times [\mathbbm{1},\gamma_4] \times
[\mathbbm{1},\vect{\gamma}\vect{q},\vect{\gamma}\vect{p},(\vect{\gamma}\vect{q})(\vect{\gamma}\vect{p})]$ accounts
for all 32 different possibilities.} Although by using the transversality of Landau gauge these can be reduced
to 8 (vacuum) or 24 (medium) structure, it remains a highly complicated object. The longitudinal part of the vertex satisfies a Slavnov-Taylor identity (STI), which has been solved approximately in the vacuum, see 
\cite{Aguilar:2018epe} and references therein. At finite temperature, however, these are not yet available. 
Similarly, exploratory lattice calculations of the vertex only exist at zero temperature and chemical 
potential \cite{Skullerud:2003qu}. The Dyson-Schwinger equation for the quark-gluon vertex has been explored 
in quite some detail in the vacuum using highly elaborate truncation schemes, 
see e.g. \cite{Braun:2014ata,Williams:2015cvx}, the review \cite{Eichmann:2016yit} and references therein. 
At finite temperature and chemical potential, however,
explicit calculations are still only on an exploratory level with first results discussed in
Ref.~\cite{Welzbacher:2016,Huber:2016xbs,Contant:2018zpi} (see Ref.~\cite{Huber:2013yqa} for corresponding exploratory results 
for the ghost-gluon vertex at finite temperature). Despite its complexity, however, the vertex-DSE is an 
important source of additional information. In fact, following Refs.~\cite{Fischer:2007ze,Eichmann:2015kfa}, 
we will now argue that there are important effects related to the physics of the Columbia plot that can 
be identified.

 \begin{figure}[t]
 \begin{subfigure}[t]{\textwidth}
 \centering\includegraphics[width=1.0\textwidth]{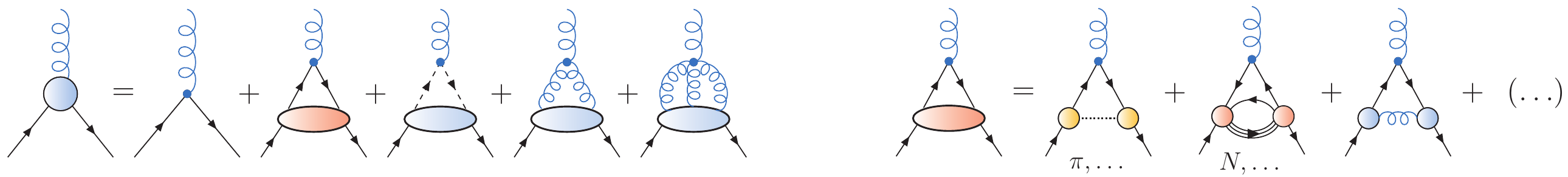}
 \caption{Left: Dyson-Schwinger equation for the quark-gluon vertex \cite{Marciano:1977su}.
  All internal propagators are fully dressed. Dashed lines with arrows denote ghost propagators, curly lines gluons 
  and solid lines quarks.
 Right: Expansion of one of the diagrams in terms of hadronic and non-hadronic contributions 
 to the quark-antiquark scattering kernel. The dotted line denotes mesons and the triple line baryons.}
 \label{fig:Vertexdse}
 \end{subfigure}
 \begin{subfigure}[t]{\textwidth}
 \centering\includegraphics[width=0.90\textwidth]{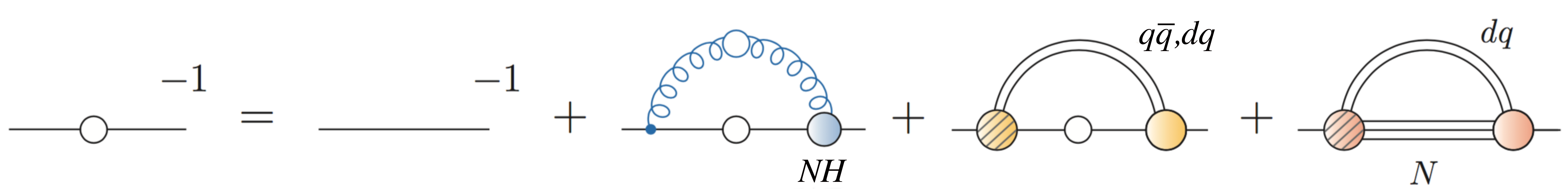}
 \caption{Resulting quark DSE with gluon, meson/diquark and baryon loop. In these loops the right
 vertices (circles) are Bethe-Salpeter amplitudes whereas the left vertices
 (hatched circles) are effective ones, well approximated by bare ones.}
 \label{fig:quark-dse-3}
 \end{subfigure}
 \caption{Meson and baryon effects in the quark-DSE.}\label{fig:mesonbaryon} 
 \end{figure}

The exact DSE for the quark-gluon vertex is shown in the left part of Fig.~\ref{fig:Vertexdse}. It contains 
a two-loop and three one-loop diagrams 
with fully dressed quarks (solid), ghosts (dashed) and gluon lines (curly) running through the loops and attached to the 
external gluon by a corresponding bare vertex. Consider now the highlighted diagram with the internal four-quark Green's 
function. The right part of Fig.~\ref{fig:Vertexdse} shows a skeleton expansion of this diagram in terms of Bethe-Salpeter 
vertices and propagators of mesons as well as Faddeev-type vertices for baryons. 
Why is such an expansion meaningful ? 
Suppose we would evaluate the diagram on the left hand side for time like momenta running through the quark legs.
Internally, these can be routed through the four-point function. Whenever these time-like momenta are evaluated at the
location of meson bound states or resonances, the quark four-point functions features a corresponding singularity. In
the vicinity of the singularities the four-point function is well approximated by the propagator of the bound state or
resonance and two corresponding Bethe-Salpeter amplitudes. Thus around these meson poles, a representation of the left hand
side of this equation in terms of the first diagram of the right hand side is an excellent approximation. In this context
it is important to keep in mind that these mesons are not introduced as new elementary fields; they are rather composite 
objects of a quark and an antiquark that are described (at least on-shell) by their Bethe-Salpeter equation (BSE). 
In principle, mesons with all quantum numbers contribute to this diagram including ground states and radial excitations
and it depends on the values of the external quark momenta, which of these are most important. In this review we are interested
in equilibrium phenomena at finite temperature and chemical potential and, as we will see below, the quark momenta
relevant to determine quantities such as order parameters are purely space-like. Thus we will never evaluate the vertex
for momenta where the exchanged mesons are on-shell, but will mostly be sensitive to the tails of the singularities caused by 
the mesons with lowest mass. In the $N_f=2$-theory these are the pseudoscalar pions and the scalar sigma meson, in the $N_f=3$ 
theory this includes kaons, the $\eta$-meson, and, depending on the fate of the $U_A(1)$ symmetry, also a light or heavy $\eta'$.
In addition to meson exchange, there are also diagrams which feature internal baryons. The first baryon exchange diagram 
shows up as a two-loop diagram involving the baryon's Faddeev amplitude, the second diagram on the right hand side of
the equation. Naturally, these diagrams are expected to contribute less than the mesons due to larger masses of the baryons. 
Furthermore we display a representative non-resonant contribution due to dressed one-gluon exchange.\footnote{Note 
that double-counting is trivially avoided in this combined expansion in elementary and effective degrees of freedom 
due to different quantum numbers in the exchange channel.} 

By plugging the DSE for the quark-gluon vertex into the quark-DSE and making diagrammatic rearrangements it 
has been argued \cite{Eichmann:2015kfa} that the effects of the diagrams involving hadron exchange can be represented 
on the level of the quark-DSE by the one-loop diagrams displayed in Fig.~\ref{fig:quark-dse-3}. 
 In both hadronic diagrams the vertex appearing on the right are proper Bethe-Salpeter amplitudes,
once for a meson/diquark and once for a baryon in quark-diquark approximation. The hatched
vertices on the left carry the same quantum numbers as their counterparts on the right but represent 
effective vertices, which can be argued to be well represented by bare vertices \cite{Eichmann:2015kfa}.
The non-hadronic part of the quark-gluon vertex (denoted by 'NH' in
Fig.~\ref{fig:quark-dse-3}) represents all contributions from the quark-gluon vertex which cannot be
rewritten in hadronic diagrams, i.e. all ghost and gluon loop contributions in the left equation of
Fig.~\ref{fig:Vertexdse} and the non-resonant parts of the expansion in the right equation.  

The important physics that is made apparent in such a construction is at least three-fold.
First, as discussed 
above in section \ref{gen:columbia} we expect the chiral two-flavour theory to feature a second order phase transition
in the O(4) universality class provided the $U_A(1)$ remains anomalously broken. This physics should be
reflected in the scaling properties of the chiral order parameter as e.g. the chiral condensate or the scalar 
quark dressing function. Indeed, as will be reviewed in section \ref{results:chiral} this can be achieved
in the DSE-framework using explicit meson degrees of freedom as displayed in Fig.~\ref{fig:quark-dse-3}. 
We will actually find, that (only) in the close vicinity of the phase transition the meson loop dominates 
the quark-DSE, which is precisely what we expect from universality. Second, such an explicit construction 
allows to assess the impact of baryons on the chiral order parameters, which will be important at large 
densities. Third, taking into account hadronic contributions will improve the quantitative accuracy of 
the framework with respect to the search for the critical end-point of QCD. This will be the topic of 
section \ref{sec:mesons}.

\subsubsection{Exploring the Columbia plot: a truncation including the Yang-Mills sector explicitly}\label{fulltrunc}

In the subsection above we discussed general expectations on the physics that can be extracted from DSEs 
and corresponding truncation strategies. In this subsection and the following one we explicate two
such truncation schemes, that have been employed in the past decade. 

We start with a scheme that takes the Yang-Mills sector, i.e. the DSE for the gluon propagator as shown in
Fig.~\ref{fig:DSE-prop}, explicitly into account. In the vacuum, this DSE has been subject to intense scrutiny 
over the past 20 years and we will review some of the results in section \ref{DSE:vacuum_quarks}. The temperature 
dependence of the gluon propagator, however, is difficult to explore in the DSE framework, see e.g. 
\cite{Maas:2005ym,Maas:2011se} for reviews. In the FRG framework, great efforts have been involved to extract the
temperature dependence of the magnetic and electric part of the gluon propagator in pure Yang-Mills theory
\cite{Fister:2011uw,Cyrol:2017qkl} with partial success: corresponding results of lattice simulations have been
reproduced for temperatures below and above the first order phase transition. However, the truncations were still
not rich enough to capture the drastic changes around the critical temperature. 
The authors of \cite{Fischer:2009wc,Fischer:2009gk,Fischer:2010fx,
Fischer:2011mz,Fischer:2012vc,Fischer:2014ata,Fischer:2014vxa,Eichmann:2015kfa} therefore explored the idea 
to replace the Yang-Mills self-energy diagrams of the gluon DSE with lattice data for the quenched propagator. 
In Fig.~\ref{fig:DSE-prop2} these are precisely the ghost and gluon diagrams that have been abbreviated by
the inverse dressed propagator on the right hand side of the equation (yellow dot). By replacing this symbol
with the quenched lattice data one misses unquenching (quark-loop) effects in the Yang-Mills self-energies.
These can be shown to be subleading in an $1/N_c$ expansion, whereas the leading quark-loop contributions
are the ones explicitly taken into account by the quark-loops shown in Fig.~\ref{fig:DSE-prop2}. At zero 
temperature, the effects of this approximation can be explicitly determined using the framework of 
Ref.~\cite{Fischer:2003rp} and are found to be well below the five percent level for the resulting gluon 
dressing functions. 

With the quenched lattice input, the resulting DSEs for the quark and gluon propagators read
\begin{eqnarray}
\left[S^{f}(p)\right]^{-1} &=& Z^f_{2}\left[S_0^{f}(p)\right]^{-1} 
+ C_{F}\,Z^f_{1F} \, g^2 \,T\sum_n \int\frac{d^3l}{(2\pi)^3}\, 
\gamma_\mu \,S^f(l)\, \Gamma^f_\nu(l,p;q)\, D_{\mu\nu}(q), \label{DSEs-1} \\
\left[D_{\mu\nu}(p)\right]^{-1} &=& \left[D_{\mu\nu}^{qu.}(p)\right]^{-1} - \sum_{f}^{N_f}\,Z_{1F}^f\,\frac{g^2}{2}\,
T\sum_n \int\frac{d^3l}{(2\pi)^3}\, \Tr\left[ \gamma_\mu \,S^{f}(l)\, \Gamma^f_\nu(l,q;p)\,
 S^{f}(q)\right], \label{DSEs-2}
\end{eqnarray}
where $q=(p-l)$, $S^f$ is the quark propagator for one specific flavour $f \in \{u,d,s,c\}$, 
$C_F=\frac{N_C^2-1}{2N_C}$ is the Casimir operator and $\Gamma_\nu$ the dressed quark-gluon vertex.
The (inverse) dressed quark propagator has been given already in Eq.~(\ref{quark}), its bare counterpart is given by
\beq
\left[S_0^{f}(p)\right]^{-1} = i \gamma \cdot p + Z^f_m m^f\,,
\eeq
and contains the renormalized quark mass $m^f$ from the Lagrangian of QCD. 
The quark-gluon vertex and quark wave function and mass renormalization constants are denoted by $Z_{1F}^f$, $Z_2^f$ and $Z_m^f$; 
for the running coupling they used $\alpha=g^2/(4\pi)=0.3$. The details of the fit functions for the quenched
gluon are given in the appendix of Ref.~\cite{Eichmann:2015kfa}, the corresponding lattice results 
\cite{Fischer:2010fx,Maas:2011ez} are discussed later on in section \ref{results:heavy}. 

The remaining quantity to be determined in the coupled system of DSEs (\ref{DSEs-1}) and (\ref{DSEs-2}) is the 
dressed quark-gluon vertex $\Gamma_\nu$. Here they used a convolution of the first term of the Ball-Chiu vertex, 
satisfying the 
Abelian WTI, multiplied with an infrared enhanced function $\Gamma(p^2,k^2,q^2)$ that accounts 
for the non-Abelian effects in the STI of the vertex and its correct ultraviolet running with momentum; see
the appendix of Ref.~\cite{Eichmann:2015kfa} for more details. The resulting expression reads
\begin{eqnarray}\label{vertex}
\Gamma_\mu^f(l,p;q) &=& \gamma_\mu\cdot\Gamma(l^2,p^2,q^2) \cdot 
\left(\delta_{\mu,4}\frac{C^f(l)+C^f(p)}{2} + \delta_{\mu,i}\frac{A^f(l)+A^f(p)}{2} \right)\,, \label{vertex1}\\ 
\Gamma(l^2,p^2,q^2) &=& \frac{d_1}{d_2+x} \!
 + \!\frac{x}{\Lambda^2+x}
\left(\frac{\beta_0 \alpha(\mu)\ln[x/\Lambda^2+1]}{4\pi}\right)^{2\delta}\,, \label{vertex2}
\end{eqnarray}
where $l$ and $p$ are fermionic momenta and $q$ is the gluon momentum. The vertex
features an implicit temperature, chemical potential and quark mass/flavor dependence via the 
quark dressing functions $A^f$ and $C^f$ in agreement with the Abelian part of the STI. The second term in 
Eq.(\ref{vertex2}) ensures the correct logarithmic running of the loops in the quark and gluon-DSE. 
Both scales $\Lambda= 1.4$ GeV and $d_2 = 0.5$ GeV$^2$ are fixed such that they match the 
corresponding scales in the gluon lattice data. The anomalous dimension is $\delta=\frac{-9Nc}{44N_c - 8N_f}$ 
and $\beta_0=\frac{11N_c-2N_f}{3}$. The only free parameter of the interaction is the vertex strength 
$d_1$ which has been adapted to (pseudo-)critical temperatures determined on the lattice. This results in 
$d_1 = 4.6 \,\mbox{GeV}^2$ for the quenched theory \cite{Fischer:2010fx} and $d_1 = 7.5 \,\mbox{GeV}^2$ for the
theory with $N_f=2+1$ quark flavours \cite{Fischer:2012vc}. We come back to this
point when we discuss the results in this truncation in section \ref{results}. 

The squared momentum variable $x$ is identified with the gluon momentum $q^2$ in the quark DSE 
and with the sum of the two squared quark momenta $l^2+p^2$ in the quark loops of the gluon DSE. 
This different treatment of the momentum dependence is necessary to maintain multiplicative 
renormalizability of the gluon-DSE \cite{Fischer:2003rp}. Details of the renormalization procedure 
of the gluon-DSE have been discussed in \cite{Fischer:2012vc}.

The resulting truncation scheme includes the leading back-coupling effects of four quark flavours 
onto the Yang-Mills sector of QCD. By varying the quark masses $m_f$ one can navigate 
in the Columbia plot and $m_f \rightarrow \infty$ completely switches off the corresponding flavour.
This truncation is therefore suited to explore the quenched limit of QCD with massive test quarks 
\cite{Fischer:2009wc,Fischer:2009gk}, the heavy quark deconfinement transition \cite{Fischer:2014vxa}
and the QCD phase diagram for $N_f=2$, $N_f=2+1$ and $N_f=2+1+1$ quarks with physical masses \cite{Fischer:2011mz,Fischer:2012vc,Fischer:2014ata}.
Corresponding results will be discussed in section \ref{results}. The truncation is, however, not yet
elaborate enough to discuss the critical behaviour of the two-flavour theory, i.e. the putative
second order phase transition in the upper left corner of the Columbia plot. To this end, as already indicated
above, one needs to incorporate meson effects in the quark-gluon vertex explicitly. This will be detailed
in section \ref{results:chiral}, where we summarise corresponding results. Furthermore, the back-reaction 
effects of baryons onto the quarks have been explored along the lines discussed in section \ref{general};
corresponding results will be shown in section \ref{results:baryons}.

\subsubsection{Rainbow-ladder and beyond rainbow-ladder truncations using a model for the gluon}\label{sec:RL}

We now come to the second class of truncations, those that model the temperature effects of the gluon 
without an explicit back-coupling of the quarks. In the literature, these models appear in various stages 
of sophistication. The advantage of these models is their simplicity and the reduced
CPU-time needed to perform the calculations. This allowed to include and explore the effect of additional 
tensor structures in the quark-gluon vertex. It also allowed to determine quantities like the pressure and
quark number susceptibilities and fluctuations that have not yet been addressed in the truncation scheme
detailed in the previous section. The drawback of these models is the lack of a well-defined 
quark flavour number, i.e. it is not possible to address the physics of the Columbia plot. Moreover,
the non-trivial temperature and chemical potential dependence of the gluon, discussed in sections 
\ref{results:gluon} and \ref{results:2p1}, may not be represented well in the models. Nevertheless, the
results obtained from the model truncations, presented in section \ref{results:thermo} are both, interesting
on their own and they serve as benchmarks for more complete truncation schemes. 

In the notation fixed in the quark-DSE, Eq.~(\ref{DSEs-1}), the model approach does not take into
account the DSE for the gluon but instead uses an ansatz for the magnetic part $D_T(\vect{k},\ok)$ and 
the electric part $D_L(\vect{k},\ok)$ of the gluon propagator. To make the evolution of the various models
transparent we use the following notation
\begin{align}
D_{T}(s) & = \mathcal{D}(s,0) \,, \\
D_{L}(s) & = \mathcal{D}(s,m_g) \,,
\end{align}
with $s= k^2 + m_g^2 = \vect{k}^2 + \ok^2 + m_g^2$. Thus $m_g=0$ indicates that $D_T = D_L$, whereas 
$m_g^2 = 16/5(T^2 + 6 \mu^2/(5\pi^2))$ introduces an electric gluon screening mass in agreement with 
a leading order hard thermal loop calculation \cite{Haque:2012my}. The most important
part of the gluon models is a term which provides interaction strength at low momenta that is 
large enough to trigger dynamical chiral symmetry breaking. To this end, simple exponentials
such as  
\begin{align}\label{RL:watson}
g^2 \mathcal{D}(s,m_g) &= 
\frac{4 \pi^2 D}{\omega^6} \, s \, e^{-s/\omega^2}
\end{align}
and variants thereof have been explored in the literature. The form (\ref{RL:watson}) has been used 
in \cite{Qin:2010nq} together with $m_g=0$ (i.e. $D_T = D_L$)  and in \cite{Jiang:2011ke} with non-zero 
$m_g^2$ as given above. Slight variations have been employed in Refs.~\cite{Shi:2014zpa,Shi:2016koj}.
The parameters $\omega = 0.5$ GeV and $D = (0.8 \,\mbox{GeV})^3/\omega$ distribute the strength 
of the interaction over a range of momenta dependent on the values of the parameters. 
More elaborate versions of the model include the logarithmic running of the interaction at large momenta 
together with different versions of the exponential that incorporate a potential variation $D(T,\mu)$ 
of the strength of the interaction with temperature and chemical potential via  
\begin{align}
g^2 \mathcal{D}(s,m_g) &= \left\{
\begin{array}{l}
\frac{4 \pi^2 D(T,\mu)}{\omega^6} \, s \, e^{-s/\sigma^2} \\
\frac{8 \pi^2 D(T,\mu)}{\omega^4} e^{-s/\omega^2}
\end{array}
\right\} 
          + \frac{8 \pi^2 \gamma_m}{\ln[e^2-1+(1+s/\Lambda^2_{QCD})^2]}\frac{1-e^{-s/4m^2_t}}{s}\,,\label{RL:MT}\\
D(T,\mu) & = \left\{ 
\begin{array}{ll} 
D,										& T<T_p \\
\frac{a}{b(\mu)+\ln[T'/\Lambda_{QCD}]}, & T \ge T_p
\end{array}\right.   \label{RL:D}         
\end{align}
Here $m_t=0.5$ GeV, $\gamma_m = 12/25$, $\Lambda_{QCD} = 0.234$ GeV throughout the literature, whereas 
the most common values for 
$\omega = 0.5$ GeV and $\sigma D = (0.8 \,\mbox{GeV})^3$ are sometimes slightly varied. 
The parameters $a$, $b$ and $T'$ are used to model potential screening
effects in the interaction that appear for temperatures larger than $T_p$. These parameters have been
adjusted differently in different works. 
In Refs.~\cite{Xu:2015jwa,Gao:2016qkh} this interaction has been used together with $m_g=0$ (i.e. $D_T = D_L$), 
and constant $D(T,\mu) = D$. In Ref.~\cite{Gao:2015kea} a non-zero 
$m_g^2$ as given above has been employed together with $T_p(\mu)=T_c(\mu)$, $(T')^2= T^2 + 6 \mu^2/[5\pi^2]$ 
and $a=0.029, b=0.47$. The same choices
have been used in Ref.~\cite{Gao:2016hks} but $b=0.432$. Finally, in \cite{Xin:2014ela} a non-zero $m_g^2$
has been used together with $T_p = 1.3T_c$, $T'=T$ and two implicit conditions for $a$ and $b$. Here, also 
both variants of the exponential have been compared and several variations of $\omega$ have been studied.

Together with the models for the gluon propagator, three different truncations for the quark-gluon vertex
have been used and in some publications directly compared. These are either the bare vertex
\cite{Qin:2010nq,Xin:2014ela,Xu:2015jwa,Gao:2015kea,Gao:2016hks}
\beq
\Gamma_{\mu} = \gamma_\mu
\eeq
or a Ball-Chiu construction generalised to finite temperature and chemical potential given by
\begin{align}
\Gamma_{\mu}(p,q) &= \gamma^T_\mu \Sigma_A 
                  + \gamma^L_\mu \Sigma_C
+ (\tilde{p}+\tilde{q})_\mu
\left[\frac{1}{2}\gamma^T_\alpha \,(\tilde{p}+\tilde{q})_\alpha \Delta_A
      \frac{1}{2}\gamma^L_\alpha \,(\tilde{p}+\tilde{q})_\alpha \Delta_C + \Delta_B \right]\\
\Sigma_{F \in \{A,B,C\}} &=\frac{F(q)+F(p)}{2} \\
\Delta_{F \in \{A,B,C\}} &=\frac{F(q)-F(p)}{\tilde{q}^2-\tilde{p}^2} 
\end{align}
with $\tilde{p} = (\vect{p},\op+i\mu)$ and $\tilde{q} = (\vect{q},\oq+i\mu)$ \cite{Qin:2010nq,Jiang:2011ke}. 
In \cite{Gao:2016qkh} the 
Ball-Chiu vertex was even supplemented by a construction for the transverse parts of the vertex 
that has been taken from the vacuum physics studies of Ref.~\cite{Chang:2010hb}. Since this construction
is quite elaborate we refrain from giving the details here and refer the interested reader to Ref.~\cite{Gao:2016qkh}.

For completeness we wish to mention that there are a number of studies in the Dyson-Schwinger framework 
that used even simpler truncations than the rainbow-ladder models discussed above. These have been reviewed
comprehensively in \cite{Roberts:2000aa}; later results can e.g. be found in
Refs.~\cite{Horvatic:2007wu,Horvatic:2007qs,Horvatic:2010md}. 

\subsection{Brief overview on selected vacuum results: gluons and quarks}\label{DSE:vacuum_quarks}

In the following we give a brief summary on vacuum results for the most basic correlation functions of QCD, 
the propagators of the gluon and the quark, since these will play a major role below when we discuss results
at finite temperature and chemical potential. Many more details can be found in a number of review articles
focusing on different aspects
\cite{Roberts:1994dr,Alkofer:2000wg,Roberts:2000aa,Maris:2003vk,Fischer:2006ub,Binosi:2009qm,Holt:2010vj,Maas:2011se,
Cloet:2013jya,Aguilar:2015bud,Eichmann:2016yit,Sanchis-Alepuz:2017jjd,Huber:2018ned}.

\subsubsection{The gluon}\label{sec:gluon}
The exploration of the content of the gluon DSE started already in the 70ies using simplified truncations
schemes which neglected the ghost contributions to the gluon \cite{Mandelstam:1979xd,Cornwall:1981zr,Brown:1988bn}.
At the end of the nineties, the importance of these contributions have been realised 
\cite{vonSmekal:1997ohs,vonSmekal:1997ern,Atkinson:1997tu,Atkinson:1998zc} and subsequently full numerical
solutions of the coupled ghost and gluon DSE have been obtained \cite{Fischer:2002hna} and expanded to also 
include the quark-DSE in a self-consistent scheme \cite{Fischer:2003rp}. A particular focus in the vacuum 
continuum explorations of the gluon and ghost-DSEs has been the infrared behaviour of the gluon and the ghost. 
In the small momentum regime, $p \ll 100 \,\mbox{MeV}$, the DSEs can be solved analytically and exact solutions 
without any truncations 
are possible \cite{Zwanziger:2001kw,Lerche:2002ep,Alkofer:2004it,Fischer:2006vf,Fischer:2009tn}. This has been
corroborated also in the corresponding tower of functional renormalization group equations (FRGs) 
\cite{Pawlowski:2003hq,Fischer:2006vf,Fischer:2009tn}. In both towers, two qualitatively different solutions 
have been found named 'scaling' and 'decoupling'. Whereas the scaling 
solution consists of infrared power laws for all Green's functions with an infrared vanishing gluon propagator 
and an infrared divergent ghost, the decoupling solution \cite{Aguilar:2008xm,Boucaud:2008ky,Dudal:2008sp,Alkofer:2008jy,Aguilar:2015bud}
is characterized by an infrared finite gluon propagator and a finite ghost dressing function. Self-consistent
numerical solutions of both types have been discussed in \cite{Fischer:2008uz,Huber:2012kd}. 
Current lattice calculations on 
very large volumes clearly favour the decoupling type of solutions \cite{Cucchieri:2008fc,Bogolubsky:2009dc}; 
potentially significant effects from different gauge fixing strategies in the deep infrared have been discussed,
see e.g. \cite{Cucchieri:2007rg,vonSmekal:2008ws,Sternbeck:2008mv,Cucchieri:2009zt,Maas:2009ph,Maas:2009se,
Sternbeck:2012mf,Dudal:2014rxa,Cucchieri:2016qyc} and references therein. The existence of a family of decoupling
type solutions together with the scaling limit may be connected to the principal problem of incomplete gauge fixing 
in Landau gauge and the Gribov copy problem as discussed e.g. in \cite{Fischer:2008uz,Maas:2009ph,Maas:2009se,Sternbeck:2012mf}.
Since in this review we are concerned with the QCD phase diagram and the associated deconfinement transition, the 
most important question related to the deep infrared behaviour of the gluon propagator is its relation to confinement
as indicated by the order parameters for center symmetry breaking discussed in section \ref{sec:deconf}. It is
therefore reassuring that both types of solutions, scaling and decoupling, satisfy a confinement criterion based 
on the Polyakov-loop potential derived in Ref.~\cite{Braun:2010cy}. Furthermore, the relevant scales for all studies
at finite temperature and chemical potential are much larger than the scaling/decoupling region: well below the chiral
transition the most important scales are $\Lambda_{QCD} \approx 250 \,\mbox{MeV}$ and the dynamically generated 
quark masses $M \approx 400 \,\mbox{MeV}$; at larger temperatures and chemical potential $2\pi T$ and $\mu_q$ are
important landmarks. Thus for the subjects discussed in this review, the deep infrared behaviour of the ghost and 
gluon propagators is not relevant. 

Numerical solutions for the coupled system of ghost and gluon-DSEs are available in the vacuum, see e.g.  
\cite{Fischer:2002hna,Aguilar:2008xm,Fischer:2008uz,Hopfer:2014zna,Aguilar:2015bud,Aguilar:2017dco} and Refs. therein
even including the non-perturbative gluonic two-loop diagrams \cite{Huber:2017txg}. In recent years, several groups
have begun to explore in addition the DSEs for the ghost-gluon, three-gluon, four-gluon and quark-gluon vertices,
see the end of section 3.2. in Ref.~\cite{Eichmann:2016yit} for a short overview and guidance to further literature.  

Technically, there are a number of issues for both, the gluon and the quark DSE that have to be carefully taken into
account in order to obtain reasonable and meaningful solutions. One is multiplicative renormalizability (MR). This exact 
property of QCD has to be maintained by any reasonable truncation claiming to respect the properties of the theory.
Technically, it is easy to show that the exact equations are MR, see appendix \ref{renorm} for an upshot, and therefore
it is not too difficult to devise truncation schemes that maintain this property. Much more elaborated are issues
related to transversality of the gluon propagator and the appearance of artificial quadratic divergences in numerical
treatments of the gluon-DSE. These have to be eliminated carefully, see e.g. \cite{Fischer:2002hna,Huber:2014tva} and
Refs. therein for discussions of this problem and practical solutions.

\begin{figure}[t]
        \begin{center}
        \includegraphics[width=0.48\textwidth]{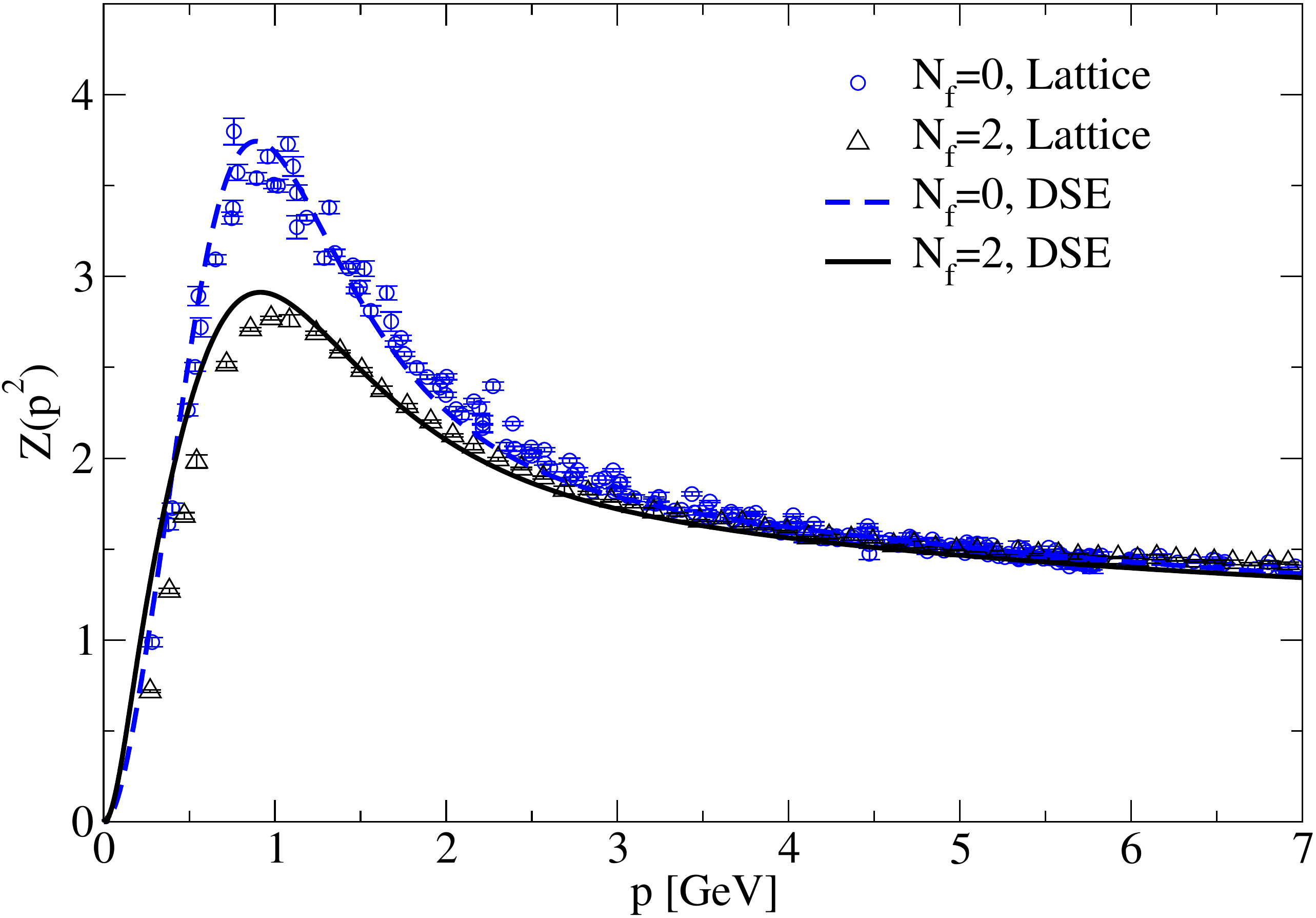}\hfill
        \includegraphics[width=0.48\textwidth]{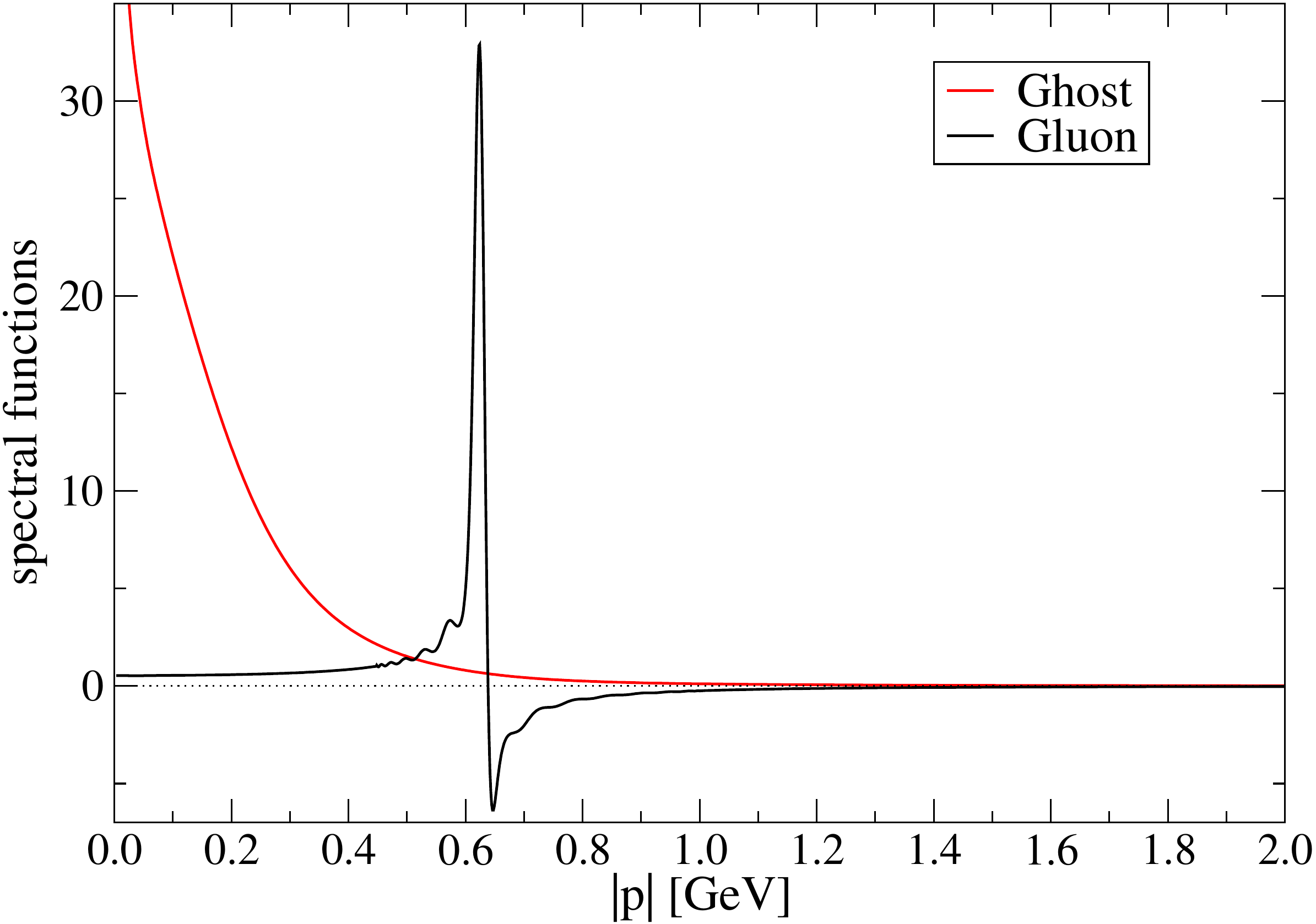}
        \caption{\textit{Left:} Gluon dressing function $Z(p^2)$ for $N_f=0$ and $N_f=2$ calculated from DSEs~\cite{Williams:2015cvx} and compared to lattice calculations~\cite{Sternbeck:2005tk,Sternbeck:2016}.
 \textit{Right:} Spectral functions of the gluon and ghost propagators from a direct calculation in the complex 
       momentum plane \cite{Strauss:2012dg}; see text for explanations.}\label{fig:gluon}
        \end{center}
\end{figure}
A lot of interesting physics is hiding in the gluon propagator. In the vacuum the two dressing 
functions $Z_{T}(p)$ and $Z_{L}(p)$ of Eq.~(\ref{eq:qProp}) become degenerate and
the gluon propagator is given in terms of one function $Z_(p^2) = Z_{T}(p) = Z_{L}(p)$ 
\begin{equation}\label{eq:qPropv}
D_{\mu\nu}(p) = \left(\delta_{\mu\nu} - \frac{p_\mu p_\nu}{p^2} \right) \frac{Z(p^2)}{p^2}\,,
\end{equation}
A typical result for the gluon dressing $Z(p^2)$ in Landau gauge is shown in the left diagram of Fig.~\ref{fig:gluon}.
The DSE-solutions are taken from \cite{Williams:2015cvx} and are compared to lattice calculations~\cite{Sternbeck:2005tk,Sternbeck:2016}.
There are numerous studies of the gluon propagator on the lattice, see e.g. \cite{Cucchieri:1997dx,Boucaud:2000nd,Langfeld:2001cz,Silva:2004bv,Sternbeck:2005tk,Cucchieri:2007rg,Bowman:2007du,
Bogolubsky:2009dc,Cucchieri:2011ig,Oliveira:2012eh,Ayala:2012pb,Boucaud:2017ksi,Biddle:2018dtc,Boucaud:2018xup} 
and references therein 
as well as \cite{Maas:2011se} for a review. 
In the large momentum region the DSEs can be solved analytically and the numerical results follow the analytical solution 
\begin{eqnarray}
Z(p^2) &=& Z(\mu^2) \left[  \frac{\alpha(\mu^2) \beta_0}{4\pi} \ln\left(\frac{p^2}{\mu^2}\right)+1 \right]^\gamma  \,,
\label{gluon_uv}\\
\end{eqnarray}
in accordance with resummed perturbation theory. Here $\mu^2$ denotes the renormalisation point, 
$\beta_0 \alpha(\mu^2)/(4\pi) = (11N_c - 2N_f)/3$ and the leading order anomalous dimension of the gluon
reads $\gamma = (-13N_c +4N_f)/(22N_c-4N_f)$. A detailed account of the analytical ultraviolet 
analysis is given e.g. in \cite{Fischer:2003zc}. In the mid-momentum region one finds a characteristic bump 
in the gluon dressing function. This bump has interesting properties. First, lattice studies suggest that
on the level of individual gluon field configurations this bump is mostly generated by center 
vortices \cite{Gattnar:2004bf,Biddle:2018dtc}. To demonstrate this, lattice ensembles of gauge fields have 
been generated where all vortex content has been removed, see \cite{Montero:1999by,Faber:1999sq} for technical 
details. This reveals an interesting correlation: the full configurations of the pure gauge theory deliver 
a gluon dressing function as the one shown in Fig.~\ref{fig:gluon} (for $N_f=0$) together with a static quark-antiquark potential 
that is linearly rising for large inter-quark distances. The vortex-removes ensembles, however, deliver a drastically 
reduced (though not vanishing) bump in the dressing function and a constant potential which is no longer confining.
It therefore seems as if the gauge field configurations with vortex content are responsible for both, confinement
in the pure gauge theory and the appearance of a sizeable bump in the gluon dressing function. 

As mentioned above, the DSEs can also be solved analytically in the small momentum 
region and one finds either a scaling solution, $Z(p^2) \sim (p^2)^{(2\kappa)}$ with the precise value of the 
exponent $\kappa > 0.5$ depending on the truncation, or one finds decoupling which means $Z(p^2) \sim p^2$ and 
consequently the propagator dressing function $D(p^2) = Z(p^2)/p^2 \sim const.$ approaches a constant, 
which can be associated with a mass of the gluon. Since this 'decoupling' solution is the one also favoured 
by the lattice, we focus on this type in the following. The apparent massive behaviour of the decoupling 
solution cannot be produced by a mass in the ordinary sense attached to the physics of a particle that can be 
detected. This can be seen in many ways and has been reviewed for example in 
\cite{Maas:2011se,Aguilar:2015bud}, we therefore give only a short summary here. 

The simplest possibility for
the propagator of a massive particle is given by 
\beq \label{simpleprop}
D(p^2) \sim \frac{1}{p^2 + m^2}\,,
\eeq
which is indeed constant at small momenta, respects positivity and shows a single pole at time-like 
momentum $p^2=-m^2$. The Landau gauge 
gluon propagator cannot have this form, simply because a constant mass is excluded already at the level of the 
QCD Lagrangian due to gauge invariance. Thus the 'mass' contribution to the gluon needs to be momentum dependent,
as indeed encoded in the dressing function $Z(p^2)$ shown above. Although its infrared behaviour generates a 'mass-like' 
behaviour similar to (\ref{simpleprop}), its ultraviolet behaviour is different in accordance with the tree-level
theory and perturbative corrections. Furthermore, as discussed above, the momentum dependence at intermediate 
momenta contains important non-perturbative physics (the 'bump') that cannot be captured by (\ref{simpleprop}).
Consequently, also the analytic structure of the gluon propagator in the complex $p^2$-momentum plane is different 
than the one of (\ref{simpleprop}). Formally, this can be seen also from the Oehme-Zimmermann superconvergence 
relation \cite{Oehme:1979ai,Nishijima:1993fq,Oehme:1994hf,Nishijima:1995ie} (see also \cite{Alkofer:2000wg} for 
a summary)
\beq
0 = \int ds \rho(s)\,,
\eeq   
where $\rho(s) \sim \mbox{Im}(D(s))$ and the integration is on the time-like momentum axis. This sum rule shows,
that the gluon propagator necessarily has to contain negative norm contributions in its spectral function; in fact
the presence of these can be shown already in perturbation theory.

The analytic structure of the fully dressed gluon propagator has been studied intensely in the past years
with methods ranging from explicit solutions of DSEs in the complex momentum plane \cite{Strauss:2012dg}, 
analytic continuations using MEM and related methods \cite{Gattnar:2004bf,Dudal:2013yva,Cyrol:2018xeq}, 
considerations based on axiomatic field theory \cite{Lowdon:2017uqe}, studies of the Schwinger function 
\cite{Alkofer:2003jj} up to guided fits to lattice data \cite{Dudal:2010tf,Cucchieri:2011ig,Cucchieri:2016jwg}. 
While some of these results point towards a pair of complex conjugate singularities in the squared momentum plane, 
others indicate a cut along the time-like momentum axis together with potential further structure on the second
Riemann sheet. In the right diagram of Fig.~\ref{fig:gluon} we show the result for the ghost and gluon spectral
functions from a direct calculation in the complex momentum plane \cite{Strauss:2012dg}. A strict derivation of
boundary conditions for the spectral function at small momenta (in disagreement with the non-zero gluon spectral 
function of the gluon seen in Fig.~\ref{fig:gluon}) has been presented recently in Ref.~\cite{Cyrol:2018xeq}. 
At finite momenta, the spectral reconstruction performed in \cite{Cyrol:2018xeq} agrees qualitatively with 
the explicit calculation.  

The physical interpretation of a cut-structure in the gluon propagator is straight forward: one unphysical 
particle (the gluon) is splitting into other unphysical particles (two or three gluons or a ghost-antighost pair). 
While the last word is not yet spoken on this issue, a simple particle interpretation of the gluon along 
the lines of Eq.(\ref{simpleprop}) is certainly ruled out.

\subsubsection{The quark}\label{sec:quark}

\begin{figure}[t]
        \begin{center}
        \includegraphics[width=0.48\textwidth]{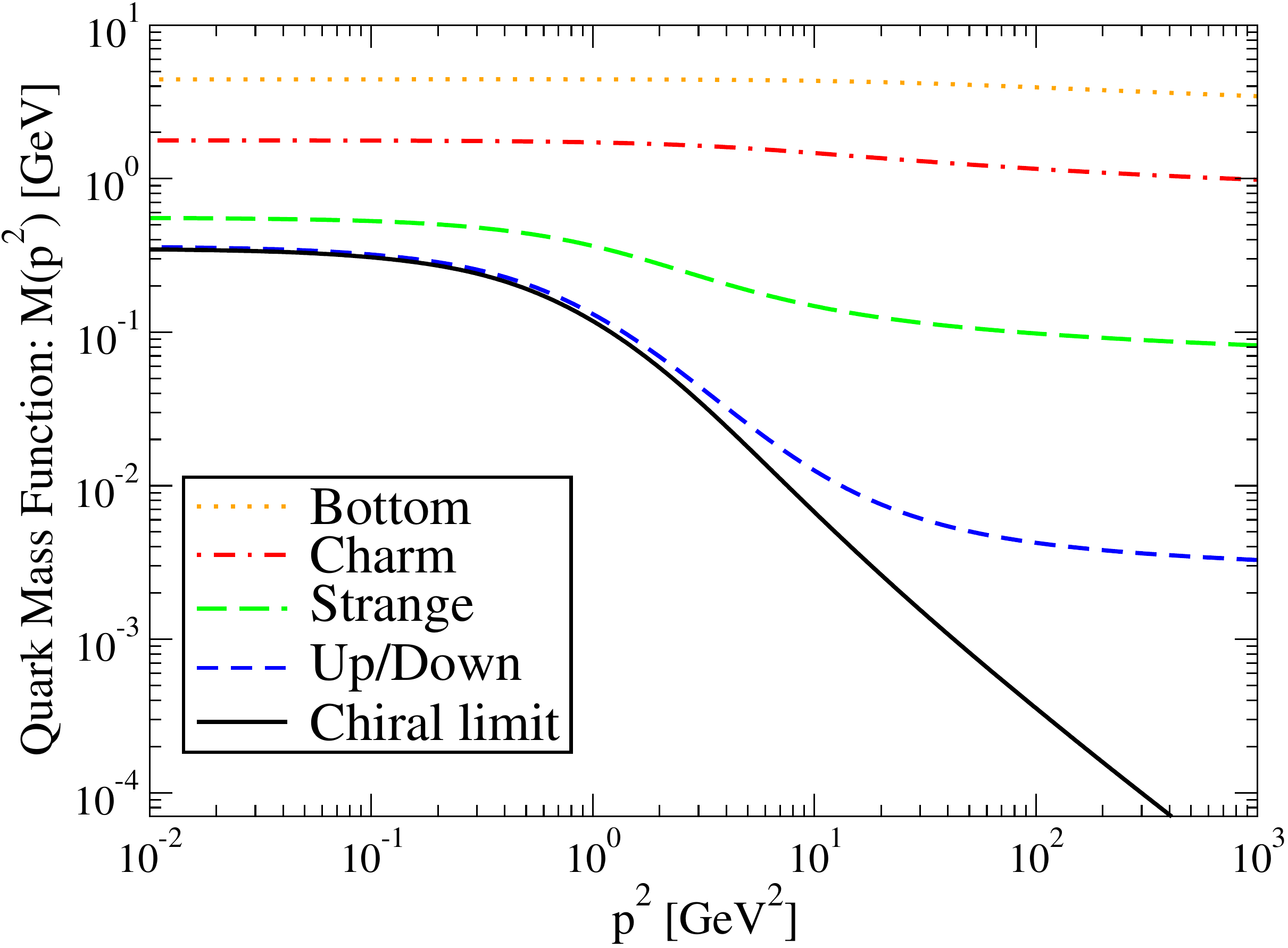}\hfill
        \includegraphics[width=0.48\textwidth]{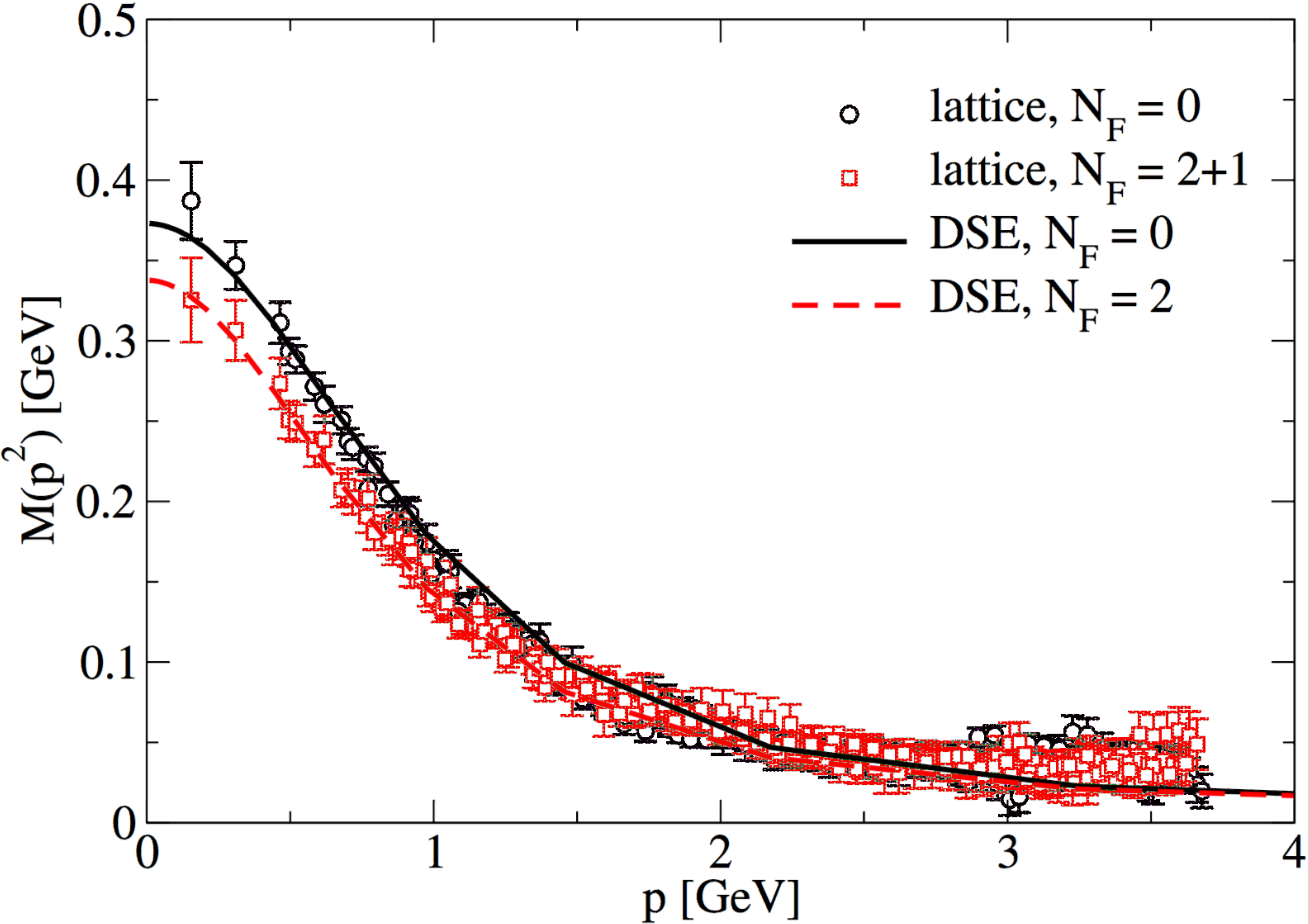}
        \caption{\textit{Left:} Typical DSE solutions for the quark mass function; 
                          figure adapted from the Reviews \cite{Fischer:2006ub,Eichmann:2016yit}.
 \textit{Right:} Quark mass function for an up/down quark from a realistic truncation of DSEs \cite{Williams:2015cvx} 
                 compared to quenched ($N_f=0$) and unquenched ($N_f=2+1$) lattice data \cite{Bowman:2005vx}. Figure adapted from \cite{Williams:2015cvx}.}\label{fig:quark}
        \end{center}
\end{figure}

Similar to the gluon DSE, also the quark DSE has been explored already in the 70ies starting with \cite{Fukuda:1976zb}.
Important milestones were the classification of the ultraviolet asymptotic behaviour in the presence of 
dynamical chiral symmetry breaking \cite{Miransky:1984ef,Miransky:1986ib} and the numerical exploration 
of gauge invariance at the beginning of the nineties, see \cite{Roberts:1994dr} for an early review. The 
importance of the quark-DSE as the central equation to study dynamical chiral symmetry breaking on the level
of the microscopic degrees of freedom of QCD has been repeatedly pointed out, c.f. the review articles mentioned
at the beginning of section \ref{DSE:vacuum_quarks}.  

The inverse dressed quark propagator at zero temperature and chemical potential can be parametrised by
\beq\label{eq:quarkvac}
S^{-1}(p) = i \pslash A(p^2) + B(p^2) = \frac{1}{Z_f(p^2)} \left( i \pslash + M(p^2) \right)
\eeq
either by the vector and scalar dressing functions $A(p^2)$ and $B(p^2)$ or, equivalently, by the quark mass function 
$M(p^2)=B(p^2)/A(p^2)$ and the quark wave function $Z_f(p^2) = 1/A(p^2)$. The latter representation has the advantage
that one of the functions, the mass function, is independent of the renormalization point, cf. \ref{renorm}. 
When comparing the vacuum expression (\ref{eq:quarkvac}) with the one for finite temperature given earlier
in Eq.~(\ref{quark}) we find that in the zero temperature limit the functions $C(\vect{p},\omega_p)$ and
 $A(\vect{p},\omega_p)$ 
become degenerate and are combined into one function $A(p^2)$. The two different Dirac tensor structures in the 
quark propagator, 
$\pslash$ and 1, behave different under chiral symmetry transformations. The vector part $\pslash$ is invariant, 
whereas the scalar part is not. A non-vanishing function $B(p^2)$ (or equivalently $M(p^2)$) thus signals unambiguously 
that chiral symmetry is broken and therefore may serve as an order parameter as discussed in section \ref{chiral}. 
Typical results for the mass functions of different quark flavours are shown in the left diagram of Fig.~\ref{fig:quark},
whereas in the right diagram we display results for the quenched and unquenched theory compared to lattice data. 

Similar to the gluon-DSE also the one for the quark propagator can be solved analytically in the large momentum 
region \cite{Miransky:1984ef,Miransky:1986ib}, featuring the asymptotic behaviour 
\beq
M(p^2) = \frac{2 \pi^2 \gamma_m}{3}
\frac{-\langle \bar{\Psi}\Psi\rangle}{p^2 \left(\frac{1}{2} \ln(p^2/\Lambda^2_{QCD})\right)^{1-\gamma_m}} 
+ M(\mu^2) \left[\alpha(\mu^2) \beta_0 \ln\left(\frac{p^2}{\mu^2}\right)+1\right]^{-\gamma_m}\,.
\label{chiral-M_UV}
\eeq
Here $\gamma_m = \frac{12}{11N_c-2N_f}$ is the anomalous dimension of the quark and 
$\langle \bar{\Psi}\Psi\rangle$ denotes the renormalisation point independent quark condensate in the chiral limit. 
In this limit, this condensate is related to the renormalisation point dependent one already discussed
in Eq.~(\ref{eq:condensate}) by a simple logarithmic factor 
\beq
\langle \bar{\Psi}\Psi\rangle(\mu^2) = \left(\frac{1}{2}\ln(\mu^2/\Lambda^2_{QCD})\right)^{\gamma_m}
\langle \bar{\Psi}\Psi\rangle \,, 
\label{ch-loop}
\eeq
provided the renormalisation point $\mu^2$ is taken large enough. In the chiral limit, the second term in Eq.~(\ref{chiral-M_UV})
is absent and the chiral condensate is a well-defined and convergent quantity. The quark mass function then vanishes
like a power law for large momenta as can be seen in the left plot of Fig.~\ref{fig:quark}. For a finite quark mass in the 
Lagrangian the logarithmic term in Eq.~(\ref{chiral-M_UV}) dominates at large momenta as is also visible in Fig.~\ref{fig:quark}.  
This logarithmic term, however, is directly responsible for the divergence of the quark condensate as extracted from the 
trace of the quark propagator, Eq.~(\ref{eq:condensate}). Introducing a hard cut-off $\Lambda$ in the integration in the 
vacuum version of Eq.~(\ref{eq:condensate}) it can be shown on dimensional grounds that this divergence is proportional 
to $m \Lambda^2$, with bare quark mass $m$. This explains, why the expression in Eq.(\ref{eq:cond_renorm}) remains 
finite.

In the infrared momentum region one clearly sees the sizeable effect of dynamical mass generation as one consequence of
dynamical chiral symmetry breaking. At zero momentum, the quarks acquire dynamical masses of the order of 350-400 MeV.
These dynamical effects play together with the presence of the (renormalized) current quark masses into continuous functions
connecting a 'constituent quark' mass at small momenta with the running 'current' quark mass at large momenta. Thus the notions
of the quark model and the findings of deep inelastic scattering are naturally connected. As can be seen in the right diagram
of Fig.~\ref{fig:quark}, the size of the dynamically generated quark mass decreases when quark-loops are present in the unquenched theory.
This is similar to the gluon propagator, where screening effects due to the presence of quark-loops also lead to a decrease 
of the bump, cf. Fig.~\ref{fig:gluon}. 

The quark-DSE is a coupled system of two (vacuum) or three (finite T) equations for the quark dressing functions. The 
sizeable dynamical effects discussed for the quark mass function therefore also lead to drastic modifications of the 
wave function $Z_f(p^2)$. In advanced truncation schemes, these can be extracted reliably in agreement with the
results of lattice gauge theory, see \cite{Williams:2015cvx} for details. 

The analytic structure of the quark propagator in the complex momentum plane in an open issue. Generic rainbow-ladder truncations
lead to complex conjugate singularities in the complex squared momentum plane, see e.g. Fig.~3.9 in Ref.~\cite{Eichmann:2016yit}
for a graphic representations of the situation or Ref.~\cite{Windisch:2016iud} for a heroic exploration beyond the leading
singularity structures. Although this type of structure seems to persists in more complete truncations beyond rainbow 
ladder \cite{Fischer:2008sp,Williams:2015cvx} there is also evidence for the principal possibility of a leading singularity
on the real axis \cite{Alkofer:2003jj}. The latter situation is realised in QED\footnote{Indeed, numerical evidence 
presented in \cite{Alkofer:2003jj} suggests, that vertices that satisfy the Abelian Ward-Takahashi identity lead to a
singularity on the time-like momentum axis accompanied by a cut. In QED this is the natural structure one would expect for 
a fermion with a photon cloud. Abelianised truncations of QCD deliver similar results.}, but may not be very plausible 
in the
non-Abelian theory. In any case, the analytic structure of the quark propagator plays an important role when it comes to 
the possibilities and limitations of determining hadronic observables in the functional framework. This is discussed in 
detail in the reviews Ref.~\cite{Eichmann:2016yit,Sanchis-Alepuz:2017jjd}.   

Finally, we wish to mention that the solution displayed in Fig.\ref{fig:quark} is not the only one possible for
the quark-DSE. To see this, consider the quark-DSE in the chiral limit, i.e. with $m \rightarrow 0$. The equation
for the scalar quark dressing function is then homogeneous and always admits an additional solution with $B(p^2)=0$
for all momenta. In contrast to the Nambu-Goldstone solution $B(p^2)\ne 0$ with broken chiral symmetry, this 
Wigner-Weyl solution is chirally symmetric. At not too large finite quark masses, this solution persists and  
has even been used to extract a finite quark condensate away in the chiral limit by an alternative procedure 
than Eq.(\ref{eq:cond_renorm}) \cite{Chang:2006bm,Williams:2006vva}. Details of the Wigner solution beyond 
rainbow-ladder have been explored in \cite{Fischer:2008sp}. We will briefly come back to the Wigner solution
in section \ref{results:spectral}, when we discuss quark spectral functions at large temperature.

\subsection{Brief overview on selected vacuum results: mesons and baryons}\label{DSE:vacuum_hadrons}

In this section we give a very brief overview on selected results for meson and baryon spectra obtained in the
Dyson-Schwinger approach using bound state Bethe-Salpeter equations. Detailed recent reviews on this and many 
more topics such as spectra, electromagnetic properties, decays, etc. are available
\cite{Holt:2010vj,Bashir:2012fs,Cloet:2013jya,Horn:2016rip,Eichmann:2016yit}. Therefore, we keep the 
discussion brief and focus only on results that will play a role later on in the presentation 
of the finite temperature and chemical potential results. 

\subsubsection{Mesons}\label{sec:mesons}

Chiral symmetry, its breaking and its restoration at large temperatures and chemical potential is one of the main topics
in this review, therefore we will focus on this aspect first. The dynamical breaking of the $SU_A(3)$ part of the flavour
chiral symmetry of the QCD Lagrangian leads to the appearance of a multiplet of eight pseudoscalar Goldstone bosons in
the chiral limit, which become massive due to additional explicit symmetry breaking effects by finite bare quark masses.
These pseudoscalar
states are actually both, Goldstone bosons and bound states of quarks and antiquarks. This dichotomy has been explored in
great detail in the framework of Dyson-Schwinger and Bethe-Salpeter equations, see e.g.
\cite{Maris:1997hd,Maris:1997tm,Horn:2016rip,Eichmann:2016yit}. 
There are a number of exact results that can be shown analytically in this
framework such as the Goldstone boson nature of the pseudo-scalars, the Gell-Mann-Oakes-Renner relation 
\beq
f_\pi^2 m_\pi^2 = -2 m_q \langle \bar{\Psi} \Psi \rangle_0 
\eeq
with pion decay constant $f_\pi$, pion mass $m_\pi$, light quark mass $m_q$ and the light quark condensate in the chiral limit.
Furthermore in the chiral limit $m_q \rightarrow 0$ there is the exact relation
\beq\label{Boverf}
\Gamma_0(p;P) = \gamma_{5} \frac{B(p^2)}{f_\pi}
\eeq
between the scalar part $B(p^2)$ of the quark propagator and the leading component $\Gamma_0$ of the pion's Bethe-Salpeter 
vertex, thus manifesting the tight relation between signatures of dynamical mass generation and the vertex of the Goldstone 
boson. For a pedagogical derivation of these results see section 3.4 and 4.2 of Ref.~\cite{Eichmann:2016yit}.
The full Bethe-Salpeter vertex of any pseudoscalar quark-antiquark bound state can be decomposed into four Dirac-structures,
\beq
\hspace*{-5mm}
\Gamma^{PS}(p;P) \;=\; \gamma_{5} \left[\Gamma_0(p;P) - \imath \Pslash \Gamma_1(p;P)
-\imath \pslash \Gamma_2(p;P) - \left[\Pslash,\pslash \right]\Gamma_3(p;P) \right]\,,
\eeq 
with total momentum $P$ of the bound state and relative momentum $p$ between the internal quark-antiquark pair. For
the sake of brevity we suppressed the flavour and colour structure of the vertex.

Pseudoscalar bound states of a quark and an antiquark are described by the 
homogeneous Bethe-Salpeter equation (BSE) which can be written schematically
as
\beq
\Gamma^{PS}(p;P)\;=\;\int \frac{d^4k}{(2 \pi)^4}\, K(p,k;P)\, S(k_+)\, \Gamma^{PS}(k;P)\, S(k_-).
\label{eq:bse}
\eeq
Here $K(p,k;P)$ is a 2PI quark line irreducible Bethe-Salpeter kernel, which describes the interaction of the
quark and the antiquark. The momentum arguments  $k_+=k+\xi P$ and $k_-=k+(\xi-1)P$ of the two quark 
propagators are defined such that the total momentum of the pion is given by $P=k_+-k_-$. All physical 
results are independent of the momentum partitioning $\xi=[0,1]$ between the quark and the antiquark.
Bethe-Salpeter equations with similar structure have been written down and solved also for scalar, (axial-)vector
mesons and tensor mesons with total spin $J=2$ \cite{Krassnigg:2010mh} and $J=3$ \cite{Fischer:2014xha}.

The crucial link between the meson bound states and their quark and gluon constituents is provided by the axial vector 
Ward-Takahashi identity (axWTI). Abbreviating the quark DSE by
\beq
S^{-1}(p)= S_0^{-1}(p) - \Sigma(p)
\eeq
one can write the axial vector Ward-Takahashi identity as
\beq
-\left[\Sigma(p_+)\gamma_5+\gamma_5\Sigma(p_-)\right] =
\int\frac{d^4k}{(2\pi)^4} K(p,k;P)
\left[\gamma_5 S(k_-) + S(k_+) \gamma_5 \right],
\label{eq:axwti}
\eeq
where again all flavour and spinor indices have been omitted. We see that this identity demands a tight relation between 
the quark self-energy $\Sigma$ and the Bethe-Salpeter kernel $K$. It is therefore at the heart of the proof of the 
above mentioned analytical results. In addition, analytic proofs have been established \cite{Bicudo:2001jq,Bicudo:2003fp} 
that Weinberg's low energy theorems for $\pi-\pi$ scattering, the Goldberger Treiman relation and the Adler zero in the 
chiral limit hold in all approximation schemes for the quark DSE and the Bethe-Salpeter kernel $K$ that satisfy the axWTI. 
This is guaranteed in all rainbow-ladder type schemes, but also in beyond rainbow-ladder (BRL) truncations, see 
Ref.~\cite{Eichmann:2016yit} for an overview. 

           \begin{figure}[t]
            \centering
            \includegraphics[width=0.55\textwidth]{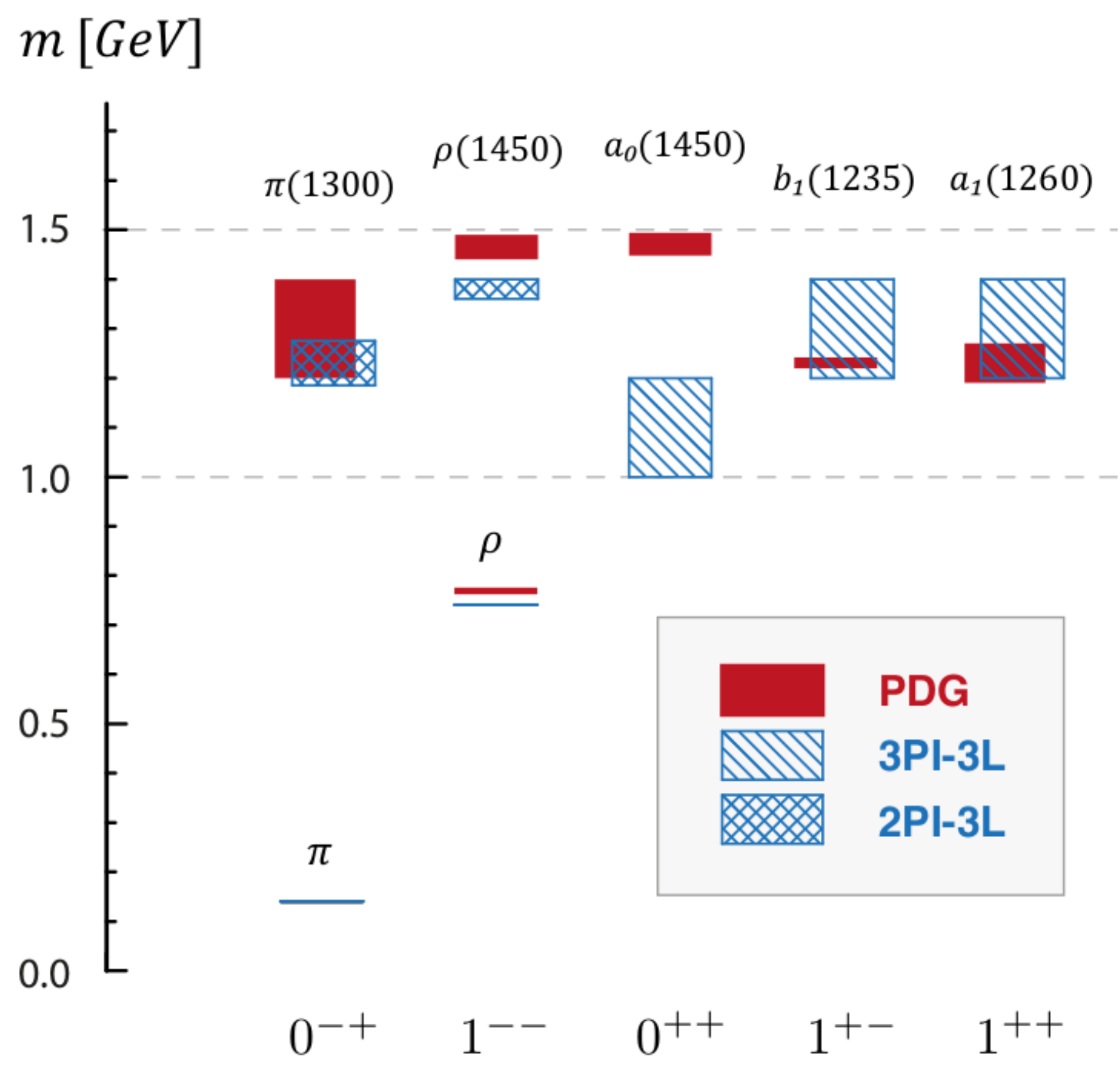}
            \caption{The isovector meson spectrum for light quarks and total angular momentum $J=0,1$ beyond rainbow-ladder,
            obtained with 2PI and (where available) 3PI truncations \cite{Williams:2015cvx}. Figure adapted from \cite{Eichmann:2016yit}.}
            \label{fig:mesons}
            \end{figure}

In Fig.~\ref{fig:mesons} we show numerical results for ground and excited state light mesons with total 
angular momentum $J=0$ and $J=1$ arising in such a BRL truncation \cite{Williams:2015cvx}. 
In comparison with the PDG results, most results for the ground
and even for the excited states below 1.5 GeV are in good agreement with experiment. For the pseudoscalar and vector channels,
results of similar quality can also be obtained in simpler rainbow-ladder schemes. However, these fail badly in the scalar 
and axial-vector channels due to missing tensor structures in the quark-gluon interaction. This deficiency is mostly remedied when 
effects beyond the rainbow are taken into account \cite{Chang:2010hb,Williams:2015cvx}. An interesting exception is the scalar meson
channel: typical rainbow-ladder results for the lowest mass scalar bound state are of the order of 650 MeV. In beyond rainbow
ladder calculations this mass is increased above 1 GeV as can be seen in Fig.~\ref{fig:mesons}. 
At the same time, DSE/BSE studies of scalar tetraquark states find a multiplet in qualitative agreement with the lowest
scalar meson states seen in experiment, i.e. the $f_0$(500) and its cousins \cite{Heupel:2012ua,Eichmann:2015cra}. This indeed
suggests the identification of the lowest scalar nonet with tetraquarks as discussed frequently in the literature, see
e.g. \cite{Jaffe:1976ig,Amsler:2004ps,Giacosa:2006tf,Ebert:2008id,Parganlija:2012fy,Pelaez:2015qba}
and references therein. The quark antiquark bound state seen in Fig.~\ref{fig:mesons} is then at roughly the right scale
to be identified with one of the scalar states seen experimentally in this region.
Finally, we wish to mention that very recently also the problem of the dynamical decay of mesons has been considered 
and solved for the case of the rho-meson \cite{Williams:2018adr}. 
 
\subsubsection{Baryons}\label{sec:baryons}

           \begin{figure}[t]
            \centering
            \begin{subfigure}[t]{0.98\textwidth}
            \includegraphics[width=0.98\textwidth]{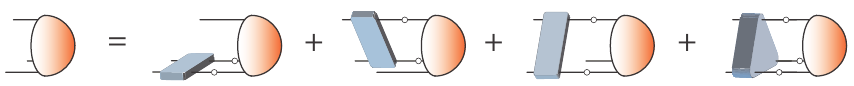}
            \caption{Three-quark Faddeev equation.}
            \label{fig:faddeev}
            \end{subfigure}\vspace*{4mm}
            \begin{subfigure}[t]{0.98\textwidth}
            \includegraphics[width=0.98\textwidth]{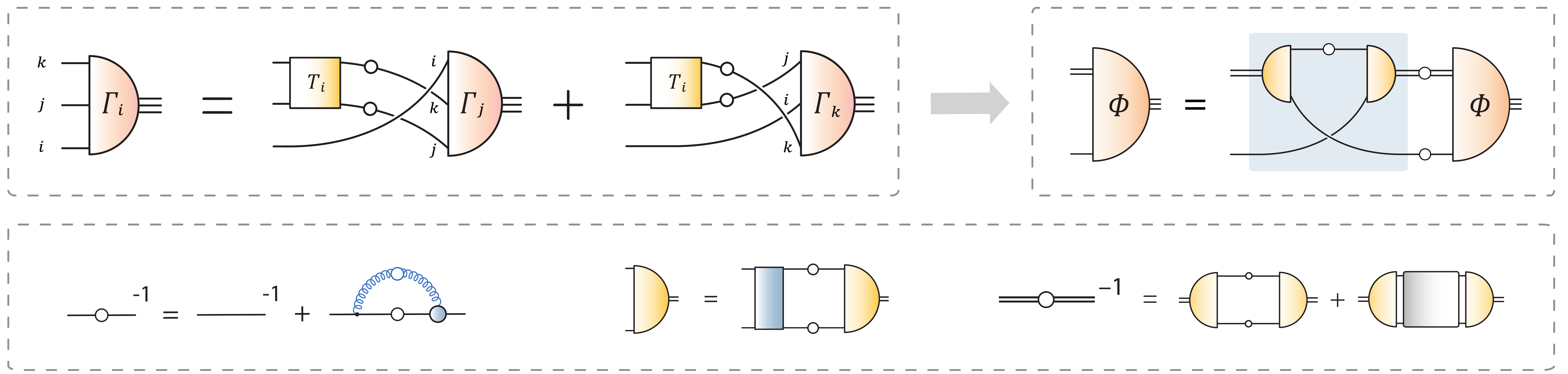}
            \caption{Simplification of the Faddeev equation in Fig.~\ref{fig:faddeev}
                     to the quark-diquark Bethe-Salpeter equation (\textit{upper panel}).
                     The lower panel shows the ingredients that enter in the equation and are calculated
                     beforehand: the quark propagator, diquark Bethe-Salpeter amplitudes and diquark propagators.}
            \label{fig:quark-diquark}
            \end{subfigure}
            \caption{Baryons treated in the DSE/BSE framework once as three-quark systems (a), once in the quark-diquark approximation (b).
            Figures adapted from \cite{Eichmann:2016yit}.}
            \end{figure}

The masses and wave functions of baryons can be extracted from their Faddeev equation, shown diagrammatically in 
Fig.~\ref{fig:faddeev} (the Faddeev amplitudes are illustrated by the shaded half-spheres). Its ingredients are 
the fully dressed quark propagator (solid line with open circle) as well as the quark two-body and irreducible 
three-body interactions. The wave functions (Faddeev amplitudes) of baryons are much more complicated than the ones
for mesons discussed above. Besides their colour and flavour parts, the fully relativistic Faddeev amplitudes contain 
a substantial number of Dirac tensor structures (64 for $J=\frac{1}{2}$ and 128 for $J=\frac{3}{2}$ baryons) with 
dressing functions dependent on one total and two relative momenta of the quarks inside the baryons.

\begin{figure*}[t]
        \centering
        \includegraphics[width=\textwidth]{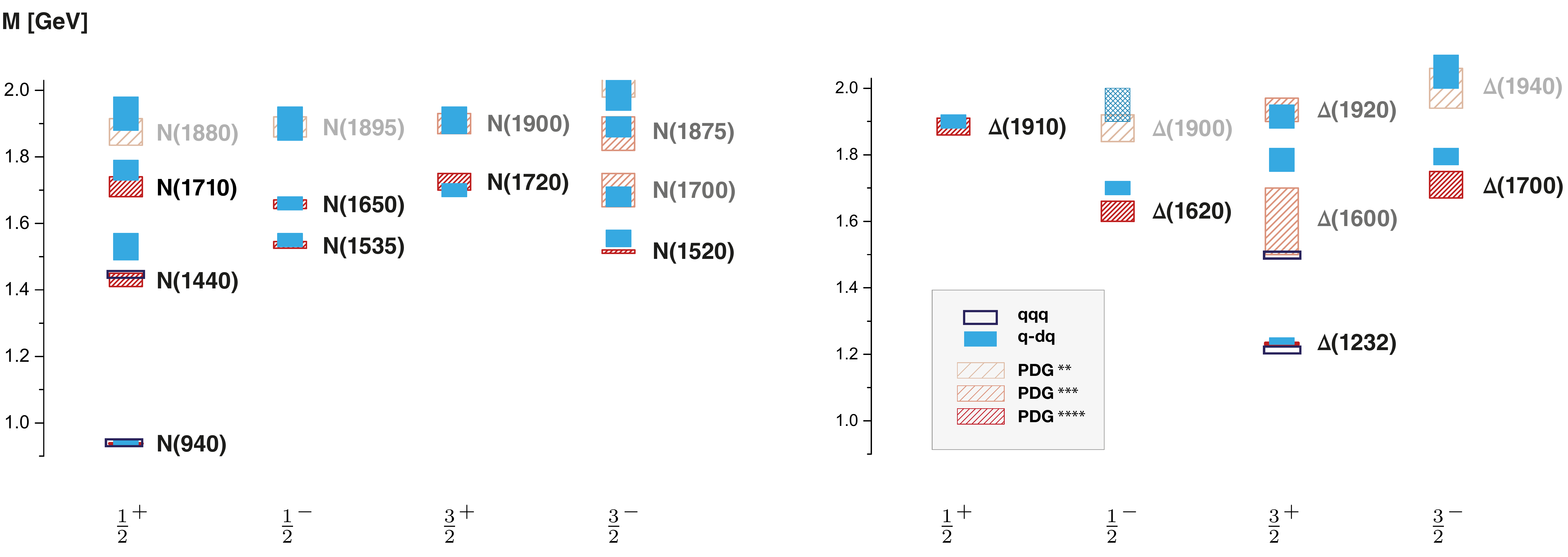}
        \caption{Nucleon and $\Delta$ baryon spectrum for $J^P=1/2^\pm$ and $3/2^\pm$ states determined within
                 a rainbow-ladder truncation \cite{Eichmann:2016hgl,Eichmann:2017inp}.
                 The three-body results (open boxes) are shown along with results from a quark-diquark approximation
                 with full diquark content (filled boxes) and with the PDG values \cite{Patrignani:2016xqp}. The 
                 width of the symbols for the PDG results represent their experimental uncertainties, the widths 
                 of the DSE results represent an estimate of (part of) the systematic error, see \cite{Eichmann:2016hgl} for details.
                 } \label{spectrum}
\end{figure*}

While the treatment of the three-body Faddeev equation has become well feasible in recent years, it is instructive to 
compare its results with those in a quark-diquark approximation, shown in the upper panel of Fig.~\ref{fig:quark-diquark}, 
which simplifies matters considerably. Its main ingredients are again the dressed quark propagator, together 
with the diquark propagator (double line with open circle) as well as the diquark Bethe-Salpeter amplitude. 
In turn, the latter needs to be calculated from the diquark Bethe-Salpeter equation (bottom centre diagram of 
Fig.~\ref{fig:quark-diquark} with similar structure than the BSE for mesons). It turns out that for $J^P=1/2^+$ 
octet and $J^P=3/2^+$ decuplet baryons it is sufficient to take scalar and axial-vector diquarks into account \cite{Oettel:1998bk},
whereas for the remaining states contributions from pseudoscalar and vector diquarks cannot be left out \cite{Eichmann:2016jqx}.
The three-body and quark-diquark approximation can only be systematically compared when the same underlying
quark-gluon interaction is chosen. This is not possible in quark-diquark models that utilize ansaetze for
the quark propagator and the diquark wave functions without contact to the underlying
QCD dynamics \cite{Oettel:1998bk,Oettel:2000jj,Segovia:2015hra}. 
Such a comparison has been discussed in \cite{Eichmann:2016hgl,Eichmann:2017inp} and we show the corresponding results 
from the three-body calculation (open boxes) \cite{Eichmann:2009qa,SanchisAlepuz:2011jn}, the quark-diquark
approximation (filled boxes) \cite{Eichmann:2016hgl} in Fig.~\ref{spectrum} compared them with the two-, three- and 
four-star states given by the PDG~\cite{Patrignani:2016xqp}.

Let us first concentrate on the nucleon channel. There is good agreement of the three-body with the quark-diquark
approach for the ground-state nucleon as well as the first radially excited state. The mass of the latter is in the ballpark 
of the Breit-Wigner mass of the Roper shown as a red shaded box.\footnote{The mass evolution of the Roper with varying pion mass
is discussed in~\cite{Eichmann:2016hgl} and compared with results from lattice QCD.} The next two excited states have been
determined only in the quark-diquark framework with masses close to the PDG's N(1710) and N(1880). In the $1/2^-$ channel 
there is again good agreement of the ground and excited states with experiment. In particular the N(1535) is where it should be
and the experimental level ordering as compared to the positive parity channel is correctly reproduced. The rest-frame Faddeev 
amplitudes of these states can be analysed in terms of angular momentum and interesting results have been found: In agreement with 
the quark model and a recent quark-diquark model calculation~\cite{Chen:2017pse}, the negative parity states are dominated 
by contributions with one unit of angular momentum (p waves), although other contributions (s-wave and d-wave) are also present. 
In contrast, in the positive-parity channel the ground-state nucleon is s-wave dominated, whereas the first excited state, 
the Roper, is dominated by p-wave components which are absent in the quark model. This points towards the extraordinary nature 
of this state which still needs to be explored further. A similar picture is observed in the various $\Delta$ channels. The 
quantitative agreement between the quark-diquark and the three-body approach is somewhat less pronounced than in the nucleon 
case but still satisfied on a semi-quantitative level. In general, not only the first radial excitations but also the second 
and third ones are close to experimentally identified states. The DSE/BSE framework therefore delivers a consistent and 
quantitative description of the light baryon spectrum below 2 GeV in terms of quark degrees of freedom.

It has been emphasised, however, that this does not mean that pion-cloud or coupled-channel effects are
absent~\cite{Eichmann:2016hgl,Chen:2017pse}. On the contrary, these are expected to contribute substantially, 
but compete with opposite sign beyond rainbow-ladder effects from non-Abelian corrections in the quark-gluon interaction.
In the meson sector, this cancellation is indicated by the results of \cite{Fischer:2008wy,Fischer:2009jm}. In the baryon
sector these effects have already been explored for baryons on an exploratory level~\cite{Sanchis-Alepuz:2014wea,Sanchis-Alepuz:2015qra}
and work will continue in this direction. This cancellation mechanism may very well be dependent on the channel in question as
well as on the internal structure of the states in question, such that sizeable net effects may remain for some states
such as the Roper. This will need to be explored in the future.

\newpage

\section{Results under variation of the number and masses of quarks, temperature and chemical potential}\label{results}

In the preceding two sections we laid the foundations, both conceptually and technically, for the main part of this 
review, the discussion of recent results for QCD at finite temperature and chemical potential. 
In the following subsections we will walk through the Columbia plot. We start in the upper right corner, i.e.
the pure gauge theory and the heavy quark physics associated with the first order region and second order
surface of the deconfinement transition in section \ref{results:heavy}, continue to the chiral two flavour
theory, i.e. the upper left corner of the Columbia plot in section \ref{results:chiral} and end up at the 
physical point and the quest for the critical end point at finite chemical potential in section \ref{results:CEP}.
Along the way we discuss properties of mesons (\ref{results:chiral}) and baryons (\ref{results:baryons}) at finite 
temperature as studied in the DSE/BSE approach. In section \ref{results:thermo} we deal with issues of thermodynamics,
transport properties and quark number susceptibilities and digress to the issue of quark spectral functions 
in section \ref{results:spectral}. We finish this chapter with a discussion of results for the colour superconducting
phases in section \ref{results:colorSC}.

\subsection{Heavy quark limit: (De-)confinement and the Roberge-Weiss transition}\label{results:heavy}

The physics of the pure gauge/heavy quark region of the Columbia plot is dominated by the gauge theory, the
associated deconfinement phase transition and the Roberge-Weiss point at imaginary chemical potential. In order 
to address these issues within the DSE/BSE framework it is mandatory to take the Yang-Mills sector, i.e. at least
the DSE for the gluon propagator explicitly into account. This has been done in 
Refs.~\cite{Fischer:2009wc,Fischer:2009gk,Fischer:2014vxa} using the truncation scheme discussed in section \ref{fulltrunc}.
To this end input from corresponding lattice simulations has been used: numerical results for the 
transverse (magnetic) and longitudinal (electric) part of the gluon propagator have been fitted, interpolated and
incorporated into the gluon DSE as described in section \ref{fulltrunc}. We therefore proceed by first discussing
the properties of the pure gauge gluon propagator at finite temperature, before we come back to the case of heavy
(but not static) quarks.  

\subsubsection{Pure gauge gluon propagator at finite temperature}\label{results:gluon}

The gluon propagator at finite temperature is an interesting object to study, see \cite{Maas:2011se} for a comprehensive 
review. On the lattice SU(2) simulations
\cite{Cucchieri:2007ta,Bornyakov:2010nc,Cucchieri:2011di,Maas:2011ez} in two and three spatial dimensions\footnote{See 
also \cite{Cucchieri:2001tw,Cucchieri:2000cy} for early works.} have been complemented by SU(3) simulations in 
Refs.~\cite{Fischer:2010fx,Maas:2011ez,Aouane:2011fv,Bornyakov:2011jm,Silva:2013maa}. 
The interest in comparing these different cases 
comes from the different orders of the transitions: whereas the SU(2) gauge theory features a second order transition
in the universality class of the Ising model, the SU(3) theory is characterised by a (weak) first order phase transition.
In section \ref{sec:deconf} we discussed the extraction of order parameters such as the Polyakov loop potential
from the correlation functions of the theory, in particular the (gauge fixed) propagators. Therefore one may expect 
that the critical behaviour of the theory is also encoded at least in some of the two-point functions and can be extracted.
Indeed from an analytic analysis of the correlators in a background gauge formulation it has been argued in Ref.~\cite{Maas:2011ez} 
that the electric part of the gluon propagator should be sensitive to critical physics, whereas the magnetic part as 
well as the ghost propagator should not. A suitable candidate for critical behaviour is the electric screening mass
\beq
m_L = \frac{1}{\sqrt{D_L(0)}}\,,
\eeq
defined from the zero energy and momentum behaviour of the electric part of the propagator, cf. Eq.(\ref{eq:qProp}), 
and its associated susceptibility
\beq
\chi_T = \frac{\partial m_L}{\partial T}\,.
\eeq
As discussed in detail in \cite{Maas:2011ez} it turns out to be extremely involved to separate the temperature dependent
screening mass and its expected critical behaviour from the (temperature independent) 'gluon mass' already present at zero
momentum as discussed in section \ref{sec:gluon}. In the vicinity of a second order phase transition the latter adds a constant
offset to a part scaling with temperature resulting in a total of
\beq
m_L(t) = m_{|_{T=0}} + a_{\pm} |t|^{\gamma/2}\,.
\eeq\label{scaling}
Here $t=T/T_c-1$ is the reduced temperature, the critical anomalous dimension characteristic for the universality 
class is denoted by $\gamma$ and there are non-universal coefficients $a_-$ for $t<0$ and $a_+$ for $t>0$. 

\begin{figure}[t]
\begin{subfigure}[t]{0.9\textwidth}
        \begin{center}
        \includegraphics[width=0.48\textwidth]{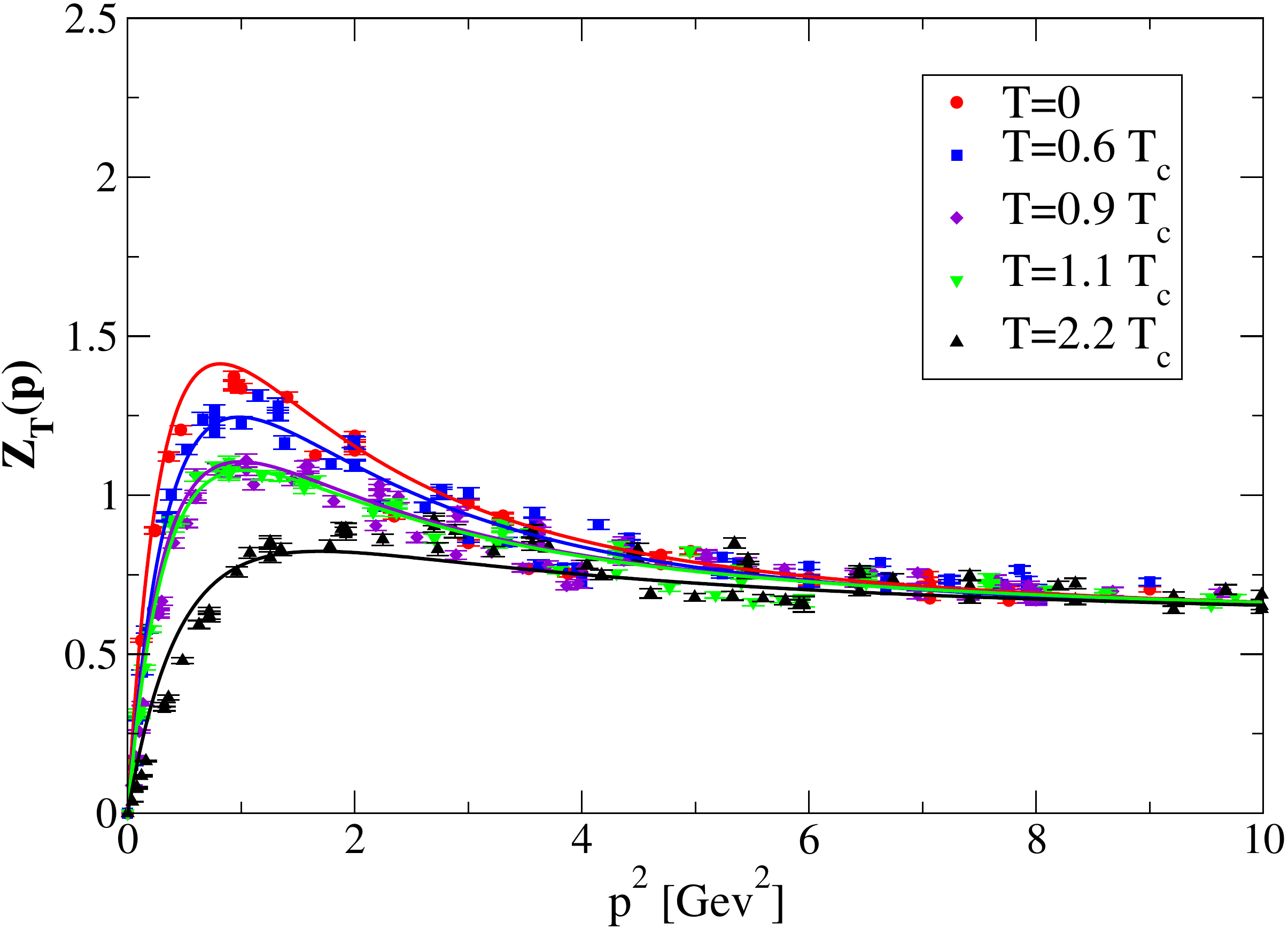}\hfill
        \includegraphics[width=0.48\textwidth]{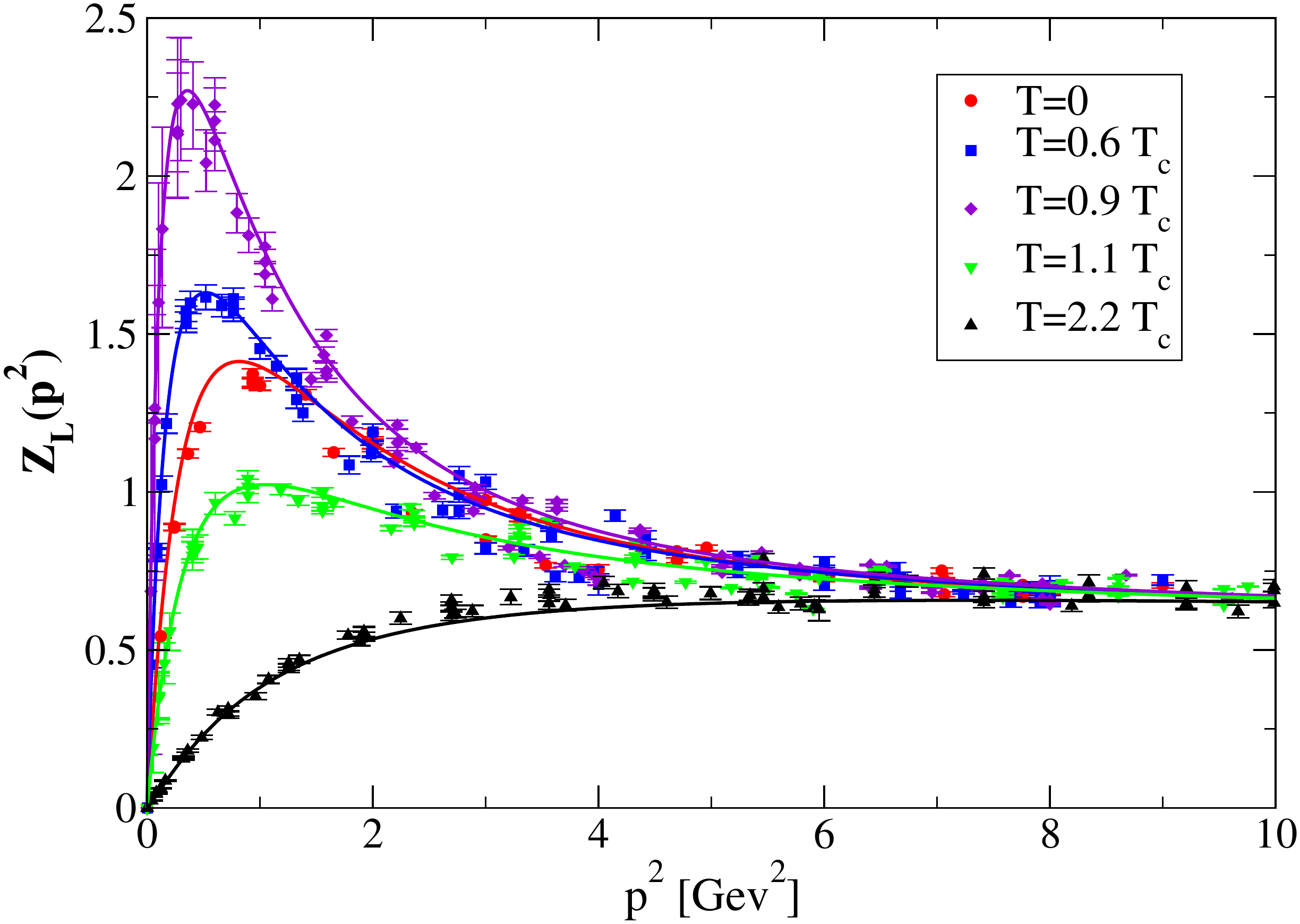}
        \caption{\textit{Left:} Transversal (magnetic) gluon dressing function $Z_T(\vec{p}^2,0)$ 
        [pure SU(3) gauge theory, zeroth Matsubara sum]
        for several temperatures compared to fits. Data taken from \cite{Fischer:2010fx,Maas:2011ez}.
 \textit{Right:} Corresponding longitudinal (electric) gluon dressing $Z_L(\vec{p}^2,0)$.}\label{fig:gluonT}
        \end{center}
\end{subfigure}\vspace*{5mm}
\begin{subfigure}[t]{0.9\textwidth}
        \begin{center}
        \includegraphics[width=0.48\textwidth]{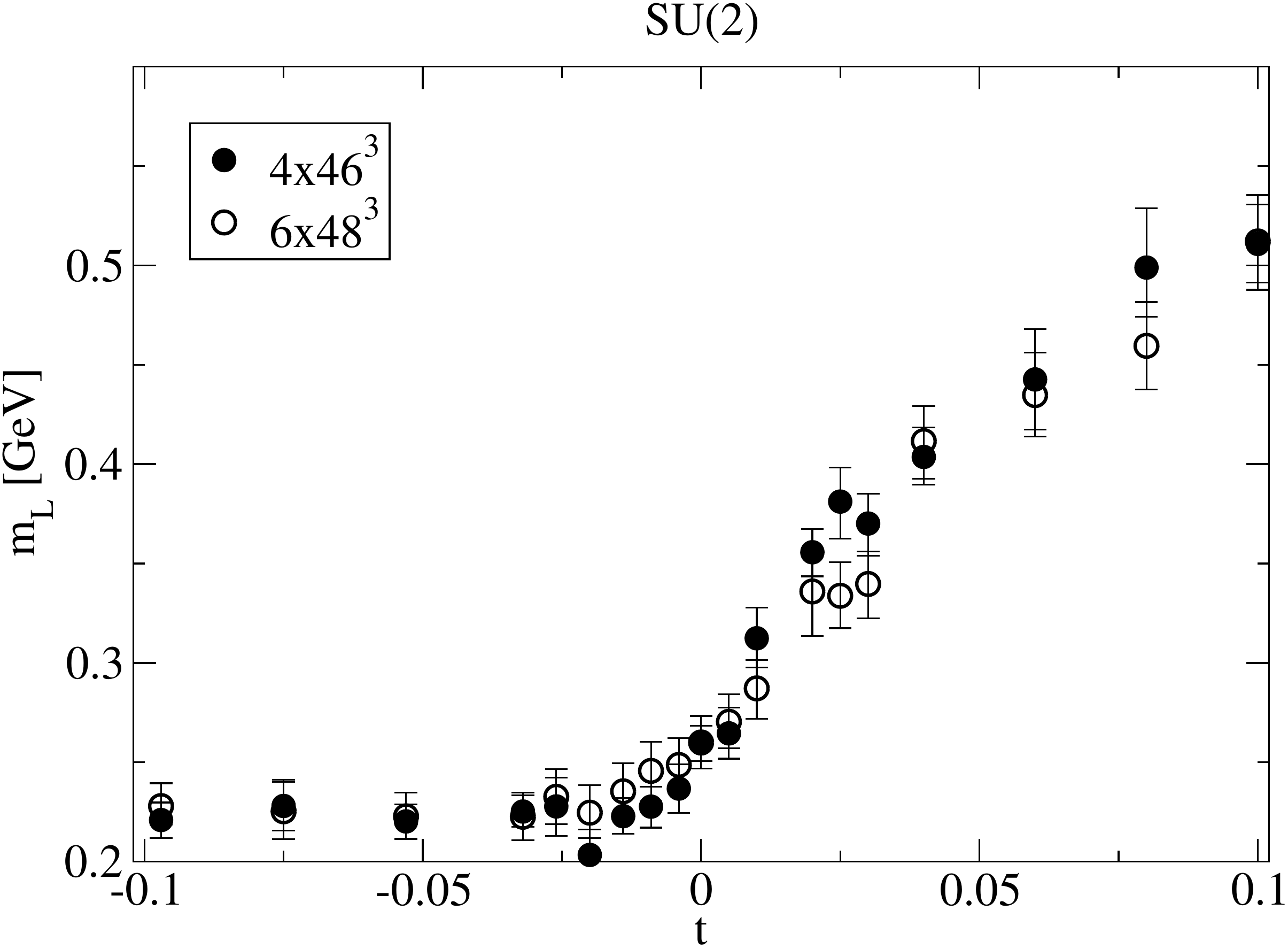}\hfill
        \includegraphics[width=0.48\textwidth]{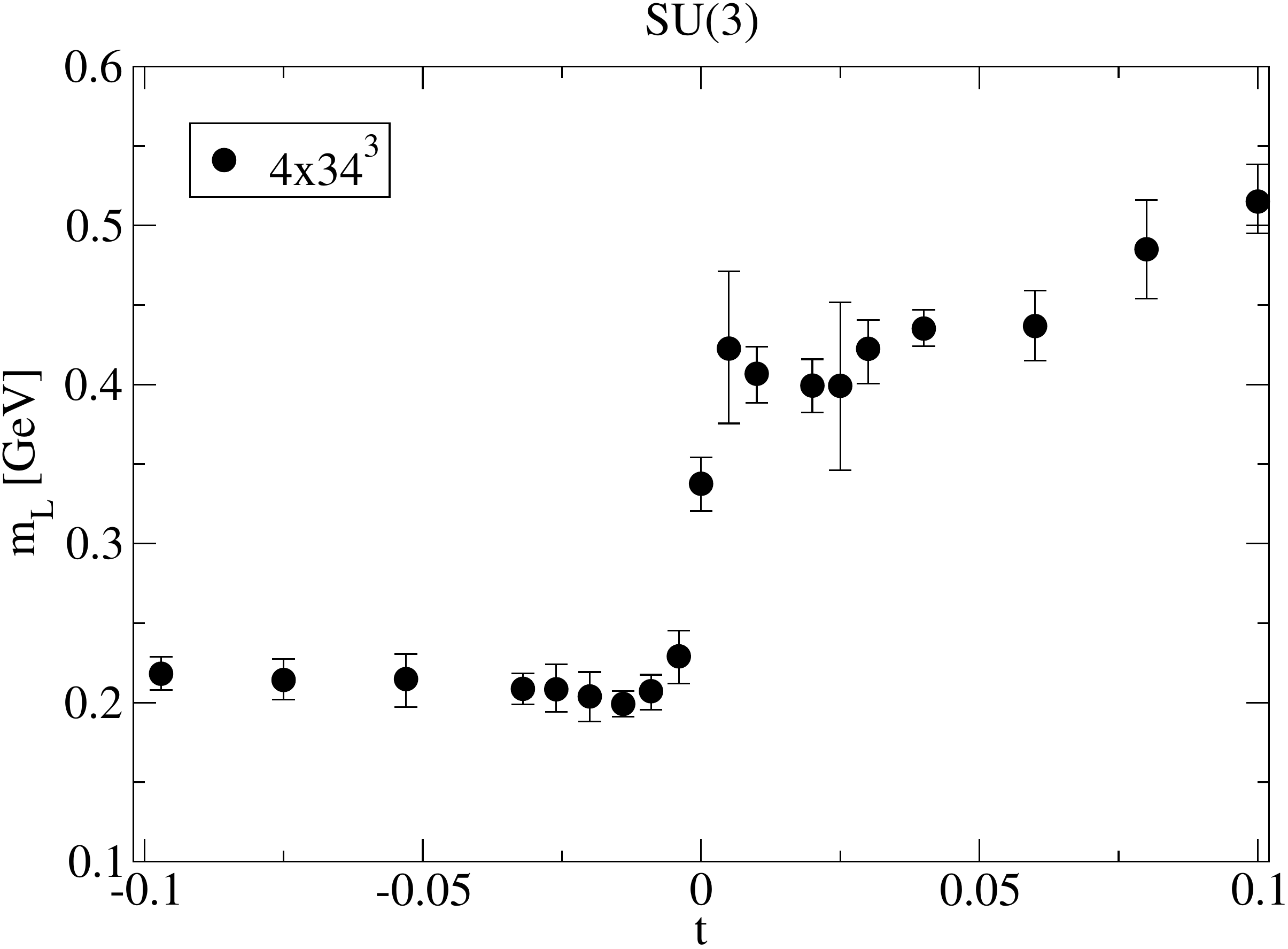}
        \caption{\textit{Left:}  Screening mass extracted from the electric part of the 
        SU(2) gluon propagator \cite{Maas:2011ez}.
 \textit{Right:} Same quantity extracted from the SU(3) gluon propagator. Figures adapted from \cite{Maas:2011ez}.}\label{fig:screeningmass}
        \end{center}
\end{subfigure}
\caption{Dressing functions and screening masses of the gluon propagator at finite temperature.}
\end{figure}

Lattice data for the magnetic and electric gluon dressing functions in pure SU(3) gauge theory are shown in Fig.~\ref{fig:gluonT} 
together with the above mentioned fits. Whereas the running of both dressing functions at large momenta is mainly unaffected
by temperature effects (as expected), for low momenta at and around the bump at roughly 1 GeV temperature effects are drastic
and different for both dressing functions. The magnetic part of the propagator continuously 'melts', i.e. the low momentum
bump discussed in section \ref{sec:gluon} decreases continuously with temperature. In contrast, the electric part of the
propagator shows a markedly different behaviour: the bump first increases by a sizeable amount and then decreases dramatically
around the critical temperature. This decrease then continues into the high temperature phase. Whereas there has been some 
debate on the influence of technical lattice parameters associated with volume and cut-off effects on the size of change
in the electric part of the propagator \cite{Cucchieri:2011di,Maas:2011ez}, there is general agreement that the 
difference between the magnetic and electric parts is genuine and contains important physics. 

The general temperature behaviour of the $SU(3)$ lattice gluon propagators for temperatures below and above the 
first order transition have been reproduced in the FRG framework \cite{Fister:2011uw,Cyrol:2017qkl}. However,
although an impressive amount of technical efforts have been involved, the truncations were still
not rich enough to capture the drastic changes around the critical temperature. Potential reasons and solutions
have been discussed in \cite{Cyrol:2017qkl} and await their implementation in the future. In the Hamilton approach,
the ghost and gluon propagator at two different temperatures above and below the critical one have been studied in 
Ref.~\cite{Quandt:2015aaa}; corresponding thermodynamic quantities are presented in \cite{Quandt:2017poi}. In the
Gribov-Zwanziger approach, thermodynamic quantities of pure Yang-Mills theory have been extracted in \cite{Fukushima:2013xsa}.

The electric screening masses for the SU(2) and SU(3) gauge theory associated with the lattice results 
are shown in Fig.~\ref{fig:screeningmass} as
a function of the reduced temperature $t$. The corresponding critical temperature $T_c$ in both cases is extracted from the 
string tension. This critical temperature coincides with the one where the screening masses display an interesting change 
of behaviour. Below the critical temperature in both cases the masses are constant within error 
bars. Above the critical temperature both masses rise with temperature. For the SU(3) case a fit to a $\sqrt{t}$-behaviour 
has been given in \cite{Maas:2011ez} (a linear fit is also possible within error bars). The marked difference between
the SU(2) and SU(3) case occurs in the close vicinity of the critical temperature. Whereas the SU(3) data clearly signal
a pronounced jump associated with the discontinuity expected for a first order phase transition, there is a smooth
transition in the SU(2) data that can be explained by the scaling law Eq.~(\ref{scaling}) of a second order transition. 
These findings substantiated earlier indications discussed in \cite{Cucchieri:2007ta,Fischer:2010fx} that the electric
screening mass of the gluon indeed serves as an order parameter for the deconfinement phase transition. Unfortunately,
with the volumes and lattice spacings available at the time it was not possible to extract a clean value for 
the anomalous dimension $\gamma$ \cite{Maas:2011ez}. This task is left for future studies. Further interesting
results for the electric and magnetic gluon susceptibilities of the SU(2) theory have been found in a background 
field approach inspired by the Curci-Ferrari model in Refs.~\cite{Reinosa:2014zta,Reinosa:2016iml}. 

\begin{figure}[t]
        \begin{center}
        \includegraphics[width=0.54\textwidth]{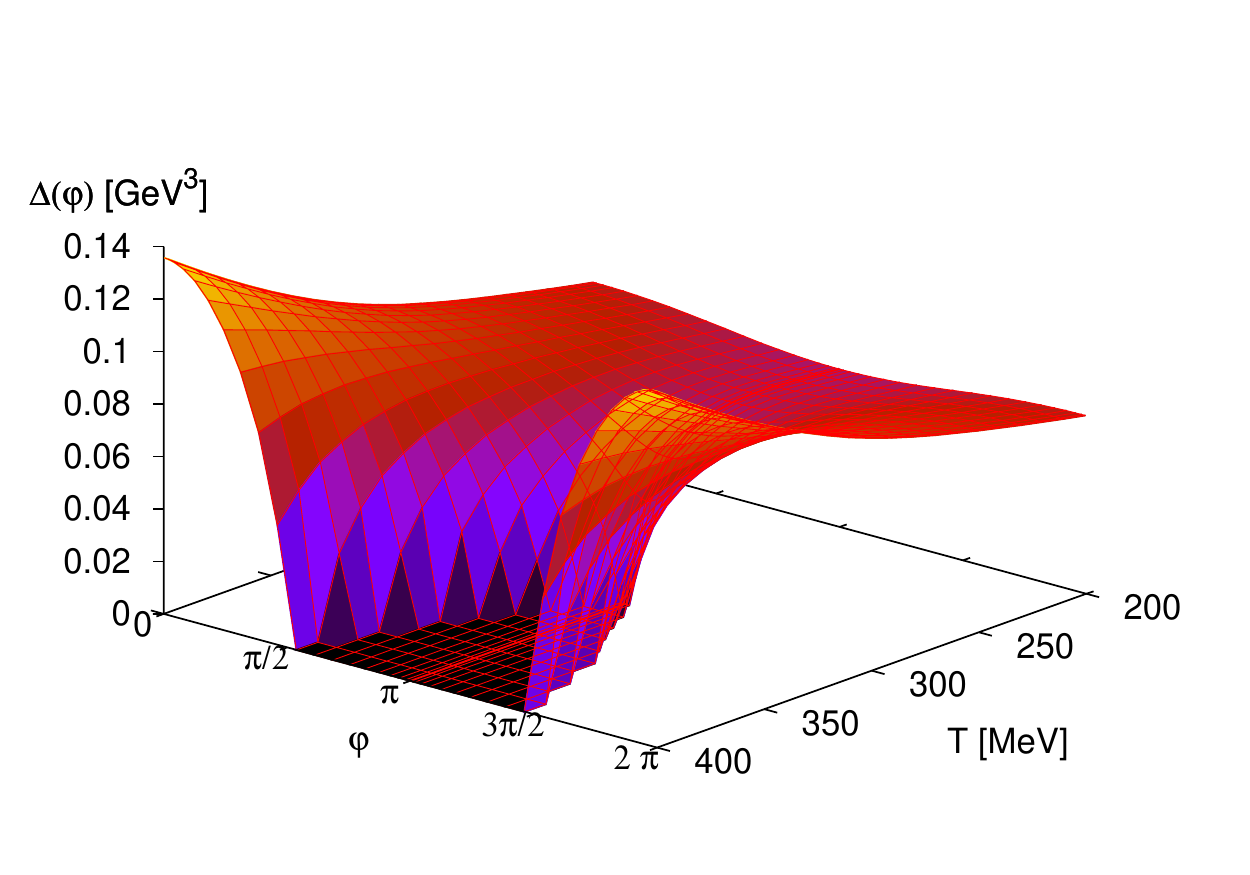}\hfill
        \includegraphics[width=0.46\textwidth]{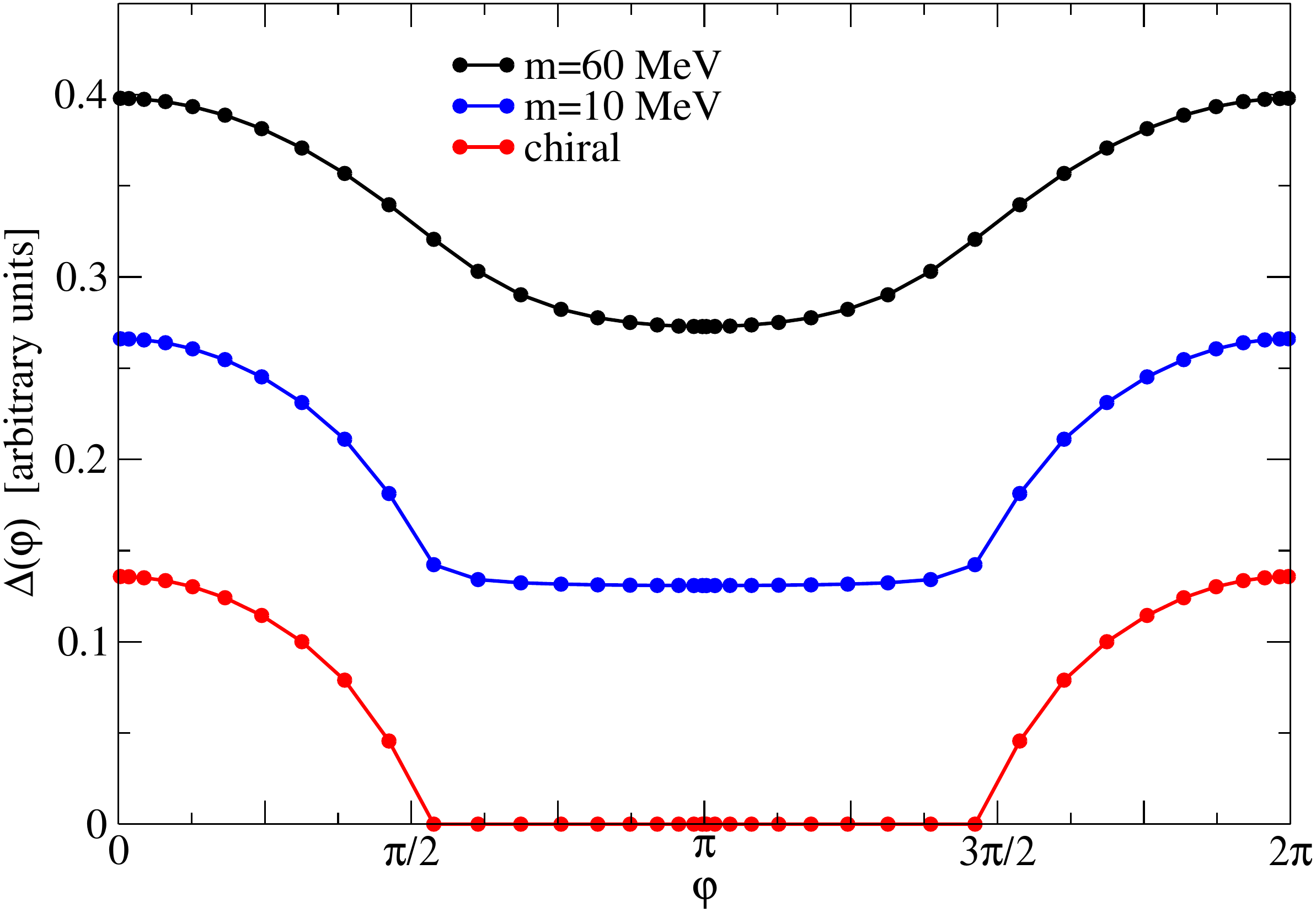}
        \caption{\textit{Left:} Dependence of the chiral condensate
                $\Delta(\varphi) \equiv \langle \bar{\Psi} \Psi \rangle_\varphi$ on boundary angle and temperature.
                \textit{Right:} Dependence of the quark condensate 
                 $\Delta(\varphi)$ on the boundary angle $\varphi$ for 
                 three different values of the test quark mass at $T=400$ MeV.
                Both figures are adapted from \cite{Fischer:2009gk}.} 
                \label{fig:dual}
        \end{center}
\end{figure}

Since signatures of the deconfinement transition are encoded in the gluon propagator it is interesting to study the 
deconfinement order parameters discussed in section \ref{sec:deconf} that either can be determined from the gluon 
and the ghost propagators of the pure Yang-Mills theory (Polyakov loop potential) or from a test quark in the quenched
theory (dressed Polyakov loop, ...). Indeed,
within the FRG-approach the Polyakov-loop potential has been determined in Ref.~\cite{Braun:2007bx,Marhauser:2008fz}
and delivered a clean distinction between the SU(2) second order and the SU(3) first order transition. The same has
been found for the dressed Polyakov loop determined in the DSE-approach in Ref.~\cite{Fischer:2009gk,Fischer:2010fx}
using the truncation described in section \ref{fulltrunc}. Since we will have a closer look at the Polyakov loop 
potential in the next subsection, here we discuss briefly the dressed Polyakov loop $\Sigma_1$. 

To this end, consider again Eq.~(\ref{dual}). In the following we discuss results for the first order 
deconfinement transition in the SU(3) gauge theory at $T\approx 270$ MeV; similar results are found for the case of SU(2). 
Below the critical temperature the condensate does not depend on the variation of the boundary condition with $\varphi$. 
Within numerical error this can be seen in the left diagram of Fig.~\ref{fig:dual}, where we plot the
dependence of the chiral condensate on $\varphi$ and $T$. As a consequence the Fourier transform in 
Eq.~(\ref{dual}) integrates to zero and the order parameter $\Sigma_1=0$ in this region. Above the critical 
temperature this behaviour changes drastically and the dressed Polyakov loop develops non-zero values. 
The precise form of this variation depends on the mass of the test quark that is exposed to the Yang-Mills 
gluon propagator. For large test quark masses,
the condensate develops a smooth variation with respect to $\varphi$, whereas for small test quark masses this
variation becomes stronger until it develops a zero plateau with a derivative discontinuity at two finite values 
of $\varphi$. This mass dependence can be readily understood from the loop-expansion, Eq.(\ref{loop}): each chain
of gauge links winding around the compactified time direction is weighted by the quark mass to the power of the 
number of links involved in the loop. Thus at sufficiently large quark masses longer loops are heavily suppressed.
As a result mostly loops winding only once around the torus contribute and the resulting angular behaviour of
the condensate is approximately proportional to the smooth $\cos(\varphi)$. Indeed, this is seen for the large 
quark mass in the left diagram of Fig.~\ref{fig:dual}. For small quark masses more and more loops winding $n$ times 
around the torus contribute and the additional $\cos(n\varphi)$-terms generate the flat behaviour in the region 
around $\varphi=\pi$. In the chiral limit, the expansion Eq.(\ref{loop}) finally breaks down and becomes meaningless. 
Note, however, that Eq.~(\ref{dual}) is still valid and the dressed Polyakov loop can be used as order parameter 
even for test quarks in the chiral limit, see \cite{Fischer:2009gk,Braun:2009gm,Fischer:2010fx} for 
details. In the Hamilton approach, the dressed Polyakov loop has been determined in Ref.~\cite{Reinhardt:2016pfe}.

\subsubsection{Phase structure of QCD for heavy quarks}\label{res:heavy}

After the successful description of the first order phase transition in the pure gauge theory discussed in the 
last section, the authors of \cite{Fischer:2014vxa} studied the upper right corner of the Columbia plot, i.e. 
the region where the first order deconfinement transition turns into a crossover separated by a second order 
critical line at critical quark masses $m_c$. This critical line becomes a critical surface $m_c(m_{u/d},m_s,\mu)$ 
when extended to real and imaginary chemical potential as discussed in section \ref{Columbia_ext}. 
To study this critical surface the authors of \cite{Fischer:2014vxa} employed the truncation described 
in section \ref{fulltrunc} with $N_f=2+1$ heavy but dynamical quarks back-coupled to the Yang-Mills sector. 
The order parameter studied has been the Polyakov loop potential evaluated using the DSE of Fig.~\ref{fig:DSE-A} 
in the presence of a constant background field ${A}_4$. Such a constant field can always be rotated
in the Cartan sub-algebra of the SU(3) colour group and decomposed into  
\begin{equation}
{A}_4 = \frac{2\pi T}{g}\left(\varphi_3 \frac{\lambda_3}{2} + \varphi_8 \frac{\lambda_8}{2}\right),
\label{eq:bfDecomposition}
\end{equation}
with Gell-Mann matrices $\lambda_a$. The introduction of a constant background field allows one to evaluate the 
order parameter $L[\langle A_4 \rangle] = L[\bar A_4]$ discussed around Eq.~(\ref{PL2}) and one obtains  
\begin{equation}
L[\bar A_4]
  = \frac{1}{3}\left[e^{-i\frac{2\pi\varphi_8}{\sqrt{3}}}
	+ 2e^{-i\frac{\pi\varphi_8}{\sqrt{3}}}\cos(\pi\varphi_3)\right]\,.
\end{equation}
In general this is a complex function which becomes real for $\varphi_8=0$. The minima of the Polyakov-loop potential
determine the values of $\varphi_3$ and $\varphi_8$.

\begin{figure}[t]
\begin{subfigure}[t]{0.9\textwidth}
        \begin{center}
        \includegraphics[width=0.48\textwidth]{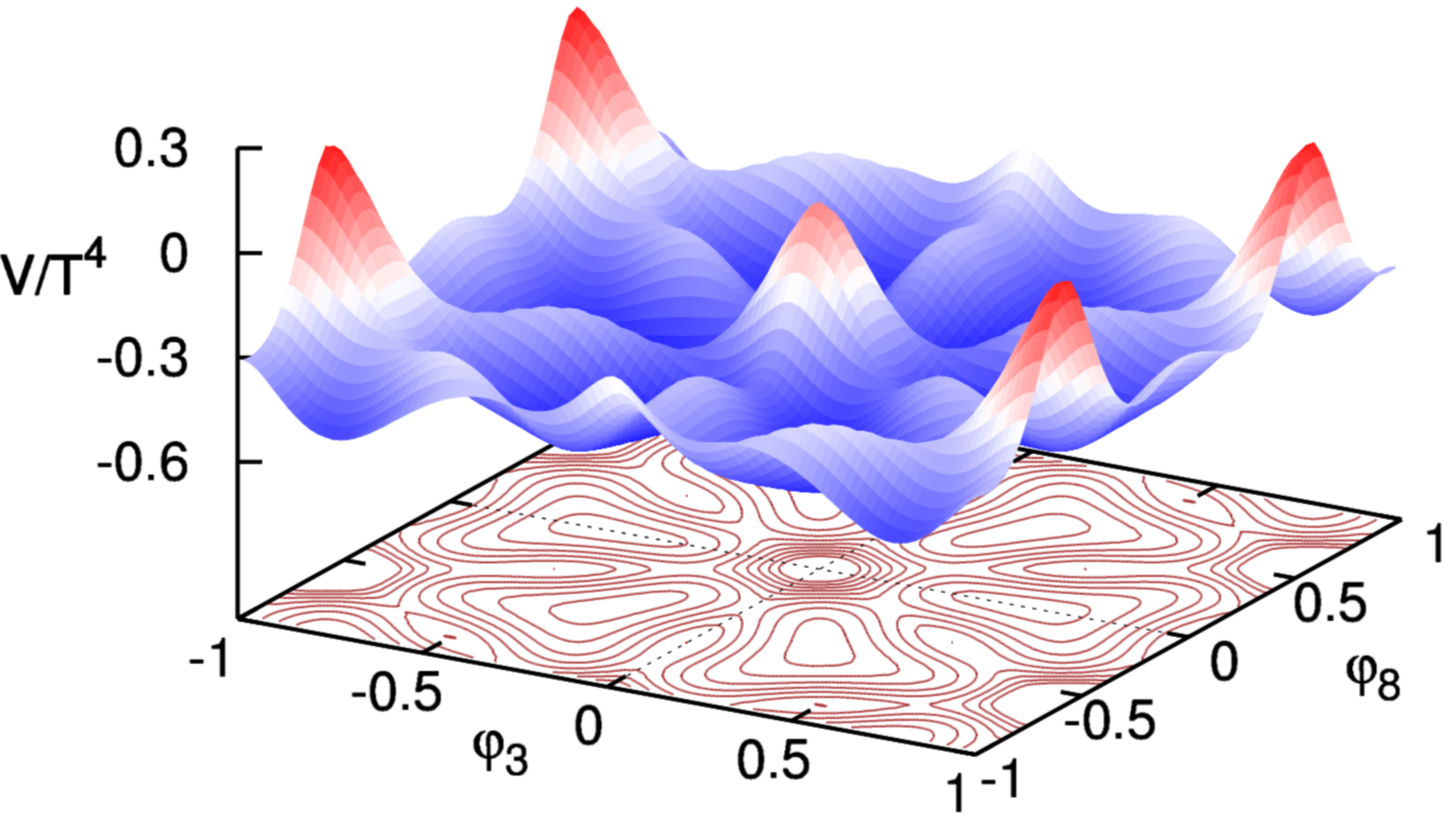}\hfill
        \includegraphics[width=0.48\textwidth]{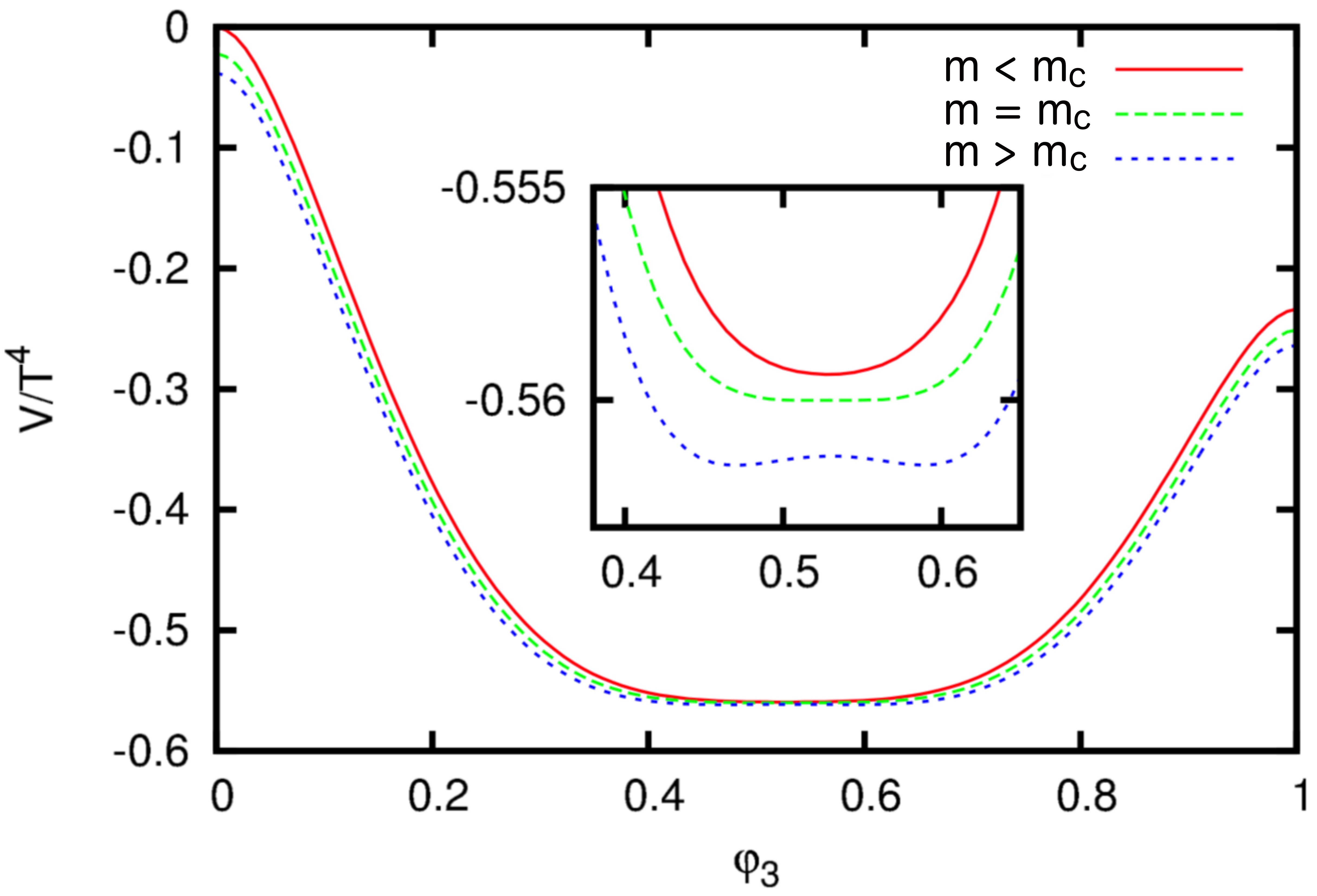}
        \caption{\textit{Left:} Polyakov loop potential at zero chemical potential in the approximately 
        center symmetric phase.
        \textit{Right:} Polyakov loop potential at $T_c$ ($\mu=0$) for quark masses above, equal and below 
        the critical mass of the $N_f=1$ theory (shifted up and down by arbitrary values for better visibility).}\label{fig:heavy_a}
        \end{center}
\end{subfigure}\vspace*{5mm}
\begin{subfigure}[t]{0.9\textwidth}
        \begin{center}
        \includegraphics[width=0.48\textwidth]{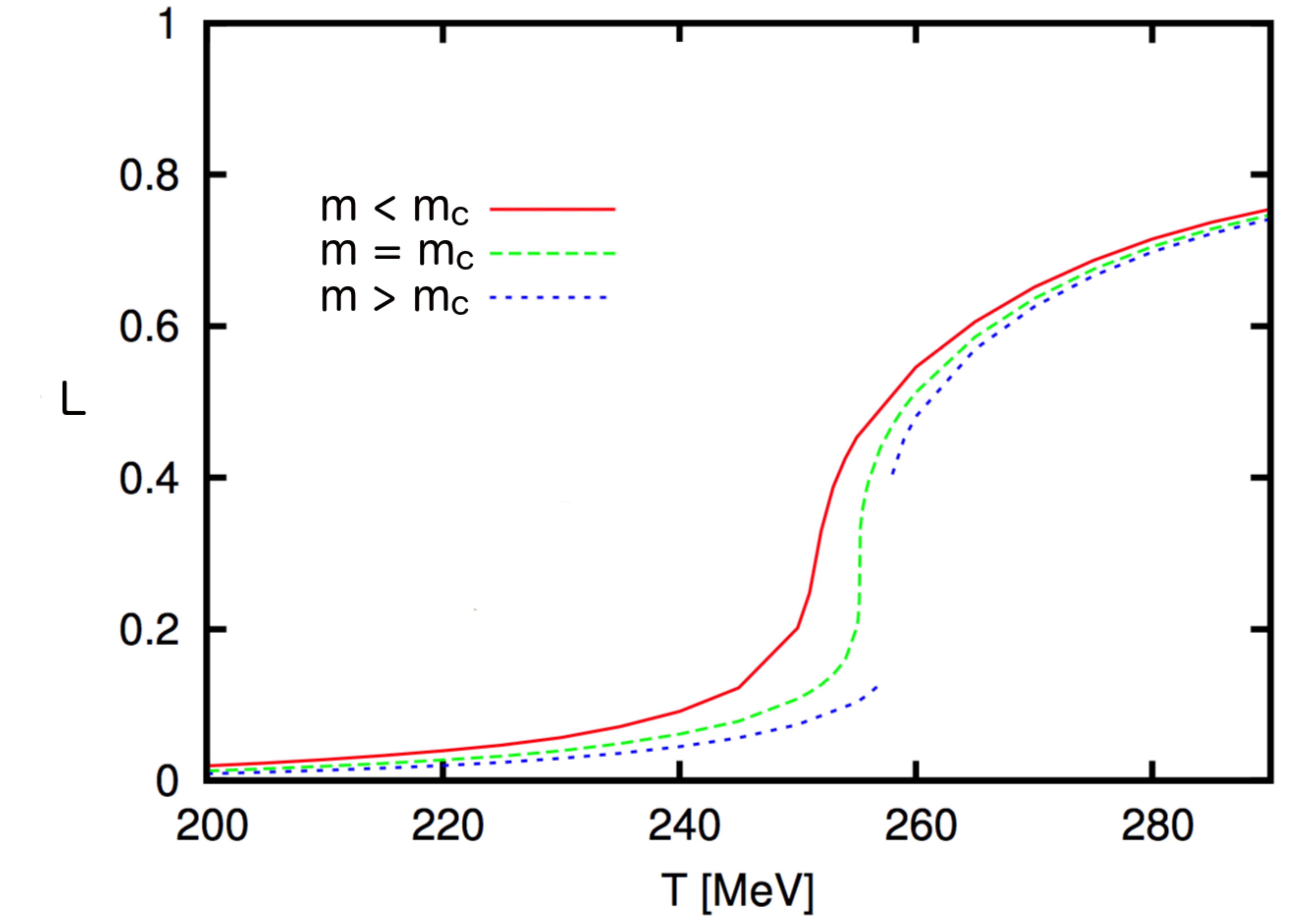}\hfill
        \includegraphics[width=0.48\textwidth]{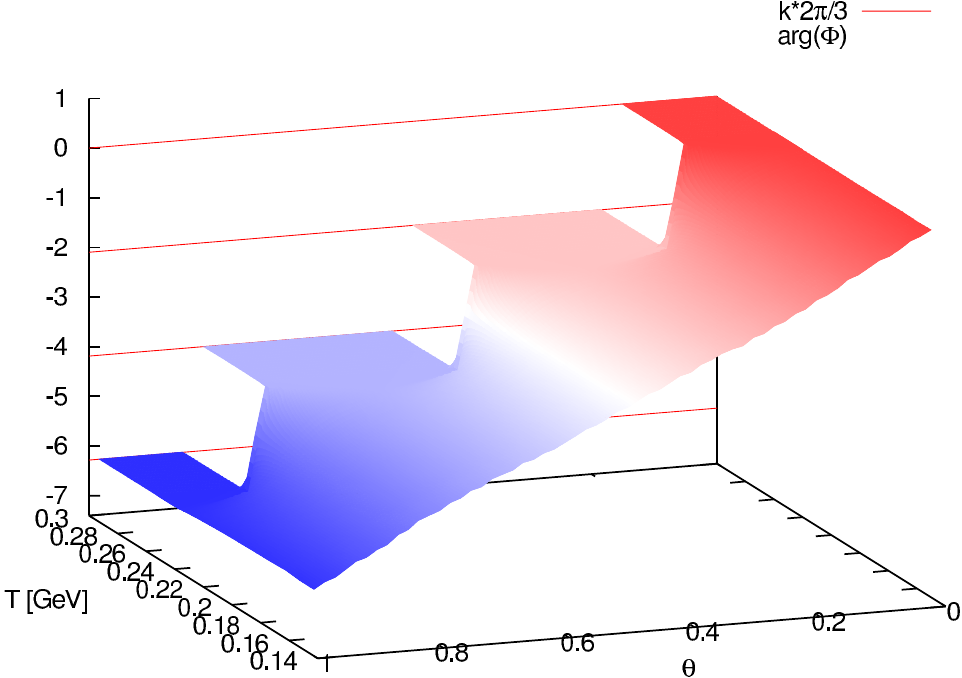}
        \caption{\textit{Left:} Polyakov loop as a function of temperature ($\mu=0$) for masses above, equal 
        and below the critical mass of the $N_f=1$ theory.
        \textit{Right:} Argument of the Polyakov loop as a function of the imaginary part $\theta=\mu_I/(2\pi T)$ of the
        chemical potential.}\label{fig:heavy_b}
        \end{center}
\end{subfigure}
         \caption{Results from DSEs for the deconfinement transition at heavy quark masses. 
         All figures adapted from \cite{Fischer:2014vxa}.}
\end{figure}
\begin{figure}[t]
        \begin{center}
        \includegraphics[width=0.48\textwidth]{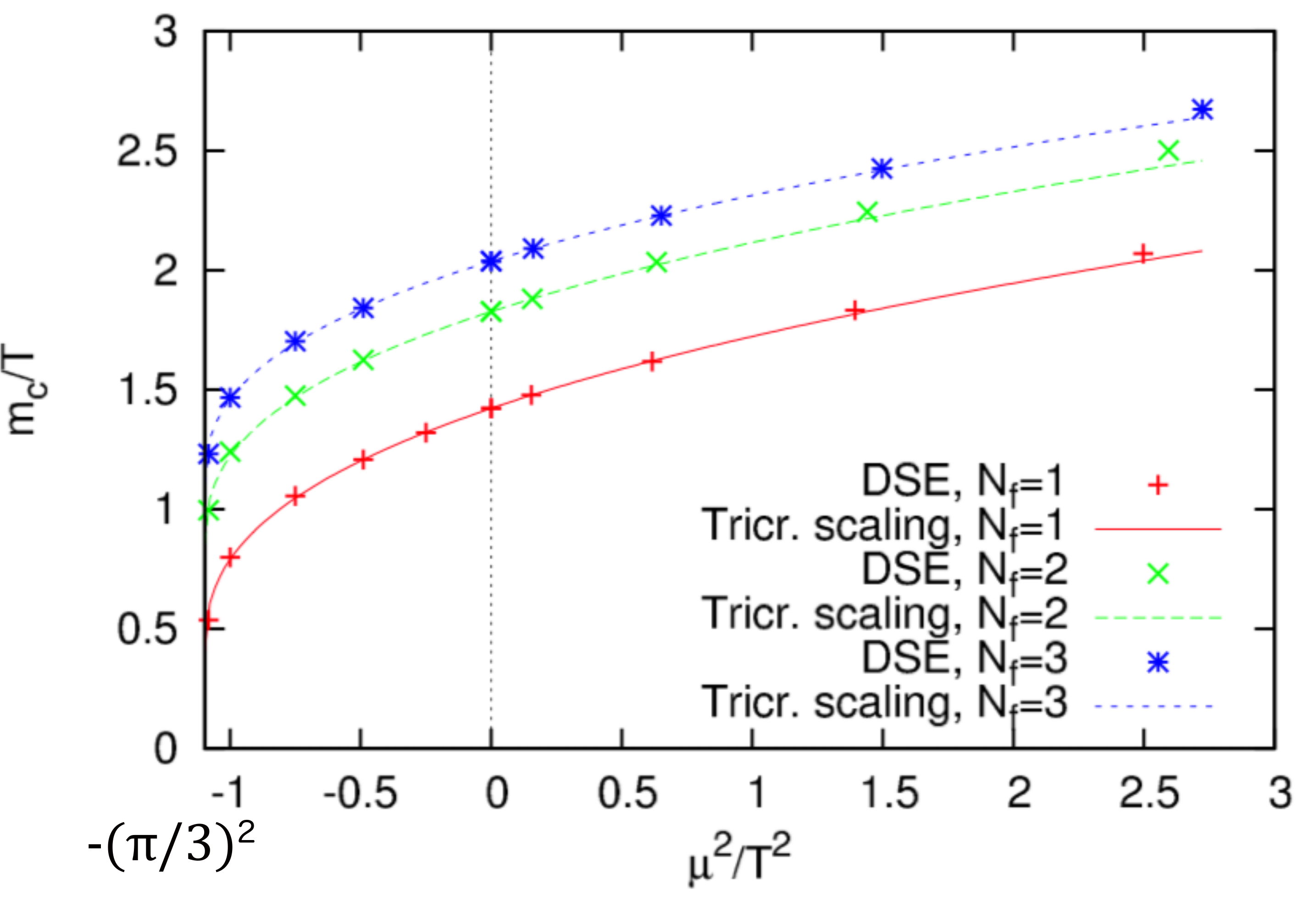}\hfill
        \includegraphics[width=0.38\textwidth]{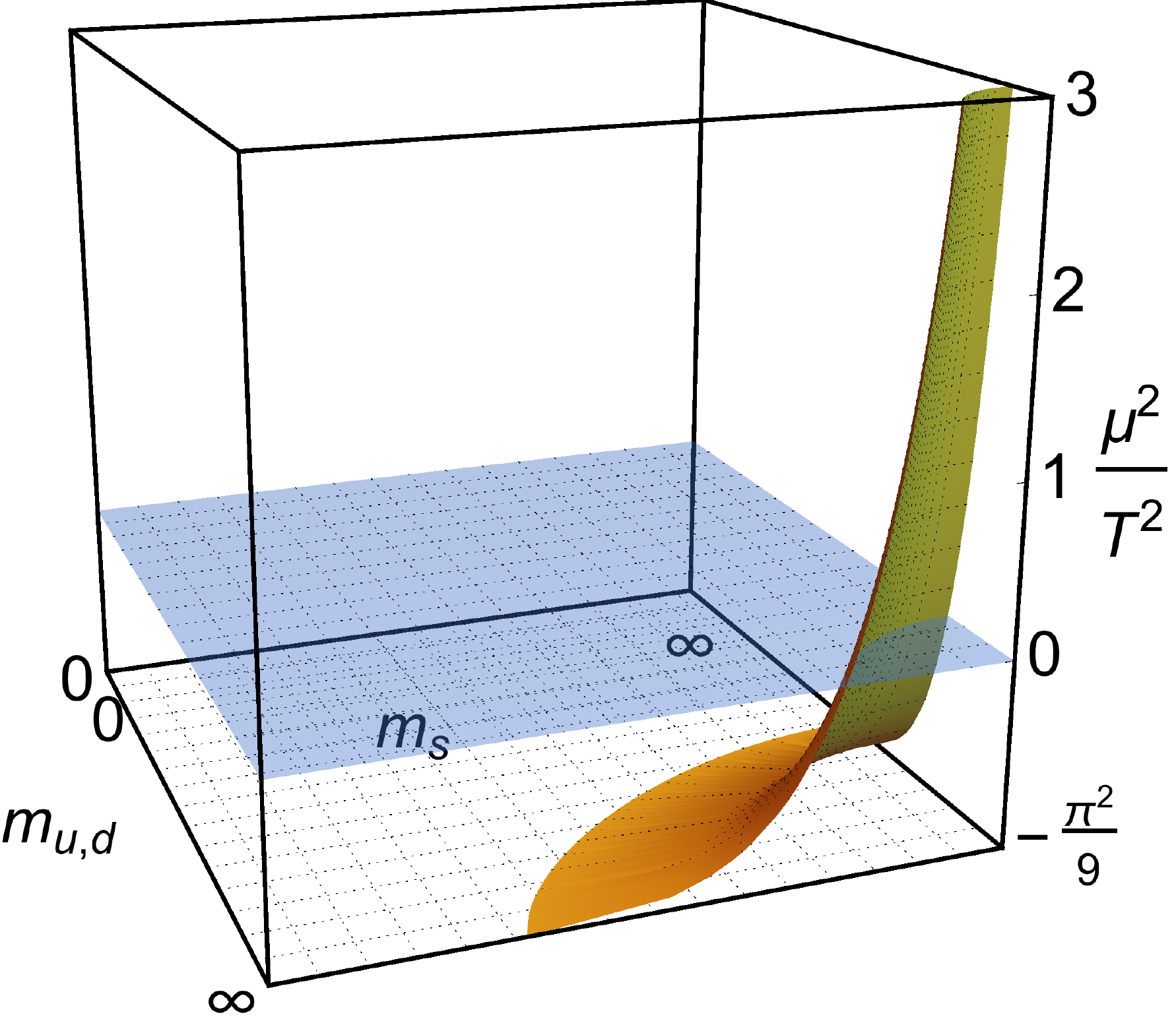}
        \caption{\textit{Left:} Critical quark mass as a function of $(\mu/T)^2$ together with the scaling laws
        Eq.~(\ref{scaling}). 
        \textit{Right:} Sketch of the Columbia plot extended to real and imaginary chemical potential.
        All figures adapted from \cite{Fischer:2014vxa}.}\label{fig:heavy_c}
        \end{center}
\end{figure}

In the left plot of Fig.~\ref{fig:heavy_a} the Polyakov-loop potential at zero chemical potential
is shown as a function $\varphi_3$ and $\varphi_8$ for a quark mass below $m_c$ and a temperature
where the system is in the approximate center symmetric low temperature phase. The potential then 
has six degenerate minima located close to $(\varphi_3,\varphi_8) = (\pm 2/3,0)$ and
$(\varphi_3,\varphi_8) = (\pm 1/3, \pm 1/\sqrt{3})$ and consequently the order parameter 
$L[\bar A_4]\approx0$ is approximately zero. In the right plot of Fig.~\ref{fig:heavy_a} the
behaviour of the potential around the minimum at $(\varphi_3,\varphi_8) = (2/3,0)$ is shown
as a function of $\varphi_3$ for quark masses below, at and above the critical mass $m_c$,
all tuned to the corresponding critical temperatures. From the inlay magnifying the region around 
$\varphi_3 = 0.5$ one can clearly see the emergence of a second order and a first order phase transition
when the quark masses are increased across the critical one: For $m<m_c$ one finds only one minimum 
away from the confining value $\varphi_3=\frac{2}{3}$, at $m=m_c$ the potential is flat and for $m>m_c$ 
one has two degenerate minima. This structure of the potential translates to the behaviour 
of the Polyakov loop shown in the left plot of Fig.~\ref{fig:heavy_b} as a function of $T$ for the 
same quark masses as used for the potential. Again, one can clearly distinguish the crossover for 
$m<m_c$ from the weak first order transition at $m>m_c$ and the second order phase transition at $m=m_c$.
For the critical exponents it turned out that the approximation to the Polyakov loop potential used 
in \cite{Fischer:2014vxa} was not sufficient to obtain critical scaling beyond mean field, i.e. they 
found $L[\bar A_4] \sim |T - T_c|^\beta$ in the vicinity of $T_c$ with a mean field exponent $\beta=1/2$. 
In order to go beyond mean field one would need to incorporate the effects of the background field 
also into the coupled system of DSEs for the propagators, which has not yet been 
done.\footnote{In the vacuum there is a large body of successful work available on DSEs in the background 
field formalism, see \cite{Binosi:2009qm,Aguilar:2015bud} for reviews.} 
In general, the Matsubara modes of the quark propagator are shifted by both, the background gauge field 
and the chemical potential. For imaginary chemical potential $\mu = i \mu_I = i 2\pi T \theta$ both shifts 
become real with
\beq
\tilde{\omega} = \omega + g {A}_4 + 2 \pi T \theta \,.
\eeq
Roberge-Weiss symmetry, discussed in section \ref{Columbia_ext}, results in the fact that a shift in imaginary 
chemical potential by $\theta \rightarrow \theta + k/3$ with integer $k$ can be absorbed by a shift 
in the Matsubara sum and a center transformation of the background gauge field $A_4$. Correspondingly,
the Polyakov loops are shifted by the center transformation and so are the minima of the effective potential.
For the argument of the Polyakov loop this shift is displayed in the right plot of Fig.~\ref{fig:heavy_b},
which needs to be compared with Fig.~\ref{fig:RW} in section \ref{Columbia_ext}. One clearly identifies
the structure of the Roberge-Weiss transitions, where the Polyakov loop changes its phase discontinuously
into the next center sector. These transitions become continuous for temperatures below the Roberge-Weiss
transition temperature. The situation shown in Fig.~\ref{fig:heavy_b} corresponds to the crossover region
of the Columbia plot, i.e. point $D$ of Fig.~\ref{fig:RW}). The other situations corresponding to the
points $A-C$ of Fig.~\ref{fig:RW}) have also been verified in \cite{Fischer:2014vxa}.

The main result of Ref.~\cite{Fischer:2014vxa} is shown in the left plot of Fig.~\ref{fig:heavy_c} and incorporated
into a sketch of the extended Columbia plot in the right diagram of Fig.~\ref{fig:heavy_c}. The 
plots show the critical surface $m_c(m_{u/d},m_s,\mu)$ as a function of squared chemical potential
for $N_f\in\{1,2,3\}$ starting at the plane $\mu/T=i\pi/3$ of the first Roberge-Weiss transition. These
results are compared to the expected tricritical scaling from the Roberge-Weiss endpoint, given by
\begin{equation}
\frac{m_c}{T} = \frac{m_{tric}}{T} + K\left[\left(\frac{\pi}{3}\right)^2+
\left(\frac{\mu}{T}\right)^2\right]^{2/5}\,, \label{eq:tricScaling}
\end{equation}
and displayed in the left plot of Fig.~\ref{fig:heavy_c}.
Here $m_{tric}$ is the quark mass on the tricritical surface $\mu/T=i\pi/3$
and $K$ is a parameter that has been determined by fitting. The agreement of the numerical results with
the expected scaling law is excellent even up to large real chemical potential, i.e. the Roberge-Weiss end-point
exercises its influence far beyond the zero chemical potential plane. Only at very large $\mu^2$ one finds 
slight deviations from the scaling behaviour. These results have been found previously also in lattice gauge 
theory \cite{Fromm:2011qi}. The critical quark masses have been addressed recently also in a perturbative study 
in the background field Curci-Ferrari inspired approach of Refs.~\cite{Reinosa:2015oua,Maelger:2017amh} and in 
a very recent publication \cite{Maelger:2018vow} universal aspects of the critical quark mass have been studied. 

We come back to the Polyakov loop potential at finite chemical potential in section \ref{results:CEP}, 
when we discuss the $N_f=2+1$ theory at physical quark masses.  

\subsection{$N_f=2$: Critical scaling in the chiral limit and temperature dependence of meson masses}\label{results:chiral}

Assuming the $U_A(1)$ symmetry remains anomalously broken across the critical temperature, two-flavour QCD
is expected to exhibit a second order phase transition in the $O(4)$ universality class of the Heisenberg
anti-ferromagnet \cite{Pisarski:1983ms,Rajagopal:1992qz}. The critical physics at reduced temperature 
$t=(T-T_c)/T_c$ is characterised by six critical exponents
\begin{align}
\alpha &= 2 - d \nu \,,\nonumber\\ 
\beta  &= \frac{\nu}{2} (d-2+\eta)\,, \\
\gamma &= (2-\eta) \nu\,, \\
\delta &= \frac{d+2-\eta}{d-2+\eta} \,,
\end{align}
and the relevant dimension is $d=3$. The six critical exponents are expressed in terms of the two independent 
quantities $\eta$ and $\nu$, which describe the scaling relation for the inverse correlation length (here the
mass of the scalar sigma meson) and the order parameter (here the chiral condensate):
\be
m_\sigma\sim t^{\nu}\;,\quad\quad \langle \bar{\psi}\psi \rangle\sim t^{\frac{\nu}{2}(1+\eta)}\,.\label{eq:mscaling}
\ee
The values for the exponents of the $O(4)$-universality class are given by $\nu\approx 0.73$ and 
$\eta\approx 0.03$, see e.g. \cite{Baker:1977hp,Rajagopal:1992qz}. Mean field scaling corresponds to
$\nu = 0.5$ and $\eta = 0$.

In the DSE approach to the chiral limit two-flavour theory a second order phase transition has been seen 
already at a very early stage \cite{Alkofer:1986bm}. Results obtained in the last millennium have been 
discussed by Roberts
and Schmidt in their review Ref.~\cite{Roberts:2000aa}. They pointed out that all truncations
falling in the class of the rainbow-ladder models discussed in section \ref{sec:RL} will show critical
scaling on the mean field level only, simply because meson correlators that are expected to develop long-range
correlations are not explicitly taken into account. This notion has been corroborated also in the
systematic study of Ref.~\cite{Blank:2010bz}. In the Hamilton variational approach, first results for the
chiral transition have been presented in \cite{Quandt:2018bbu}, but no attempts have been made to extract the
critical exponents of the second-order transition.

\begin{figure}[t]
        \begin{center}
        \includegraphics[width=0.70\textwidth]{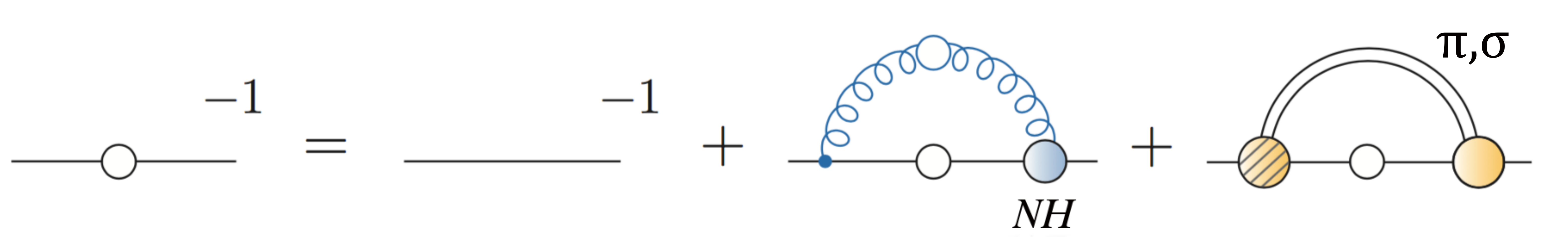}
        \caption{DSE for the quark propagator with non-hadronic contributions and
        hadronic contributions from pion and sigma mesons.}\label{fig:DSE-pion}
        \end{center}
\end{figure}

What is needed in a more complete approach? First of all note again that dynamical chiral symmetry 
breaking and the associated formation of a chiral condensate is driven by the Yang-Mills sector of QCD.
The interaction strength of QCD is largest in the quenched theory and receives unquenching corrections of order $1/N_c$
from quark loops which decrease the amount of dynamical mass generation as discussed already in section
\ref{sec:quark}, see Fig.~\ref{fig:quark}. These corrections are of the order of 10-20 \%. In the quark-DSE,
these corrections stem from two different sources: they occur in the DSE for the gluon 
propagator \cite{Fischer:2003rp,Fischer:2005en} and they contribute to the dressed quark gluon vertex.
In section \ref{general} we argued that the latter contributions can be cast into diagrammatic representations
that involve (off-shell) hadronic propagators and wave functions \cite{Fischer:2008wy,Fischer:2008sp}. 
With respect to the discussion here it is particularly important that meson exchange diagrams appear,
which contain the degrees of freedom relevant for the problem at hand, namely an iso-triplet of pions
and the iso-singlet scalar correlator which we call the sigma meson in the following. In the $N_f=2$ theory 
these are the dominating hadronic contributions with the smallest masses. In Fig.~\ref{fig:DSE-pion} we
show the resulting quark-DSE, which is an approximation of the one shown in Fig.~\ref{fig:mesonbaryon}.
In the vacuum and for all temperatures sufficiently far away from the critical one, the mesonic contributions
to the quark-DSE are subleading as compared to the gluonic diagram. For temperatures
close to the critical one, however, universality tells us that the long range fluctuations take over and
the microscopic interaction becomes irrelevant. In fact this is the very notion of universality. In the quark-DSE
this means that the dressing diagrams with the mesons should then dominate and lead to a self-consistent scaling
law for the order parameters, i.e. for the chiral condensate or, equivalently, the scalar quark dressing function
$B(0)$ evaluated e.g. at lowest Matsubara frequency and zero spatial momentum. This has been shown analytically 
and numerically in Ref.~\cite{Fischer:2011pk}, which we summarise in the following.

In order to analyse the quark-DSE of Fig.~\ref{fig:DSE-pion}, we first need to specify the pion propagator
and wave functions appearing in the second diagram. Following Ref.~\cite{Son:2001ff}, we write the (real part 
of the) Euclidean in-medium pion propagator as
\begin{equation}\label{pionprop}
 D_\pi = \frac{1}{\omega_p^2 + u^2 (\vp^2 + m_\pi^2)}\,.
\end{equation}
denoting by $u$ the pion velocity at $m_\pi=0$. The pion screening mass $m_\pi$ is defined by the zero
of the dispersion relation at vanishing energy and the pion pole mass $\omega_p = iE_\pi = iu m_\pi$ by the zero of the
dispersion relation at zero spatial momentum. At finite temperature there are also two distinct pion decay 
constants denoted by $f_s$ transverse to the heat bath and $f_t$ longitudinal to the heat bath \cite{Pisarski:1996mt}
with $f_t=f_s=f_\pi$ in the zero temperature limit. The ratio of these is equal to the pion velocity
\beq\label{vel}
u^2 = \frac{f^2_s}{f^2_t}\,,
\eeq
and the generalized Gell-Mann-Oakes-Renner relation at finite temperature is given by 
\beq
f_s^2 m_\pi^2 = -2m_q \langle \bar{\Psi} \Psi \rangle_0\,.
\eeq
The pion decay constants are static quantities and are in principle calculable 
in a thermodynamic equilibrium approach \cite{Son:2001ff}. 
Using the abbreviation $\tilde{P}_{\mu}=(u\vect{P},\omega_P)$ for the total four-momentum of the pion 
one finds 
\begin{equation}
 \tilde{P}_{\mu}f_t
=\,3\,\textrm{\large{tr}}_{D}T\sum_{n_q}\int\frac{d^3q}{(2\,\pi)^3}\,\Gamma_\pi(q,P)\,
S(q+P)\gamma_5\gamma_{\mu}S(q) \,,\label{pi_decay}
\end{equation}
valid on the pion mass shell. In the chiral limit, $m_\pi^2 \rightarrow 0$ and $P_\mu\rightarrow 0$, one obtains 
$f_t$ from the time component of Eq.~(\ref{pi_decay}), whereas the transverse decay constant  
$f_s=u f_t$ can be extracted from the spatial components of the equation. The pion Bethe-Salpeter vertex $\Gamma_\pi$
can be extracted from its Bethe-Salpeter equation which is discussed below in section \ref{sec:mesons}.
In the chiral limit and in the vacuum the leading part of the vertex satisfies Eq.~(\ref{Boverf}), which 
generalises to 
\beq \label{BoverfT}
\Gamma_0(p;P) = \gamma_{5} \frac{B(p)}{f_t} \,,
\eeq
at finite temperature.

The scaling behaviour of the pion velocity and decay constants close to the critical temperature
has been obtained from a matching of an effective theory with QCD at the scale $m_\sigma$ in Ref.~\cite{Son:2001ff}. 
This matching results in 
\beq \label{scal1}
u\sim f_s\sim t^{\nu/2} \,,
\eeq
which means that the pion velocity vanishes at the critical temperature. Using the pion Bethe-Salpeter vertex
(\ref{BoverfT}), the scaling law (\ref{scal1}), the pion velocity (\ref{vel}) and the pion 
propagator (\ref{pionprop}) the authors of Ref.~\cite{Fischer:2011pk} found that at the critical 
temperature (and only there) the meson dressing loop (third diagram on the right hand side in Fig.~\ref{fig:DSE-pion})
dominates over the one with the gluon. The scalar quark dressing function then satisfies the self-consistent scaling law
\beq\label{chiralscaling}
B(t) \sim t^{\nu/2}\,.
\eeq
This demonstrates universality. As a direct consequence, the chiral condensate satisfies a similar scaling law
which (up to small corrections due to $\eta$) agrees with Eq.~(\ref{eq:mscaling}). Plugging these scaling
laws into Eq.(\ref{pi_decay}) furthermore leads to ${f_s} \sim t^{\nu/2}$ in agreement with Eq.~(\ref{scal1}).
This renders the scaling analysis self-consistent.

\begin{figure}[t]
        \begin{center}
        \includegraphics[width=0.45\textwidth]{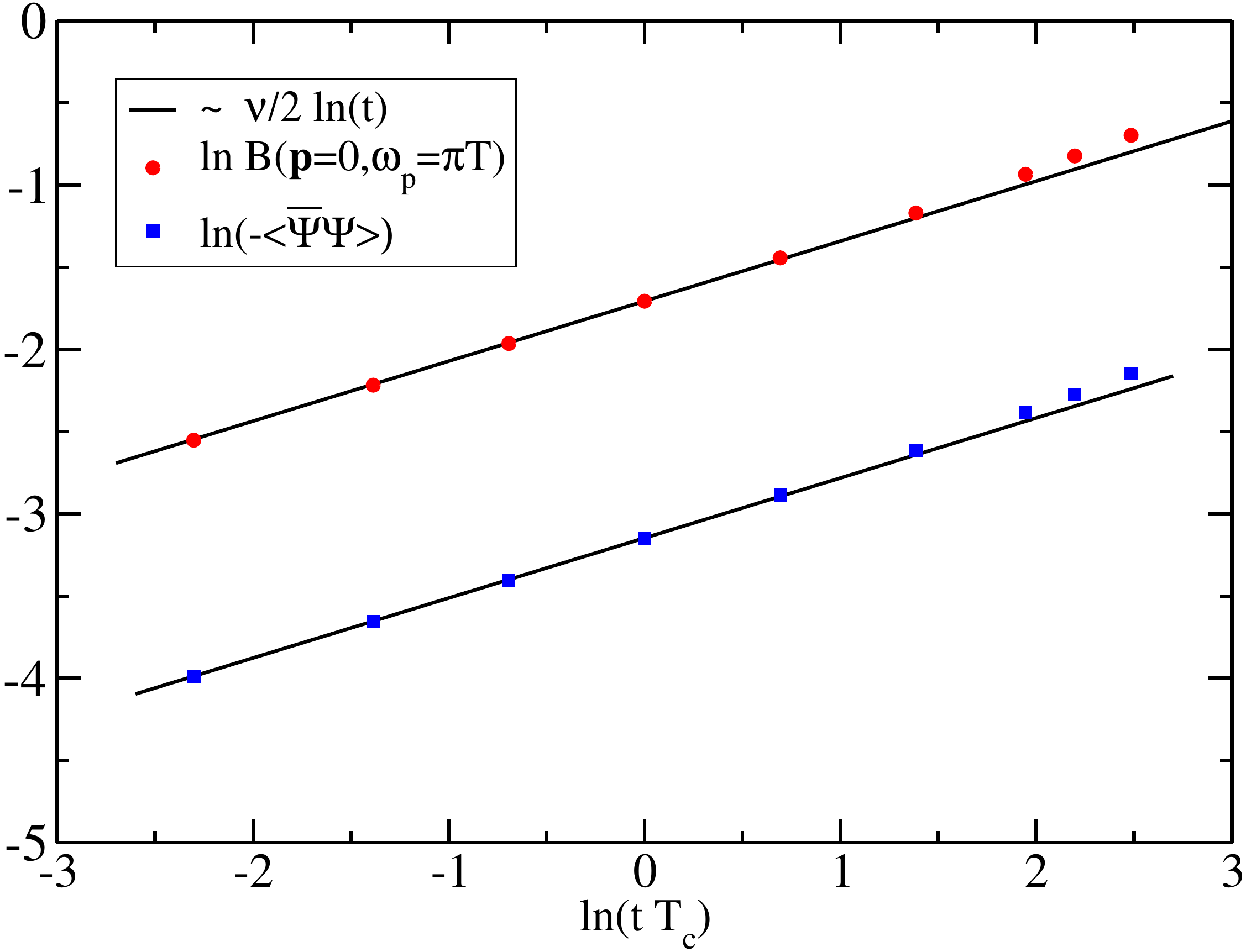}\hfill
        \includegraphics[width=0.50\textwidth]{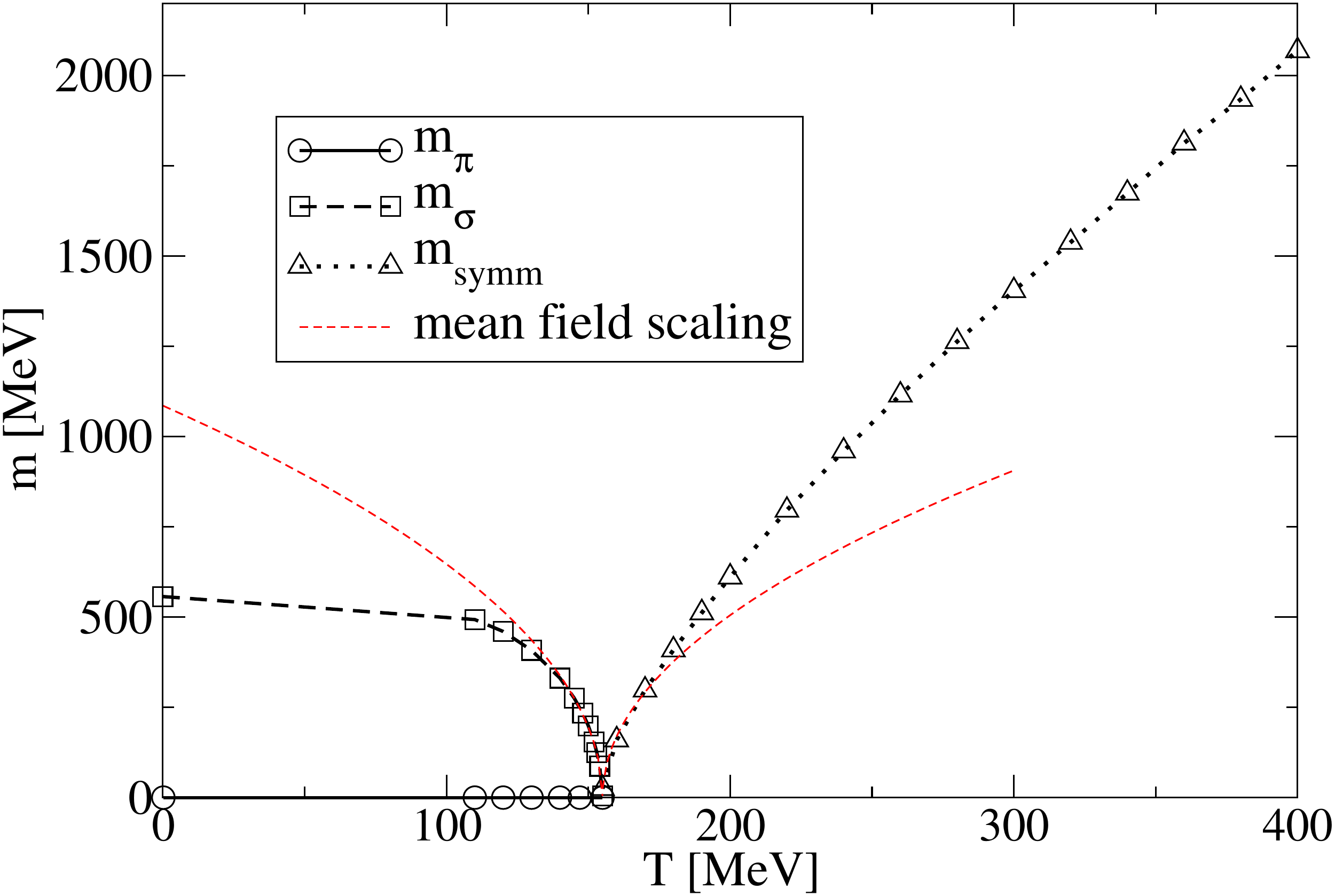}
        \caption{\textit{Left:} Critical behaviour of the quark mass function and the quark condensate 
        as a function of reduced temperature together with the scaling law Eq.~(\ref{chiralscaling}). 
        Figure adapted from \cite{Fischer:2011pk}.
        \textit{Right:} Masses of the pion and the sigma meson as a function of temperature together with
        the mean field scaling laws. Figure adapted from \cite{Fischer:2014vxa}.}\label{fig:chiral}
        \end{center}
\end{figure}

These analytical results have been verified also numerically in Ref.~\cite{Fischer:2011pk}. Employing two
different forms of the truncation for the gluonic dressing loop, the Yang-Mills back-coupled truncation 
described in section \ref{fulltrunc} and a rainbow-ladder type of truncation with completely different
infrared behaviour, they verified that in both cases universality kicks in close to $T_c$ and the gluonic
dressing loop is overwhelmed by the meson diagram. Furthermore, using the scaling ansatz Eq.~(\ref{scal1}) 
for the pion decay constant they showed that the scalar quark dressing function and the chiral condensate 
scale in a temperature region of about $10$ MeV below the chiral critical temperature, see
the left plot in Fig.~\ref{fig:chiral}. Despite this progress, 
however, the exploratory study of Ref.~\cite{Fischer:2011pk} did not succeed in producing the O(4) critical 
exponents self-consistently. To this end the truncation used was still not rich enough. One would need to 
include the meson Bethe-Salpeter equation, its normalisation condition and the explicit equations for the 
pion decay constants in the analysis. This has been left for future work. 

Using a rainbow-ladder framework,
the study of Ref.~\cite{Maris:2000ig} already long ago managed to determine the masses of the pion and the 
sigma meson as a function of temperature in the chiral limit. The results, also discussed in the 
review \cite{Roberts:2000aa}, can be seen in the right plot of Fig.~\ref{fig:chiral}. As expected, 
in the chiral symmetry preserving scheme the pion remains a massless Goldstone boson up to the critical
temperature of the second order phase transition. The sigma meson starts out massive, but becomes massless
at $T_c$ with the proper mean field scaling law indicated by the red dashed line. Above $T_c$ the masses of the
two mesons are degenerate and follow again the mean field scaling law until the screening masses approach the
expected behaviour proportional to $T$ for large temperatures. 

As a final remark to the critical physics associated with the upper left corner on the Columbia plot (and
a putative critical end point) let us briefly compare the inherent strengths and problems of approaches 
that are used to study this issue.
As we have seen, in the Dyson-Schwinger approach it is a highly non-trivial problem to self-consistently include
the degrees of freedom that generate the critical physics at and around $T_c$. In effective chiral theories, 
such as the the PQM model, these degrees of freedom are built in from the start and therefore the critical
physics is extracted easily \cite{Schaefer:2004en,Schaefer:2006sr} and even the question of the fate of the
U$_A$(1) anomaly can be discussed \cite{Resch:2017vjs}. On the other hand, the non-universal
physics of the Yang-Mills sector is readily accessible in the DSE approach, whereas it has to be included indirectly
and without control over the back-reaction effects of the quarks onto the Yang-Mills sector in the PQM approach. 
Since this back-reaction drives the interaction strength of the theory, its proper inclusion is mandatory to
obtain quantitative results on a putative critical end point in the QCD phase diagram. In the DSE approach this
is straightforward; corresponding results are discussed below in section \ref{results:CEP}.
On the other hand, if one is interested in the details of the critical universal behaviour 
of the theory at the CEP, the PQM model may have distinct advantages since the putative Z(2) universality class 
of the CEP is associated with a massless sigma that is readily accessible in the PQM. 
This interplay of different approaches, together with the common contact to lattice gauge theory at small 
chemical potential, seems interesting and fruitful for future research.

The theory with $N_f=2$ quark flavours is not only interesting with respect to the problems discussed above,
it also serves as a test bed for the investigation of systematic aspects of phase transitions without having
to deal with the numerical complexity of the $N_f=2+1$ theory. This has been exploited in \cite{Contant:2017gtz},
where the back-coupled truncation scheme described in section \ref{fulltrunc} has been explored for three different
gauge theories, SU(2), SU(3) and G$_2$ and the expected types of transition in the quenched and unquenched
theories have been found. Furthermore, this has been exploited also in the FRG approach in Ref.~\cite{Braun:2009gm}, 
where the chiral limit of the two-flavour theory has been studied and general arguments concerning the connection of 
quark confinement and chiral symmetry breaking derived from the dressed Polyakov loop and other dual quantities 
have been given.

\subsection{Physical quark masses: The critical end point for $N_f=2+1$ and $N_f=2+1+1$}\label{results:CEP}
In our walk through the Columbia plot we now come to the most interesting point: the one with physical quark 
masses. In order to study this point with DSEs in the back-coupled truncation of section \ref{fulltrunc} one 
needs to solve eight coupled integral equations self-consistently: one for the magnetic and one for the electric
part of the gluon propagator, three equations for the dressing functions of the isospin symmetric and therefore
degenerate up and down quarks and three equations for the dressing functions of the strange quark. This formidable
numerical task has been performed first in Ref.~\cite{Fischer:2012vc} and extended to include the effects of the
charm quark in Ref.\cite{Fischer:2014ata}. Finally, baryonic effects on the CEP in the $N_f=2+1$-theory have been 
studied in \cite{Eichmann:2015kfa}. In the following we will discuss the corresponding results in turn. 

\subsubsection{The QCD phase diagram for $N_f=2+1$}\label{results:2p1}

In section \ref{general} we discussed general aspects of truncations of DSEs. We argued that the physics 
of the Columbia plot can only be explored if the truncation is rich enough that back-reaction effects on the
Yang-Mills sector are taken into account appropriately. This then also allows to study
temperature effects in the gluon propagator and the associated generation of a temperature dependent
electric screening mass. Furthermore, we identified important physics encoded in the non-perturbative coupling
of the gluon to the quark. Although much can be done in this respect from DSEs alone, at some point it is
crucial to assess the quality of a given truncation not only by the richness of the physics it is supposed to 
contain, but also by direct comparison with the results from other approaches. In this respect, results from
lattice gauge theory offer extremely valuable guidance: on the one hand, gauge invariant quantities like 
the quark condensate can be compared but also gauge dependent quantities like the gluon propagator can be
extracted from gauge fixed lattice simulations.

\begin{figure}[t]
        \begin{center}
        \includegraphics[width=0.48\textwidth]{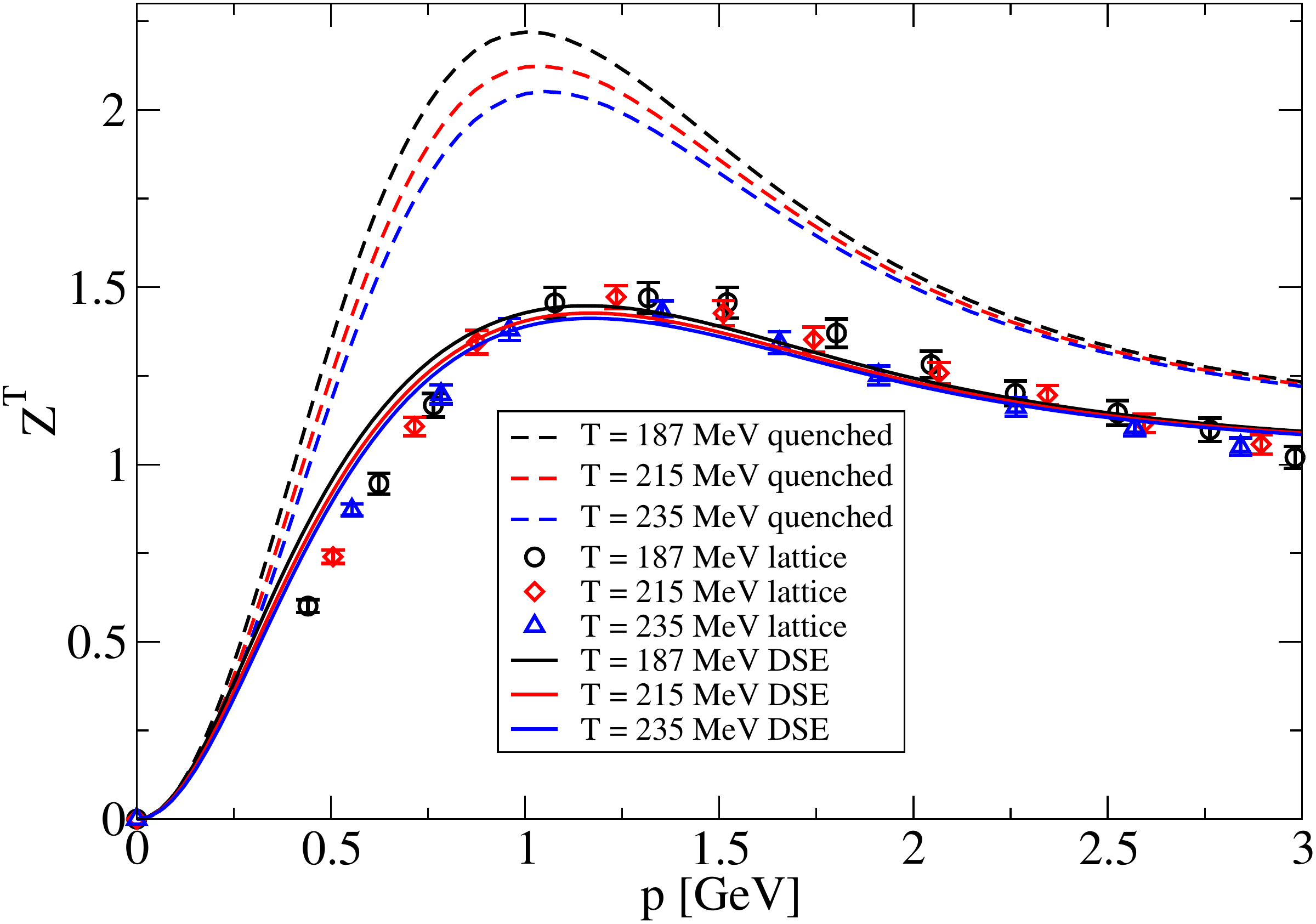}\hfill
        \includegraphics[width=0.48\textwidth]{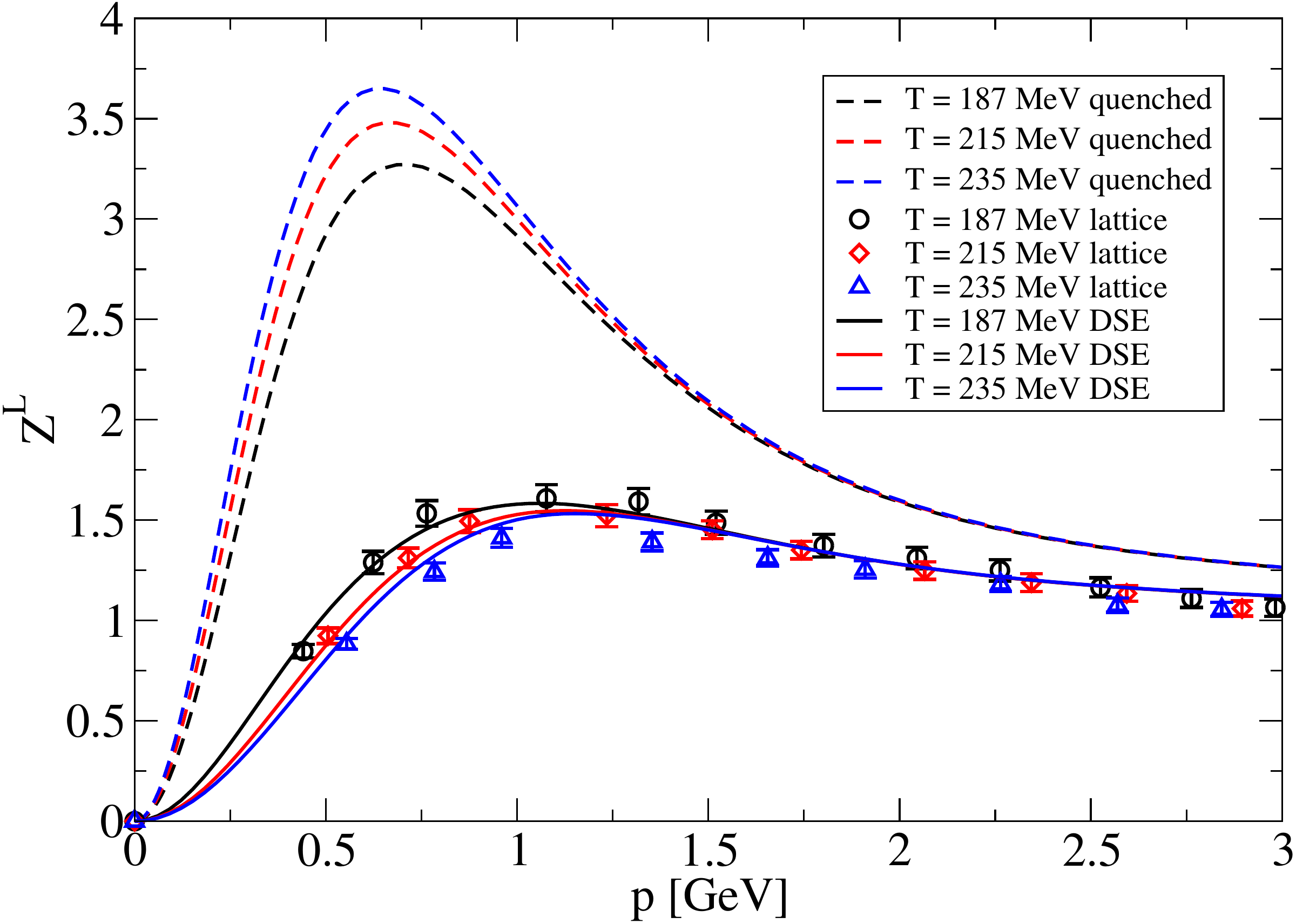}
        \caption{DSE results for the unquenched magnetic (left) and electric (right) gluon dressing
        function. Compared are fits to corresponding quenched lattice data of Ref.~\cite{Fischer:2010fx,Maas:2011ez} 
        (cf. Fig.~\ref{fig:gluonT}) for three different temperatures in dashed lines to unquenched DSE-results \cite{Fischer:2012vc} and unquenched lattice data \cite{Aouane:2012bk}. All results are for $N_f=2$
        and a pion mass of $m_\pi=316$ MeV. Figure adapted from Ref.~\cite{Fischer:2013eca}.}\label{fig:unqgluon}
        \end{center}
\end{figure}

Using the truncation outlined in section \ref{fulltrunc}, the (iterated) back-reaction of the quarks 
onto the gluons in the $N_f=2$ and $N_f=2+1$ theory has been determined first in Ref.~\cite{Fischer:2012vc}
for physical quark masses. The first unquenched lattice results for the temperature dependence of the gluon
propagator became available somewhat later in a simulation with $N_f=2$ quark flavours and quark masses 
corresponding to $m_\pi = 316$ MeV. In \cite{Fischer:2013eca} these two sets of results have been compared
directly, whereas in \cite{Fischer:2014ata} an update of this comparison has been performed with DSE quark 
masses adapted to the large quark masses on the lattice. It turned out that the changes of the unquenched 
gluon due to the different up/down quark masses in the DSE result were smaller than the (statistical) error 
bars of the lattice data, such that both comparisons have been meaningful and delivered a similar result,
shown in Fig.~\ref{fig:unqgluon}. One finds large unquenching effects in both, the magnetic and electric 
part of the gluon propagator. These affect the momentum dependence of the gluon with a large reduction of 
the size of the bump in the non-perturbative moment region. Furthermore, the quark loop effects even invert 
the temperature dependence of the electric gluon dressing function $Z_L$: for the temperatures shown the 
bump in the quenched dressing function increases with $T$ \cite{Fischer:2010fx}, whereas it decreases in 
the unquenched case. This effect is seen in both, the DSE approach and on the lattice. In general, the 
quantitative agreement between the two approaches is excellent. Considering that the DSE results have
been predicting the lattice results, this provides a non-trivial quality check 
for the truncation scheme outlined in section \ref{fulltrunc}.  
\begin{figure}[t]
\begin{subfigure}[t]{0.9\textwidth}
        \begin{center}
        \includegraphics[width=0.48\textwidth]{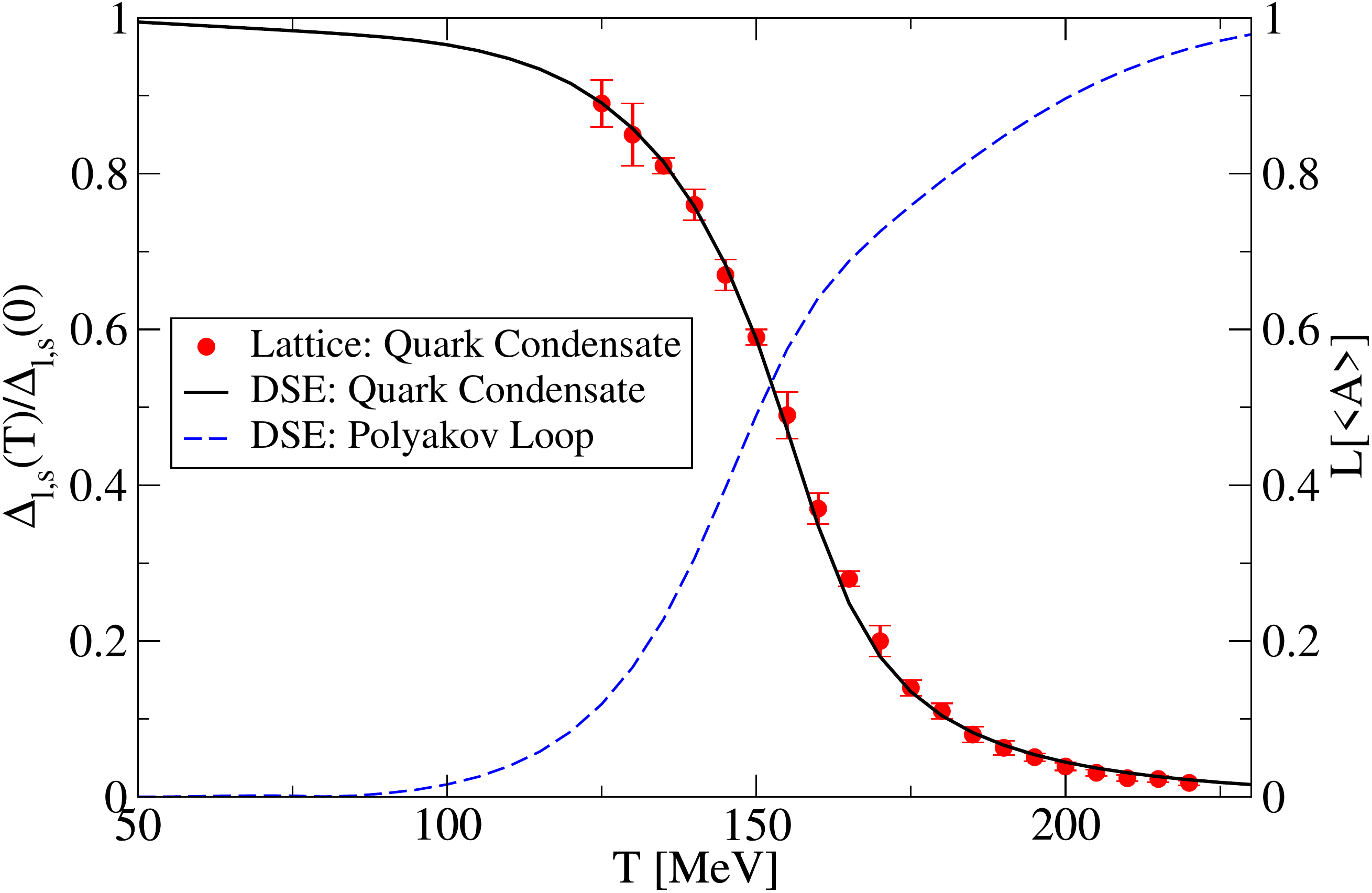}\hfill
        \includegraphics[width=0.48\textwidth]{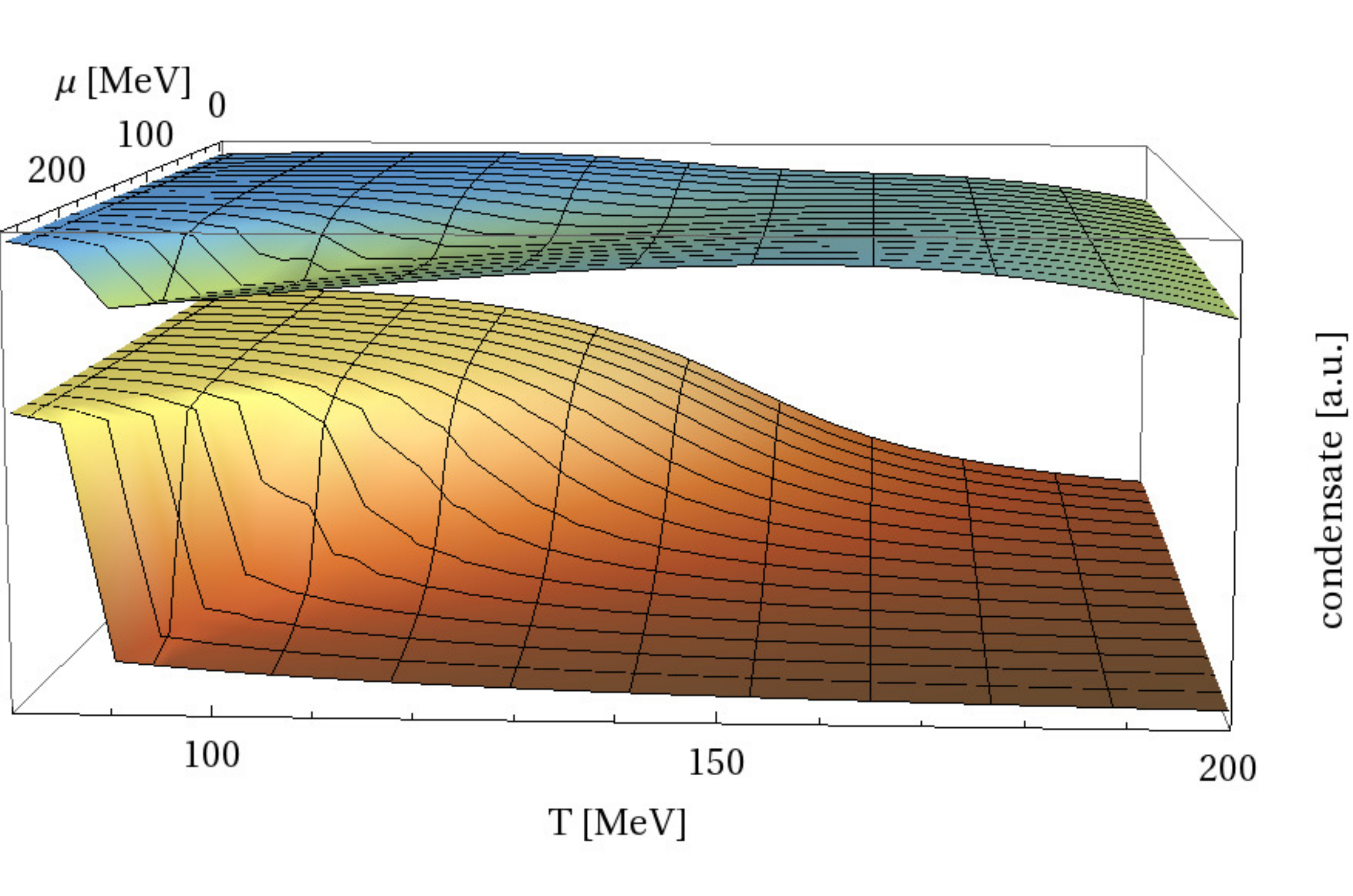}
        \caption{\textit{Left:} Regularized quark condensate and the Polyakov loop for $N_f=2+1$ quark flavours
        as a function of temperature at zero chemical potential. Figure adapted from Ref.~\cite{Fischer:2014ata};
        lattice results are from Ref.~\cite{Borsanyi:2010bp}.
        \textit{Right:} Regularized quark condensate for light (lower surface) and strange (upper surface)
        quarks as a function of temperature and chemical potential. 
        Figure adapted from Ref.~\cite{Fischer:2012vc}.}\label{fig:2+1a}
        \end{center}
\end{subfigure}\vspace*{5mm}
\begin{subfigure}[t]{0.9\textwidth}
        \begin{center}
        \includegraphics[width=0.48\textwidth]{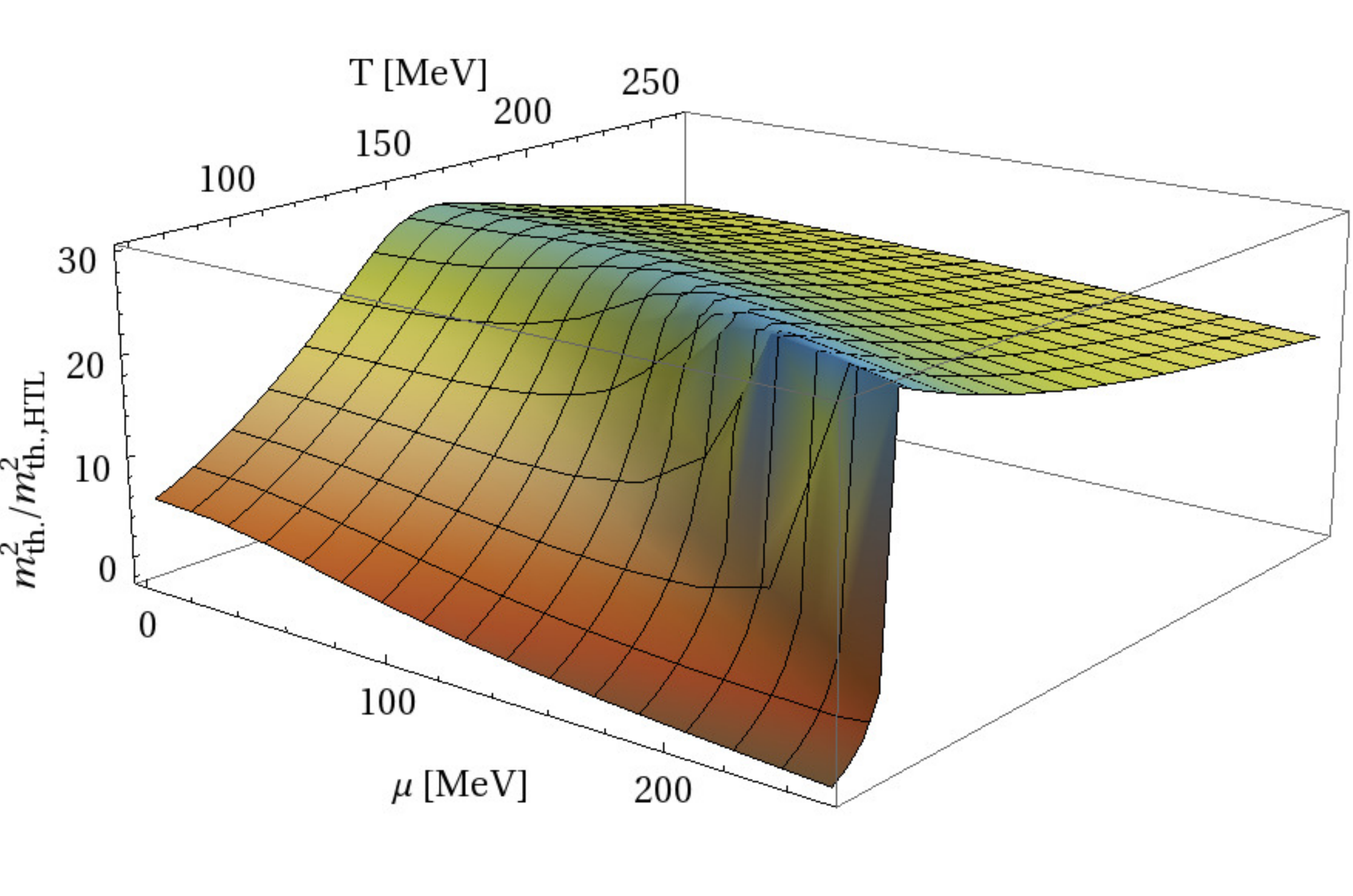}\hfill
        \includegraphics[width=0.48\textwidth]{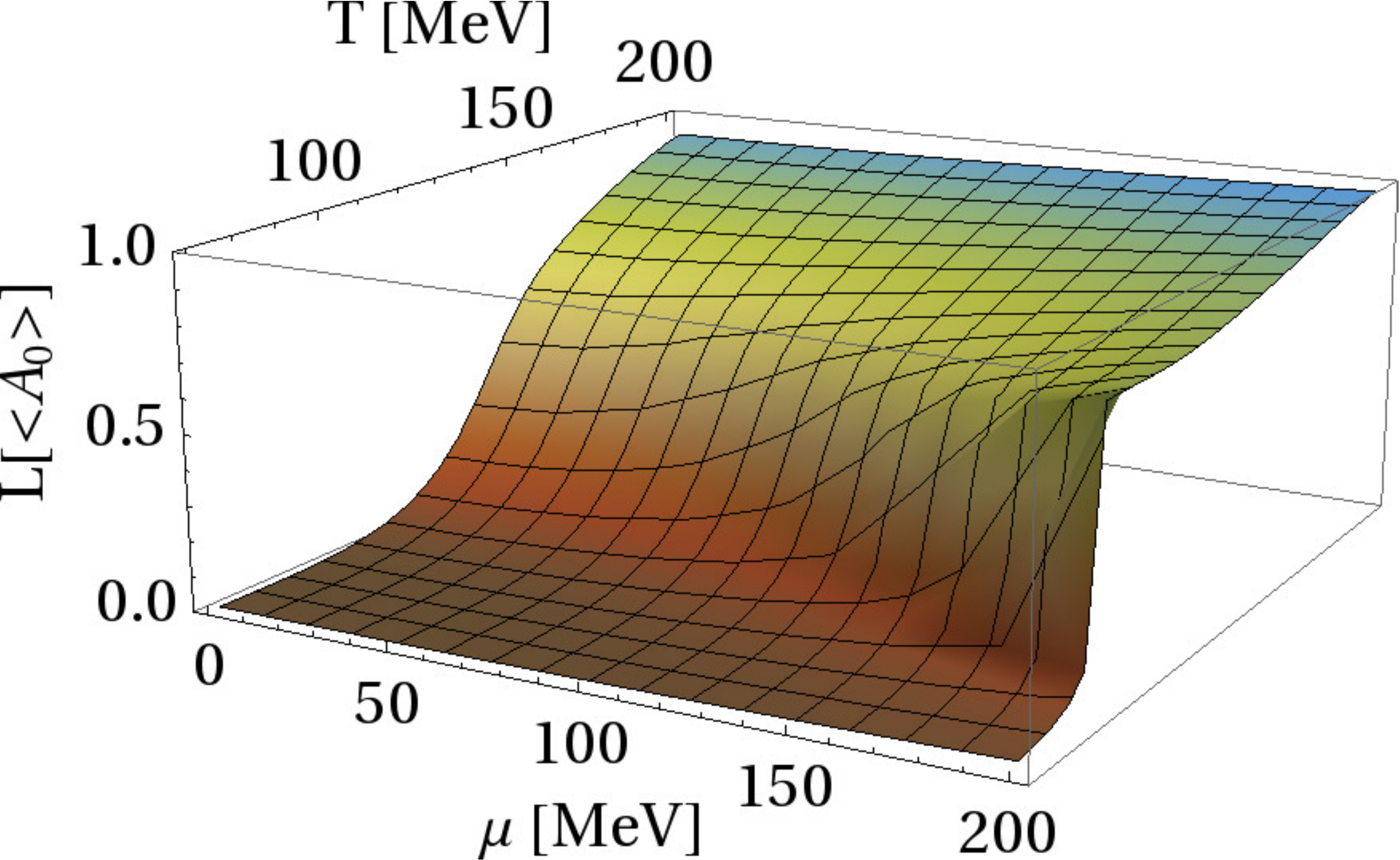}
        \caption{\textit{Left:} Electric screening mass of the gluon normalized by its asymptotic (HTL) behaviour as a function of temperature and chemical potential. Figure adapted from Ref.~\cite{Fischer:2012vc}.
        \textit{Right:} Polyakov loop $L[\langle A_4 \rangle]$ as a function of temperature and chemical potential.
        Figure adapted from Ref.~\cite{Fischer:2013eca}.}\label{fig:2+1b}
        \end{center}
\end{subfigure}
         \caption{Results from DSEs for $N_f=2+1$ at finite temperature and chemical potential.}
\end{figure}
 
The resulting temperature behaviour of the (normalized) light quark condensate \cite{Fischer:2014ata}
is shown in the left diagram of Fig.~\ref{fig:2+1a} and extended to finite chemical potential in the 
right diagram\footnote{Note that Ref.~\cite{Fischer:2012vc} reported results for the light quark condensate
that have been obtained with up/down quark masses about a factor of two too large, resulting in a mismatch
with the lattice data at large temperatures. This has been corrected in \cite{Fischer:2014ata}.}. There we also show the 
corresponding condensate of the strange quark. The melting of the condensate with temperature seen in the 
left plot is nicely matched by corresponding lattice results taken from Ref.~\cite{Borsanyi:2010bp}. The 
pseudo-critical temperature of the chiral crossover has been determined from the chiral susceptibility and 
from the inflection point of the light-quark condensate resulting in slightly different values 
\begin{eqnarray}
\left.T_c\right|_{\frac{d\langle\bar\psi\psi\rangle}{dm}} &= 160.2 \,\mbox{MeV}\,, \nonumber \\ 
\left.T_c\right|_{\frac{d\langle\bar\psi\psi\rangle}{dT}} &= 155.6 \,\mbox{MeV}.
\end{eqnarray}
These reproduce the corresponding transition temperatures from different lattice groups
\cite{Borsanyi:2010bp,Bazavov:2011nk,Bonati:2015bha,Bazavov:2017dus,Bonati:2018nut} within error bars.
As discussed in \cite{Fischer:2014ata}, this agreement is not a result of the DSE framework, but has been 
achieved by an adjustment of the strength parameter $d_1$ in the quark-gluon vertex Eq.~(\ref{vertex}).
Less trivial is the excellent agreement of the steepness of the chiral transition with the lattice data. 
This has been achieved as well in effective models such 
as the Polyakov loop quark-meson model \cite{Herbst:2013ufa,Mitter:2013fxa,Rennecke:2016tkm}. 
The effect of non-vanishing chemical potential
on the chiral transition can be seen in the right diagram of Fig.~\ref{fig:2+1a}. The condensates are not 
normalized and shown on different scales for the z-axis for the sake of clarity. Clearly, the chiral crossover
becomes steeper and steeper with chemical potential until we reach a critical point and subsequently the
gap of a first order phase transition opens. A corresponding behaviour can be seen in the strange quark
condensate; here it is the back-coupling of the effects of the light quark onto the strange quarks via 
the gluon propagator that causes the opening of a gap at the same location as for the light quark condensate.

Corresponding plots for order parameters of the deconfinement phase transition can be seen in Fig.~\ref{fig:2+1b}.
In the left diagram we show the electric screening mass of the gluon propagator \cite{Fischer:2012vc} 
multiplied (but not normalized) with the asymptotic form evaluated in the hard thermal loop formalism, 
$m^2_{th,HTL} \sim T^2 + 3 \mu^2/\pi^2$. For very large temperatures this behaviour is reproduced.
Around $T \approx 150$ MeV one clearly sees the continuous change of a crossover for small 
chemical potential. For larger chemical potential the transition becomes steeper until it becomes discontinuous 
in the vicinity of the critical endpoint of the chiral transition. A similar behaviour can be seen from the 
Polyakov loop $L[\langle A_4 \rangle]$ \cite{Fischer:2013eca} shown in the right diagram of Fig.~\ref{fig:2+1b}, 
see also the zero chemical potential curve shown in the left plot of Fig.~\ref{fig:2+1a}. Both deconfinement
order parameters react to the chiral transition and develop a critical end point at the same location.
The direct calculation presented in Ref.~\cite{Fischer:2013eca} made the Polyakov loop potential available
at finite chemical potential for the first time. These results have been further refined and their relation to the
effective potentials used in model approaches such as the PQM and the PNJL model have been discussed in detail in 
Ref.~\cite{Herbst:2015ona}. 

\begin{figure}[t]
        \begin{center}
        \includegraphics[width=0.48\textwidth]{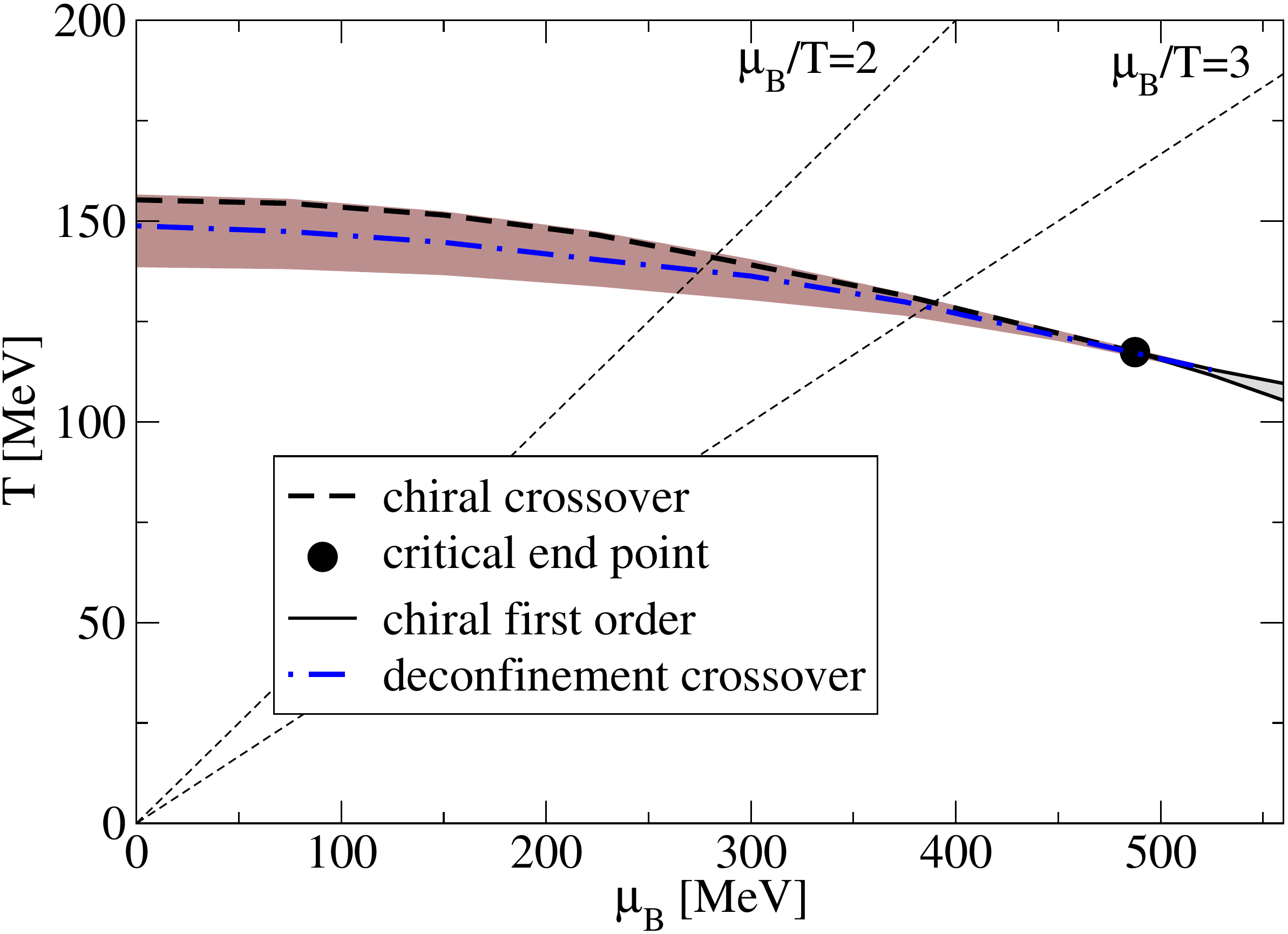}\hfill
        \includegraphics[width=0.48\textwidth]{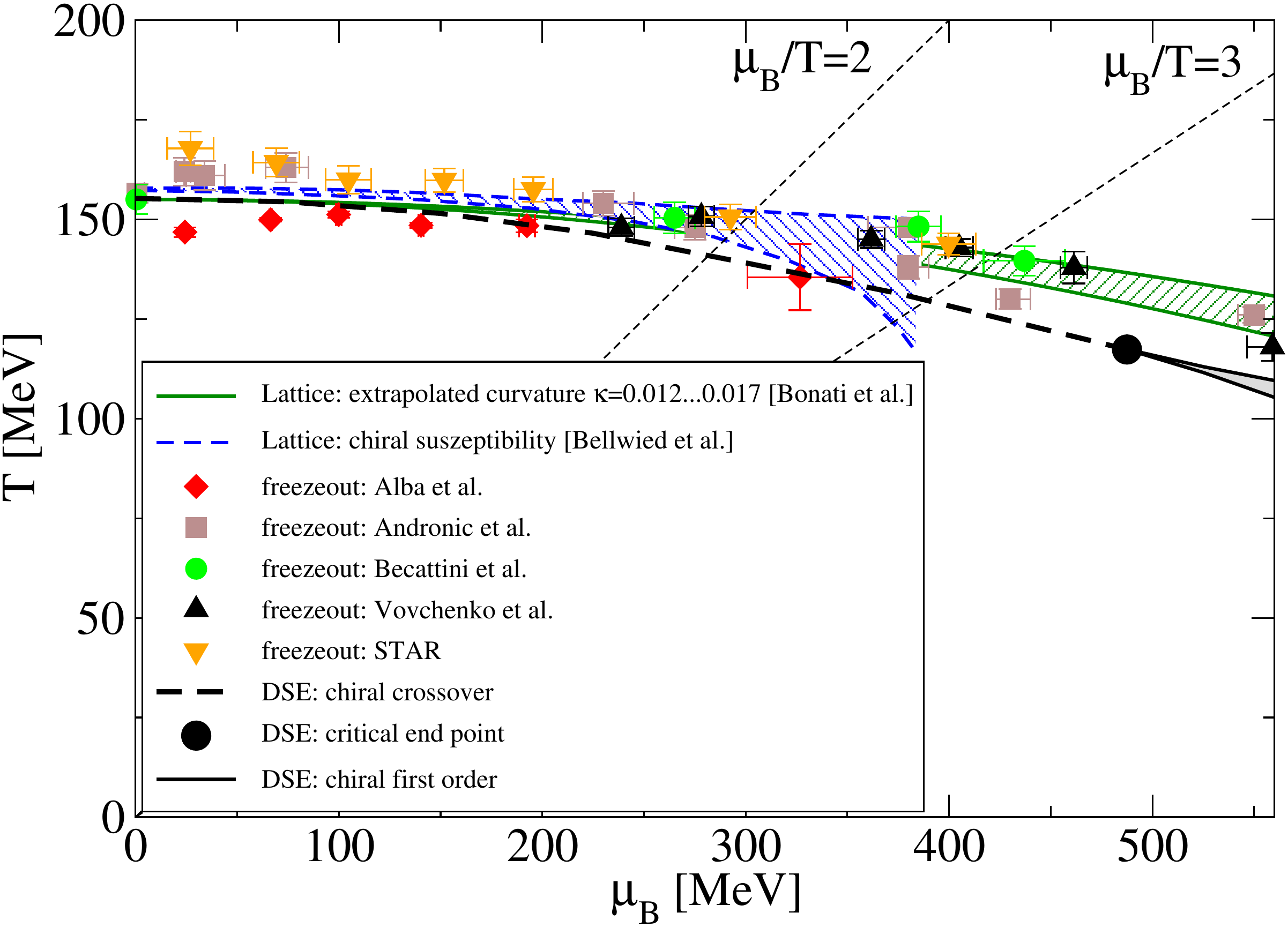}
        \caption{\textit{Left:} QCD phase diagram from DSEs \cite{Fischer:2014ata}. 
        Shown is the chiral transition (inflection point) and the deconfinement transition from the 
        Polyakov loop (inflection point). The brown shaded area shows the width of the
        deconfinement cross-over defined by 80 \% of the inflection point.
        \textit{Right:} QCD phase diagram from DSEs including: the region of extrapolated curvatures 
        extracted from lattice QCD (green band), see \cite{Bonati:2018nut} and references therein; 
        region of chiral crossover from lattice QCD (blue band) \cite{Bellwied:2015rza} 
        (see also \cite{Borsanyi:2010bp,Bazavov:2011nk}); freeze-out points 
        from heavy ion collisions extracted by different methods/groups
        \cite{Alba:2014eba,Becattini:2016xct,Vovchenko:2015idt,Adamczyk:2017iwn,Andronic:2016nof,Andronic:2017pug}; 
        and the DSE results of Ref.~\cite{Fischer:2014ata}.}\label{fig:phasediag}
        \end{center}
\end{figure}

The resulting phase diagram is shown in the left diagram of Fig.~\ref{fig:phasediag}. The chiral cross over line indicates
the chiral transition extracted from the inflection point of the quark condensate. It turns into a critical end-point
at \cite{Fischer:2014ata}\footnote{In Ref.~\cite{Fischer:2014ata} accidentally two slightly different values for the 
CEP have been given which stem from numerical runs with different precision: In the figure, high precision data have 
been used, whereas the number given in the main text has been obtained with somewhat lower accuracy. Here, we only include
the high accuracy result.}
\begin{equation}\label{eq:CEP}
(T^{CEP},\mu_B^{CEP})=(117,488) \,\mbox{MeV}\,, 
\end{equation}
which corresponds to a ratio $\mu_B^{CEP}/T^{CEP} = 4.2$, i.e. large chemical potential. The deconfinement crossover line
extracted from the inflection point of the Polyakov-loop $L[\langle A_4 \rangle]$ is a couple of MeV below the 
chiral transition line at zero chemical potential, but joins the chiral transition at and beyond the critical 
end-point. Thus in this region of the phase diagram there is no quarkyonic phase in the sense of 
Ref.~\cite{McLerran:2007qj}.
The (brown) shaded area indicates the width of the deconfinement cross-over defined by a $\pm 20$ \% range around the 
inflection point. This range includes the chiral transition and deconfinement transition lines from other order
parameters such as the dressed Polyakov loop. Since all these quantities test different properties of the quark
and gluon propagators this agreement underlines the consistency of the approach \cite{Fischer:2013eca}. 

As an aside, note that the location of the CEP is hardly affected by the choice of $\mu_s$. In the calculation
shown in Fig.~\ref{fig:phasediag} $\mu_s$ has been set to zero, but it has been checked \cite{Welzbacher:2016}
that the implementation of strangeness neutrality discussed in section \ref{sec:fluct} does not change the 
location of the CEP within the numerical error of the calculation, which is below five MeV in each direction. 
Other quantities such as isentropes are more sensitive to $\mu_s$, as explored in the PNJL and PQM models 
see e.g. \cite{Fukushima:2009dx,Fu:2018qsk,Fu:2018swz}.

Furthermore, note that the presence of the critical end-point at finite chemical potential indicates 
a bending of the corresponding 2nd order critical surface in the three-dimensional Columbia plot as 
indicated in Fig.~\ref{fig:columbia_ext_real}. In an exploratory calculation, the positive slope of
the critical surface with respect to growing light quark masses has been explicitly verified for three 
fixed strange quark masses in Ref.~\cite{Welzbacher:2016}. This bending of the chiral critical surface 
is opposite to the behaviour of the deconfinement critical surface discussed above around Fig.~\ref{fig:heavy_c}.

In the right diagram of Fig.~\ref{fig:phasediag} we compare the transition line from the DSE-approach with 
corresponding lattice calculations and freeze-out points from heavy ion collision experiments. Within errors,
there is no tension between all results. Let us compare in turn. The lattice results for the chiral transition
\cite{Bellwied:2015rza} (blue dashed band) have been obtained by analytic continuation from imaginary chemical 
potential. The transition temperature at zero chemical potential is $T_c = 157$ MeV. The systematic error of 
the calculation is represented by the width of the band and is claimed to be under control up to a chemical potential
of $\mu_B=300$ MeV, i.e. up to a ratio $\mu_B/T = 2$. For larger chemical potentials the error accumulates rapidly
as can be seen in the plot. Similar results for the chiral crossover at finite $\mu_B$ have been extracted
in \cite{Bazavov:2017dus} using Taylor expansion methods. Thus the combined evidence from different lattice 
techniques clearly disfavours the existence of a CEP for $\mu_B/T \le 2$. This result confirms the predictions 
from functional methods \cite{Fischer:2012vc,Fischer:2013eca,Pawlowski:2010ht,Pawlowski:2014aha}.

The green band in the right diagram of Fig.~\ref{fig:phasediag} represents the results of Ref.~\cite{Bonati:2018nut} 
for the curvature of the chiral transition using the Taylor expansion technique. As explained in section
\ref{sec:curvature}, the curvature $\kappa$ can be extracted from the expansion Eq.~(\ref{eq:kappa}) at small 
chemical potential. With $T_c = 155$ MeV, the green band represents a continuation of Eq.~(\ref{eq:kappa}) to 
the chemical potentials shown in the plot for the values of $\kappa = 0.0145 (25)$ given in \cite{Bonati:2018nut}. 
This range is in line with the result $\kappa=0.0149 (21)$ of Ref.~\cite{Bellwied:2015rza} and similar results from 
other works already discussed in table \ref{tab:curvature}.
Compared to the lattice results, the DSE cross-over line has a somewhat larger curvature, given by $\kappa = 0.0238$,
which results in a shift of the CEP towards lower temperatures of about 10-15 MeV as compared to the green band. 
This shift may serve as indication for the size of the systematic error of the DSE calculations. We come back 
to this point in the next section. Finally, when comparing the lattice and DSE-results with the freeze-out points 
extracted by various methods from heavy ion experiments it seems that at least in this region of the phase diagram 
the chiral transition temperatures are not much larger than the freeze-out ones. Thus, if the CEP is confirmed to 
be located in the region indicated by the DSE, there may be hope to detect its signals in experimental observables.

\subsubsection{The QCD phase diagram for $N_f=2+1+1$}\label{results:2p1p1}
\begin{figure}[t]
        \begin{center}
        \includegraphics[width=0.48\textwidth]{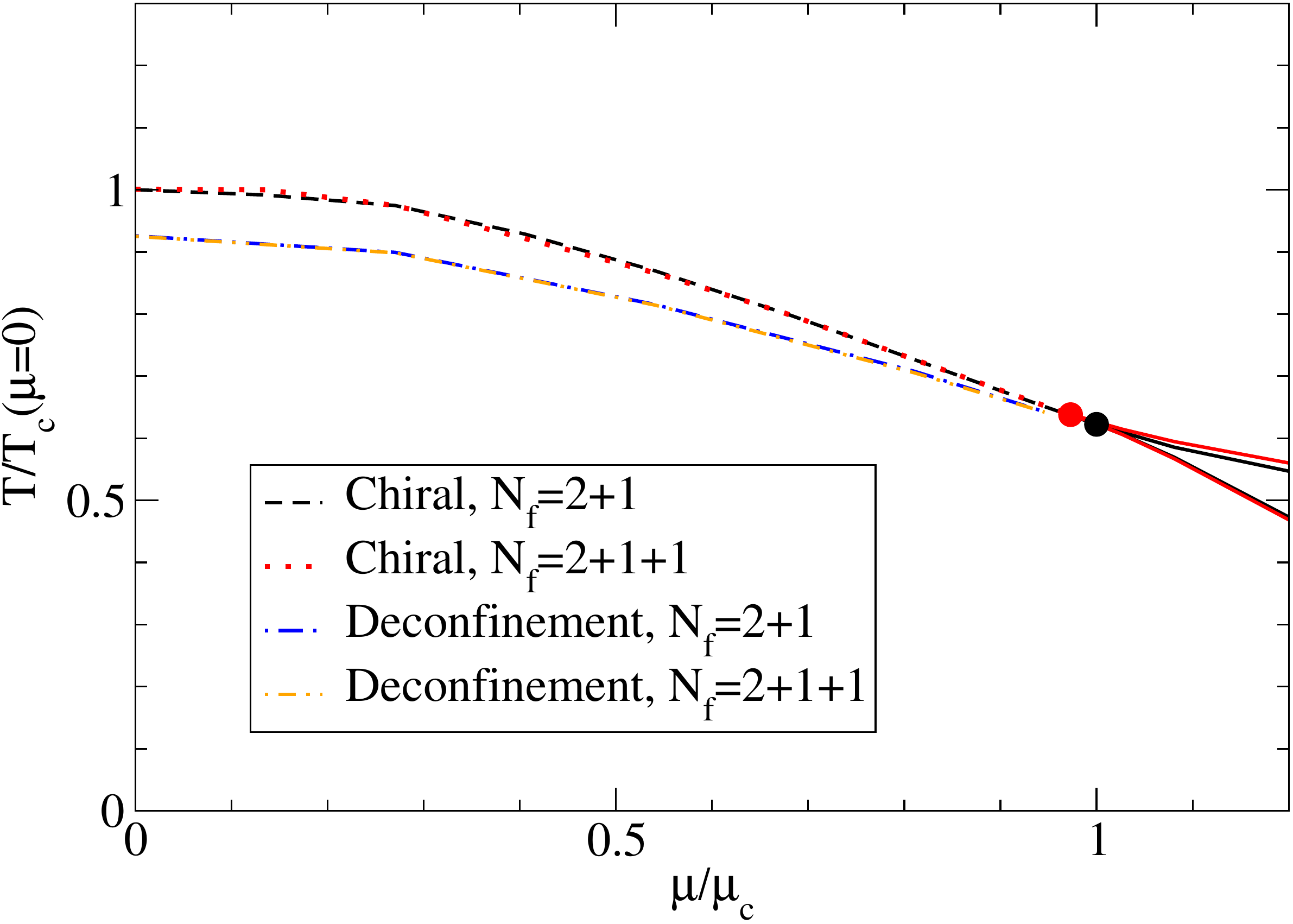}
        \caption{Changes in the QCD phase diagram due to the additional presence of the charm quark. The black dot
        signals the CEP for $N_f=2+1$, whereas the red dot is the corresponding CEP for $N_f=2+1+1$
        Figure adapted from Ref.~\cite{Fischer:2014ata}.}\label{fig:charm}
        \end{center}
\end{figure}

In the last subsection we discussed results for the phase diagram of QCD with $N_f=2+1$ obtained from DSEs in the
truncation discussed in section \ref{fulltrunc}. It turned out that this truncation can be expanded to accommodate 
more quark flavours in a systematic fashion without conceptual difficulties. This has been done for QCD
with $N_f=2+1+1$ in Ref.~\cite{Fischer:2014ata}. The only new element, studied thoroughly in \cite{Fischer:2014ata},
is the problem of setting the scale. For the $N_f=2+1$ calculations the scale has been inherited by corresponding
lattice calculations via the fixing of the strength parameter $d_1$ in the quark-gluon vertex, which guaranteed
agreement with the lattice pseudo-critical temperature at zero chemical potential. Since there were no lattice data 
on the chiral transition available for $N_f=2+1+1$, the authors of \cite{Fischer:2014ata} probed a different 
strategy: they fixed the scales in the $N_f=2+1$ and $N_f=2+1+1$ theory both from vacuum physics. This procedure allows
for a systematic comparison of the two theories and delivers the overall effect of the inclusion of the charm quark
onto the QCD phase diagram. The results were two-fold. Firstly, the different strategy to fix the parameters of the
interaction by vacuum physics led to a reduction of the pseudo-critical temperature at zero chemical potential by
$\delta T = 23$ MeV, i.e. 15 \% also in the $N_f=2+1$ case. This has been interpreted as a measure of the systematic
error of the truncation scheme \cite{Fischer:2014ata}. This estimate is of the same order as the one discussed in the
previous section, comparing shifts in temperature along the chiral transition lines. 
Secondly, with scales fixed as described above, the influence
of the charm quark onto the phase diagram is almost negligible, see Fig.~\ref{fig:charm}. This does not mean that the
charm has no effect at all on the system. On the contrary, the authors noted that the presence of the charm quark affected 
the momentum dependence of the gluon propagator on the level of 15-20 \% in the mid and ultraviolet momentum region.
However, for low momenta at scales of temperatures relevant for the chiral transition, the gluon propagator remained
essentially unchanged such that the chiral transition temperature remained the same within their numerical 
uncertainty of 1--2 MeV. Finite chemical potential did not change this situation such that the location of the 
critical end point is hardly affected by the charm. This effect, first established in \cite{Fischer:2014ata}, now
awaits confirmation from other approaches.

\subsubsection{Baryon effects on the CEP}\label{results:baryons}

Above, we have seen that the critical endpoint found from DSEs is at rather large quark chemical
potential. Since this result relies on a truncation of the quark-gluon interaction which is far 
from complete, it is an important task to quantify its systematic error. At zero chemical potential,
justification can be obtained by good agreement with lattice data on the quark condensate and the
unquenched gluon propagator as discussed above. However, effects at non-zero chemical potential cannot 
be tested in this way and may provide for sizeable quantitative corrections. The authors of 
Ref.~\cite{Eichmann:2015kfa} therefore focused on a particular class of such corrections, namely 
vertex corrections that can be parametrized in terms of (off-shell) baryons. As outlined in section 
\ref{general} baryonic back-reaction effects onto the quark propagator provide a direct mechanism 
how the quark condensate may be influenced by changes in the baryon's wave functions such as the 
one inflicted e.g. by the nuclear liquid-gas transition at very small temperatures. These back-reaction 
effects, however, may very well decrease in size for growing temperatures and it is an interesting question 
whether they are still important at the location of the CEP.\footnote{In a two-color version of QCD this influence 
has been studied in Refs.~\cite{Strodthoff:2011tz,Strodthoff:2013cua,Khan:2015puu} using functional methods 
and found to be crucial to an extent that not only the location but even the very existence of a CEP is affected. 
SU(3), however, may very well be an entirely different matter.} 
\begin{figure}[t]
        \begin{center}
        \includegraphics[width=0.48\textwidth]{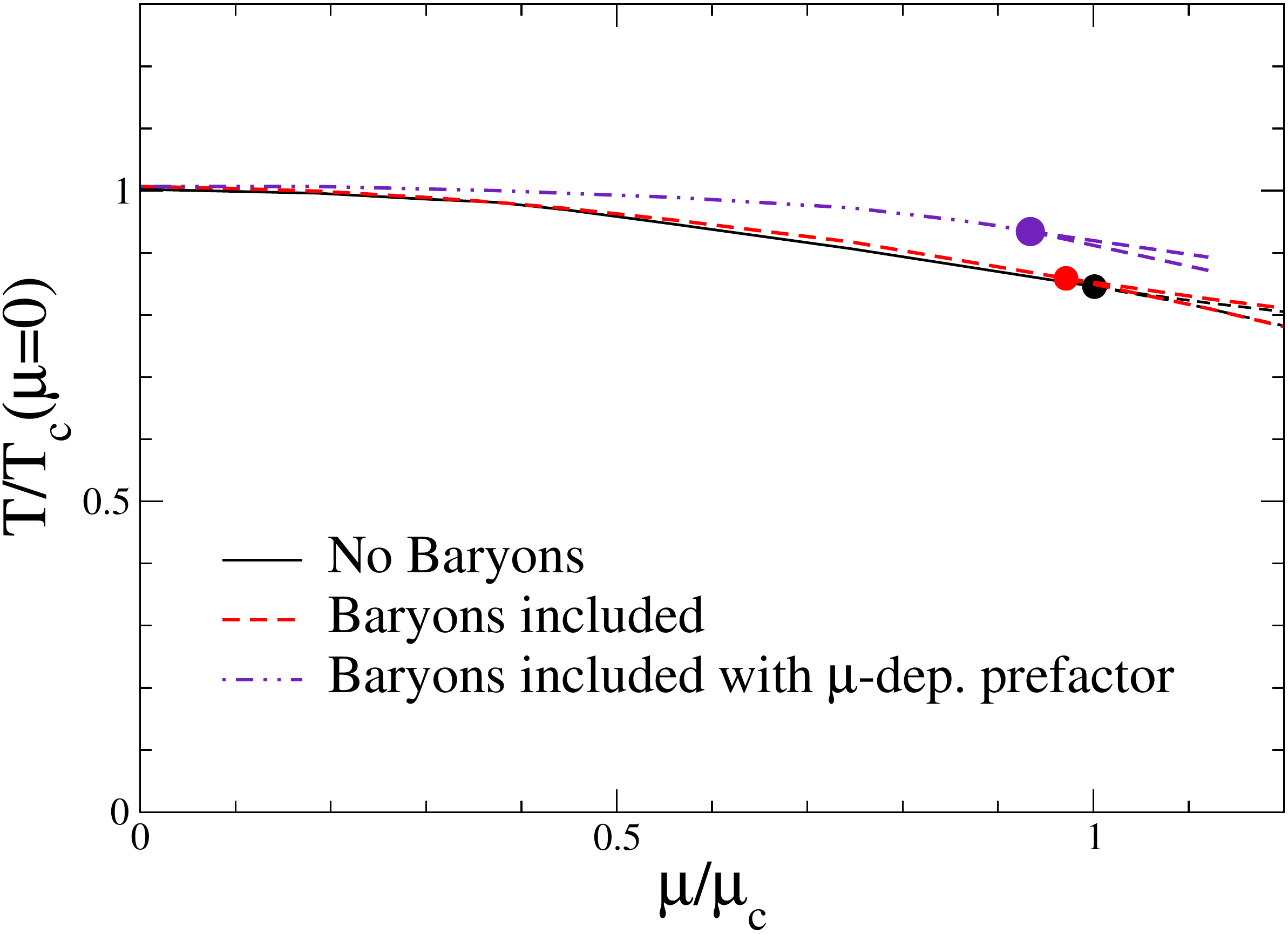}
        \caption{Changes of the location of the CEP when baryons are included (see text for details). 
        Figure adapted from Ref.~\cite{Eichmann:2015kfa}.}\label{fig:phasediag_baryon}
        \end{center}
\end{figure}

To this end the authors of \cite{Eichmann:2015kfa} considered the quark-DSE in an approximation where the
explicit contributions from diquarks and baryons become apparent, see Fig.\ref{fig:mesonbaryon} but without
the meson contributions. The diquark and baryon wave functions have been parametrized from explicit solutions
of the baryon's Faddeev equation, discussed in section \ref{sec:baryons}, see Ref.~\cite{Eichmann:2015kfa}
for details. In order to come to terms with the numerical effort of this exploratory study, the authors
furthermore chose $N_f=2$. The resulting pseudo-critical temperature for the chiral transition at zero chemical
potential is at $T_c^{N_f=2} = 211$ MeV, i.e. larger than in the $N_f=2+1$ theory. For $N_f=2$ the ground state
baryons that have been taken into account are the nucleon with quantum numbers $J_P = 1/2^+$ and its 
parity partner with $J_P = 1/2^-$. The rationale for this choice was that the effect of excited states 
with mass $m_B$ is suppressed compared to the nucleon with mass $m_N$ by powers of $m_N^2/m_B^2$. The parity 
partner, however, although initially heavier than the nucleon becomes (approximately) mass-degenerate
once chiral symmetry is restored, i.e. in the high temperature/density phase. In fact, due to different
Dirac structure leading to different signs, the effects of mass-degenerate parity partners onto the quark 
cancel out in the quark mass function. 

What is not yet known in detail is the temperature and chemical potential dependence of the baryons wave 
function.\footnote{Within the NJL-model results for the temperature and chemical potential dependence 
of the nucleon masses have been discussed in \cite{Mu:2012zz,Wang:2013wk}. Lattice results on this issue
are presented in \cite{Aarts:2017rrl}.} 
Ideally these need to be determined consistently from their BSEs evaluated at finite $T$ and $\mu_c$, but
this formidable numerical task is yet to be performed (see section \ref{results:chiral} for first results on mesons).
In order to evaluate the potential impact of such changes, the authors of \cite{Eichmann:2015kfa} considered
two cases: (i) they used wave functions from the vacuum thus neglecting all $T$- and $\mu$-effects in the 
wave functions (they have been taken into account, though, in the baryon propagator); (ii) they modelled
the $\mu$-dependence of the wave function with a strength function $f(\mu)$. Both results are shown in 
Fig.~\ref{fig:phasediag_baryon}. Whereas the changes imposed by baryonic corrections on the CEP are minuscule
using the vacuum wave functions for the baryons, they become somewhat larger when a chemical potential dependence
has been assumed. The result shown in the curve has been obtained at the expense of a sizeable modification of the strength
of the baryon loop by more than 50 \%. Whether such a variation of the baryon wave function and masses with 
chemical potential is realistic or not needs to be investigated in the future. In any case, it is interesting to note
that in principle, such variations are capable to drive the curvature of the cross-over obtained in DSE results
to smaller values matching those obtained in lattice QCD: the result shown in Fig.~\ref{fig:phasediag_baryon}
corresponds to $\kappa = 0.0149$.    

\subsection{Thermodynamics and quark number susceptibilities}\label{results:thermo}
\begin{figure}[t]
        \begin{center}
        \includegraphics[width=0.48\textwidth]{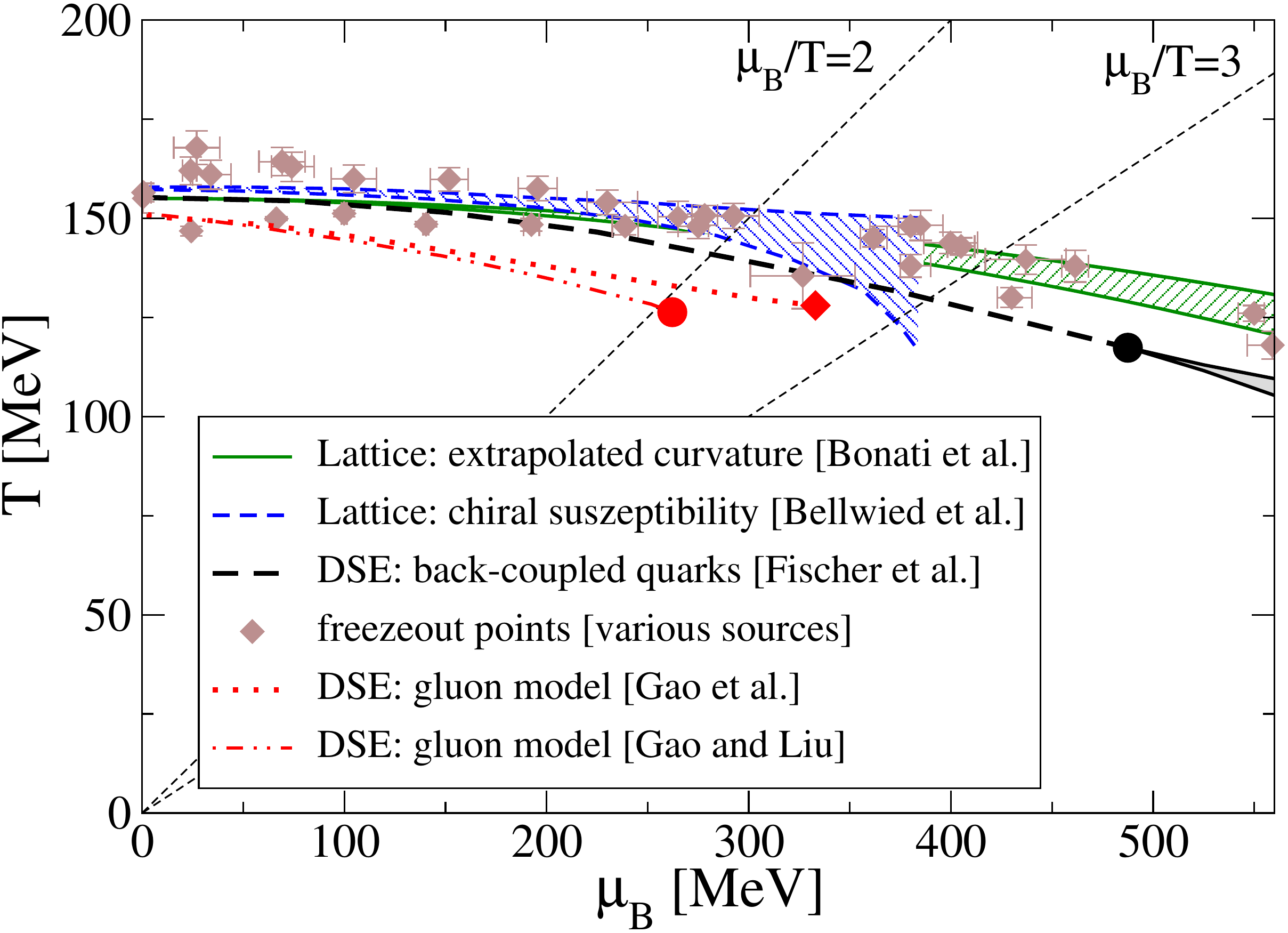}\hfill
        \includegraphics[width=0.48\textwidth]{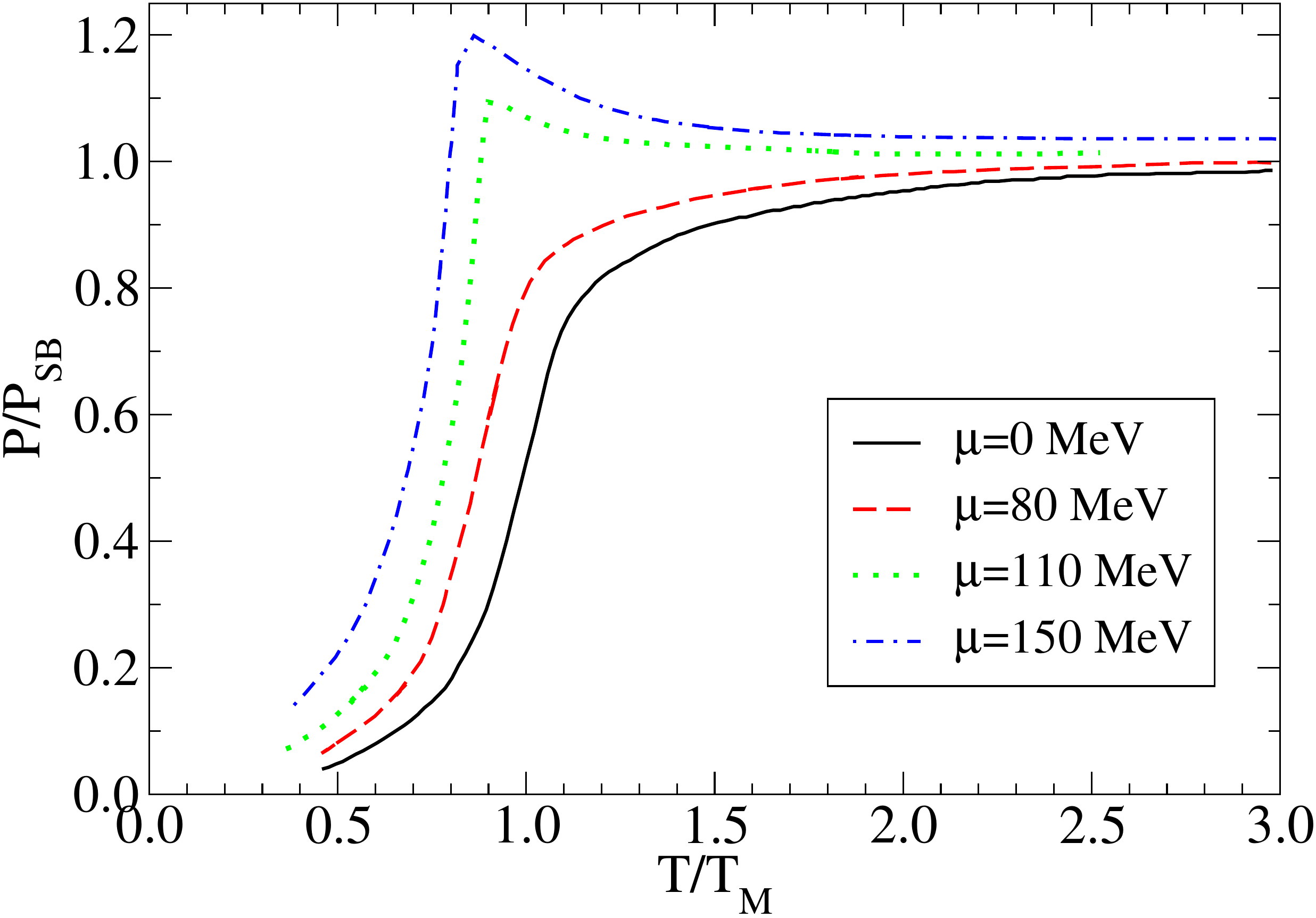}\\\vspace*{3mm}
        \includegraphics[width=0.50\textwidth]{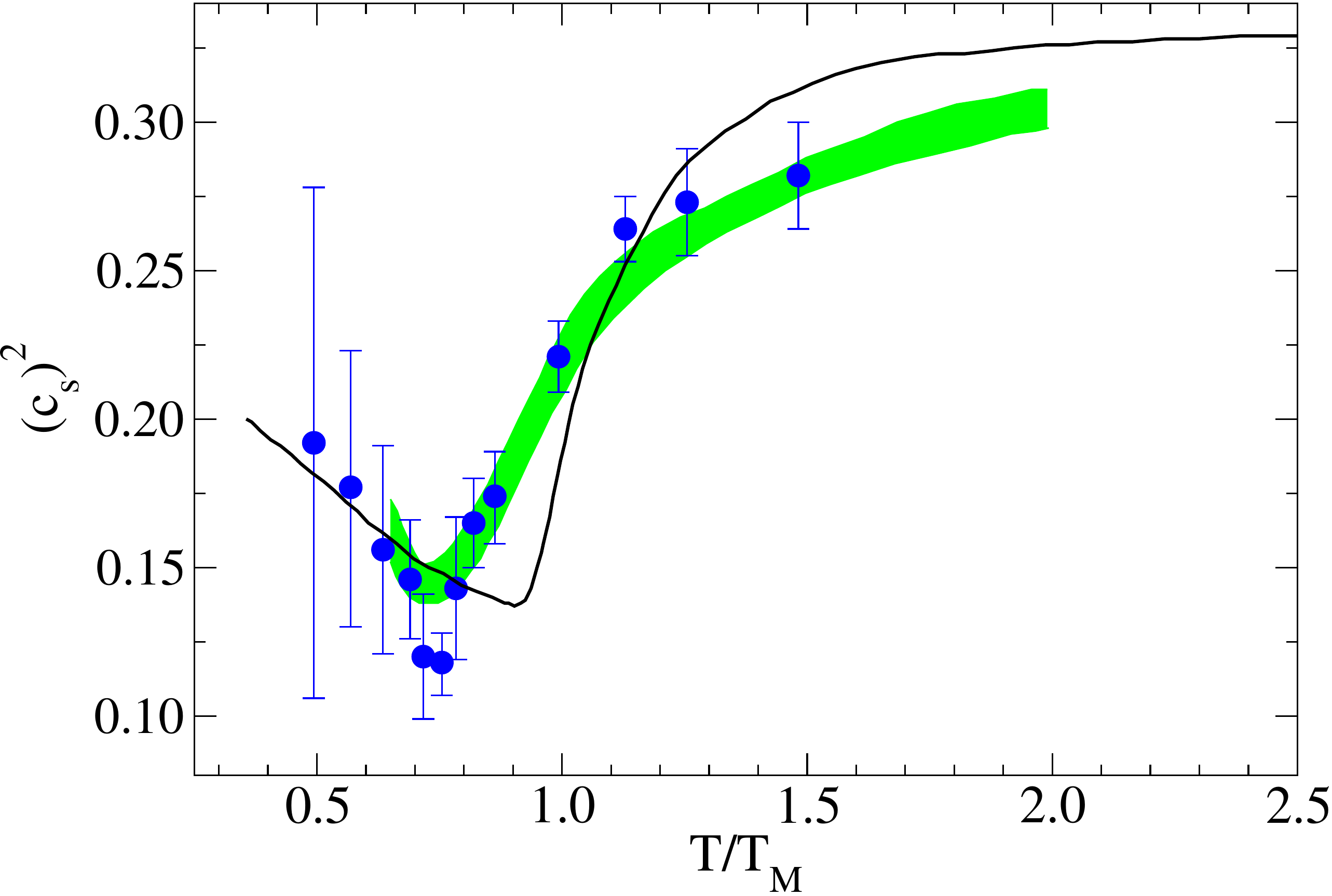}\hfill
        \includegraphics[width=0.48\textwidth]{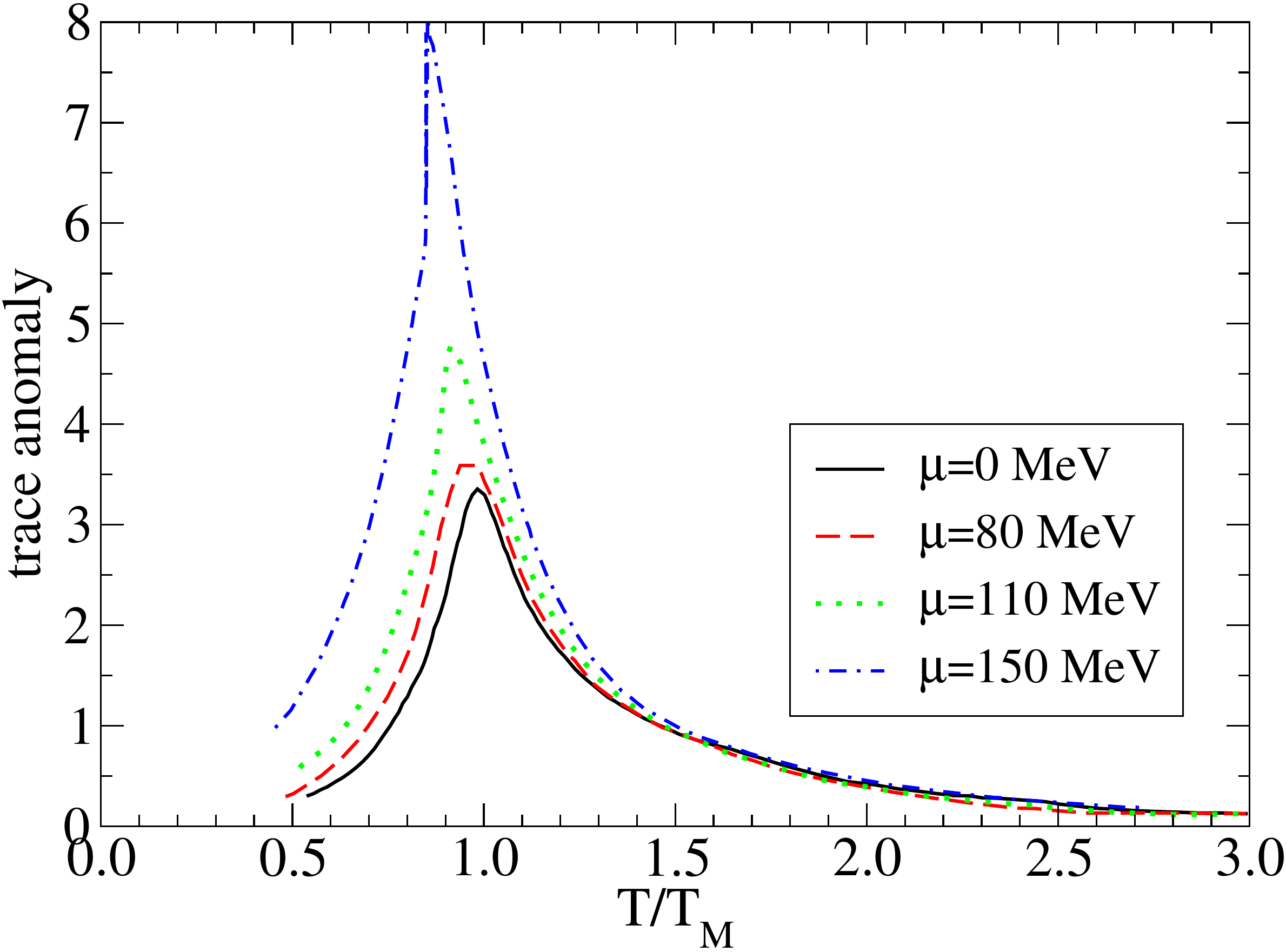}
        \caption{\textit{Upper left diagram:} Phase diagram obtained from DSEs with modelled gluons
        \cite{Gao:2015kea,Gao:2016qkh} (red curve with dotted and dash-dotted lines) compared to the 
        DSE \cite{Fischer:2014ata} (black curve with dashed line) and lattice results 
        \cite{Bellwied:2015rza,Bonati:2018nut} already displayed in Fig.~\ref{fig:phasediag}. 
        \textit{Upper right and lower right diagrams:} Normalised pressure $\mathrm{P}/\mathrm{P}_{\mathrm{SB}}$ 
        and trace anomaly $\mathcal{I}$ from \cite{Gao:2015kea} evaluated for different values of quark chemical 
        potential $\mu$. The temperature is normalised to $T_M$, the location of the peak of the trace anomaly at $\mu=0$;
        see \cite{Gao:2015kea} for details. 
        \textit{Lower left diagram:} Speed of sound from \cite{Gao:2015kea} (black curve) compared to results
        from lattice QCD \cite{Borsanyi:2012cr} (blue dots) and \cite{Bazavov:2014pvz} (green band).}  \label{fig:phasediag_RL}
        \end{center}
\end{figure}

All results in the previous subsections have been obtained in the truncation scheme outlined in section \ref{fulltrunc}
and generalisations thereof, i.e. including the back-reaction of the quarks onto the gluon explicitly. We now
focus on results using the gluon models detailed in section \ref{sec:RL}. Since quark-loop effects alone determine
the number of quark-flavours, $N_f$ is not a well defined quantity in these truncation schemes. In the chiral
limit these truncations lead to a (mean-field) second order transition at $\mu=0$ and consequently they have been
interpreted as models for the two-flavour theory \cite{Gao:2016qkh}. On the other hand, the parameters are tuned 
such that the chiral transition temperatures are much more in line with the ones typical for the $N_f=2+1$-theory.
In the following, we adopt the latter interpretation and interpret the single quark-DSE in these models as the 
one for the light up/down quark with interaction strength generated from a model gluon that incorporates the effects of 
two light and one strange quark loops. 
This allows for a direct comparison with lattice calculations and DSE-results for $N_f=2+1$ (as done anyway 
e.g. in \cite{Gao:2016qkh}). The location of the CEP in the QCD phase diagram has been explored in a broad range
of variants of the truncations discussed in section \ref{sec:RL}
\cite{Qin:2010nq,Shi:2014zpa,Xin:2014ela,Gao:2015kea,Shi:2016koj,Gao:2016qkh}. 
The resulting phase diagram for the set-up with the most detailed gluon model \cite{Gao:2015kea}
and the one with the richest quark-gluon vertex \cite{Gao:2016qkh} is shown in Fig.~\ref{fig:phasediag_RL}. 
For completeness we also show results already discussed in Fig.~\ref{fig:phasediag}. Comparing
the locations of the CEP in the different DSE truncations 
one finds roughly the same critical temperatures, but quite different values for the 
critical chemical potential:
\begin{align}
\mbox{back-coupled quarks, dressed vertex \,\cite{Fischer:2014ata}:}\,\,\,\, (T^{CEP},\mu_B^{CEP}) &= (117,488) \,\mbox{MeV} \nonumber\\
\mbox{gluon model, bare vertex \,\cite{Gao:2015kea}:}\,\,\,\, (T^{CEP},\mu_B^{CEP}) &= (128,333) \,\mbox{MeV} \nonumber\\
\mbox{gluon model, dressed vertex \,\cite{Gao:2016qkh}:}\,\,\,\, (T^{CEP},\mu_B^{CEP}) &= (126,262) \,\mbox{MeV} \nonumber
\end{align}
Both versions of truncations with gluon model find a CEP at rather moderate chemical potential, far lower than 
the result obtained with the back-coupled truncation including the gluon-DSE of Ref.~\cite{Fischer:2014ata}. 
In \cite{Gao:2016qkh} other choices of parameters and vertex truncations are discussed and locations of the
CEP have been determined (although no data for the respective crossover lines are given). None of these choices 
result in as large a chemical potential as the truncation with back-coupling. It also seems as if the details of
different truncations for the quark-gluon vertex do not have a material impact on the location of the CEP
(a conclusion also stated in \cite{Gao:2015kea} and in line with the discussion of section \ref{results:baryons}), 
whereas the treatment of the Yang-Mills sector obviously has.\footnote{This conclusion is also supported by the observation 
of a sizeable shift in the location of the critical end-point when comparing the fully back-coupled truncation scheme
of \cite{Fischer:2012vc} with the one of Ref.~\cite{Fischer:2011mz} where an HTL-approximation for the quark-loop
has been used.}   

Also it seems as if the gluon models tend to produce much too large values for the curvature of the phase boundary 
marked by the cross-over line: The authors of \cite{Gao:2016qkh} report $\kappa=0.038$ from a fit of the whole 
cross-over line from $\mu_B=0$ up to the CEP. However, this value increases to $\kappa = 0.100 (10)$ if only the 
interval $\mu_B \in [0,100]$ is considered in the fit. Such large values seem to be generic for all versions of 
the gluon model that roughly reproduce the (pseudo-)critical temperature $T_c \approx 155$ MeV from the lattice 
at zero chemical potential, see e.g. \cite{Gao:2015kea,Shi:2016koj}. In particular, similarly large curvatures 
are obtained in versions of the gluon model truncation with and without vertex dressing.

On the other hand, there are a number of studies of interesting quantities using the gluon model truncations, that
have not yet been able in the back-coupled truncation scheme. These include thermodynamic quantities and fluctuations
of conserved charges, which we will discuss in the following.

In rainbow-ladder truncations of the quark-DSE the pressure (and subsequently the thermodynamics)
of the system can be obtained via 
\beq
P(S) = \frac{T}{V} \ln Z = \frac{T}{V} \left(\mbox{Tr} \ln[T^{-1}S^{-1}] + \frac{1}{2} \mbox{Tr}[\Sigma S] \right) 
\eeq
where $Z$ denotes the generating functional, $S$ is the fully dressed quark propagator and 
$\Sigma = S_0^{-1}-S^{-1}$ the quark self-energy. This expression can be derived from a 2PI effective 
action at the stationary point, but neglecting contributions from the Yang-Mills sector. 
Ultraviolet divergences are taken care of by a subtraction scheme
described in \cite{Gao:2015kea,Gao:2016hks}. From the pressure one can also determine the entropy density
$s = \partial P/\partial T$, the energy density $\varepsilon = -P + Ts + \mu n$, the 
trace anomaly $\mathcal{I}=\varepsilon - 3P$ and the speed of sound $c_s^2 = \partial P/\partial \varepsilon$.

The results of Ref.~\cite{Gao:2015kea} for the normalised pressure and trace anomaly can be 
seen in the two diagrams on the right of Fig.~\ref{fig:phasediag_RL}. At zero chemical potential,
and for appropriate rescalings, it has been argued in \cite{Gao:2015kea} that the trace anomaly is in qualitative
agreement with results from lattice QCD. At non-zero chemical potential one clearly sees a change
of behaviour for the pressure once the chemical potential approaches the critical one, $\mu_B = 3 \mu = 330$ MeV, 
and beyond: the continuous growth with temperature at small chemical potential turns into one with a maximum 
at the location of the CEP/first order phase transition. For similar values of chemical potential one also observes 
a drastic increase of the trace anomaly. The results for the speed of sound at $\mu=0$ are shown in the bottom
left diagram of Fig.~\ref{fig:phasediag_RL} and compared to results of lattice QCD 
\cite{Borsanyi:2012cr,Bazavov:2014pvz}, again with qualitative and partly quantitative agreement. 
It is furthermore interesting, to consider the thermodynamic details of the first order phase transition for 
chemical potentials larger than the critical one. This has been discussed in Ref.~\cite{Gao:2016hks}. In order 
to study both, the transition from the hadronic to the quark-gluon plasma phase and vice versa, the authors 
included not only bulk entropy contributions but also contributions from interfaces between the bubbles of
different phases in the coexistence region of the phase diagram. These proofed vital to ensure that the 
entropy increases in both directions of the phase transitions.   

\begin{figure}[t]
        \begin{center}
        \includegraphics[width=0.44\textwidth]{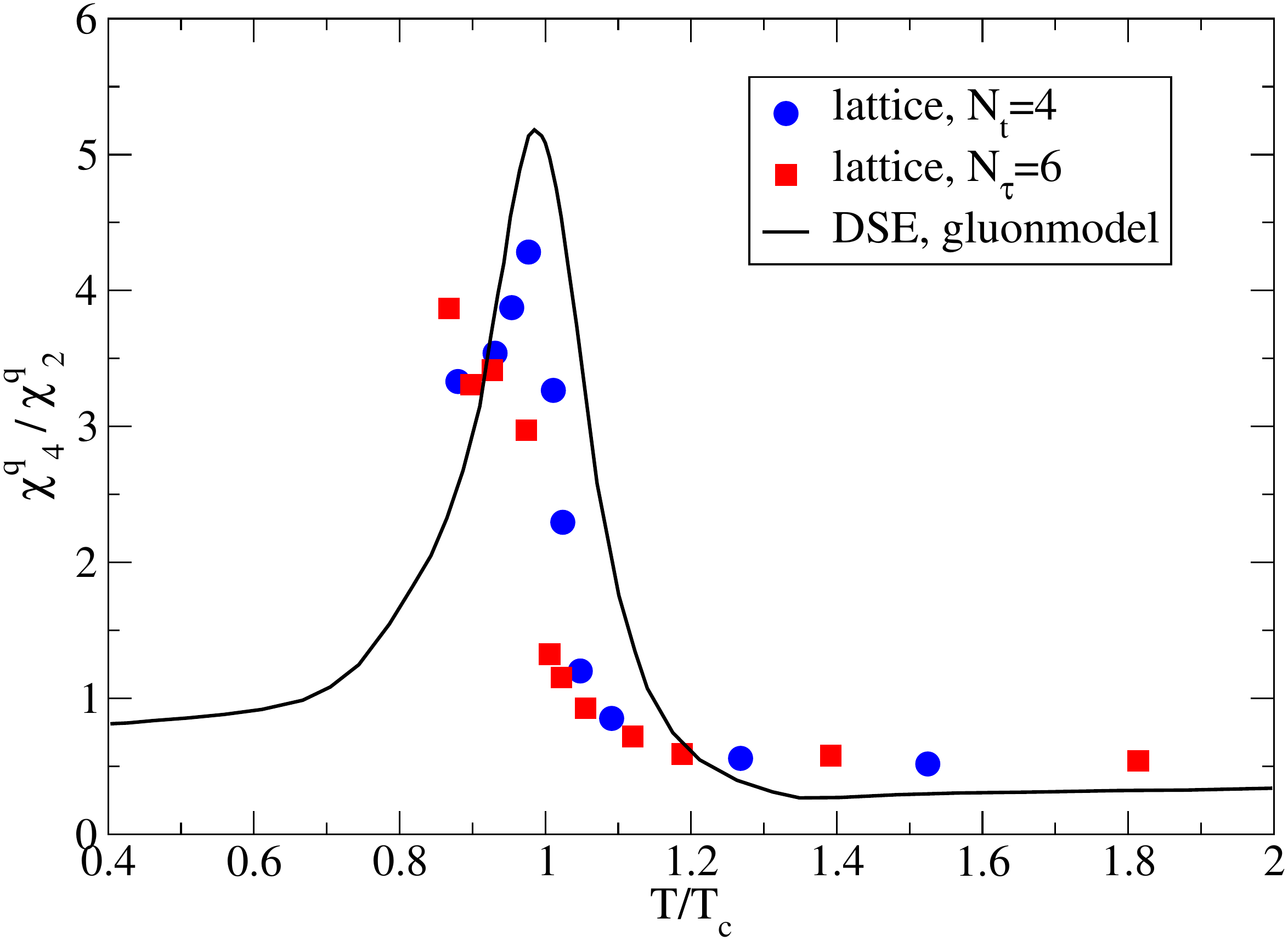}\hfill
        \includegraphics[width=0.54\textwidth]{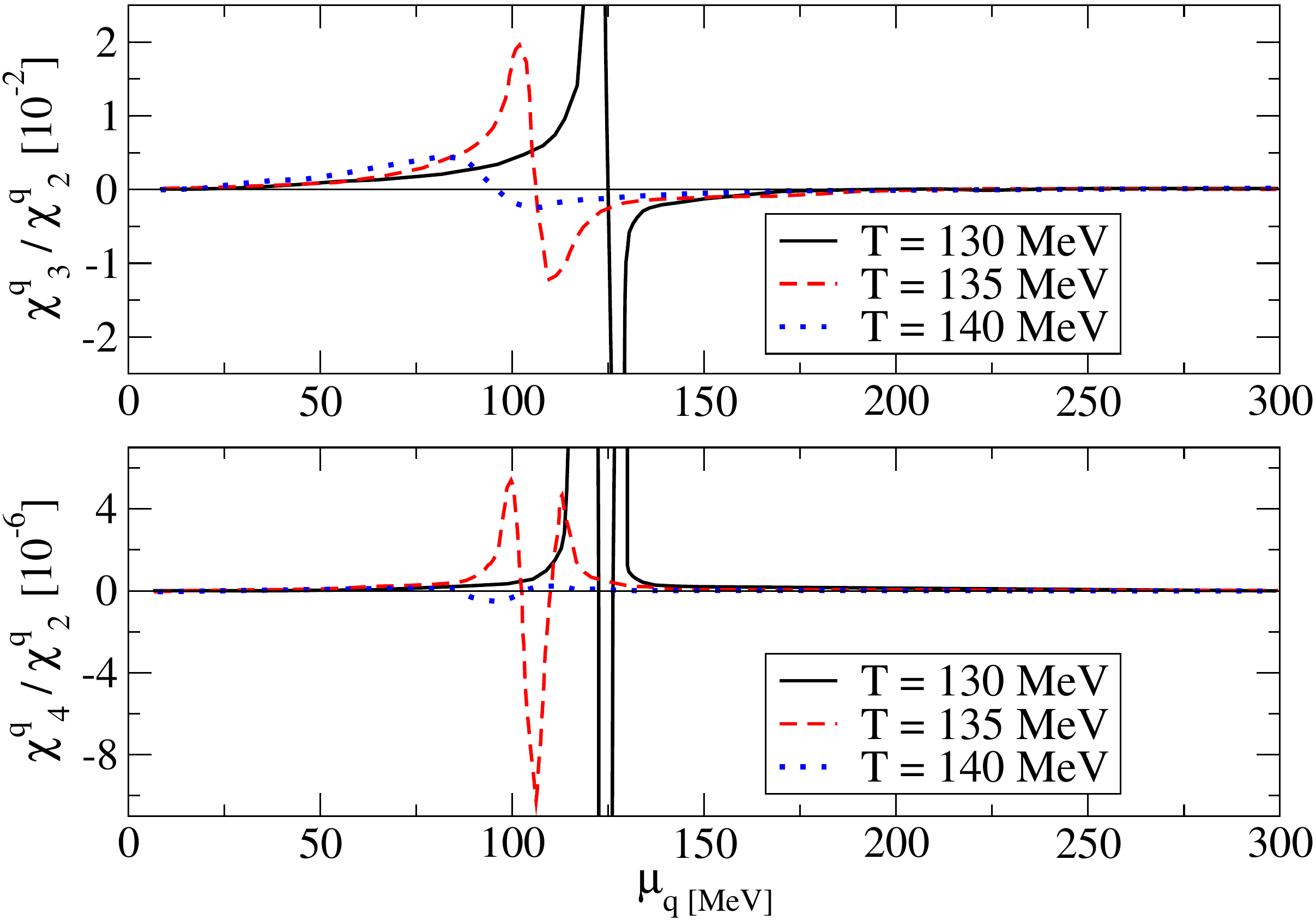}
        \caption{
        \textit{Left diagram:} Temperature dependence of the ratio $\chi_4^q/\chi_2^q$ proportional to the kurtosis
        determined from DSEs (model gluon) compared to the lattice results of \cite{Cheng:2008zh}.  
        \textit{Right diagram:} The quark chemical potential dependence of ratios of generalised susceptibilities.
        Both figures are adapted from Ref.~\cite{Xin:2014ela}. 
        }\label{fig:fluctuations}
        \end{center}
\end{figure}

Finally we summarise results for the quark number susceptibilities and fluctuations, obtained in the model set-up
of Ref.~\cite{Xin:2014ela} (see also \cite{Xu:2015jwa}) and shown in Fig.\ref{fig:fluctuations}. These have been 
determined with the same
chemical potential for both light quarks, i.e. $\mu_u=\mu_d \equiv \mu_q$. In the left diagram the ratio
\beq
\frac{\chi_4^q}{\chi_2^q} = \kappa_q \sigma_q^2
\eeq
proportional to the kurtosis $\kappa_q$ is shown. Results from the gluon model in the lower equation of  
Eq.~(\ref{RL:MT}) are displayed together with the data from the
lattice simulation of Ref.~\cite{Cheng:2008zh}. The qualitative agreement is reasonable, although 
in magnitude one notes a sizeable overshoot of the DSE-results and a shift towards larger temperatures.
With this in mind one can study the behaviour of such ratios at different values of chemical potential.
The results for two ratios are shown in the right plot of Fig.\ref{fig:fluctuations}. Analysing the behaviour 
of the ratios the authors of \cite{Xin:2014ela} were able to extract the corresponding critical end 
point $(T^{CEP},\mu_q^{CEP}) = (129,124)$ MeV in this model in agreement with the one determined from the chiral 
susceptibilities. This clearly demonstrates the presence of signals of the CEP in the fluctuations.  
For corresponding recent studies in the PQM model see \cite{Fu:2016tey,Almasi:2017bhq}. 

\subsection{Quark spectral functions and positivity restoration}\label{results:spectral}

\begin{figure}[t]
        \begin{center}\hspace*{-5mm}
        \includegraphics[width=0.39\textwidth]{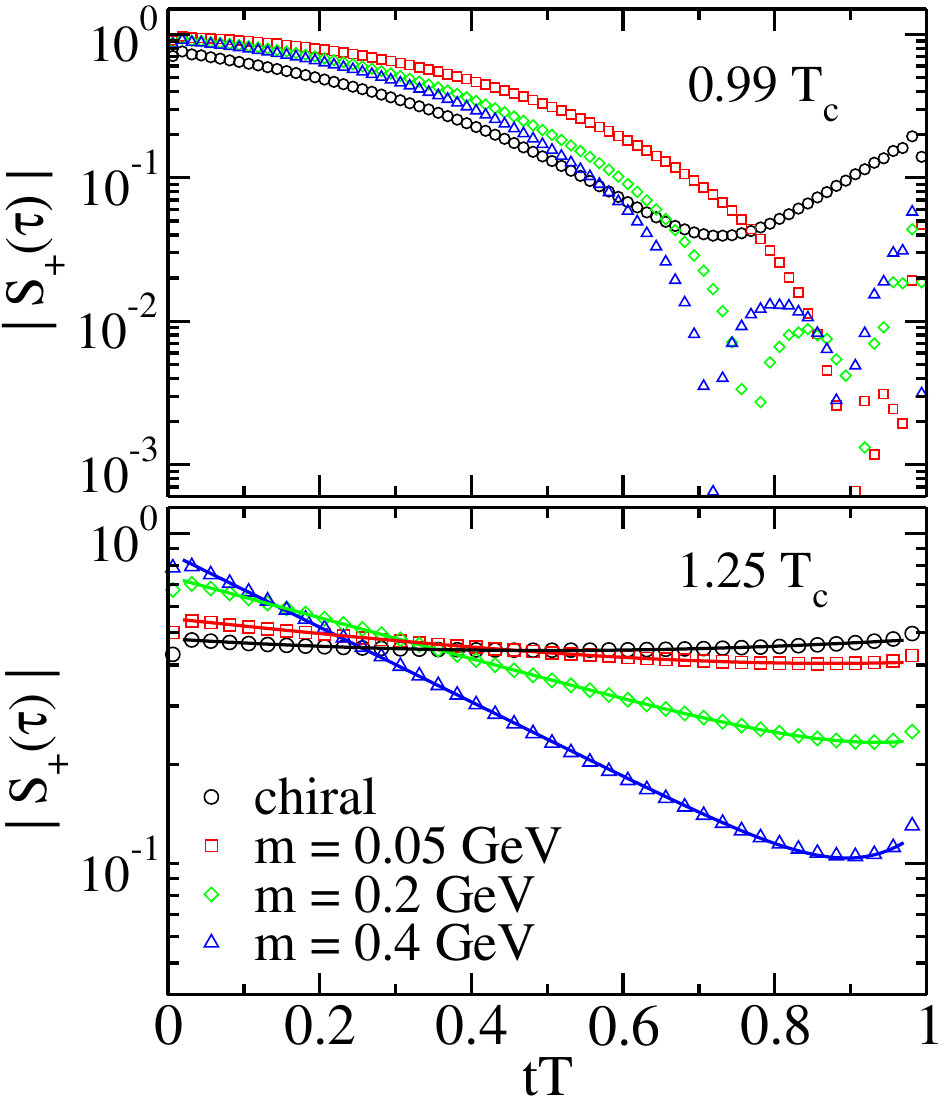}\hspace*{1cm}
        \includegraphics[width=0.41\textwidth]{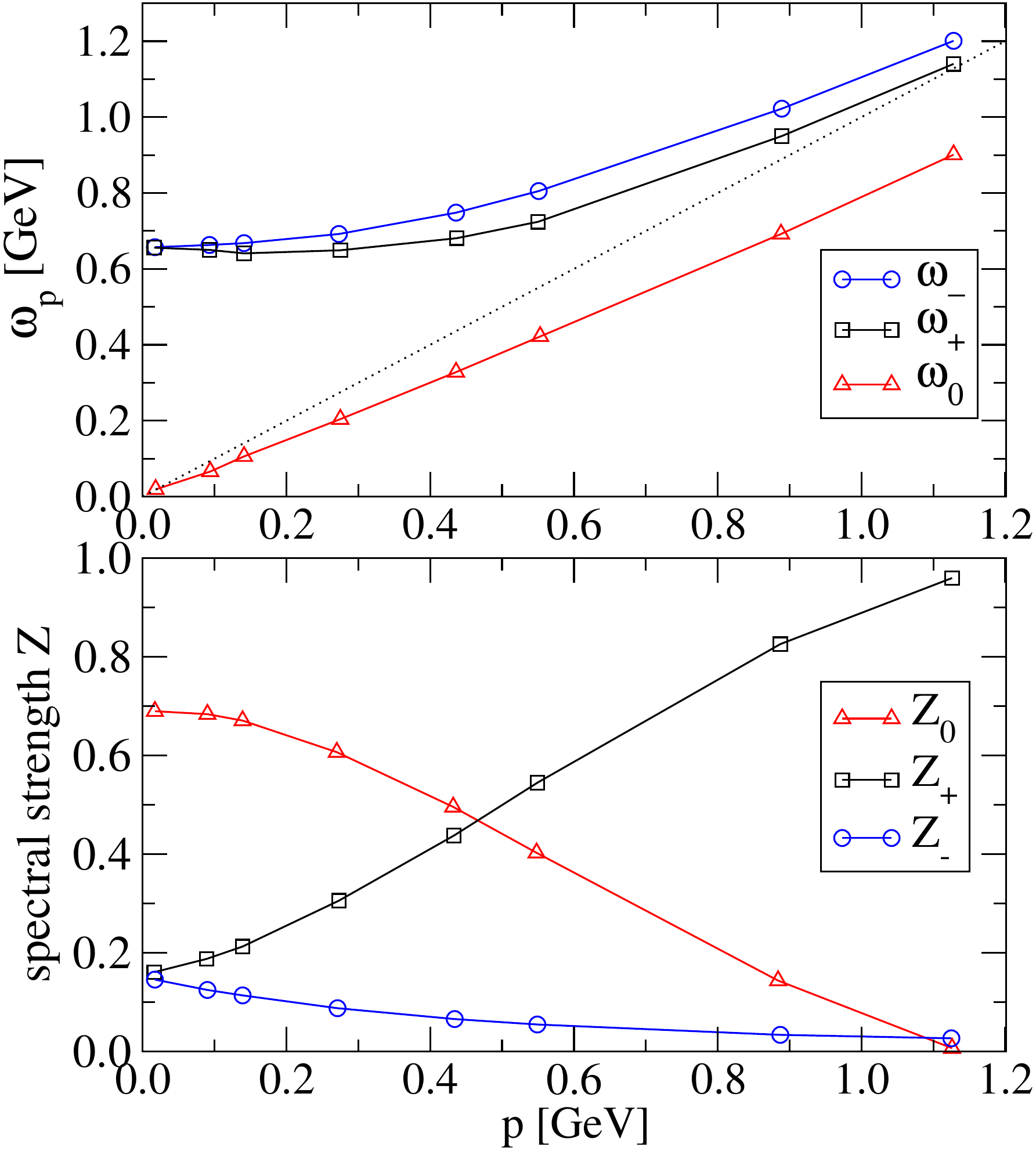}\\\vspace*{3mm}
        \includegraphics[width=0.40\textwidth]{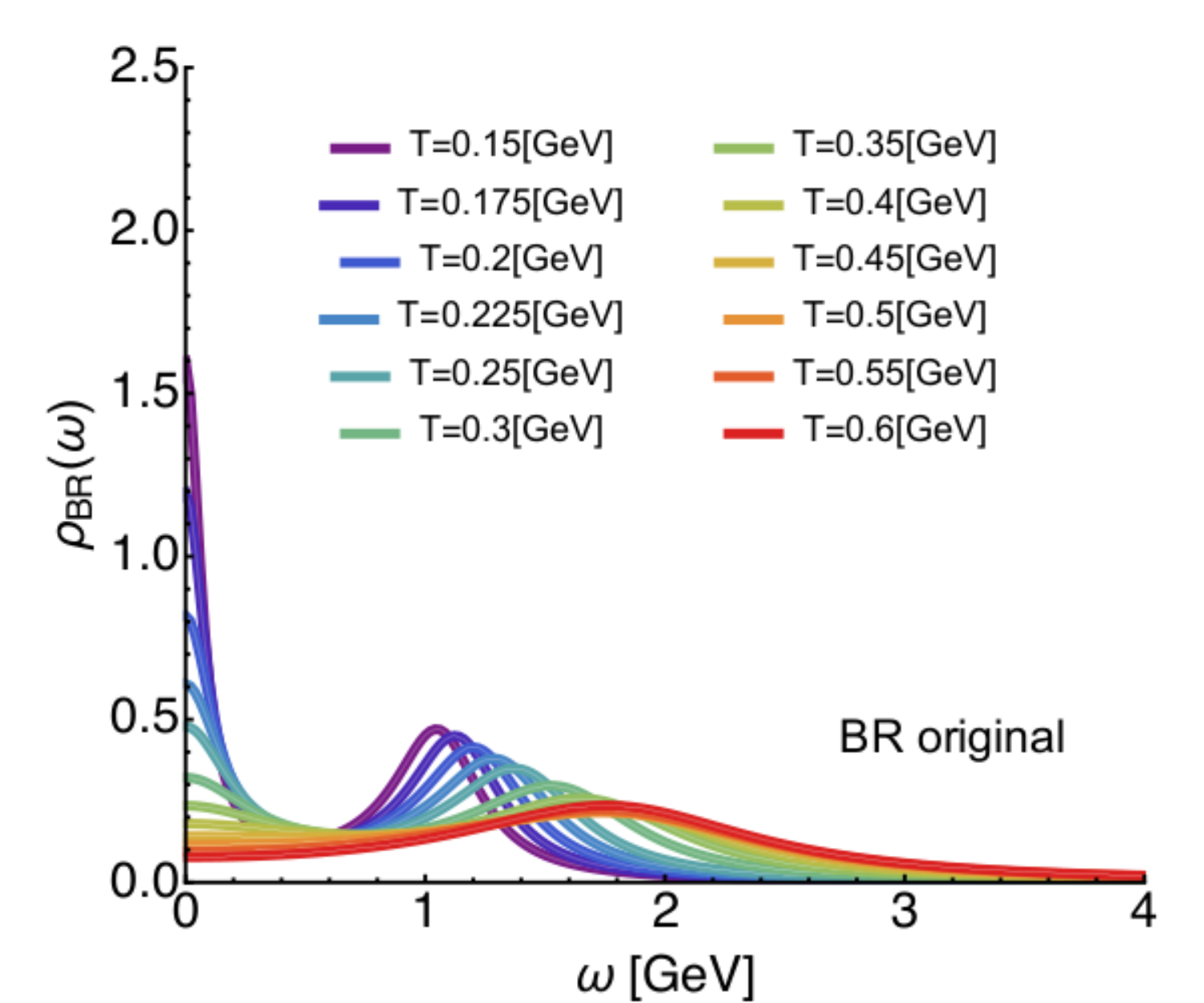}\hspace*{1cm}
        \includegraphics[width=0.40\textwidth]{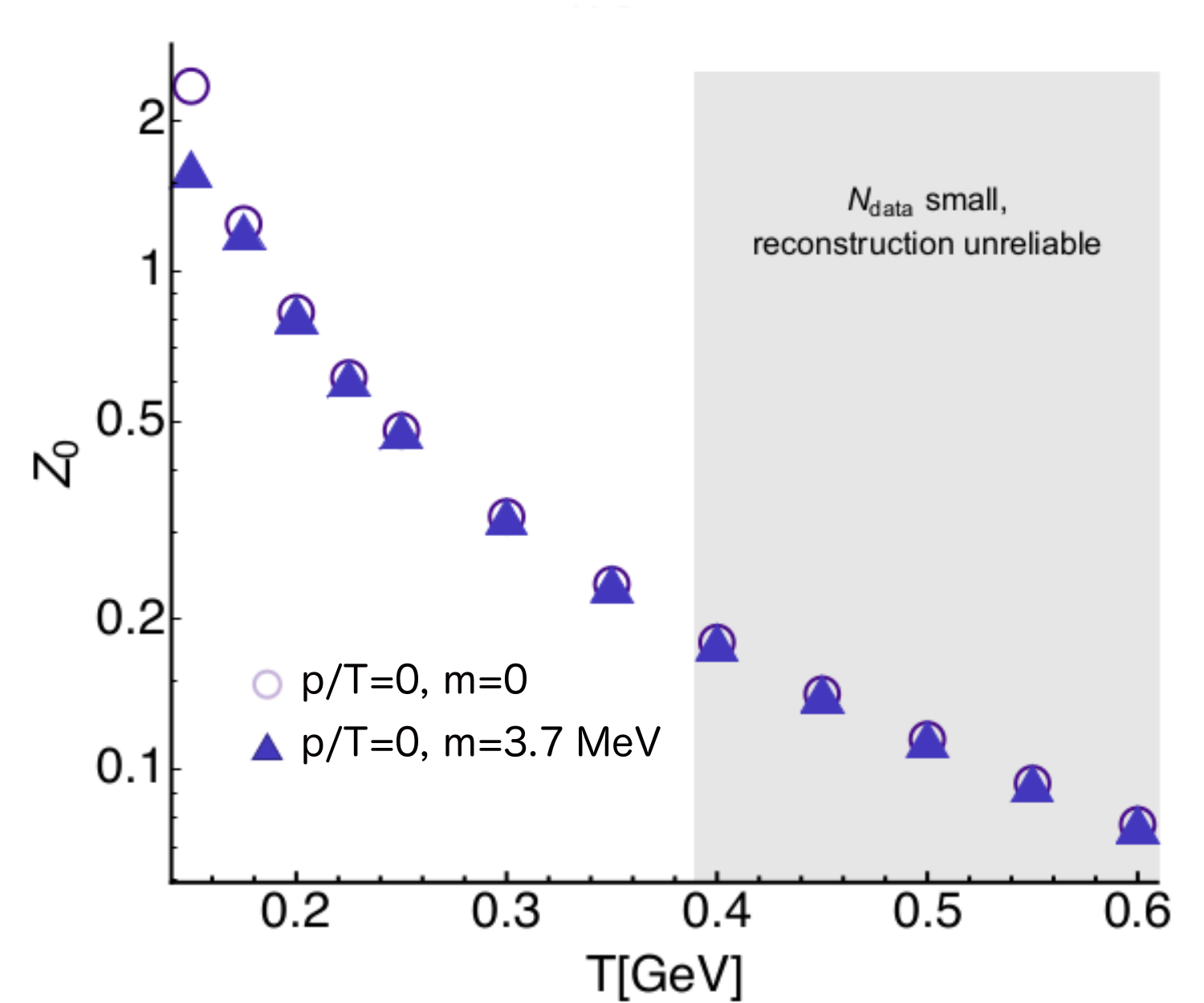}
        \caption{\textit{Upper left diagram:} Absolute value of the Schwinger function $S_+(\tau)$ as a function of 
        time $\tau$ in units of inverse temperature for two temperatures below and above $T_c$ \cite{Fischer:2009gk}.  
        \textit{Upper right diagram:} Dispersion relation of the quark propagator and spectral strength 
        showing three different branches
        at $T=1.1 T_c$; figure adapted from \cite{Gao:2014rqa}.  
        \textit{Lower left diagram:} Spectral function for a range of temperatures above $T_c=142$ MeV 
        (gluon model) \cite{Fischer:2017kbq}.
        \textit{Lower right diagram:} Strength of the zero frequency peak \cite{Fischer:2017kbq}.}  \label{fig:spectral}
        \end{center}
\end{figure}

In order to connect the properties of quarks and gluons in the high temperature quark-gluon plasma phase with
observable quantities it is mandatory to determine their properties at time-like momenta. Moreover,
as discussed in section \ref{sec:positivity}, the analytic structure of quarks and gluon in the complex momentum
plane might give vital clues why we are not able to observe these as asymptotic states. To this end, let us first
discuss the quenched theory which features the first order deconfinement transition of pure SU(3) Yang-Mills theory,
c.f. Figs.\ref{fig:screeningmass} and \ref{fig:heavy_c}. The corresponding Schwinger function of the test quark
has been determined in Ref.~\cite{Fischer:2009gk} using the truncation scheme of section \ref{fulltrunc}; earlier 
results in Ref.~\cite{Bender:1996bm} were obtained using a very simple model. The 
results of \cite{Fischer:2009gk} are shown in the upper left diagram of Fig.~\ref{fig:spectral}. 
There are clear signals for a qualitative change in the Schwinger function at $T_c$.
Above the critical temperature $S_+(\tau)$ is positive and found to be convex. Chiral symmetry restoration for
massless quarks furthermore translates into $S_+(\omega_n)=-S_+(-\omega_n)$ and $S_+(\tau)=S_+(1/T-\tau)$.
Indeed, this symmetry emerges when the current quark mass $m$ of the test quark is decreased.
Below the critical temperature, the Schwinger function changes it behaviour: it becomes concave and for some
quark masses even negative for larger times. As discussed in section \ref{sec:positivity} this entails a 
non-positive spectral function. Similar results have been found in quenched lattice calculations \cite{Karsch:2009tp}
and the model calculations of Ref.~\cite{Bender:1996bm}. Within numerical accuracy, these changes in the Schwinger function
occur at the same temperature as the critical one extracted from the dressed Polyakov loop. This establishes
an interesting connection between the analytic properties of the quark propagator and center symmetry breaking,
which needs to explored in more detail. 

The positive and convex Schwinger function of the quark propagator at temperatures above the chiral transition
opens up the possibility to employ standard MEM methods to determine the spectral function of the quark. This has
been pioneered in Ref.~\cite{Fischer:2009gk} for the quenched theory, refined in
Refs.~\cite{Qin:2010pc,Qin:2013ufa,Gao:2014rqa} for the
truncation of section \ref{sec:RL} with model gluon, and in Ref.~\cite{Fischer:2017kbq} for the model gluon 
and the back-coupled
truncation of section \ref{fulltrunc}. For weak coupling, i.e. large temperatures, reliable results have been 
obtained previously in the hard-thermal loop (HTL) expansion \cite{Braaten:1989mz,Baym:1992eu,Blaizot:1993bb}. 
The quark spectral function there shows two excitations in the dispersion relation, the ordinary quark with 
a positive ratio of chirality to helicity and a collective 'plasmino' mode with a corresponding negative ratio. 
Both have thermal masses of order $gT$ and decay widths of order $g^2T$, where $g$ is the coupling constant. 
The two excitations are accompanied by a continuum contribution from a branch cut in the quark propagator due to
Landau damping, i.e. the absorption of a space-like quark by a hard gluon or hard antiquark. In addition, ultra-soft
fermionic collective excitations have been discussed frequently in the literature, see e.g.
\cite{Lebedev:1989ev,Kitazawa:2005mp,Hidaka:2011rz,Blaizot:2014hka} and references therein. 
By some authors, these have been attributed to the 
breaking of a supersymmetry of the free Lagrangian by temperature or chemical potential effects at weak coupling.
The emerging Nambu-Goldstone mode has been called quasi-goldstino.

Since the DSEs contain the weak coupling perturbative limit, it is not surprising that these structures have 
also been extracted in the MEM-approaches to the DSE-results at large temperatures. It is, however, non-trivial
that these persist in the strong coupling region, i.e. for temperatures close to the pseudo-critical one. 
In the upper right panel of Fig.~\ref{fig:spectral} we show the corresponding results of Ref.~\cite{Gao:2014rqa} 
in the gluon model truncation discussed in section \ref{sec:RL}. The quark and the plasmino branch can be seen 
in the time-like region above the free-fermion demarcation line $\omega \sim p$. Below this line, the ultra-soft 
spectral branch starting at zero frequency has been seen first in \cite{Qin:2010pc} and has been corroborated 
in Refs.~\cite{Qin:2013ufa,Gao:2014rqa,Fischer:2017kbq}. Whereas the quark and the plasmino branches mostly share
their properties with the ones obtained in the HTL-approach, it is not entirely clear whether the ultra-soft branch 
can be identified with the quasi-goldstino mentioned above.\footnote{An alternative interpretation has been given 
in \cite{Gao:2014rqa}, where it has been identified as the temperature analogue of the vacuum Wigner type solution 
of the DSEs, discussed at the end of section \ref{sec:quark}.} To this end one also needs to discuss the residues
$Z_{+,-,0}$ associated with the three quasi-particle branches, which are also displayed
in the upper right diagram of Fig.~\ref{fig:spectral} as a function of momentum. Whereas at large momenta the
ordinary quark branch dominates, at low momenta the peak of the zero mode is clearly the largest.
This behaviour has been confirmed in a different formulation of MEM that allows for a systematic error control
in Ref.~\cite{Fischer:2017kbq}. From the spectral function shown in the lower left plot of Fig.~\ref{fig:spectral}
we again see that for temperatures not too far above $T_c=142$ MeV the dominating part of the spectral function
is the peak at zero frequency. For larger temperatures, however, this peak becomes smaller and its fate at very 
large temperatures (where it could be identified with the pseudo-goldstino) remains an open question. 
Whereas in \cite{Qin:2010pc,Qin:2013ufa,Gao:2014rqa} the peak
disappeared around $T=1.4 T_c$, the authors of \cite{Fischer:2017kbq} found evidence for the persistence
of this peak for much larger temperatures, as can bee seen in the lower right digram of Fig.~\ref{fig:spectral}
for two different current quark masses $m$. Since the region of very high temperatures above $T=400$ MeV is not 
reliably captured due to the sparseness of the grid of Matsubara frequencies, it is not clear whether the 
zero peak vanishes at all at some temperature. In any case, the existence of this zero peak has been confirmed
not only in truncations using the model function for the gluon but also in the truncation of section \ref{fulltrunc}
including back-coupling effects in the Yang-Mills sector \cite{Fischer:2017kbq}. Thus it seems to be a stable, 
truncation independent feature of the quark for temperatures larger than the (pseudo-)critical one. 

A zero frequency mode with different properties as the one discussed above has been extracted analytically
from the quark propagator in the Gribov-Zwanziger approach of Ref.~\cite{Su:2014rma}. It has the interesting
properties that for small momenta the dispersion relation behaves similar as the one seen in the upper right 
diagram of Fig.~\ref{fig:spectral}, whereas at large momenta it approaches the free fermion line. Moreover,
its residue is negative and approaches zero for large as well as vanishing momenta, in marked contrast to the one
discussed above. In the set-up of Ref.~\cite{Su:2014rma}, the properties of this mode have been traced back 
to complex conjugate poles in the gluon propagator of the Gribov-Zwanziger framework. Due to its negativity, 
the resulting quark mode has been interpreted not as a physical excitation but as a remnant of positivity 
violation in the quark propagator also for temperatures above the deconfinement transition. Whether this result 
is in contradiction with the positive and convex Schwinger function observed in the DSE-approaches as discussed
above has to be explored in more detail.  

Finally, we wish to mention that finite temperature gluon spectral functions in the quenched theory have been
determined in the FRG approach in Ref.~\cite{Haas:2013hpa} using a MEM approach that has been adjusted for 
non-positive definite spectral functions. As a result they provided gluonic spectral functions for 
$0.4\, T_c \le T \le 4.5\, T_c$, with a zero temperature extrapolation that agrees with the result from explicit
DSE-calculations discussed in section \ref{sec:gluon}. These gluon spectral functions have then been used to 
compute the viscosity over entropy ratio in this temperature range. In agreement with other approaches, they
found a minimum of the ratio at temperatures slightly above $T_c$, which is close but above the lower bound
$\eta/s = 1/(4\pi)$ derived from the AdS-CFT correspondence \cite{Kovtun:2004de}. An interpretation of the results
in terms of a glueball resonance gas at low temperature and a high temperature behaviour consistent with HTL-resummed
perturbation theory has been given in \cite{Christiansen:2014ypa} together with a first estimate for $\eta/s$
in full QCD. A very recent application to heavy-ion collision phenomenology within the framework of hydrodynamics 
is reported in \cite{Dubla:2018czx}. For $N_f=2+1+1$ quark flavours, results for the gluon spectral functions
from lattice QCD have been discussed in \cite{Ilgenfritz:2017kkp}. In the Gribov-Zwanziger approach, shear 
and bulk viscosities have been determined in \cite{Florkowski:2015dmm}.

\subsection{Colour superconductivity}\label{results:colorSC}

We conclude this chapter with a brief summary of results obtained in the low temperature and high density region of the
QCD phase diagram. As outlined in section \ref{gen:sketch}, this is the region where we expect to find a
first order transition for the chiral restoration followed by a region where Cooper pairs of quarks condense and
form the $SU(3)$ analogue of superconductivity. The study of this form of matter using the DSE-approach has been
pioneered in a series of four papers \cite{Nickel:2006vf,Nickel:2006kc,Marhauser:2006hy,Nickel:2008ef}. Subsequently,
it has been extended to the truncation with back-coupling of the quarks onto the gluons: In a first step, this has 
been done in a hard 
thermal and dense loop approximation \cite{Muller:2013pya} (similar to \cite{Fischer:2011mz}) and in a second step 
\cite{Muller:2016fdr} in the fully back-coupled truncation discussed in section \ref{fulltrunc}. In the following we
discuss the results of the most advanced truncation scheme of Ref.~\cite{Muller:2016fdr}.   

In order to study superconducting phases in the DSE framework one has to gain access to the associated 
condensates. To this end the Nambu-Gorkov formalism has been used, generalising the quark-propagator $S$ and the
quark self-energy $\Sigma$ to the forms 
\beq
S(p) = \left(
\begin{array}{ll}
S^+(p) & T^-(p) \\
T^+(p) & S^-(p) 
\end{array}\right)\,,
\hspace*{2cm}
\Sigma(p) = \left(
\begin{array}{ll}
\Sigma^+(p) & \phi^-(p) \\
\phi^+(p)   & \Sigma^-(p) 
\end{array}\right)\,.
\eeq
The normal components $S^\pm$ and $\Sigma^\pm$ correspond to particle and charge conjugate particle propagators
and self-energies, whereas the off-diagonal components are related to the colour-superconducting condensates.
The colour and flavour components of the propagator have to be adjusted to the situation under scrutiny. In
the colour-flavour locked phase (CFL) all flavour and colour components are arranged in a one-to-one correspondence;
this form of pairing is known to prevail in the limit of very high densities accessible in hard-dense-loop perturbation
theory \cite{Alford:1998mk}. In this limit, the mass differences between the up/down and the strange quark are
negligible as compared to the scale set by the chemical potential. However, at lower chemical potential this
mass difference can become important and the two-flavour colour superconducting state (2SC) may be energetically
favoured. In this state only the up/down quarks (cross-)pair leaving the strange quarks as 
spectators.\footnote{Taking boundary conditions such as local neutrality and beta equilibrium into account even
this type of pairing might be disfavoured as compared to pairings of same quarks in each flavour channel. Symmetries
then suggest Copper pairs in spin 1 channels with a potential preference to the so-called colour-spin locked 
phase (CSL) \cite{Schafer:2000tw,Alford:2002rz,Schmitt:2004et}. In the DSE-approach, this possibility has been explored 
in Ref.~\cite{Marhauser:2006hy}.}. It is of potential great interest in connection to the physics in the interior of
neutron stars to explore the phase boundary between the CFL and the 2SC states. This has been one of the main points
of the studies in the DSE approach. In contrast to NJL-model studies \cite{Buballa:2003qv}, the early DSE calculations
indicated that the CFL phase is dominant for chemical potentials down to the chiral phase transition
\cite{Nickel:2006vf,Nickel:2006kc}. This result has been revised in the more sophisticated truncation schemes 
of Refs.~\cite{Muller:2013pya,Muller:2016fdr} reintroducing stable 2SC phase(s) directly after the chiral restoration.

\begin{figure}[t]
        \begin{center}
        \includegraphics[width=0.50\textwidth]{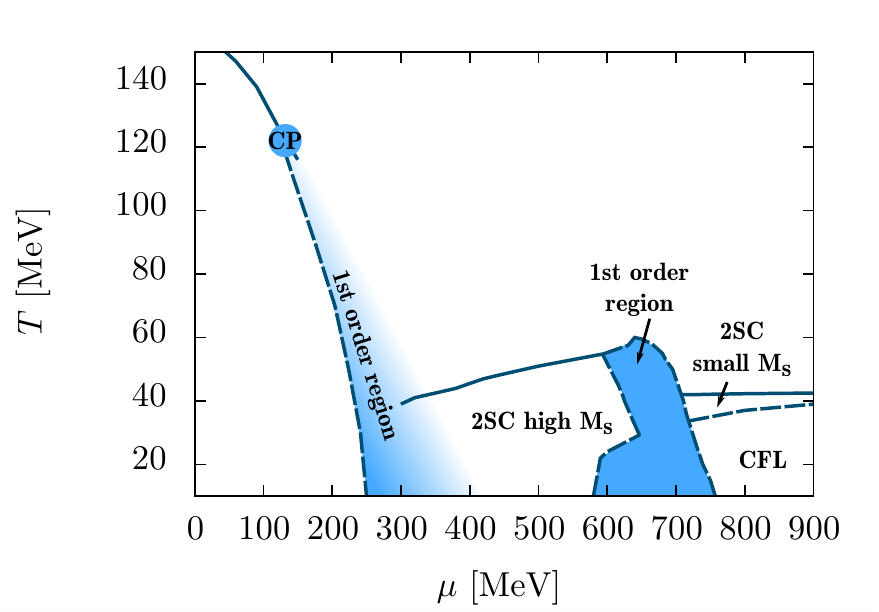}\hfill
        \includegraphics[width=0.46\textwidth]{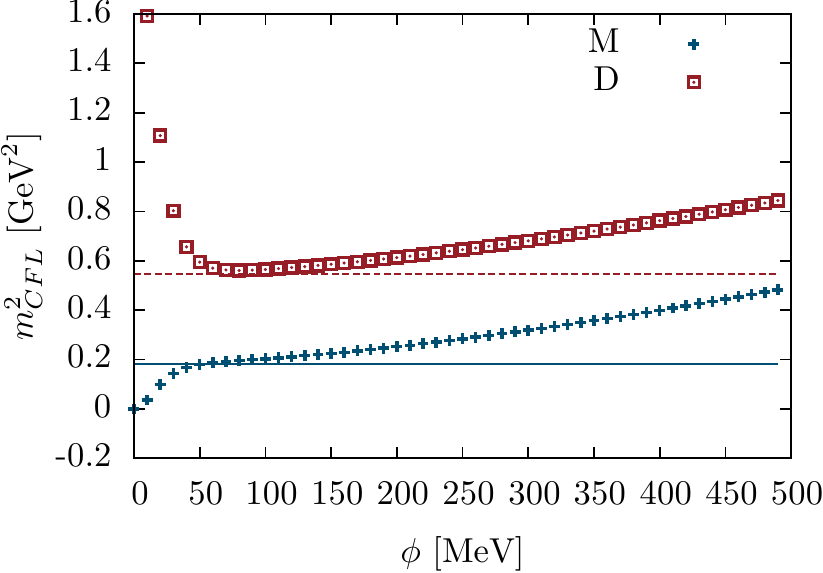}
        \caption{
        \textit{Left diagram:} QCD phase diagram as a function of temperature 
        and quark chemical potential including phases of colour-superconductivity. 
        \textit{Right diagram:} Meissner and Debye masses (dots) as function of the gap parameter $\phi$
        in comparison with weak-coupling results 
        \cite{Rischke:2000ra} (lines) for the CL phase. 
        Both diagrams taken from Ref.~\cite{Muller:2016fdr}.  
        }\label{fig:super}
        \end{center}
\end{figure}
The results of Ref.~\cite{Muller:2016fdr} for the QCD phase diagram with $N_f=2+1$ flavours is shown 
in the left diagram of Fig.~\ref{fig:super}. Let us first compare the location of the critical end-point with the 
result of Ref.~\cite{Fischer:2014ata} discussed in section \ref{results:2p1}, Fig.~\ref{fig:phasediag}. Compared
to Eq.~(\ref{eq:CEP}) the critical end-point $(T^{CEP},\mu_B^{CEP})=(120,400) \,\mbox{MeV}$ seen in Fig.~\ref{fig:super} occurs
at roughly the same temperature but smaller chemical potential. This can be attributed to the omission of the
Ball-Chiu dressing of the quark-gluon vertex in \cite{Muller:2016fdr} that affects the back-coupling of the quarks
onto the gluon sector. As a consequence, also the location of the chiral first order transition at zero temperature
is shifted to smaller values than expected, cf. the discussion in section \ref{sec:curvature}. In the superconducting
phase, however, the truncation used in Ref.~\cite{Muller:2016fdr} unfolds its full power even including non-diagonal
terms in the quark-gluon vertex that are important for the CSC phases. Consequently, the phase structure seen in the 
left diagram of Fig.~\ref{fig:super} is the most elaborate extracted from DSEs (and arguably also in general). 
In the region of large chemical potential one finds the usual CFL-phase in accordance with the hard-dense loop results.
At zero temperature, when the chemical potential is lowered this phase turns into a 2SC-state which persists down 
to the region of the chiral first order transition. 
Due to the elaborate construction of the quark-gluon vertex used in \cite{Muller:2016fdr} the
authors did not have a thermodynamic potential readily at their disposal. Consequently, only first order 
coexistence regions 
can be shown bounded by spinodals. An interesting second region of 2SC-favoured pairing occurs at finite temperatures
and large chemical potential above the CFL-region. In contrast to the 2SC region at smaller chemical potentials, this
region is characterised by small values of the screening mass of the strange quark defined by 
\beq
M_s(\vect{p},\omega_p) = \frac{B_s(\vect{p},\omega_p)}{C_s(\vect{p},\omega_p)}_{|_{\vect{p}=0,\omega_p=\pi T}}
\eeq
with scalar dressing function $B$ and temporal dressing function $C$ of the strange quark propagator, 
cf. Eq.(\ref{quark}). The two 2SC phases are not continuously connected but separated by a large spinodal region 
(dark blue in the plot) which borders at four different phases - the CFL-phase, the two 2SC phases and the 
quark-gluon plasma phase. 
Within this region, all phase transitions between these four phases are first order. Outside this region,
the transitions between the 2SC phases and the quark-gluon plasma phase are found to be second order and take
place between $T=40-60$ MeV. 

In the right diagram of Fig.~\ref{fig:super} we display the quark contribution to the gluon Debye (electric)
and Meissner (magnetic) screening masses in the weak coupling limit defined by
\begin{align}
(m^{ab}_{D,M})^2 &= \lim_{\vect{p}\rightarrow 0} \Pi^{ab}_{L,T}(\vect{p},\omega_p=0)\,,
\end{align} 
where $\Pi^{ab}_{L,T}$ denote the transverse and longitudinal quark-loop contributions to the gluon self-energy.
These masses are shown as a function of the gap parameter $\phi$, which enters the quark self energies
via $\phi^+ = \gamma_5 \phi_i M_i$. Here the matrices $M_i$ reflect the colour-flavour structure of the
superconducting phase and for the comparison with the weak coupling limit $\phi$ is taken as constant, 
see \cite{Muller:2016fdr} for technical details.  
The electric and magnetic masses agree with the weak coupling results in the region $ T \ll \phi \ll \mu$, 
where one expects the weak
coupling expansion to hold. This is an important cross-check of the calculation. In other regions sizeable deviations
occur indicating the strong coupling non-perturbative nature of the problem. For more detailed results on the
screening masses we refer the interested reader to Ref.~\cite{Muller:2016fdr}.  

A further interesting problem, addressed within the DSE approach in Ref.~\cite{Nickel:2008ef} and recently followed 
up in Ref.~\cite{Alford:2017ale} is the one of charge neutrality. For the CFL phase it has been argued in
Ref.~\cite{Rajagopal:2000ff} that charge neutrality is satisfied exactly. Thus no electrons would be allowed in this 
phase; it would be an insulator. This has been confirmed in self-consistent NJL-type calculations
\cite{Ruester:2005jc,Abuki:2005ms} which produces energy independent gap functions. In contrast, the DSE-approach
takes into account the full energy and momentum dependence of the gap functions which in turn opens the possibility
of deviations from charge neutrality. Indeed, this has been found in Ref.~\cite{Nickel:2008ef}. 
While at the time it was not clear whether this finding is robust enough to persist in even more elaborate truncations,
it has recently gained support from the study of Ref.~\cite{Alford:2017ale} in a different framework.

\newpage

\section{Outlook}\label{sec:sum}
With the ongoing experimental program at ALICE/CERN and the beam energy scan at RHIC/BNL as well as 
the future dedicated programs at NICA and HADES/CBM/FAIR, it remains a major task and challenge for 
the theoretical approaches of QCD to make solid qualitative and quantitative predictions for the
potentially rich structure of the QCD phase diagram and the associated physics of strongly interacting 
matter. One of these approaches, the Dyson-Schwinger equations of QCD, has been reviewed in this article.
In principle DSEs can be used with two essentially different goals in mind: (i) they may serve as devices 
for model building with a prime focus on qualitative features of the strong interaction and the 
exploration of phenomenological effects with low numerical costs; or (ii) they can be used as heavy 
duty machinery to study the physics of QCD in a systematic expansion in terms of n-point functions.
In this review we presented results that originated in the past decade from first steps in the second 
direction.  
   
Contemporary truncation 
schemes of DSEs at finite temperature and chemical potential are advanced to a point which enables us to 
explore the physics of the Columbia plot and its extensions to real and imaginary chemical potential as well as
the structure of the QCD phase diagram at realistic quark masses in a systematic and meaningful way. Since 
the back-coupling of the quarks 
onto the gluons is handled explicitly, variations of the number of active quarks can be assessed systematically.
The physics of deconfinement for heavy quark masses is accessible via order parameters such as 
the Polyakov loop potential and the dressed Polyakov loop. The associated change of the analytic structure
of the quark propagator has been studied in some detail. At physical quark masses, results at zero temperature
can be compared with available lattice data. This shows very good agreement for quantities such as the 
temperature behaviour of the chiral order parameter, the quark condensate, and the unquenched gluon. Furthermore,
the extrapolations of lattice QCD to finite chemical potential are in agreement with the results from DSEs.
Based on the combined evidence from functional methods and lattice gauge theory, the appearance of a critical
end point for chemical potentials $\mu_B/T \le 2$ is clearly disfavoured. Instead, the DSEs find a CEP at
much large values but still close to the extrapolated curve of the chiral transition found on the lattice.
The further improvement of the truncations used to extract this CEP is an important and necessary task for
the future. To this end, results using other functional approaches such as the functional renormalisation group
that corroborate (or reject) this finding are highly desirable.

The very existence of a CEP is potentially called into question by the possibility of more complicated phases
such as the ones featuring inhomogeneous condensates. Although many calculations including 
inhomogeneous phases merely replace the CEP with a Lifshitz point (thereby rendering searches for the location 
of the CEP not including this possibility still meaningful) this may not be the case in general and needs to 
be explored further.

An important task for the next years is to intensify contact with experimental heavy ion physics in all possible
respects. Important interfaces are the calculation of fluctuations and ratios thereof, the calculation of 
thermodynamic quantities and transport coefficients and the access to spectral functions not only of quarks
and gluons but also of light and heavy mesons. Promising first steps in this direction have been discussed in 
this review and will be expanded in the future.

\section*{Acknowledgements}

It is a great pleasure to thank Reinhard Alkofer, Jacqueline Bonnet, Romain Contant,
Gernot Eichmann, Leonard Fister, 
Pascal Gunkel, Markus Q. Huber, Philipp Isserstedt, Jan Luecker, Axel Maas, Jens Mueller, Niklas Mueller, 
Dominik Nickel, Alexander Rothkopf, 
Helios Sanchis-Alepuz, Jan M. Pawlowski, Christian A. Welzbacher and Richard Williams for pleasant and fruitful
collaborations on topics discussed in this review. Furthermore I would like to thank
Gert Aarts, Szabols Borsanyi, Jens Braun, Michael Buballa, Wolfgang Cassing, Tetyana Galatyuk, Jeff Greensite,
Kenji Fukushima, 
Frithjof Karsch, Volker Koch, Yu-Xin Liu, Swagato Mukherjee, Joannis Papavassiliu, Rob Pisarski, Owe Philipsen, 
Hugo Reinhardt, Craig D. Roberts, Bernd-Jochen Schaefer, Sebastian M. Schmidt, Lorenz von Smekal and Nu Xu 
for inspiring discussions.
I am indebted to Reinhard Alkofer and Bernd-Jochen Schaefer for a critical reading of the manuscript.   
This work was supported by the Helmholtz International Center for FAIR within the LOEWE program of the
State of Hesse and the BMBF under contracts No.~05P15RGFCA and No.~05P18RGFCA.

\appendix
\renewcommand*{\thesection}{\Alph{section}}

\section{Multiplicative renormalizability in the DSEs for the propagators}\label{renorm}
In the following we discuss the renormalization of the Dyson-Schwinger equations for 
the ghost, gluon and quark propagators at zero temperature and chemical potential.
The case of finite temperature and chemical potential is completely analogue, since 
no new divergences may appear at finite temperature and chemical potential \cite{Das:1997}. 
The discussion extends the one on the DSE for the ghost propagator given in Ref.~\cite{Fischer:2008uz} 
to include the DSE of the gluon propagator and the one for the quark.

Let us start with the Yang-Mills sector. Multiplicative renormalizability of QCD implies 
the following relations between unrenormalized and renormalized ghost, gluon and quark 
dressing functions as well as the ones for the dressing functions of the 
ghost-gluon $\Gamma_\mu^{ghgl}$, three-gluon $\Gamma_{\mu\nu\sigma}^{3g}$ 
and quark-gluon $\Gamma_\mu^{qg}$ vertices.  
\begin{eqnarray} G(p^2,\mu^2)
  \widetilde{Z}_3(\mu^2,\Lambda^2) &=& G^0(p^2,\Lambda^2)\,, \\
  Z(p^2,\mu^2) Z_3(\mu^2,\Lambda^2) &=& Z^0(p^2,\Lambda^2)\,, \label{glueZ}\\
  A(p^2,\mu^2) Z_2^{-1}(\mu^2,\Lambda^2) &=& A^0(p^2,\Lambda^2)\,, \label{quarkZA}\\
  B(p^2,\mu^2) Z_2^{-1}(\mu^2,\Lambda^2) &=& B^0(p^2,\Lambda^2)\,, \label{quarkZB}\\
  g(\mu^2) Z_g(\mu^2,\Lambda^2) &=& g^0(\Lambda^2)\,, \label{couplingZ}\\
  \Gamma^{ghgl}_\mu(p,q,\mu^2) \widetilde{Z}_1^{-1}(\mu^2,\Lambda^2)&=&  \Gamma_\mu^{ghgl,0}(p,q,\Lambda^2)\,,\\ 
  \Gamma^{3g}_{\mu\nu\sigma}(p,q,\mu^2)  {Z}_1^{-1}(\mu^2,\Lambda^2)&=&  \Gamma_{\mu\nu\sigma}^{3g,0}(p,q,\Lambda^2)\,,\\ 
  \Gamma^{qg}_\mu(p,q,\mu^2){Z}_{1f}^{-1}(\mu^2,\Lambda^2)&=&  \Gamma_\mu^{qg,0}(p,q,\Lambda^2)\,.  
  \label{eq1} 
\end{eqnarray}
Here, besides momentum unrenormalized dressing functions depend on an ultraviolet cut-off 
$\Lambda$ and are denoted with superscript zero, whereas the renormalized dressing 
functions depend on the renormalization point $\mu^2$ instead. The renormalization factors
have been introduced in Eq.~(\ref{rescaling}).

Note that the right hand sides of Eqs.~(\ref{eq1}) are independent of
the renormalization point. Thus a finite re-normalization, i.\ e., a
change in the renormalization point from $\mu^2$ to $\nu^2$ is described by
\begin{eqnarray}
G(p^2,\nu^2)  &=& G(p^2,\mu^2) 
\frac{\widetilde{Z}_3(\mu^2,\Lambda^2)}{\widetilde{Z}_3(\nu^2,\Lambda^2)}  \label{eq2}
\end{eqnarray}
for the ghost dressing function and similar relations for all other dressing functions.
Furthermore the identities (\ref{Zsti}) between the renormalization factors are extremely useful. 

In the following we demonstrate explicitly, that the Dyson-Schwinger equations for the gluon and
quark propagators satisfy multiplicative renormalizability. A corresponding demonstration for the
ghost-DSE can be found in the appendix of Ref.~\cite{Fischer:2008uz}.
Consider the DSE for the gluon propagator in symbolic notation, where we kept all Lorenz indices as
well as (almost) all dependencies on momenta implicit and instead highlight the dependencies of the
dressing functions and renormalization factors on the renormalization 
point $\mu^2$ (e.g.$\Gamma^{3g}_{\mu\nu\sigma}(p,q,\mu^2) \rightarrow \Gamma^{3g}(\mu^2)$; all symmetry
factors colour factors, multiple factors of $2\pi$ etc. are absorbed in the integration symbol $\int$). 
\begin{align}
\frac{1}{Z(\mu^2)} = {Z}_3(\mu^2) 
&- \widetilde{Z}_1(\mu^2) g^2(\mu^2) \int G(\mu^2) G(\mu^2) \Gamma^{gh}(\mu^2) \nonumber\\
&-           {Z}_1(\mu^2) g^2(\mu^2) \int Z(\mu^2) Z(\mu^2) \Gamma^{3g}(\mu^2) \nonumber\\
&-        {Z}_{1F}(\mu^2) g^2(\mu^2) \int \frac{A(\mu^2)}{q^2 A^2(\mu^2) + B^2(\mu^2)} 
                                                            \frac{A(\mu^2)}{q^2 A^2(\mu^2) + B^2(\mu^2)}
                                                            \Gamma^{qgl}(\mu^2)
\label{app:gluon}
\end{align} 
This equation is invariant under a change of the renormalization point, since we have 
\begin{align} 
\frac{1}{Z(\nu^2)} &= \frac{1}{Z(\mu^2)}\frac{Z_3(\nu^2)}{Z_3(\mu^2)} \nonumber\\
                   &= {Z}_3(\nu^2) 
- \frac{Z_3(\nu^2)}{Z_3(\mu^2)}\widetilde{Z}_1(\mu^2) g^2(\nu^2) \left(\frac{Z_g(\nu^2)}{Z_g(\mu^2)}\right)^2 
\int G(\nu^2) G(\nu^2) \left(\frac{\widetilde{Z}_3(\nu^2)}{\widetilde{Z}_3(\mu^2)}\right)^2  
\Gamma^{ghgl}(\nu^2) \frac{\widetilde{Z}_1(\mu^2)}{\widetilde{Z}_1(\nu^2)} \nonumber\\
&\hspace*{1.6cm}- \frac{Z_3(\nu^2)}{Z_3(\mu^2)}          {Z}_1(\mu^2) g^2(\nu^2) \left(\frac{Z_g(\nu^2)}{Z_g(\mu^2)}\right)^2 
\int Z(\nu^2) Z(\nu^2) \left(\frac{{Z}_3(\nu^2)}{{Z}_3(\mu^2)}\right)^2
\Gamma^{3g}(\nu^2) \frac{{Z}_1(\mu^2)}{{Z}_1(\nu^2)}\nonumber\\
&\hspace*{1.6cm}- \frac{Z_3(\nu^2)}{Z_3(\mu^2)}       {Z}_{1F}(\mu^2) g^2(\nu^2) \left(\frac{Z_g(\nu^2)}{Z_g(\mu^2)}\right)^2
\int \frac{A(\nu^2)}{q^2 A^2(\nu^2) + B^2(\nu^2)}\frac{A(\nu^2)}{q^2 A^2(\nu^2) + B^2(\nu^2)}\times \nonumber\\
&\hspace{8cm}\times \left(\frac{{Z}_2(\nu^2)}{{Z}_2(\mu^2)}\right)^2
\Gamma^{qgl}(\nu^2) \frac{{Z}_{1F}(\nu^2)}{{Z}_{1F}(\mu^2)} \label{app:gluon2}\\
&= {Z}_3(\nu^2) 
 - \widetilde{Z}_1(\nu^2) g^2(\nu^2) \int G(\nu^2) G(\nu^2) \Gamma^{gh}(\nu^2) \nonumber\\
&\hspace*{1.6cm}-           {Z}_1(\nu^2) g^2(\nu^2) \int Z(\nu^2) Z(\nu^2) \Gamma^{3g}(\nu^2) \nonumber\\
&\hspace*{1.6cm}-        {Z}_{1F}(\nu^2) g^2(\nu^2) \int \frac{A(\nu^2)}{q^2 A^2(\nu^2) + B^2(\nu^2)} 
                        \frac{A(\nu^2)}{q^2 A^2(\nu^2)+B^2(\nu^2)} \Gamma^{qgl}(\nu^2)\,,
\label{app:gluon3} 
\end{align}
where in the first line of Eq.(\ref{app:gluon2}) we have used the STI $\tilde{Z}_1=Z_g \tilde{Z}_3 Z_3^{1/2}$,
in the second line the identity $Z_1 = Z_g Z_3^{3/2}$ and in the last line $Z_{1F} = Z_g Z_3^{1/2} Z_2$.

Since the equation is invariant, the change in the renormalization point $\mu^2 \rightarrow \nu^2$ does not 
affect the momentum dependence of the gluon dressing function $Z(p^2,\mu^2)$. The solutions $Z(p^2,\mu^2)$ and
$Z(p^2,\nu^2)$ of both equations (\ref{app:gluon}) and (\ref{app:gluon3}) are uniquely related by the momentum 
independent ratio $Z_3(\mu^2)/Z_3(\nu^2)$. To our mind it is mandatory for any reasonable truncation of DSEs that 
these relations are not spoiled.

Along completely analogous lines one can show the same property for the DSE of the quark propagator. Since model
building often takes place in this equation, we show this property explicitly and detail, how a rainbow-ladder
model is able to cope with this. The renormalized Dyson-Schwinger equation for the inverse dressed quark propagator 
$S^{-1}(p) = -i \pslash A(p^2,\mu^2) + B(p^2,\mu^2)$ is given in symbolic notation by 
\beq
S^{-1}(\mu^2) = Z_2(\mu^2) \, S^{-1}_0 + g^2(\mu^2)\, Z_{1F}(\mu^2)\, \int \,
S(\mu^2) \,\Gamma^{qg}(\mu^2) \,Z(\mu^2) \,,
\eeq
where again we suppressed the notation of all momentum dependencies and only made the dependence on the renormalization
point $\mu^2$ explicit. Note that the $\mu^2$ dependence of the bare propagator term with $S^{-1}_0 = -i \pslash + m_0$ 
and renormalization point independent quark mass $m_0=Z_4(\mu^2) m(\mu^2)$ is completely carried by the quark
renormalization factor $Z_2$. From Eqs.~(\ref{quarkZA}) and (\ref{quarkZB}) we infer that $S^{-1}(\mu^2) \sim Z_2(\mu^2)$.
This entails that separately, $A(p^2,\mu^2) \sim Z_2(\mu^2)$ and $B(p^2,\mu^2) \sim Z_2(\mu^2)$, and therefore 
the quark mass function $M(p^2) = B(p^2,\mu^2)/A(p^2,\mu^2)$ is independent of the renormalization scale. 
Again, we change the renormalization point from $\mu^2$ to $\nu^2$,
\begin{align}
S^{-1}(\nu^2) &= S^{-1}(\mu^2) \frac{Z_2(\nu^2)}{Z_2(\mu^2)} \nonumber\\
&=  Z_2(\nu^2) \, S^{-1}_0 + \frac{Z_2(\nu^2)}{Z_2(\mu^2)} g^2(\nu^2) \left(\frac{Z_g(\nu^2)}{Z_g(\mu^2)}\right)^2 
\, Z_{1F}(\mu^2)\, \int \,
 S(\nu^2)\frac{Z_2(\nu^2)}{Z_2(\mu^2)} \,\Gamma^{qg}(\nu^2) \frac{Z_{1F}(\mu^2)}{Z_{1F}(\nu^2)} 
\,Z(\nu^2) \frac{{Z}_3(\nu^2)}{{Z}_3(\mu^2)} \nonumber\\
&=  Z_2(\nu^2) \, S^{-1}_0 + g^2(\nu^2) \, Z_{1F}(\nu^2)\, \int \,
 S(\nu^2) \,\Gamma^{qg}(\nu^2) \,Z(\nu^2)
\end{align}
where in the last line we have used the STI $Z_{1F} = Z_g Z_3^{1/2} Z_2$. The equation is form-invariant under 
a change of the renormalization point in agreement with multiplicative renormalizability. 

An often employed truncation scheme in the DSE/BSE framework is a rainbow-ladder truncation together with an 
effective running coupling. A convenient way to introduce this truncation and keep track of multiplicative 
renormalizability is to replace the fully dressed quark gluon vertex $\Gamma^{qg}(p,q,\nu^2)$ with quark
momenta $p$ and $q$ with the expression
\beq
\Gamma^{qg}(p,q,\nu^2) \rightarrow Z_{1F}(\mu^2) \gamma_\mu \Gamma^{qg}(k^2) 
\eeq   
where the renormalization factor $Z_{1F}(\mu^2)$ keeps track of all the dependence of the vertex
on the renormalization point and the tensor structure $\gamma_\mu$ is dressed by a function $\Gamma^{qg}(k^2)$
depending on the gluon momentum only. In the DSE we then use the STI $Z_{1F}=\widetilde{Z}_1 Z_2/\widetilde{Z}_3$ 
(cf. Eq.(\ref{Zsti})) and replace the expression
\beq
\left(\frac{\widetilde{Z}_1(\mu^2)}{\widetilde{Z}_3(\mu^2)}\right)^2 
\frac{g^2(\mu^2)}{4\pi} Z(k^2,\mu^2) \Gamma^{qg}(k^2) \rightarrow \alpha(k^2) \label{alpha} 
\eeq
with the so-called effective coupling $\alpha(k^2)$ to arrive at
\beq
S^{-1}(\mu^2) = Z_2(\mu^2) \, S^{-1}_0 + Z_2^2(\mu^2)\, \int \, \gamma S(\mu^2) \gamma \,\alpha(k^2) \,,
\label{quark2}
\eeq
where we made the resulting two bare quark-gluon vertices $\gamma$ explicit (still suppressing Lorenz indices).
It is easy to check that (\ref{quark2}) is still multiplicatively renormalizable, provided that
the effective coupling $\alpha(k^2)$ is a renormalization group invariant. That this is indeed the case is apparent
from Eq.~(\ref{alpha}) using $g(\mu^2) \sim 1/Z_g(\mu^2)$, $Z(k^2,\mu^2) \sim 1/Z_3(\mu^2)$ from
Eqs.~(\ref{glueZ}),(\ref{couplingZ}) and $\tilde{Z}_1=Z_g \tilde{Z}_3 Z_3^{1/2}$. The frequently used rainbow-ladder
truncation therefore indeed satisfies multiplicative renormalizability, provided careful track of all renormalization
factors is kept. In particular, the appearance of the factor $Z_2^2$ (instead of $Z_2$) in front of the integral of the
resulting quark-DSE is mandatory.

\section*{References}
\bibliography{phase_v2}

\providecommand{\href}[2]{#2}\begingroup\raggedright\begin{thebibliography}{100}
  \setlength{\itemsep}{0mm}

\bibitem{Aoki:2006we}
Y.~Aoki, G.~Endrodi, Z.~Fodor, S.~D. Katz, and K.~K. Szabo, ``\textit{{The
  Order of the quantum chromodynamics transition predicted by the standard
  model of particle physics}},''
  \href{http://dx.doi.org/10.1038/nature05120}{{\em Nature} {\bf 443} (2006)
  675--678} \href{http://arxiv.org/abs/hep-lat/0611014}{{ $\bullet$}}
\href{http://inspirehep.net/search?p=find+eprint+hep-lat/0611014}{{$
  \triangleright $}}

\bibitem{Aoki:2009sc}
Y.~Aoki, S.~Borsanyi, S.~Durr, Z.~Fodor, S.~D. Katz, S.~Krieg, and K.~K. Szabo,
  ``\textit{{The QCD transition temperature: results with physical masses in
  the continuum limit II.}},''
  \href{http://dx.doi.org/10.1088/1126-6708/2009/06/088}{{\em JHEP} {\bf 06}
  (2009)  088} \href{http://arxiv.org/abs/0903.4155}{{ $\bullet$}}
\href{http://inspirehep.net/search?p=find+eprint+0903.4155}{{$ \triangleright
  $}}

\bibitem{Borsanyi:2010bp}
{\bf Wuppertal-Budapest} Collaboration: S.~Borsanyi, Z.~Fodor, C.~Hoelbling,
  S.~D. Katz, S.~Krieg, C.~Ratti, and K.~K. Szabo, ``\textit{{Is there still
  any $T_c$ mystery in lattice QCD? Results with physical masses in the
  continuum limit III}},''
  \href{http://dx.doi.org/10.1007/JHEP09(2010)073}{{\em JHEP} {\bf 09} (2010)
  073} \href{http://arxiv.org/abs/1005.3508}{{ $\bullet$}}
\href{http://inspirehep.net/search?p=find+eprint+1005.3508}{{$ \triangleright
  $}}

\bibitem{Bazavov:2011nk}
A.~Bazavov {\em et al.}, ``\textit{{The chiral and deconfinement aspects of the
  QCD transition}},'' \href{http://dx.doi.org/10.1103/PhysRevD.85.054503}{{\em
  Phys. Rev.} {\bf D85} (2012)  054503} \href{http://arxiv.org/abs/1111.1710}{{
  $\bullet$}}
\href{http://inspirehep.net/search?p=find+eprint+1111.1710}{{$ \triangleright
  $}}

\bibitem{Bhattacharya:2014ara}
T.~Bhattacharya {\em et al.}, ``\textit{{QCD Phase Transition with Chiral
  Quarks and Physical Quark Masses}},''
  \href{http://dx.doi.org/10.1103/PhysRevLett.113.082001}{{\em Phys. Rev.
  Lett.} {\bf 113} (2014) no.~8, 082001}
  \href{http://arxiv.org/abs/1402.5175}{{ $\bullet$}}
\href{http://inspirehep.net/search?p=find+eprint+1402.5175}{{$ \triangleright
  $}}

\bibitem{Bazavov:2014pvz}
{\bf HotQCD} Collaboration: A.~Bazavov {\em et al.}, ``\textit{{Equation of
  state in ( 2+1 )-flavor QCD}},''
  \href{http://dx.doi.org/10.1103/PhysRevD.90.094503}{{\em Phys. Rev.} {\bf
  D90} (2014)  094503} \href{http://arxiv.org/abs/1407.6387}{{ $\bullet$}}
\href{http://inspirehep.net/search?p=find+eprint+1407.6387}{{$ \triangleright
  $}}

\bibitem{Fukushima:2003fw}
K.~Fukushima, ``\textit{{Chiral effective model with the Polyakov loop}},''
  \href{http://dx.doi.org/10.1016/j.physletb.2004.04.027}{{\em Phys. Lett.}
  {\bf B591} (2004)  277--284} \href{http://arxiv.org/abs/hep-ph/0310121}{{
  $\bullet$}}
\href{http://inspirehep.net/search?p=find+eprint+hep-ph/0310121}{{$
  \triangleright $}}

\bibitem{Megias:2004hj}
E.~Megias, E.~Ruiz~Arriola, and L.~L. Salcedo, ``\textit{{Polyakov loop in
  chiral quark models at finite temperature}},''
  \href{http://dx.doi.org/10.1103/PhysRevD.74.065005}{{\em Phys. Rev.} {\bf
  D74} (2006)  065005} \href{http://arxiv.org/abs/hep-ph/0412308}{{ $\bullet$}}
\href{http://inspirehep.net/search?p=find+eprint+hep-ph/0412308}{{$
  \triangleright $}}

\bibitem{Ratti:2005jh}
C.~Ratti, M.~A. Thaler, and W.~Weise, ``\textit{{Phases of QCD: Lattice
  thermodynamics and a field theoretical model}},''
  \href{http://dx.doi.org/10.1103/PhysRevD.73.014019}{{\em Phys. Rev.} {\bf
  D73} (2006)  014019} \href{http://arxiv.org/abs/hep-ph/0506234}{{ $\bullet$}}
\href{http://inspirehep.net/search?p=find+eprint+hep-ph/0506234}{{$
  \triangleright $}}

\bibitem{Schaefer:2007pw}
B.-J. Schaefer, J.~M. Pawlowski, and J.~Wambach, ``\textit{{The Phase Structure
  of the Polyakov--Quark-Meson Model}},''
  \href{http://dx.doi.org/10.1103/PhysRevD.76.074023}{{\em Phys. Rev.} {\bf
  D76} (2007)  074023} \href{http://arxiv.org/abs/0704.3234}{{ $\bullet$}}
\href{http://inspirehep.net/search?p=find+eprint+0704.3234}{{$ \triangleright
  $}}

\bibitem{Skokov:2010wb}
V.~Skokov, B.~Stokic, B.~Friman, and K.~Redlich, ``\textit{{Meson fluctuations
  and thermodynamics of the Polyakov loop extended quark-meson model}},''
  \href{http://dx.doi.org/10.1103/PhysRevC.82.015206}{{\em Phys. Rev.} {\bf
  C82} (2010)  015206} \href{http://arxiv.org/abs/1004.2665}{{ $\bullet$}}
\href{http://inspirehep.net/search?p=find+eprint+1004.2665}{{$ \triangleright
  $}}

\bibitem{Herbst:2010rf}
T.~K. Herbst, J.~M. Pawlowski, and B.-J. Schaefer, ``\textit{{The phase
  structure of the Polyakov?quark?meson model beyond mean field}},''
  \href{http://dx.doi.org/10.1016/j.physletb.2010.12.003}{{\em Phys. Lett.}
  {\bf B696} (2011)  58--67} \href{http://arxiv.org/abs/1008.0081}{{
  $\bullet$}}
\href{http://inspirehep.net/search?p=find+eprint+1008.0081}{{$ \triangleright
  $}}

\bibitem{Drews:2016wpi}
M.~Drews and W.~Weise, ``\textit{{Functional renormalization group studies of
  nuclear and neutron matter}},''
  \href{http://dx.doi.org/10.1016/j.ppnp.2016.10.002}{{\em Prog. Part. Nucl.
  Phys.} {\bf 93} (2017)  69} \href{http://arxiv.org/abs/1610.07568}{{
  $\bullet$}}
\href{http://inspirehep.net/search?p=find+eprint+1610.07568}{{$ \triangleright
  $}}

\bibitem{Fukushima:2017csk}
K.~Fukushima and V.~Skokov, ``\textit{{Polyakov loop modeling for hot QCD}},''
  \href{http://dx.doi.org/10.1016/j.ppnp.2017.05.002}{{\em Prog. Part. Nucl.
  Phys.} {\bf 96} (2017)  154--199} \href{http://arxiv.org/abs/1705.00718}{{
  $\bullet$}}
\href{http://inspirehep.net/search?p=find+eprint+1705.00718}{{$ \triangleright
  $}}

\bibitem{Roberts:2000aa}
C.~D. Roberts and S.~M. Schmidt, ``\textit{{Dyson-Schwinger equations: Density,
  temperature and continuum strong QCD}},''
  \href{http://dx.doi.org/10.1016/S0146-6410(00)90011-5}{{\em Prog. Part. Nucl.
  Phys.} {\bf 45} (2000)  S1--S103}
  \href{http://arxiv.org/abs/nucl-th/0005064}{{ $\bullet$}}
\href{http://inspirehep.net/search?p=find+eprint+nucl-th/0005064}{{$
  \triangleright $}}

\bibitem{Borsanyi:2010cj}
S.~Borsanyi, G.~Endrodi, Z.~Fodor, A.~Jakovac, S.~D. Katz, S.~Krieg, C.~Ratti,
  and K.~K. Szabo, ``\textit{{The QCD equation of state with dynamical
  quarks}},'' \href{http://dx.doi.org/10.1007/JHEP11(2010)077}{{\em JHEP} {\bf
  11} (2010)  077} \href{http://arxiv.org/abs/1007.2580}{{ $\bullet$}}
\href{http://inspirehep.net/search?p=find+eprint+1007.2580}{{$ \triangleright
  $}}

\bibitem{Borsanyi:2013bia}
S.~Borsanyi, Z.~Fodor, C.~Hoelbling, S.~D. Katz, S.~Krieg, and K.~K. Szabo,
  ``\textit{{Full result for the QCD equation of state with 2+1 flavors}},''
  \href{http://dx.doi.org/10.1016/j.physletb.2014.01.007}{{\em Phys. Lett.}
  {\bf B730} (2014)  99--104} \href{http://arxiv.org/abs/1309.5258}{{
  $\bullet$}}
\href{http://inspirehep.net/search?p=find+eprint+1309.5258}{{$ \triangleright
  $}}

\bibitem{Ding:2015ona}
H.-T. Ding, F.~Karsch, and S.~Mukherjee, ``\textit{{Thermodynamics of
  strong-interaction matter from Lattice QCD}},''
  \href{http://dx.doi.org/10.1142/S0218301315300076}{{\em Int. J. Mod. Phys.}
  {\bf E24} (2015) no.~10, 1530007} \href{http://arxiv.org/abs/1504.05274}{{
  $\bullet$}}
\href{http://inspirehep.net/search?p=find+eprint+1504.05274}{{$ \triangleright
  $}}

\bibitem{Bazavov:2017dus}
A.~Bazavov {\em et al.}, ``\textit{{The QCD Equation of State to O($\mu_B^6$)
  from Lattice QCD}},''
  \href{http://dx.doi.org/10.1103/PhysRevD.95.054504}{{\em Phys. Rev.} {\bf
  D95} (2017) no.~5, 054504} \href{http://arxiv.org/abs/1701.04325}{{
  $\bullet$}}
\href{http://inspirehep.net/search?p=find+eprint+1701.04325}{{$ \triangleright
  $}}

\bibitem{Asakawa:1989bq}
M.~Asakawa and K.~Yazaki, ``\textit{{Chiral Restoration at Finite Density and
  Temperature}},''
\href{http://dx.doi.org/10.1016/0375-9474(89)90002-X}{{\em Nucl. Phys.} {\bf
  A504} (1989)  668--684}

\bibitem{Stephanov:1998dy}
M.~A. Stephanov, K.~Rajagopal, and E.~V. Shuryak, ``\textit{{Signatures of the
  tricritical point in QCD}},''
  \href{http://dx.doi.org/10.1103/PhysRevLett.81.4816}{{\em Phys. Rev. Lett.}
  {\bf 81} (1998)  4816--4819} \href{http://arxiv.org/abs/hep-ph/9806219}{{
  $\bullet$}}
\href{http://inspirehep.net/search?p=find+eprint+hep-ph/9806219}{{$
  \triangleright $}}

\bibitem{Stephanov:1999zu}
M.~A. Stephanov, K.~Rajagopal, and E.~V. Shuryak, ``\textit{{Event-by-event
  fluctuations in heavy ion collisions and the QCD critical point}},''
  \href{http://dx.doi.org/10.1103/PhysRevD.60.114028}{{\em Phys. Rev.} {\bf
  D60} (1999)  114028} \href{http://arxiv.org/abs/hep-ph/9903292}{{ $\bullet$}}
\href{http://inspirehep.net/search?p=find+eprint+hep-ph/9903292}{{$
  \triangleright $}}

\bibitem{Stephanov:2004wx}
M.~A. Stephanov, ``\textit{{QCD phase diagram and the critical point}},''
  \href{http://dx.doi.org/10.1142/S0217751X05027965}{{\em Prog. Theor. Phys.
  Suppl.} {\bf 153} (2004)  139--156}
  \href{http://arxiv.org/abs/hep-ph/0402115}{{ $\bullet$}}
\href{http://inspirehep.net/search?p=find+eprint+hep-ph/0402115}{{$
  \triangleright $}}

\bibitem{Halasz:1998qr}
A.~M. Halasz, A.~D. Jackson, R.~E. Shrock, M.~A. Stephanov, and J.~J.~M.
  Verbaarschot, ``\textit{{On the phase diagram of QCD}},''
  \href{http://dx.doi.org/10.1103/PhysRevD.58.096007}{{\em Phys. Rev.} {\bf
  D58} (1998)  096007} \href{http://arxiv.org/abs/hep-ph/9804290}{{ $\bullet$}}
\href{http://inspirehep.net/search?p=find+eprint+hep-ph/9804290}{{$
  \triangleright $}}

\bibitem{Berges:1998rc}
J.~Berges and K.~Rajagopal, ``\textit{{Color superconductivity and chiral
  symmetry restoration at nonzero baryon density and temperature}},''
  \href{http://dx.doi.org/10.1016/S0550-3213(98)00620-8}{{\em Nucl. Phys.} {\bf
  B538} (1999)  215--232} \href{http://arxiv.org/abs/hep-ph/9804233}{{
  $\bullet$}}
\href{http://inspirehep.net/search?p=find+eprint+hep-ph/9804233}{{$
  \triangleright $}}

\bibitem{Pisarski:1983ms}
R.~D. Pisarski and F.~Wilczek, ``\textit{{Remarks on the Chiral Phase
  Transition in Chromodynamics}},''
\href{http://dx.doi.org/10.1103/PhysRevD.29.338}{{\em Phys. Rev.} {\bf D29}
  (1984)  338--341}

\bibitem{Schaefer:2006ds}
B.-J. Schaefer and J.~Wambach, ``\textit{{Susceptibilities near the QCD
  (tri)critical point}},''
  \href{http://dx.doi.org/10.1103/PhysRevD.75.085015}{{\em Phys. Rev.} {\bf
  D75} (2007)  085015} \href{http://arxiv.org/abs/hep-ph/0603256}{{ $\bullet$}}
\href{http://inspirehep.net/search?p=find+eprint+hep-ph/0603256}{{$
  \triangleright $}}

\bibitem{Karsch:2003jg}
F.~Karsch and E.~Laermann, ``\textit{{Thermodynamics and in medium hadron
  properties from lattice QCD}},''
  \href{http://arxiv.org/abs/hep-lat/0305025}{{\tt hep-lat/0305025[hep-lat]}}
  \href{http://arxiv.org/abs/hep-lat/0305025}{{ $\bullet$}}
\href{http://inspirehep.net/search?p=find+eprint+hep-lat/0305025}{{$
  \triangleright $}}

\bibitem{deForcrand:2006pv}
P.~de~Forcrand and O.~Philipsen, ``\textit{{The Chiral critical line of N(f) =
  2+1 QCD at zero and non-zero baryon density}},''
  \href{http://dx.doi.org/10.1088/1126-6708/2007/01/077}{{\em JHEP} {\bf 01}
  (2007)  077} \href{http://arxiv.org/abs/hep-lat/0607017}{{ $\bullet$}}
\href{http://inspirehep.net/search?p=find+eprint+hep-lat/0607017}{{$
  \triangleright $}}

\bibitem{Kaczmarek:2011zz}
O.~Kaczmarek, F.~Karsch, E.~Laermann, C.~Miao, S.~Mukherjee, P.~Petreczky,
  C.~Schmidt, W.~Soeldner, and W.~Unger, ``\textit{{Phase boundary for the
  chiral transition in (2+1) -flavor QCD at small values of the chemical
  potential}},'' \href{http://dx.doi.org/10.1103/PhysRevD.83.014504}{{\em Phys.
  Rev.} {\bf D83} (2011)  014504} \href{http://arxiv.org/abs/1011.3130}{{
  $\bullet$}}
\href{http://inspirehep.net/search?p=find+eprint+1011.3130}{{$ \triangleright
  $}}

\bibitem{Endrodi:2011gv}
G.~Endrodi, Z.~Fodor, S.~D. Katz, and K.~K. Szabo, ``\textit{{The QCD phase
  diagram at nonzero quark density}},''
  \href{http://dx.doi.org/10.1007/JHEP04(2011)001}{{\em JHEP} {\bf 04} (2011)
  001} \href{http://arxiv.org/abs/1102.1356}{{ $\bullet$}}
\href{http://inspirehep.net/search?p=find+eprint+1102.1356}{{$ \triangleright
  $}}

\bibitem{Bellwied:2015rza}
R.~Bellwied, S.~Borsanyi, Z.~Fodor, J.~G\"unther, S.~D. Katz, C.~Ratti, and
  K.~K. Szabo, ``\textit{{The QCD phase diagram from analytic continuation}},''
  \href{http://dx.doi.org/10.1016/j.physletb.2015.11.011}{{\em Phys. Lett.}
  {\bf B751} (2015)  559--564} \href{http://arxiv.org/abs/1507.07510}{{
  $\bullet$}}
\href{http://inspirehep.net/search?p=find+eprint+1507.07510}{{$ \triangleright
  $}}

\bibitem{Fodor:2001pe}
Z.~Fodor and S.~D. Katz, ``\textit{{Lattice determination of the critical point
  of QCD at finite T and mu}},''
  \href{http://dx.doi.org/10.1088/1126-6708/2002/03/014}{{\em JHEP} {\bf 03}
  (2002)  014} \href{http://arxiv.org/abs/hep-lat/0106002}{{ $\bullet$}}
\href{http://inspirehep.net/search?p=find+eprint+hep-lat/0106002}{{$
  \triangleright $}}

\bibitem{Fodor:2004nz}
Z.~Fodor and S.~D. Katz, ``\textit{{Critical point of QCD at finite T and mu,
  lattice results for physical quark masses}},''
  \href{http://dx.doi.org/10.1088/1126-6708/2004/04/050}{{\em JHEP} {\bf 04}
  (2004)  050} \href{http://arxiv.org/abs/hep-lat/0402006}{{ $\bullet$}}
\href{http://inspirehep.net/search?p=find+eprint+hep-lat/0402006}{{$
  \triangleright $}}

\bibitem{Datta:2012pj}
S.~Datta, R.~V. Gavai, and S.~Gupta, ``\textit{{The QCD Critical Point :
  marching towards continuum}},''
  \href{http://dx.doi.org/10.1016/j.nuclphysa.2013.02.156}{{\em Nucl. Phys.}
  {\bf A904-905} (2013)  883c--886c} \href{http://arxiv.org/abs/1210.6784}{{
  $\bullet$}}
\href{http://inspirehep.net/search?p=find+eprint+1210.6784}{{$ \triangleright
  $}}

\bibitem{Fischer:2014ata}
C.~S. Fischer, J.~Luecker, and C.~A. Welzbacher, ``\textit{{Phase structure of
  three and four flavor QCD}},''
  \href{http://dx.doi.org/10.1103/PhysRevD.90.034022}{{\em Phys. Rev.} {\bf
  D90} (2014) no.~3, 034022} \href{http://arxiv.org/abs/1405.4762}{{
  $\bullet$}}
\href{http://inspirehep.net/search?p=find+eprint+1405.4762}{{$ \triangleright
  $}}

\bibitem{Eichmann:2015kfa}
G.~Eichmann, C.~S. Fischer, and C.~A. Welzbacher, ``\textit{{Baryon effects on
  the location of QCD's critical end point}},''
  \href{http://dx.doi.org/10.1103/PhysRevD.93.034013}{{\em Phys. Rev.} {\bf
  D93} (2016) no.~3, 034013} \href{http://arxiv.org/abs/1509.02082}{{
  $\bullet$}}
\href{http://inspirehep.net/search?p=find+eprint+1509.02082}{{$ \triangleright
  $}}

\bibitem{Cohen:1991nk}
T.~D. Cohen, R.~J. Furnstahl, and D.~K. Griegel, ``\textit{{Quark and gluon
  condensates in nuclear matter}},''
\href{http://dx.doi.org/10.1103/PhysRevC.45.1881}{{\em Phys. Rev.} {\bf C45}
  (1992)  1881--1893}

\bibitem{Cohen:2004qp}
T.~D. Cohen, ``\textit{{QCD functional integrals for systems with nonzero
  chemical potential}},'' \href{http://arxiv.org/abs/hep-ph/0405043}{{\tt
  hep-ph/0405043[hep-ph]}} \href{http://arxiv.org/abs/hep-ph/0405043}{{
  $\bullet$}}
\href{http://inspirehep.net/search?p=find+eprint+hep-ph/0405043}{{$
  \triangleright $}}

\bibitem{Fromm:2012eb}
M.~Fromm, J.~Langelage, S.~Lottini, M.~Neuman, and O.~Philipsen,
  ``\textit{{Onset Transition to Cold Nuclear Matter from Lattice QCD with
  Heavy Quarks}},''
  \href{http://dx.doi.org/10.1103/PhysRevLett.110.122001}{{\em Phys. Rev.
  Lett.} {\bf 110} (2013) no.~12, 122001}
  \href{http://arxiv.org/abs/1207.3005}{{ $\bullet$}}
\href{http://inspirehep.net/search?p=find+eprint+1207.3005}{{$ \triangleright
  $}}

\bibitem{Muller:2016fdr}
D.~M\"uller, M.~Buballa, and J.~Wambach, ``\textit{{Dyson-Schwinger Approach to
  Color-Superconductivity: Effects of Selfconsistent Gluon Dressing}},''
  \href{http://arxiv.org/abs/1603.02865}{{\tt 1603.02865[hep-ph]}}
  \href{http://arxiv.org/abs/1603.02865}{{ $\bullet$}}
\href{http://inspirehep.net/search?p=find+eprint+1603.02865}{{$ \triangleright
  $}}

\bibitem{Fukushima:2013rx}
K.~Fukushima and C.~Sasaki, ``\textit{{The phase diagram of nuclear and quark
  matter at high baryon density}},''
  \href{http://dx.doi.org/10.1016/j.ppnp.2013.05.003}{{\em Prog. Part. Nucl.
  Phys.} {\bf 72} (2013)  99--154} \href{http://arxiv.org/abs/1301.6377}{{
  $\bullet$}}
\href{http://inspirehep.net/search?p=find+eprint+1301.6377}{{$ \triangleright
  $}}

\bibitem{Langelage:2014vpa}
J.~Langelage, M.~Neuman, and O.~Philipsen, ``\textit{{Heavy dense QCD and
  nuclear matter from an effective lattice theory}},''
  \href{http://dx.doi.org/10.1007/JHEP09(2014)131}{{\em JHEP} {\bf 09} (2014)
  131} \href{http://arxiv.org/abs/1403.4162}{{ $\bullet$}}
\href{http://inspirehep.net/search?p=find+eprint+1403.4162}{{$ \triangleright
  $}}

\bibitem{Weyrich:2015hha}
J.~Weyrich, N.~Strodthoff, and L.~von Smekal, ``\textit{{Chiral
  mirror-baryon-meson model and nuclear matter beyond mean-field
  approximation}},'' \href{http://dx.doi.org/10.1103/PhysRevC.92.015214}{{\em
  Phys. Rev.} {\bf C92} (2015) no.~1, 015214}
  \href{http://arxiv.org/abs/1504.02697}{{ $\bullet$}}
\href{http://inspirehep.net/search?p=find+eprint+1504.02697}{{$ \triangleright
  $}}

\bibitem{Alford:1997zt}
M.~G. Alford, K.~Rajagopal, and F.~Wilczek, ``\textit{{QCD at finite baryon
  density: Nucleon droplets and color superconductivity}},''
  \href{http://dx.doi.org/10.1016/S0370-2693(98)00051-3}{{\em Phys. Lett.} {\bf
  B422} (1998)  247--256} \href{http://arxiv.org/abs/hep-ph/9711395}{{
  $\bullet$}}
\href{http://inspirehep.net/search?p=find+eprint+hep-ph/9711395}{{$
  \triangleright $}}

\bibitem{Rapp:1997zu}
R.~Rapp, T.~Sch\"afer, E.~V. Shuryak, and M.~Velkovsky, ``\textit{{Diquark Bose
  condensates in high density matter and instantons}},''
  \href{http://dx.doi.org/10.1103/PhysRevLett.81.53}{{\em Phys. Rev. Lett.}
  {\bf 81} (1998)  53--56} \href{http://arxiv.org/abs/hep-ph/9711396}{{
  $\bullet$}}
\href{http://inspirehep.net/search?p=find+eprint+hep-ph/9711396}{{$
  \triangleright $}}

\bibitem{Alford:1998mk}
M.~G. Alford, K.~Rajagopal, and F.~Wilczek, ``\textit{{Color flavor locking and
  chiral symmetry breaking in high density QCD}},''
  \href{http://dx.doi.org/10.1016/S0550-3213(98)00668-3}{{\em Nucl. Phys.} {\bf
  B537} (1999)  443--458} \href{http://arxiv.org/abs/hep-ph/9804403}{{
  $\bullet$}}
\href{http://inspirehep.net/search?p=find+eprint+hep-ph/9804403}{{$
  \triangleright $}}

\bibitem{Alford:1999pa}
M.~G. Alford, J.~Berges, and K.~Rajagopal, ``\textit{{Unlocking color and
  flavor in superconducting strange quark matter}},''
  \href{http://dx.doi.org/10.1016/S0550-3213(99)00410-1}{{\em Nucl. Phys.} {\bf
  B558} (1999)  219--242} \href{http://arxiv.org/abs/hep-ph/9903502}{{
  $\bullet$}}
\href{http://inspirehep.net/search?p=find+eprint+hep-ph/9903502}{{$
  \triangleright $}}

\bibitem{Rajagopal:2000wf}
K.~Rajagopal and F.~Wilczek,
  \href{http://dx.doi.org/10.1142/9789812810458_0043}{``\textit{{The Condensed
  matter physics of QCD}},''} in {\em At the frontier of particle physics.
  Handbook of QCD. Vol. 1-3}, M.~Shifman and B.~Ioffe, Eds. pp.~2061--2151.
\newblock 2000.
\newblock
\href{http://arxiv.org/abs/hep-ph/0011333}{{\tt hep-ph/0011333[hep-ph]}}.
\newblock

\bibitem{Rischke:2003mt}
D.~H. Rischke, ``\textit{{The Quark gluon plasma in equilibrium}},''
  \href{http://dx.doi.org/10.1016/j.ppnp.2003.09.002}{{\em Prog. Part. Nucl.
  Phys.} {\bf 52} (2004)  197--296}
  \href{http://arxiv.org/abs/nucl-th/0305030}{{ $\bullet$}}
\href{http://inspirehep.net/search?p=find+eprint+nucl-th/0305030}{{$
  \triangleright $}}

\bibitem{Buballa:2003qv}
M.~Buballa, ``\textit{{NJL} model analysis of dense quark matter},''
  \href{http://dx.doi.org/10.1016/j.physrep.2004.11.004}{{\em Phys. Rept.} {\bf
  407} (2005)  205--376} \href{http://arxiv.org/abs/hep-ph/0402234}{{
  $\bullet$}}
\href{http://inspirehep.net/search?p=find+eprint+hep-ph/0402234}{{$
  \triangleright $}}

\bibitem{Alford:2007xm}
M.~G. Alford, A.~Schmitt, K.~Rajagopal, and T.~Sch\"afer, ``\textit{{Color
  superconductivity in dense quark matter}},''
  \href{http://dx.doi.org/10.1103/RevModPhys.80.1455}{{\em Rev. Mod. Phys.}
  {\bf 80} (2008)  1455--1515} \href{http://arxiv.org/abs/0709.4635}{{
  $\bullet$}}
\href{http://inspirehep.net/search?p=find+eprint+0709.4635}{{$ \triangleright
  $}}

\bibitem{Nickel:2006vf}
D.~Nickel, J.~Wambach, and R.~Alkofer, ``\textit{{Color-superconductivity in
  the strong-coupling regime of Landau gauge QCD}},''
  \href{http://dx.doi.org/10.1103/PhysRevD.73.114028}{{\em Phys. Rev.} {\bf
  D73} (2006)  114028} \href{http://arxiv.org/abs/hep-ph/0603163}{{ $\bullet$}}
\href{http://inspirehep.net/search?p=find+eprint+hep-ph/0603163}{{$
  \triangleright $}}

\bibitem{Nickel:2006kc}
D.~Nickel, R.~Alkofer, and J.~Wambach, ``\textit{{On the unlocking of color and
  flavor in color-superconducting quark matter}},''
  \href{http://dx.doi.org/10.1103/PhysRevD.74.114015}{{\em Phys. Rev.} {\bf
  D74} (2006)  114015} \href{http://arxiv.org/abs/hep-ph/0609198}{{ $\bullet$}}
\href{http://inspirehep.net/search?p=find+eprint+hep-ph/0609198}{{$
  \triangleright $}}

\bibitem{Marhauser:2006hy}
F.~Marhauser, D.~Nickel, M.~Buballa, and J.~Wambach, ``\textit{{Color-spin
  locking in a selfconsistent Dyson-Schwinger approach}},''
  \href{http://dx.doi.org/10.1103/PhysRevD.75.054022}{{\em Phys. Rev.} {\bf
  D75} (2007)  054022} \href{http://arxiv.org/abs/hep-ph/0612027}{{ $\bullet$}}
\href{http://inspirehep.net/search?p=find+eprint+hep-ph/0612027}{{$
  \triangleright $}}

\bibitem{Nickel:2008ef}
D.~Nickel, R.~Alkofer, and J.~Wambach, ``\textit{{Neutrality of the
  color-flavor-locked phase in a Dyson-Schwinger approach}},''
  \href{http://dx.doi.org/10.1103/PhysRevD.77.114010}{{\em Phys. Rev.} {\bf
  D77} (2008)  114010} \href{http://arxiv.org/abs/0802.3187}{{ $\bullet$}}
\href{http://inspirehep.net/search?p=find+eprint+0802.3187}{{$ \triangleright
  $}}

\bibitem{Muller:2013pya}
D.~M\"uller, M.~Buballa, and J.~Wambach, ``\textit{{Dyson-Schwinger approach to
  color superconductivity at finite temperature and density}},''
  \href{http://dx.doi.org/10.1140/epja/i2013-13096-5}{{\em Eur. Phys. J.} {\bf
  A49} (2013)  96} \href{http://arxiv.org/abs/1303.2693}{{ $\bullet$}}
\href{http://inspirehep.net/search?p=find+eprint+1303.2693}{{$ \triangleright
  $}}

\bibitem{Franz:2001zz}
M.~Franz and Z.~Tesanovic, ``\textit{{Algebraic Fermi Liquid from Phase
  Fluctuations: 'Topological' Fermions, Vortex 'Berryons, ' and QE D-3 Theory
  of Cuprate Superconductors}},''
\href{http://dx.doi.org/10.1103/PhysRevLett.87.257003}{{\em Phys. Rev. Lett.}
  {\bf 87} (2001)  257003}

\bibitem{Herbut:2002wd}
I.~F. Herbut, ``\textit{{Antiferromagnetism from phase disordering of a d wave
  superconductor}},''
  \href{http://dx.doi.org/10.1103/PhysRevLett.88.047006}{{\em Phys. Rev. Lett.}
  {\bf 88} (2002)  047006} \href{http://arxiv.org/abs/cond-mat/0110188}{{
  $\bullet$}}
\href{http://inspirehep.net/search?p=find+eprint+cond-mat/0110188}{{$
  \triangleright $}}

\bibitem{Franz:2002qy}
M.~Franz, Z.~Tesanovic, and O.~Vafek, ``\textit{{QED(3) theory of pairing
  pseudogap in cuprates. 1. From D wave superconductor to antiferromagnet via
  'algebraic' Fermi liquid}},''
  \href{http://dx.doi.org/10.1103/PhysRevB.66.054535}{{\em Phys. Rev.} {\bf
  B66} (2002)  054535} \href{http://arxiv.org/abs/cond-mat/0203333}{{
  $\bullet$}}
\href{http://inspirehep.net/search?p=find+eprint+cond-mat/0203333}{{$
  \triangleright $}}

\bibitem{Fischer:2004nq}
C.~S. Fischer, R.~Alkofer, T.~Dahm, and P.~Maris, ``\textit{{Dynamical chiral
  symmetry breaking in unquenched QED(3)}},''
  \href{http://dx.doi.org/10.1103/PhysRevD.70.073007}{{\em Phys. Rev.} {\bf
  D70} (2004)  073007} \href{http://arxiv.org/abs/hep-ph/0407104}{{ $\bullet$}}
\href{http://inspirehep.net/search?p=find+eprint+hep-ph/0407104}{{$
  \triangleright $}}

\bibitem{Deryagin:1992rw}
D.~V. Deryagin, D.~{\relax Yu}. Grigoriev, and V.~A. Rubakov,
  ``\textit{{Standing wave ground state in high density, zero temperature QCD
  at large N(c)}},''
\href{http://dx.doi.org/10.1142/S0217751X92000302}{{\em Int. J. Mod. Phys.}
  {\bf A7} (1992)  659--681}

\bibitem{Shuster:1999tn}
E.~Shuster and D.~T. Son, ``\textit{{On finite density QCD at large N(c)}},''
  \href{http://dx.doi.org/10.1016/S0550-3213(99)00615-X}{{\em Nucl. Phys.} {\bf
  B573} (2000)  434--446} \href{http://arxiv.org/abs/hep-ph/9905448}{{
  $\bullet$}}
\href{http://inspirehep.net/search?p=find+eprint+hep-ph/9905448}{{$
  \triangleright $}}

\bibitem{Park:1999bz}
B.-Y. Park, M.~Rho, A.~Wirzba, and I.~Zahed, ``\textit{{Dense QCD: Overhauser
  or BCS pairing?}},'' \href{http://dx.doi.org/10.1103/PhysRevD.62.034015}{{\em
  Phys. Rev.} {\bf D62} (2000)  034015}
  \href{http://arxiv.org/abs/hep-ph/9910347}{{ $\bullet$}}
\href{http://inspirehep.net/search?p=find+eprint+hep-ph/9910347}{{$
  \triangleright $}}

\bibitem{Rapp:2000zd}
R.~Rapp, E.~V. Shuryak, and I.~Zahed, ``\textit{{A Chiral crystal in cold QCD
  matter at intermediate densities?}},''
  \href{http://dx.doi.org/10.1103/PhysRevD.63.034008}{{\em Phys. Rev.} {\bf
  D63} (2001)  034008} \href{http://arxiv.org/abs/hep-ph/0008207}{{ $\bullet$}}
\href{http://inspirehep.net/search?p=find+eprint+hep-ph/0008207}{{$
  \triangleright $}}

\bibitem{Buballa:2014tba}
M.~Buballa and S.~Carignano, ``\textit{{Inhomogeneous chiral condensates}},''
  \href{http://dx.doi.org/10.1016/j.ppnp.2014.11.001}{{\em Prog. Part. Nucl.
  Phys.} {\bf 81} (2015)  39--96} \href{http://arxiv.org/abs/1406.1367}{{
  $\bullet$}}
\href{http://inspirehep.net/search?p=find+eprint+1406.1367}{{$ \triangleright
  $}}

\bibitem{Muller:2013tya}
D.~M\"uller, M.~Buballa, and J.~Wambach, ``\textit{{Dyson-Schwinger study of
  chiral density waves in QCD}},''
  \href{http://dx.doi.org/10.1016/j.physletb.2013.10.050}{{\em Phys. Lett.}
  {\bf B727} (2013)  240--243} \href{http://arxiv.org/abs/1308.4303}{{
  $\bullet$}}
\href{http://inspirehep.net/search?p=find+eprint+1308.4303}{{$ \triangleright
  $}}

\bibitem{Carignano:2014jla}
S.~Carignano, M.~Buballa, and B.-J. Schaefer, ``\textit{{Inhomogeneous phases
  in the quark-meson model with vacuum fluctuations}},''
  \href{http://dx.doi.org/10.1103/PhysRevD.90.014033}{{\em Phys. Rev.} {\bf
  D90} (2014) no.~1, 014033} \href{http://arxiv.org/abs/1404.0057}{{
  $\bullet$}}
\href{http://inspirehep.net/search?p=find+eprint+1404.0057}{{$ \triangleright
  $}}

\bibitem{McLerran:2007qj}
L.~McLerran and R.~D. Pisarski, ``\textit{{Phases of cold, dense quarks at
  large N(c)}},'' \href{http://dx.doi.org/10.1016/j.nuclphysa.2007.08.013}{{\em
  Nucl. Phys.} {\bf A796} (2007)  83--100}
  \href{http://arxiv.org/abs/0706.2191}{{ $\bullet$}}
\href{http://inspirehep.net/search?p=find+eprint+0706.2191}{{$ \triangleright
  $}}

\bibitem{Kojo:2009ha}
T.~Kojo, Y.~Hidaka, L.~McLerran, and R.~D. Pisarski, ``\textit{{Quarkyonic
  Chiral Spirals}},''
  \href{http://dx.doi.org/10.1016/j.nuclphysa.2010.05.053}{{\em Nucl. Phys.}
  {\bf A843} (2010)  37--58} \href{http://arxiv.org/abs/0912.3800}{{
  $\bullet$}}
\href{http://inspirehep.net/search?p=find+eprint+0912.3800}{{$ \triangleright
  $}}

\bibitem{Torrieri:2010gz}
G.~Torrieri and I.~Mishustin, ``\textit{{The nuclear liquid-gas phase
  transition at large $N_c$ in the Van der Waals approximation}},''
  \href{http://dx.doi.org/10.1103/PhysRevC.82.055202}{{\em Phys. Rev.} {\bf
  C82} (2010)  055202} \href{http://arxiv.org/abs/1006.2471}{{ $\bullet$}}
\href{http://inspirehep.net/search?p=find+eprint+1006.2471}{{$ \triangleright
  $}}

\bibitem{Andersen:2014xxa}
J.~O. Andersen, W.~R. Naylor, and A.~Tranberg, ``\textit{{Phase diagram of QCD
  in a magnetic field: A review}},''
  \href{http://dx.doi.org/10.1103/RevModPhys.88.025001}{{\em Rev. Mod. Phys.}
  {\bf 88} (2016)  025001} \href{http://arxiv.org/abs/1411.7176}{{ $\bullet$}}
\href{http://inspirehep.net/search?p=find+eprint+1411.7176}{{$ \triangleright
  $}}

\bibitem{Miransky:2015ava}
V.~A. Miransky and I.~A. Shovkovy, ``\textit{{Quantum field theory in a
  magnetic field: From quantum chromodynamics to graphene and Dirac
  semimetals}},'' \href{http://dx.doi.org/10.1016/j.physrep.2015.02.003}{{\em
  Phys. Rept.} {\bf 576} (2015)  1--209}
  \href{http://arxiv.org/abs/1503.00732}{{ $\bullet$}}
\href{http://inspirehep.net/search?p=find+eprint+1503.00732}{{$ \triangleright
  $}}

\bibitem{Klevansky:1989vi}
S.~P. Klevansky and R.~H. Lemmer, ``\textit{{Chiral symmetry restoration in the
  Nambu-Jona-Lasinio model with a constant electromagnetic field}},''
\href{http://dx.doi.org/10.1103/PhysRevD.39.3478}{{\em Phys. Rev.} {\bf D39}
  (1989)  3478--3489}

\bibitem{Suganuma:1990nn}
H.~Suganuma and T.~Tatsumi, ``\textit{{On the Behavior of Symmetry and Phase
  Transitions in a Strong Electromagnetic Field}},''
\href{http://dx.doi.org/10.1016/0003-4916(91)90304-Q}{{\em Annals Phys.} {\bf
  208} (1991)  470--508}

\bibitem{Bali:2011qj}
G.~S. Bali, F.~Bruckmann, G.~Endrodi, Z.~Fodor, S.~D. Katz, S.~Krieg,
  A.~Schafer, and K.~K. Szabo, ``\textit{{The QCD phase diagram for external
  magnetic fields}},'' \href{http://dx.doi.org/10.1007/JHEP02(2012)044}{{\em
  JHEP} {\bf 02} (2012)  044} \href{http://arxiv.org/abs/1111.4956}{{
  $\bullet$}}
\href{http://inspirehep.net/search?p=find+eprint+1111.4956}{{$ \triangleright
  $}}

\bibitem{Bali:2012zg}
G.~S. Bali, F.~Bruckmann, G.~Endrodi, Z.~Fodor, S.~D. Katz, and A.~Schafer,
  ``\textit{{QCD quark condensate in external magnetic fields}},''
  \href{http://dx.doi.org/10.1103/PhysRevD.86.071502}{{\em Phys. Rev.} {\bf
  D86} (2012)  071502} \href{http://arxiv.org/abs/1206.4205}{{ $\bullet$}}
\href{http://inspirehep.net/search?p=find+eprint+1206.4205}{{$ \triangleright
  $}}

\bibitem{Bruckmann:2013oba}
F.~Bruckmann, G.~Endrodi, and T.~G. Kovacs, ``\textit{{Inverse magnetic
  catalysis and the Polyakov loop}},''
  \href{http://dx.doi.org/10.1007/JHEP04(2013)112}{{\em JHEP} {\bf 04} (2013)
  112} \href{http://arxiv.org/abs/1303.3972}{{ $\bullet$}}
\href{http://inspirehep.net/search?p=find+eprint+1303.3972}{{$ \triangleright
  $}}

\bibitem{Bali:2013esa}
G.~S. Bali, F.~Bruckmann, G.~Endrodi, F.~Gruber, and A.~Schaefer,
  ``\textit{{Magnetic field-induced gluonic (inverse) catalysis and pressure
  (an)isotropy in QCD}},''
  \href{http://dx.doi.org/10.1007/JHEP04(2013)130}{{\em JHEP} {\bf 04} (2013)
  130} \href{http://arxiv.org/abs/1303.1328}{{ $\bullet$}}
\href{http://inspirehep.net/search?p=find+eprint+1303.1328}{{$ \triangleright
  $}}

\bibitem{Ilgenfritz:2013ara}
E.~M. Ilgenfritz, M.~Muller-Preussker, B.~Petersson, and A.~Schreiber,
  ``\textit{{Magnetic catalysis (and inverse catalysis) at finite temperature
  in two-color lattice QCD}},''
  \href{http://dx.doi.org/10.1103/PhysRevD.89.054512}{{\em Phys. Rev.} {\bf
  D89} (2014) no.~5, 054512} \href{http://arxiv.org/abs/1310.7876}{{
  $\bullet$}}
\href{http://inspirehep.net/search?p=find+eprint+1310.7876}{{$ \triangleright
  $}}

\bibitem{Mueller:2015fka}
N.~Mueller and J.~M. Pawlowski, ``\textit{{Magnetic catalysis and inverse
  magnetic catalysis in QCD}},''
  \href{http://dx.doi.org/10.1103/PhysRevD.91.116010}{{\em Phys. Rev.} {\bf
  D91} (2015) no.~11, 116010} \href{http://arxiv.org/abs/1502.08011}{{
  $\bullet$}}
\href{http://inspirehep.net/search?p=find+eprint+1502.08011}{{$ \triangleright
  $}}

\bibitem{Skokov:2016yrj}
V.~Koch, S.~Schlichting, V.~Skokov, P.~Sorensen, J.~Thomas, S.~Voloshin,
  G.~Wang, and H.-U. Yee, ``\textit{{Status of the chiral magnetic effect and
  collisions of isobars}},''
  \href{http://dx.doi.org/10.1088/1674-1137/41/7/072001}{{\em Chin. Phys.} {\bf
  C41} (2017) no.~7, 072001} \href{http://arxiv.org/abs/1608.00982}{{
  $\bullet$}}
\href{http://inspirehep.net/search?p=find+eprint+1608.00982}{{$ \triangleright
  $}}

\bibitem{Yamamoto:2011gk}
A.~Yamamoto, ``\textit{{Chiral magnetic effect in lattice QCD with a chiral
  chemical potential}},''
  \href{http://dx.doi.org/10.1103/PhysRevLett.107.031601}{{\em Phys. Rev.
  Lett.} {\bf 107} (2011)  031601} \href{http://arxiv.org/abs/1105.0385}{{
  $\bullet$}}
\href{http://inspirehep.net/search?p=find+eprint+1105.0385}{{$ \triangleright
  $}}

\bibitem{Braguta:2010ej}
V.~V. Braguta, P.~V. Buividovich, T.~Kalaydzhyan, S.~V. Kuznetsov, and M.~I.
  Polikarpov, ``\textit{{The Chiral Magnetic Effect and chiral symmetry
  breaking in SU(3) quenched lattice gauge theory}},''
  \href{http://dx.doi.org/10.1134/S1063778812030052}{{\em Phys. Atom. Nucl.}
  {\bf 75} (2012)  488--492} \href{http://arxiv.org/abs/1011.3795}{{
  $\bullet$}}
\href{http://inspirehep.net/search?p=find+eprint+1011.3795}{{$ \triangleright
  $}}

\bibitem{Braguta:2015owi}
V.~V. Braguta, E.~M. Ilgenfritz, A.~{\relax Yu}. Kotov, B.~Petersson, and S.~A.
  Skinderev, ``\textit{{Study of QCD Phase Diagram with Non-Zero Chiral
  Chemical Potential}},''
  \href{http://dx.doi.org/10.1103/PhysRevD.93.034509}{{\em Phys. Rev.} {\bf
  D93} (2016) no.~3, 034509} \href{http://arxiv.org/abs/1512.05873}{{
  $\bullet$}}
\href{http://inspirehep.net/search?p=find+eprint+1512.05873}{{$ \triangleright
  $}}

\bibitem{Braguta:2016aov}
V.~V. Braguta and A.~{\relax Yu}. Kotov, ``\textit{{Catalysis of Dynamical
  Chiral Symmetry Breaking by Chiral Chemical Potential}},''
  \href{http://dx.doi.org/10.1103/PhysRevD.93.105025}{{\em Phys. Rev.} {\bf
  D93} (2016) no.~10, 105025} \href{http://arxiv.org/abs/1601.04957}{{
  $\bullet$}}
\href{http://inspirehep.net/search?p=find+eprint+1601.04957}{{$ \triangleright
  $}}

\bibitem{Ruggieri:2011xc}
M.~Ruggieri, ``\textit{{The Critical End Point of Quantum Chromodynamics
  Detected by Chirally Imbalanced Quark Matter}},''
  \href{http://dx.doi.org/10.1103/PhysRevD.84.014011}{{\em Phys. Rev.} {\bf
  D84} (2011)  014011} \href{http://arxiv.org/abs/1103.6186}{{ $\bullet$}}
\href{http://inspirehep.net/search?p=find+eprint+1103.6186}{{$ \triangleright
  $}}

\bibitem{Yu:2015hym}
L.~Yu, H.~Liu, and M.~Huang, ``\textit{{Effect of the chiral chemical potential
  on the chiral phase transition in the NJL model with different regularization
  schemes}},'' \href{http://dx.doi.org/10.1103/PhysRevD.94.014026}{{\em Phys.
  Rev.} {\bf D94} (2016) no.~1, 014026}
  \href{http://arxiv.org/abs/1511.03073}{{ $\bullet$}}
\href{http://inspirehep.net/search?p=find+eprint+1511.03073}{{$ \triangleright
  $}}

\bibitem{Ruggieri:2016xww}
M.~Ruggieri, Z.~Y. Lu, and G.~X. Peng, ``\textit{{Influence of chiral chemical
  potential, parallel electric, and magnetic fields on the critical temperature
  of QCD}},'' \href{http://dx.doi.org/10.1103/PhysRevD.94.116003}{{\em Phys.
  Rev.} {\bf D94} (2016) no.~11, 116003}
  \href{http://arxiv.org/abs/1608.08310}{{ $\bullet$}}
\href{http://inspirehep.net/search?p=find+eprint+1608.08310}{{$ \triangleright
  $}}

\bibitem{Ruggieri:2016ejz}
M.~Ruggieri and G.~X. Peng, ``\textit{{Critical Temperature of Chiral Symmetry
  Restoration for Quark Matter with a Chiral Chemical Potential}},''
  \href{http://dx.doi.org/10.1088/0954-3899/43/12/125101}{{\em J. Phys.} {\bf
  G43} (2016) no.~12, 125101} \href{http://arxiv.org/abs/1602.05250}{{
  $\bullet$}}
\href{http://inspirehep.net/search?p=find+eprint+1602.05250}{{$ \triangleright
  $}}

\bibitem{Wang:2015tia}
B.~Wang, Y.-L. Wang, Z.-F. Cui, and H.-S. Zong, ``\textit{{Effect of the chiral
  chemical potential on the position of the critical endpoint}},''
\href{http://dx.doi.org/10.1103/PhysRevD.91.034017}{{\em Phys. Rev.} {\bf D91}
  (2015) no.~3, 034017}

\bibitem{Xu:2015vna}
S.-S. Xu, Z.-F. Cui, B.~Wang, Y.-M. Shi, Y.-C. Yang, and H.-S. Zong,
  ``\textit{{Chiral phase transition with a chiral chemical potential in the
  framework of Dyson-Schwinger equations}},''
  \href{http://dx.doi.org/10.1103/PhysRevD.91.056003}{{\em Phys. Rev.} {\bf
  D91} (2015) no.~5, 056003} \href{http://arxiv.org/abs/1505.00316}{{
  $\bullet$}}
\href{http://inspirehep.net/search?p=find+eprint+1505.00316}{{$ \triangleright
  $}}

\bibitem{Cui:2016zqp}
Z.-F. Cui, I.~C. Cloet, Y.~Lu, C.~D. Roberts, S.~M. Schmidt, S.-S. Xu, and
  H.-S. Zong, ``\textit{{Critical endpoint in the presence of a chiral chemical
  potential}},'' \href{http://dx.doi.org/10.1103/PhysRevD.94.071503}{{\em Phys.
  Rev.} {\bf D94} (2016)  071503} \href{http://arxiv.org/abs/1604.08454}{{
  $\bullet$}}
\href{http://inspirehep.net/search?p=find+eprint+1604.08454}{{$ \triangleright
  $}}

\bibitem{Son:2000xc}
D.~T. Son and M.~A. Stephanov, ``\textit{{QCD at finite isospin density}},''
  \href{http://dx.doi.org/10.1103/PhysRevLett.86.592}{{\em Phys. Rev. Lett.}
  {\bf 86} (2001)  592--595} \href{http://arxiv.org/abs/hep-ph/0005225}{{
  $\bullet$}}
\href{http://inspirehep.net/search?p=find+eprint+hep-ph/0005225}{{$
  \triangleright $}}

\bibitem{Kogut:2004zg}
J.~B. Kogut and D.~K. Sinclair, ``\textit{{The Finite temperature transition
  for 2-flavor lattice QCD at finite isospin density}},''
  \href{http://dx.doi.org/10.1103/PhysRevD.70.094501}{{\em Phys. Rev.} {\bf
  D70} (2004)  094501} \href{http://arxiv.org/abs/hep-lat/0407027}{{
  $\bullet$}}
\href{http://inspirehep.net/search?p=find+eprint+hep-lat/0407027}{{$
  \triangleright $}}

\bibitem{deForcrand:2007uz}
P.~de~Forcrand, M.~A. Stephanov, and U.~Wenger, ``\textit{{On the phase diagram
  of QCD at finite isospin density}},'' {\em PoS} {\bf LATTICE2007} (2007)  237
  \href{http://arxiv.org/abs/0711.0023}{{ $\bullet$}}
\href{http://inspirehep.net/search?p=find+eprint+0711.0023}{{$ \triangleright
  $}}

\bibitem{Detmold:2012wc}
W.~Detmold, K.~Orginos, and Z.~Shi, ``\textit{{Lattice QCD at non-zero isospin
  chemical potential}},''
  \href{http://dx.doi.org/10.1103/PhysRevD.86.054507}{{\em Phys. Rev.} {\bf
  D86} (2012)  054507} \href{http://arxiv.org/abs/1205.4224}{{ $\bullet$}}
\href{http://inspirehep.net/search?p=find+eprint+1205.4224}{{$ \triangleright
  $}}

\bibitem{Brandt:2017oyy}
B.~B. Brandt, G.~Endrodi, and S.~Schmalzbauer, ``\textit{{QCD phase diagram for
  nonzero isospin-asymmetry}},''
  \href{http://dx.doi.org/10.1103/PhysRevD.97.054514}{{\em Phys. Rev.} {\bf
  D97} (2018) no.~5, 054514} \href{http://arxiv.org/abs/1712.08190}{{
  $\bullet$}}
\href{http://inspirehep.net/search?p=find+eprint+1712.08190}{{$ \triangleright
  $}}

\bibitem{Splittorff:2000mm}
K.~Splittorff, D.~T. Son, and M.~A. Stephanov, ``\textit{{QCD - like theories
  at finite baryon and isospin density}},''
  \href{http://dx.doi.org/10.1103/PhysRevD.64.016003}{{\em Phys. Rev.} {\bf
  D64} (2001)  016003} \href{http://arxiv.org/abs/hep-ph/0012274}{{ $\bullet$}}
\href{http://inspirehep.net/search?p=find+eprint+hep-ph/0012274}{{$
  \triangleright $}}

\bibitem{Kamikado:2012bt}
K.~Kamikado, N.~Strodthoff, L.~von Smekal, and J.~Wambach,
  ``\textit{{Fluctuations in the quark-meson model for QCD with isospin
  chemical potential}},''
  \href{http://dx.doi.org/10.1016/j.physletb.2012.11.055}{{\em Phys. Lett.}
  {\bf B718} (2013)  1044--1053} \href{http://arxiv.org/abs/1207.0400}{{
  $\bullet$}}
\href{http://inspirehep.net/search?p=find+eprint+1207.0400}{{$ \triangleright
  $}}

\bibitem{Loewe:2002tw}
M.~Loewe and C.~Villavicencio, ``\textit{{Thermal pions at finite isospin
  chemical potential}},''
  \href{http://dx.doi.org/10.1103/PhysRevD.67.074034}{{\em Phys. Rev.} {\bf
  D67} (2003)  074034} \href{http://arxiv.org/abs/hep-ph/0212275}{{ $\bullet$}}
\href{http://inspirehep.net/search?p=find+eprint+hep-ph/0212275}{{$
  \triangleright $}}

\bibitem{Klein:2003fy}
B.~Klein, D.~Toublan, and J.~J.~M. Verbaarschot, ``\textit{{The QCD phase
  diagram at nonzero temperature, baryon and isospin chemical potentials in
  random matrix theory}},''
  \href{http://dx.doi.org/10.1103/PhysRevD.68.014009}{{\em Phys. Rev.} {\bf
  D68} (2003)  014009} \href{http://arxiv.org/abs/hep-ph/0301143}{{ $\bullet$}}
\href{http://inspirehep.net/search?p=find+eprint+hep-ph/0301143}{{$
  \triangleright $}}

\bibitem{Andersen:2015eoa}
J.~O. Andersen, N.~Haque, M.~G. Mustafa, and M.~Strickland,
  ``\textit{{Three-loop hard-thermal-loop perturbation theory thermodynamics at
  finite temperature and finite baryonic and isospin chemical potential}},''
  \href{http://dx.doi.org/10.1103/PhysRevD.93.054045}{{\em Phys. Rev.} {\bf
  D93} (2016) no.~5, 054045} \href{http://arxiv.org/abs/1511.04660}{{
  $\bullet$}}
\href{http://inspirehep.net/search?p=find+eprint+1511.04660}{{$ \triangleright
  $}}

\bibitem{Stiele:2013pma}
R.~Stiele, E.~S. Fraga, and J.~Schaffner-Bielich, ``\textit{{Thermodynamics of
  (2+1)-flavor strongly interacting matter at nonzero isospin}},''
  \href{http://dx.doi.org/10.1016/j.physletb.2013.12.053}{{\em Phys. Lett.}
  {\bf B729} (2014)  72--78} \href{http://arxiv.org/abs/1307.2851}{{
  $\bullet$}}
\href{http://inspirehep.net/search?p=find+eprint+1307.2851}{{$ \triangleright
  $}}

\bibitem{Brandt:2018bwq}
B.~B. Brandt, G.~Endrodi, E.~S. Fraga, M.~Hippert, J.~Schaffner-Bielich, and
  S.~Schmalzbauer, ``\textit{{A new class of compact stars: pion stars}},''
  \href{http://arxiv.org/abs/1802.06685}{{\tt 1802.06685[hep-ph]}}
  \href{http://arxiv.org/abs/1802.06685}{{ $\bullet$}}
\href{http://inspirehep.net/search?p=find+eprint+1802.06685}{{$ \triangleright
  $}}

\bibitem{Fulde:1964zz}
P.~Fulde and R.~A. Ferrell, ``\textit{{Superconductivity in a Strong
  Spin-Exchange Field}},''
\href{http://dx.doi.org/10.1103/PhysRev.135.A550}{{\em Phys. Rev.} {\bf 135}
  (1964)  A550--A563}

\bibitem{larkin:1964zz}
A.~I. larkin and Y.~N. Ovchinnikov, ``\textit{{Nonuniform state of
  superconductors}},''
{\em Zh. Eksp. Teor. Fiz.} {\bf 47} (1964)  1136--1146

\bibitem{Cea:2014xva}
P.~Cea, L.~Cosmai, and A.~Papa, ``\textit{{Critical line of 2+1 flavor QCD}},''
  \href{http://dx.doi.org/10.1103/PhysRevD.89.074512}{{\em Phys. Rev.} {\bf
  D89} (2014) no.~7, 074512} \href{http://arxiv.org/abs/1403.0821}{{
  $\bullet$}}
\href{http://inspirehep.net/search?p=find+eprint+1403.0821}{{$ \triangleright
  $}}

\bibitem{Bonati:2014kpa}
C.~Bonati, P.~de~Forcrand, M.~D'Elia, O.~Philipsen, and F.~Sanfilippo,
  ``\textit{{Chiral phase transition in two-flavor QCD from an imaginary
  chemical potential}},''
  \href{http://dx.doi.org/10.1103/PhysRevD.90.074030}{{\em Phys. Rev.} {\bf
  D90} (2014) no.~7, 074030} \href{http://arxiv.org/abs/1408.5086}{{
  $\bullet$}}
\href{http://inspirehep.net/search?p=find+eprint+1408.5086}{{$ \triangleright
  $}}

\bibitem{Bonati:2015bha}
C.~Bonati, M.~D'Elia, M.~Mariti, M.~Mesiti, F.~Negro, and F.~Sanfilippo,
  ``\textit{{Curvature of the chiral pseudocritical line in QCD: Continuum
  extrapolated results}},''
  \href{http://dx.doi.org/10.1103/PhysRevD.92.054503}{{\em Phys. Rev.} {\bf
  D92} (2015) no.~5, 054503} \href{http://arxiv.org/abs/1507.03571}{{
  $\bullet$}}
\href{http://inspirehep.net/search?p=find+eprint+1507.03571}{{$ \triangleright
  $}}

\bibitem{Bonati:2018nut}
C.~Bonati, M.~D'Elia, F.~Negro, F.~Sanfilippo, and K.~Zambello,
  ``\textit{{Curvature of the pseudocritical line in QCD: Taylor expansion
  matches analytic continuation}},''
  \href{http://arxiv.org/abs/1805.02960}{{\tt 1805.02960[hep-lat]}}
  \href{http://arxiv.org/abs/1805.02960}{{ $\bullet$}}
\href{http://inspirehep.net/search?p=find+eprint+1805.02960}{{$ \triangleright
  $}}

\bibitem{Braun:2011iz}
J.~Braun, B.~Klein, and B.-J. Schaefer, ``\textit{{On the Phase Structure of
  QCD in a Finite Volume}},''
  \href{http://dx.doi.org/10.1016/j.physletb.2012.05.053}{{\em Phys. Lett.}
  {\bf B713} (2012)  216--223} \href{http://arxiv.org/abs/1110.0849}{{
  $\bullet$}}
\href{http://inspirehep.net/search?p=find+eprint+1110.0849}{{$ \triangleright
  $}}

\bibitem{Pawlowski:2014zaa}
J.~M. Pawlowski and F.~Rennecke, ``\textit{{Higher order quark-mesonic
  scattering processes and the phase structure of QCD}},''
  \href{http://dx.doi.org/10.1103/PhysRevD.90.076002}{{\em Phys. Rev.} {\bf
  D90} (2014) no.~7, 076002} \href{http://arxiv.org/abs/1403.1179}{{
  $\bullet$}}
\href{http://inspirehep.net/search?p=find+eprint+1403.1179}{{$ \triangleright
  $}}

\bibitem{deForcrand:2002hgr}
P.~de~Forcrand and O.~Philipsen, ``\textit{{The QCD phase diagram for small
  densities from imaginary chemical potential}},''
  \href{http://dx.doi.org/10.1016/S0550-3213(02)00626-0}{{\em Nucl. Phys.} {\bf
  B642} (2002)  290--306} \href{http://arxiv.org/abs/hep-lat/0205016}{{
  $\bullet$}}
\href{http://inspirehep.net/search?p=find+eprint+hep-lat/0205016}{{$
  \triangleright $}}

\bibitem{Philipsen:2008gf}
O.~Philipsen, ``\textit{{Status of Lattice Studies of the QCD Phase
  Diagram}},'' \href{http://dx.doi.org/10.1143/PTPS.174.206}{{\em Prog. Theor.
  Phys. Suppl.} {\bf 174} (2008)  206--213}
  \href{http://arxiv.org/abs/0808.0672}{{ $\bullet$}}
\href{http://inspirehep.net/search?p=find+eprint+0808.0672}{{$ \triangleright
  $}}

\bibitem{SteinbrecherI}
P.~Steinbrecher, {\em PhD-thesis, University of Bielefeld} (2018)

\bibitem{Steinbrecher:2018phh}
P.~Steinbrecher, ``\textit{{The QCD crossover at zero and non-zero baryon
  densities from Lattice QCD}},'' \href{http://arxiv.org/abs/1807.05607}{{\tt
  1807.05607[hep-lat]}} \href{http://arxiv.org/abs/1807.05607}{{ $\bullet$}}
\href{http://inspirehep.net/search?p=find+eprint+1807.05607}{{$ \triangleright
  $}}

\bibitem{Luo:2017faz}
X.~Luo and N.~Xu, ``\textit{{Search for the QCD Critical Point with
  Fluctuations of Conserved Quantities in Relativistic Heavy-Ion Collisions at
  RHIC : An Overview}},''
  \href{http://dx.doi.org/10.1007/s41365-017-0257-0}{{\em Nucl. Sci. Tech.}
  {\bf 28} (2017) no.~8, 112} \href{http://arxiv.org/abs/1701.02105}{{
  $\bullet$}}
\href{http://inspirehep.net/search?p=find+eprint+1701.02105}{{$ \triangleright
  $}}

\bibitem{Asakawa:2000wh}
M.~Asakawa, U.~W. Heinz, and B.~Muller, ``\textit{{Fluctuation probes of quark
  deconfinement}},'' \href{http://dx.doi.org/10.1103/PhysRevLett.85.2072}{{\em
  Phys. Rev. Lett.} {\bf 85} (2000)  2072--2075}
  \href{http://arxiv.org/abs/hep-ph/0003169}{{ $\bullet$}}
\href{http://inspirehep.net/search?p=find+eprint+hep-ph/0003169}{{$
  \triangleright $}}

\bibitem{Jeon:2000wg}
S.~Jeon and V.~Koch, ``\textit{{Charged particle ratio fluctuation as a signal
  for QGP}},'' \href{http://dx.doi.org/10.1103/PhysRevLett.85.2076}{{\em Phys.
  Rev. Lett.} {\bf 85} (2000)  2076--2079}
  \href{http://arxiv.org/abs/hep-ph/0003168}{{ $\bullet$}}
\href{http://inspirehep.net/search?p=find+eprint+hep-ph/0003168}{{$
  \triangleright $}}

\bibitem{Koch:2005vg}
V.~Koch, A.~Majumder, and J.~Randrup, ``\textit{{Baryon-strangeness
  correlations: A Diagnostic of strongly interacting matter}},''
  \href{http://dx.doi.org/10.1103/PhysRevLett.95.182301}{{\em Phys. Rev. Lett.}
  {\bf 95} (2005)  182301} \href{http://arxiv.org/abs/nucl-th/0505052}{{
  $\bullet$}}
\href{http://inspirehep.net/search?p=find+eprint+nucl-th/0505052}{{$
  \triangleright $}}

\bibitem{Ejiri:2005wq}
S.~Ejiri, F.~Karsch, and K.~Redlich, ``\textit{{Hadronic fluctuations at the
  QCD phase transition}},''
  \href{http://dx.doi.org/10.1016/j.physletb.2005.11.083}{{\em Phys. Lett.}
  {\bf B633} (2006)  275--282} \href{http://arxiv.org/abs/hep-ph/0509051}{{
  $\bullet$}}
\href{http://inspirehep.net/search?p=find+eprint+hep-ph/0509051}{{$
  \triangleright $}}

\bibitem{Friman:2011pf}
B.~Friman, F.~Karsch, K.~Redlich, and V.~Skokov, ``\textit{{Fluctuations as
  probe of the QCD phase transition and freeze-out in heavy ion collisions at
  LHC and RHIC}},''
  \href{http://dx.doi.org/10.1140/epjc/s10052-011-1694-2}{{\em Eur. Phys. J.}
  {\bf C71} (2011)  1694} \href{http://arxiv.org/abs/1103.3511}{{ $\bullet$}}
\href{http://inspirehep.net/search?p=find+eprint+1103.3511}{{$ \triangleright
  $}}

\bibitem{Bazavov:2012vg}
A.~Bazavov {\em et al.}, ``\textit{{Freeze-out Conditions in Heavy Ion
  Collisions from QCD Thermodynamics}},''
  \href{http://dx.doi.org/10.1103/PhysRevLett.109.192302}{{\em Phys. Rev.
  Lett.} {\bf 109} (2012)  192302} \href{http://arxiv.org/abs/1208.1220}{{
  $\bullet$}}
\href{http://inspirehep.net/search?p=find+eprint+1208.1220}{{$ \triangleright
  $}}

\bibitem{Borsanyi:2013hza}
S.~Borsanyi, Z.~Fodor, S.~D. Katz, S.~Krieg, C.~Ratti, and K.~K. Szabo,
  ``\textit{{Freeze-out parameters: lattice meets experiment}},''
  \href{http://dx.doi.org/10.1103/PhysRevLett.111.062005}{{\em Phys. Rev.
  Lett.} {\bf 111} (2013)  062005} \href{http://arxiv.org/abs/1305.5161}{{
  $\bullet$}}
\href{http://inspirehep.net/search?p=find+eprint+1305.5161}{{$ \triangleright
  $}}

\bibitem{Welzbacher:2016}
C.~A. Welzbacher, {\em Quarks and gluons in the phase diagram of quantum
  chromodynamics}.
\newblock PhD thesis, University of Giessen 2016.
\newblock \href{http://geb.uni-giessen.de/geb/volltexte/2016/12250}{{$
  \triangleright $}}

\bibitem{Karsch:2010ck}
F.~Karsch and K.~Redlich, ``\textit{{Probing freeze-out conditions in heavy ion
  collisions with moments of charge fluctuations}},''
  \href{http://dx.doi.org/10.1016/j.physletb.2010.10.046}{{\em Phys. Lett.}
  {\bf B695} (2011)  136--142} \href{http://arxiv.org/abs/1007.2581}{{
  $\bullet$}}
\href{http://inspirehep.net/search?p=find+eprint+1007.2581}{{$ \triangleright
  $}}

\bibitem{Cheng:2008zh}
M.~Cheng {\em et al.}, ``\textit{{Baryon Number, Strangeness and Electric
  Charge Fluctuations in QCD at High Temperature}},''
  \href{http://dx.doi.org/10.1103/PhysRevD.79.074505}{{\em Phys. Rev.} {\bf
  D79} (2009)  074505} \href{http://arxiv.org/abs/0811.1006}{{ $\bullet$}}
\href{http://inspirehep.net/search?p=find+eprint+0811.1006}{{$ \triangleright
  $}}

\bibitem{Borsanyi:2011sw}
S.~Borsanyi, Z.~Fodor, S.~D. Katz, S.~Krieg, C.~Ratti, and K.~Szabo,
  ``\textit{{Fluctuations of conserved charges at finite temperature from
  lattice QCD}},'' \href{http://dx.doi.org/10.1007/JHEP01(2012)138}{{\em JHEP}
  {\bf 01} (2012)  138} \href{http://arxiv.org/abs/1112.4416}{{ $\bullet$}}
\href{http://inspirehep.net/search?p=find+eprint+1112.4416}{{$ \triangleright
  $}}

\bibitem{Bazavov:2012jq}
{\bf HotQCD} Collaboration: A.~Bazavov {\em et al.}, ``\textit{{Fluctuations
  and Correlations of net baryon number, electric charge, and strangeness: A
  comparison of lattice QCD results with the hadron resonance gas model}},''
  \href{http://dx.doi.org/10.1103/PhysRevD.86.034509}{{\em Phys. Rev.} {\bf
  D86} (2012)  034509} \href{http://arxiv.org/abs/1203.0784}{{ $\bullet$}}
\href{http://inspirehep.net/search?p=find+eprint+1203.0784}{{$ \triangleright
  $}}

\bibitem{Bellwied:2015lba}
R.~Bellwied, S.~Borsanyi, Z.~Fodor, S.~D. Katz, A.~Pasztor, C.~Ratti, and K.~K.
  Szabo, ``\textit{{Fluctuations and correlations in high temperature QCD}},''
  \href{http://dx.doi.org/10.1103/PhysRevD.92.114505}{{\em Phys. Rev.} {\bf
  D92} (2015) no.~11, 114505} \href{http://arxiv.org/abs/1507.04627}{{
  $\bullet$}}
\href{http://inspirehep.net/search?p=find+eprint+1507.04627}{{$ \triangleright
  $}}

\bibitem{DElia:2016jqh}
M.~D'Elia, G.~Gagliardi, and F.~Sanfilippo, ``\textit{{Higher order quark
  number fluctuations via imaginary chemical potentials in $N_f=2+1$ QCD}},''
  \href{http://dx.doi.org/10.1103/PhysRevD.95.094503}{{\em Phys. Rev.} {\bf
  D95} (2017) no.~9, 094503} \href{http://arxiv.org/abs/1611.08285}{{
  $\bullet$}}
\href{http://inspirehep.net/search?p=find+eprint+1611.08285}{{$ \triangleright
  $}}

\bibitem{Bazavov:2017tot}
{\bf HotQCD} Collaboration: A.~Bazavov {\em et al.}, ``\textit{{Skewness and
  kurtosis of net baryon-number distributions at small values of the baryon
  chemical potential}},''
  \href{http://dx.doi.org/10.1103/PhysRevD.96.074510}{{\em Phys. Rev.} {\bf
  D96} (2017) no.~7, 074510} \href{http://arxiv.org/abs/1708.04897}{{
  $\bullet$}}
\href{http://inspirehep.net/search?p=find+eprint+1708.04897}{{$ \triangleright
  $}}

\bibitem{Borsanyi:2018grb}
S.~Borsanyi, Z.~Fodor, J.~N. Guenther, S.~K. Katz, K.~K. Szabo, A.~Pasztor,
  I.~Portillo, and C.~Ratti, ``\textit{{Higher order fluctuations and
  correlations of conserved charges from lattice QCD}},''
  \href{http://arxiv.org/abs/1805.04445}{{\tt 1805.04445[hep-lat]}}
  \href{http://arxiv.org/abs/1805.04445}{{ $\bullet$}}
\href{http://inspirehep.net/search?p=find+eprint+1805.04445}{{$ \triangleright
  $}}

\bibitem{Schaefer:2011ex}
B.~J. Schaefer and M.~Wagner, ``\textit{{QCD critical region and higher moments
  for three flavor models}},''
  \href{http://dx.doi.org/10.1103/PhysRevD.85.034027}{{\em Phys. Rev.} {\bf
  D85} (2012)  034027} \href{http://arxiv.org/abs/1111.6871}{{ $\bullet$}}
\href{http://inspirehep.net/search?p=find+eprint+1111.6871}{{$ \triangleright
  $}}

\bibitem{Fu:2016tey}
W.-j. Fu, J.~M. Pawlowski, F.~Rennecke, and B.-J. Schaefer, ``\textit{{Baryon
  number fluctuations at finite temperature and density}},''
  \href{http://dx.doi.org/10.1103/PhysRevD.94.116020}{{\em Phys. Rev.} {\bf
  D94} (2016) no.~11, 116020} \href{http://arxiv.org/abs/1608.04302}{{
  $\bullet$}}
\href{http://inspirehep.net/search?p=find+eprint+1608.04302}{{$ \triangleright
  $}}

\bibitem{Almasi:2017bhq}
G.~A. Almasi, B.~Friman, and K.~Redlich, ``\textit{{Baryon number fluctuations
  in chiral effective models and their phenomenological implications}},''
  \href{http://dx.doi.org/10.1103/PhysRevD.96.014027}{{\em Phys. Rev.} {\bf
  D96} (2017) no.~1, 014027} \href{http://arxiv.org/abs/1703.05947}{{
  $\bullet$}}
\href{http://inspirehep.net/search?p=find+eprint+1703.05947}{{$ \triangleright
  $}}

\bibitem{Brown:1990ev}
F.~R. Brown, F.~P. Butler, H.~Chen, N.~H. Christ, Z.-h. Dong, W.~Schaffer,
  L.~I. Unger, and A.~Vaccarino, ``\textit{{On the existence of a phase
  transition for QCD with three light quarks}},''
\href{http://dx.doi.org/10.1103/PhysRevLett.65.2491}{{\em Phys. Rev. Lett.}
  {\bf 65} (1990)  2491--2494}

\bibitem{Yaffe:1982qf}
L.~G. Yaffe and B.~Svetitsky, ``\textit{{First Order Phase Transition in the
  SU(3) Gauge Theory at Finite Temperature}},''
\href{http://dx.doi.org/10.1103/PhysRevD.26.963}{{\em Phys. Rev.} {\bf D26}
  (1982)  963}

\bibitem{Lucini:2005vg}
B.~Lucini, M.~Teper, and U.~Wenger, ``\textit{{Properties of the deconfining
  phase transition in SU(N) gauge theories}},''
  \href{http://dx.doi.org/10.1088/1126-6708/2005/02/033}{{\em JHEP} {\bf 02}
  (2005)  033} \href{http://arxiv.org/abs/hep-lat/0502003}{{ $\bullet$}}
\href{http://inspirehep.net/search?p=find+eprint+hep-lat/0502003}{{$
  \triangleright $}}

\bibitem{deForcrand:2003vyj}
P.~de~Forcrand and O.~Philipsen, ``\textit{{The QCD phase diagram for three
  degenerate flavors and small baryon density}},''
  \href{http://dx.doi.org/10.1016/j.nuclphysb.2003.09.005}{{\em Nucl. Phys.}
  {\bf B673} (2003)  170--186} \href{http://arxiv.org/abs/hep-lat/0307020}{{
  $\bullet$}}
\href{http://inspirehep.net/search?p=find+eprint+hep-lat/0307020}{{$
  \triangleright $}}

\bibitem{deForcrand:2010he}
P.~de~Forcrand and O.~Philipsen, ``\textit{{Constraining the QCD phase diagram
  by tricritical lines at imaginary chemical potential}},''
  \href{http://dx.doi.org/10.1103/PhysRevLett.105.152001}{{\em Phys. Rev.
  Lett.} {\bf 105} (2010)  152001} \href{http://arxiv.org/abs/1004.3144}{{
  $\bullet$}}
\href{http://inspirehep.net/search?p=find+eprint+1004.3144}{{$ \triangleright
  $}}

\bibitem{Saito:2011fs}
{\bf WHOT-QCD} Collaboration: H.~Saito, S.~Ejiri, S.~Aoki, T.~Hatsuda,
  K.~Kanaya, Y.~Maezawa, H.~Ohno, and T.~Umeda, ``\textit{{Phase structure of
  finite temperature QCD in the heavy quark region}},''
  \href{http://dx.doi.org/10.1103/PhysRevD.85.079902,
  10.1103/PhysRevD.84.054502}{{\em Phys. Rev.} {\bf D84} (2011)  054502}
  \href{http://arxiv.org/abs/1106.0974}{{ $\bullet$}}
\href{http://inspirehep.net/search?p=find+eprint+1106.0974}{{$ \triangleright
  $}}

\bibitem{Fromm:2011qi}
M.~Fromm, J.~Langelage, S.~Lottini, and O.~Philipsen, ``\textit{{The QCD
  deconfinement transition for heavy quarks and all baryon chemical
  potentials}},'' \href{http://dx.doi.org/10.1007/JHEP01(2012)042}{{\em JHEP}
  {\bf 01} (2012)  042} \href{http://arxiv.org/abs/1111.4953}{{ $\bullet$}}
\href{http://inspirehep.net/search?p=find+eprint+1111.4953}{{$ \triangleright
  $}}

\bibitem{Kashiwa:2012wa}
K.~Kashiwa, R.~D. Pisarski, and V.~V. Skokov, ``\textit{{Critical endpoint for
  deconfinement in matrix and other effective models}},''
  \href{http://dx.doi.org/10.1103/PhysRevD.85.114029}{{\em Phys. Rev.} {\bf
  D85} (2012)  114029} \href{http://arxiv.org/abs/1205.0545}{{ $\bullet$}}
\href{http://inspirehep.net/search?p=find+eprint+1205.0545}{{$ \triangleright
  $}}

\bibitem{Lo:2014vba}
P.~M. Lo, B.~Friman, and K.~Redlich, ``\textit{{Polyakov loop fluctuations and
  deconfinement in the limit of heavy quarks}},''
  \href{http://dx.doi.org/10.1103/PhysRevD.90.074035}{{\em Phys. Rev.} {\bf
  D90} (2014) no.~7, 074035} \href{http://arxiv.org/abs/1406.4050}{{
  $\bullet$}}
\href{http://inspirehep.net/search?p=find+eprint+1406.4050}{{$ \triangleright
  $}}

\bibitem{Fischer:2014vxa}
C.~S. Fischer, J.~Luecker, and J.~M. Pawlowski, ``\textit{{Phase structure of
  QCD for heavy quarks}},''
  \href{http://dx.doi.org/10.1103/PhysRevD.91.014024}{{\em Phys. Rev.} {\bf
  D91} (2015) no.~1, 014024} \href{http://arxiv.org/abs/1409.8462}{{
  $\bullet$}}
\href{http://inspirehep.net/search?p=find+eprint+1409.8462}{{$ \triangleright
  $}}

\bibitem{Reinosa:2015oua}
U.~Reinosa, J.~Serreau, and M.~Tissier, ``\textit{{Perturbative study of the
  QCD phase diagram for heavy quarks at nonzero chemical potential}},''
  \href{http://dx.doi.org/10.1103/PhysRevD.92.025021}{{\em Phys. Rev.} {\bf
  D92} (2015)  025021} \href{http://arxiv.org/abs/1504.02916}{{ $\bullet$}}
\href{http://inspirehep.net/search?p=find+eprint+1504.02916}{{$ \triangleright
  $}}

\bibitem{Maelger:2017amh}
J.~Maelger, U.~Reinosa, and J.~Serreau, ``\textit{{Perturbative study of the
  QCD phase diagram for heavy quarks at nonzero chemical potential: Two-loop
  corrections}},'' \href{http://dx.doi.org/10.1103/PhysRevD.97.074027}{{\em
  Phys. Rev.} {\bf D97} (2018) no.~7, 074027}
  \href{http://arxiv.org/abs/1710.01930}{{ $\bullet$}}
\href{http://inspirehep.net/search?p=find+eprint+1710.01930}{{$ \triangleright
  $}}

\bibitem{Butti:2003nu}
A.~Butti, A.~Pelissetto, and E.~Vicari, ``\textit{{On the nature of the finite
  temperature transition in QCD}},''
  \href{http://dx.doi.org/10.1088/1126-6708/2003/08/029}{{\em JHEP} {\bf 08}
  (2003)  029} \href{http://arxiv.org/abs/hep-ph/0307036}{{ $\bullet$}}
\href{http://inspirehep.net/search?p=find+eprint+hep-ph/0307036}{{$
  \triangleright $}}

\bibitem{deForcrand:2017cgb}
P.~de~Forcrand and M.~D'Elia, ``\textit{{Continuum limit and universality of
  the Columbia plot}},'' {\em PoS} {\bf LATTICE2016} (2017)  081
  \href{http://arxiv.org/abs/1702.00330}{{ $\bullet$}}
\href{http://inspirehep.net/search?p=find+eprint+1702.00330}{{$ \triangleright
  $}}

\bibitem{Karsch:2001nf}
F.~Karsch, E.~Laermann, and C.~Schmidt, ``\textit{{The Chiral critical point in
  three-flavor QCD}},''
  \href{http://dx.doi.org/10.1016/S0370-2693(01)01114-5}{{\em Phys. Lett.} {\bf
  B520} (2001)  41--49} \href{http://arxiv.org/abs/hep-lat/0107020}{{
  $\bullet$}}
\href{http://inspirehep.net/search?p=find+eprint+hep-lat/0107020}{{$
  \triangleright $}}

\bibitem{Karsch:2003va}
F.~Karsch, C.~R. Allton, S.~Ejiri, S.~J. Hands, O.~Kaczmarek, E.~Laermann, and
  C.~Schmidt, ``\textit{{Where is the chiral critical point in three flavor
  QCD?}},'' \href{http://dx.doi.org/10.1016/S0920-5632(03)02659-8}{{\em Nucl.
  Phys. Proc. Suppl.} {\bf 129} (2004)  614--616}
  \href{http://arxiv.org/abs/hep-lat/0309116}{{ $\bullet$}}
\href{http://inspirehep.net/search?p=find+eprint+hep-lat/0309116}{{$
  \triangleright $}}

\bibitem{deForcrand:2007rq}
P.~de~Forcrand, S.~Kim, and O.~Philipsen, ``\textit{{A QCD chiral critical
  point at small chemical potential: Is it there or not?}},'' {\em PoS} {\bf
  LATTICE2007} (2007)  178 \href{http://arxiv.org/abs/0711.0262}{{ $\bullet$}}
\href{http://inspirehep.net/search?p=find+eprint+0711.0262}{{$ \triangleright
  $}}

\bibitem{Ding:2011du}
H.~T. Ding, A.~Bazavov, P.~Hegde, F.~Karsch, S.~Mukherjee, and P.~Petreczky,
  ``\textit{{Exploring phase diagram of $N_f=3$ QCD at $\mu=0$ with HISQ
  fermions}},'' {\em PoS} {\bf LATTICE2011} (2011)  191
  \href{http://arxiv.org/abs/1111.0185}{{ $\bullet$}}
\href{http://inspirehep.net/search?p=find+eprint+1111.0185}{{$ \triangleright
  $}}

\bibitem{Jin:2014hea}
X.-Y. Jin, Y.~Kuramashi, Y.~Nakamura, S.~Takeda, and A.~Ukawa,
  ``\textit{{Critical endpoint of the finite temperature phase transition for
  three flavor QCD}},''
  \href{http://dx.doi.org/10.1103/PhysRevD.91.014508}{{\em Phys. Rev.} {\bf
  D91} (2015) no.~1, 014508} \href{http://arxiv.org/abs/1411.7461}{{
  $\bullet$}}
\href{http://inspirehep.net/search?p=find+eprint+1411.7461}{{$ \triangleright
  $}}

\bibitem{Takeda:2016vfj}
S.~Takeda, X.-Y. Jin, Y.~Kuramashi, Y.~Nakamura, and A.~Ukawa,
  ``\textit{{Update on Nf=3 finite temperature QCD phase structure with
  Wilson-Clover fermion action}},'' {\em PoS} {\bf LATTICE2016} (2017)  384
  \href{http://arxiv.org/abs/1612.05371}{{ $\bullet$}}
\href{http://inspirehep.net/search?p=find+eprint+1612.05371}{{$ \triangleright
  $}}

\bibitem{Bazavov:2017xul}
A.~Bazavov, H.~T. Ding, P.~Hegde, F.~Karsch, E.~Laermann, S.~Mukherjee,
  P.~Petreczky, and C.~Schmidt, ``\textit{{Chiral phase structure of three
  flavor QCD at vanishing baryon number density}},''
  \href{http://dx.doi.org/10.1103/PhysRevD.95.074505}{{\em Phys. Rev.} {\bf
  D95} (2017) no.~7, 074505} \href{http://arxiv.org/abs/1701.03548}{{
  $\bullet$}}
\href{http://inspirehep.net/search?p=find+eprint+1701.03548}{{$ \triangleright
  $}}

\bibitem{Iwasaki:1996ya}
Y.~Iwasaki, K.~Kanaya, S.~Kaya, and T.~Yoshie, ``\textit{{Scaling of chiral
  order parameter in two flavor QCD}},''
  \href{http://dx.doi.org/10.1103/PhysRevLett.78.179}{{\em Phys. Rev. Lett.}
  {\bf 78} (1997)  179--182} \href{http://arxiv.org/abs/hep-lat/9609022}{{
  $\bullet$}}
\href{http://inspirehep.net/search?p=find+eprint+hep-lat/9609022}{{$
  \triangleright $}}

\bibitem{DElia:2004uwa}
M.~D'Elia, A.~Di~Giacomo, and C.~Pica, ``\textit{{On the order of the
  deconfining transition in N(f) = 2 QCD}},''
  \href{http://dx.doi.org/10.1142/S0217751X05028235}{{\em Int. J. Mod. Phys.}
  {\bf A20} (2005)  4579--4584} \href{http://arxiv.org/abs/hep-lat/0408011}{{
  $\bullet$}}
\href{http://inspirehep.net/search?p=find+eprint+hep-lat/0408011}{{$
  \triangleright $}}

\bibitem{DElia:2005nmv}
M.~D'Elia, A.~Di~Giacomo, and C.~Pica, ``\textit{{Two flavor QCD and
  confinement}},'' \href{http://dx.doi.org/10.1103/PhysRevD.72.114510}{{\em
  Phys. Rev.} {\bf D72} (2005)  114510}
  \href{http://arxiv.org/abs/hep-lat/0503030}{{ $\bullet$}}
\href{http://inspirehep.net/search?p=find+eprint+hep-lat/0503030}{{$
  \triangleright $}}

\bibitem{Kogut:2006gt}
J.~B. Kogut and D.~K. Sinclair, ``\textit{{Evidence for O(2) universality at
  the finite temperature transition for lattice QCD with 2 flavors of massless
  staggered quarks}},''
  \href{http://dx.doi.org/10.1103/PhysRevD.73.074512}{{\em Phys. Rev.} {\bf
  D73} (2006)  074512} \href{http://arxiv.org/abs/hep-lat/0603021}{{
  $\bullet$}}
\href{http://inspirehep.net/search?p=find+eprint+hep-lat/0603021}{{$
  \triangleright $}}

\bibitem{Dick:2015twa}
V.~Dick, F.~Karsch, E.~Laermann, S.~Mukherjee, and S.~Sharma,
  ``\textit{{Microscopic origin of $U_A(1)$ symmetry violation in the high
  temperature phase of QCD}},''
  \href{http://dx.doi.org/10.1103/PhysRevD.91.094504}{{\em Phys. Rev.} {\bf
  D91} (2015) no.~9, 094504} \href{http://arxiv.org/abs/1502.06190}{{
  $\bullet$}}
\href{http://inspirehep.net/search?p=find+eprint+1502.06190}{{$ \triangleright
  $}}

\bibitem{Philipsen:2016hkv}
O.~Philipsen and C.~Pinke, ``\textit{{The $N_f=2$ QCD chiral phase transition
  with Wilson fermions at zero and imaginary chemical potential}},''
  \href{http://dx.doi.org/10.1103/PhysRevD.93.114507}{{\em Phys. Rev.} {\bf
  D93} (2016) no.~11, 114507} \href{http://arxiv.org/abs/1602.06129}{{
  $\bullet$}}
\href{http://inspirehep.net/search?p=find+eprint+1602.06129}{{$ \triangleright
  $}}

\bibitem{Cuteri:2017gci}
F.~Cuteri, O.~Philipsen, and A.~Sciarra, ``\textit{{QCD chiral phase transition
  from noninteger numbers of flavors}},''
  \href{http://dx.doi.org/10.1103/PhysRevD.97.114511}{{\em Phys. Rev.} {\bf
  D97} (2018) no.~11, 114511} \href{http://arxiv.org/abs/1711.05658}{{
  $\bullet$}}
\href{http://inspirehep.net/search?p=find+eprint+1711.05658}{{$ \triangleright
  $}}

\bibitem{Ding:2018auz}
H.~T. Ding, P.~Hegde, F.~Karsch, A.~Lahiri, S.~T. Li, S.~Mukherjee, and
  P.~Petreczky, ``\textit{{Chiral phase transition of (2+1)-flavor QCD}},'' in
  {\em {27th International Conference on Ultrarelativistic Nucleus-Nucleus
  Collisions (Quark Matter 2018) Venice, Italy, May 14-19, 2018}}.
\newblock 2018.
\newblock
\href{http://arxiv.org/abs/1807.05727}{{\tt 1807.05727[hep-lat]}}.
\newblock

\bibitem{Lenaghan:2000kr}
J.~T. Lenaghan, ``\textit{{Influence of the U(1)(A) anomaly on the QCD phase
  transition}},'' \href{http://dx.doi.org/10.1103/PhysRevD.63.037901}{{\em
  Phys. Rev.} {\bf D63} (2001)  037901}
  \href{http://arxiv.org/abs/hep-ph/0005330}{{ $\bullet$}}
\href{http://inspirehep.net/search?p=find+eprint+hep-ph/0005330}{{$
  \triangleright $}}

\bibitem{Kovacs:2006ym}
P.~Kovacs and Z.~Szep, ``\textit{{The critical surface of the $SU(3)_L x
  SU(3)_R$ chiral quark model at non-zero baryon density}},''
  \href{http://dx.doi.org/10.1103/PhysRevD.75.025015}{{\em Phys. Rev.} {\bf
  D75} (2007)  025015} \href{http://arxiv.org/abs/hep-ph/0611208}{{ $\bullet$}}
\href{http://inspirehep.net/search?p=find+eprint+hep-ph/0611208}{{$
  \triangleright $}}

\bibitem{Fukushima:2008wg}
K.~Fukushima, ``\textit{{Phase diagrams in the three-flavor Nambu-Jona-Lasinio
  model with the Polyakov loop}},''
  \href{http://dx.doi.org/10.1103/PhysRevD.77.114028,
  10.1103/PhysRevD.78.039902}{{\em Phys. Rev.} {\bf D77} (2008)  114028}
  \href{http://arxiv.org/abs/0803.3318}{{ $\bullet$}}
\href{http://inspirehep.net/search?p=find+eprint+0803.3318}{{$ \triangleright
  $}}

\bibitem{Schaefer:2008hk}
B.-J. Schaefer and M.~Wagner, ``\textit{{The Three-flavor chiral phase
  structure in hot and dense QCD matter}},''
  \href{http://dx.doi.org/10.1103/PhysRevD.79.014018}{{\em Phys. Rev.} {\bf
  D79} (2009)  014018} \href{http://arxiv.org/abs/0808.1491}{{ $\bullet$}}
\href{http://inspirehep.net/search?p=find+eprint+0808.1491}{{$ \triangleright
  $}}

\bibitem{Mitter:2013fxa}
M.~Mitter and B.-J. Schaefer, ``\textit{{Fluctuations and the axial anomaly
  with three quark flavors}},''
  \href{http://dx.doi.org/10.1103/PhysRevD.89.054027}{{\em Phys. Rev.} {\bf
  D89} (2014) no.~5, 054027} \href{http://arxiv.org/abs/1308.3176}{{
  $\bullet$}}
\href{http://inspirehep.net/search?p=find+eprint+1308.3176}{{$ \triangleright
  $}}

\bibitem{Resch:2017vjs}
S.~Resch, F.~Rennecke, and B.-J. Schaefer, ``\textit{{Mass sensitivity of the
  three-flavor chiral phase transition}},''
  \href{http://arxiv.org/abs/1712.07961}{{\tt 1712.07961[hep-ph]}}
  \href{http://arxiv.org/abs/1712.07961}{{ $\bullet$}}
\href{http://inspirehep.net/search?p=find+eprint+1712.07961}{{$ \triangleright
  $}}

\bibitem{Braun:2009gm}
J.~Braun, L.~M. Haas, F.~Marhauser, and J.~M. Pawlowski, ``\textit{{Phase
  Structure of Two-Flavor QCD at Finite Chemical Potential}},''
  \href{http://dx.doi.org/10.1103/PhysRevLett.106.022002}{{\em Phys. Rev.
  Lett.} {\bf 106} (2011)  022002} \href{http://arxiv.org/abs/0908.0008}{{
  $\bullet$}}
\href{http://inspirehep.net/search?p=find+eprint+0908.0008}{{$ \triangleright
  $}}

\bibitem{Fischer:2011pk}
C.~S. Fischer and J.~A. Mueller, ``\textit{{On critical scaling at the QCD
  $N_f=2$ chiral phase transition}},''
  \href{http://dx.doi.org/10.1103/PhysRevD.84.054013}{{\em Phys. Rev.} {\bf
  D84} (2011)  054013} \href{http://arxiv.org/abs/1106.2700}{{ $\bullet$}}
\href{http://inspirehep.net/search?p=find+eprint+1106.2700}{{$ \triangleright
  $}}

\bibitem{Langelage:2010yr}
J.~Langelage, S.~Lottini, and O.~Philipsen, ``\textit{{Centre symmetric 3d
  effective actions for thermal SU(N) Yang-Mills from strong coupling
  series}},'' \href{http://dx.doi.org/10.1007/JHEP07(2011)014,
  10.1007/JHEP02(2011)057}{{\em JHEP} {\bf 02} (2011)  057}
  \href{http://arxiv.org/abs/1010.0951}{{ $\bullet$}}
\href{http://inspirehep.net/search?p=find+eprint+1010.0951}{{$ \triangleright
  $}}

\bibitem{Roberge:1986mm}
A.~Roberge and N.~Weiss, ``\textit{{Gauge Theories With Imaginary Chemical
  Potential and the Phases of {QCD}}},''
\href{http://dx.doi.org/10.1016/0550-3213(86)90582-1}{{\em Nucl. Phys.} {\bf
  B275} (1986)  734--745}

\bibitem{Rivers:1987hi}
R.~J. Rivers, {\em {Path Integral Methods in Quantum Field Theory}}.
\newblock Cambridge University Press
1988.
\newblock

\bibitem{Alkofer:2000wg}
R.~Alkofer and L.~von Smekal, ``\textit{{T}he {I}nfrared behavior of {QCD}
  {G}reen's functions: {C}onfinement dynamical symmetry breaking, and hadrons
  as relativistic bound states},''
  \href{http://dx.doi.org/10.1016/S0370-1573(01)00010-2}{{\em Phys. Rept.} {\bf
  353} (2001)  281} \href{http://arxiv.org/abs/hep-ph/0007355}{{ $\bullet$}}
\href{http://inspirehep.net/search?p=find+eprint+hep-ph/0007355}{{$
  \triangleright $}}

\bibitem{Faddeev:1967fc}
L.~D. Faddeev and V.~N. Popov, ``\textit{{Feynman Diagrams for the Yang-Mills
  Field}},''
\href{http://dx.doi.org/10.1016/0370-2693(67)90067-6}{{\em Phys. Lett.} {\bf
  25B} (1967)  29--30}

\bibitem{Pokorski:1987ed}
S.~Pokorski, {\em {Gauge Field Theories}}.
\newblock Cambridge University Press 2005.
\newblock
\href{http://www.cambridge.org/uk/catalogue/catalogue.asp?isbn=0521265371}{{$
  \triangleright $}}

\bibitem{Williams:2002dw}
A.~G. Williams, ``\textit{{QCD, gauge fixing, and the Gribov problem}},''
  \href{http://dx.doi.org/10.1016/S0920-5632(02)01405-6}{{\em Nucl. Phys. Proc.
  Suppl.} {\bf 109A} (2002)  141--145}
  \href{http://arxiv.org/abs/hep-lat/0202010}{{ $\bullet$}}
\href{http://inspirehep.net/search?p=find+eprint+hep-lat/0202010}{{$
  \triangleright $}}

\bibitem{Sobreiro:2005ec}
R.~F. Sobreiro and S.~P. Sorella, ``\textit{{Introduction to the Gribov
  ambiguities in Euclidean Yang-Mills theories}},'' in {\em {13th Jorge Andre
  Swieca Summer School on Particle and Fields Campos do Jordao, Brazil, January
  9-22, 2005}}.
\newblock 2005.
\newblock
\href{http://arxiv.org/abs/hep-th/0504095}{{\tt hep-th/0504095[hep-th]}}.
\newblock

\bibitem{Vandersickel:2012tz}
N.~Vandersickel and D.~Zwanziger, ``\textit{{The Gribov problem and QCD
  dynamics}},'' \href{http://dx.doi.org/10.1016/j.physrep.2012.07.003}{{\em
  Phys. Rept.} {\bf 520} (2012)  175--251}
  \href{http://arxiv.org/abs/1202.1491}{{ $\bullet$}}
\href{http://inspirehep.net/search?p=find+eprint+1202.1491}{{$ \triangleright
  $}}

\bibitem{Cucchieri:1997dx}
A.~Cucchieri, ``\textit{{Gribov copies in the minimal Landau gauge: The
  Influence on gluon and ghost propagators}},''
  \href{http://dx.doi.org/10.1016/S0550-3213(97)80016-8,
  10.1016/S0550-3213(97)00629-9}{{\em Nucl. Phys.} {\bf B508} (1997)  353--370}
  \href{http://arxiv.org/abs/hep-lat/9705005}{{ $\bullet$}}
\href{http://inspirehep.net/search?p=find+eprint+hep-lat/9705005}{{$
  \triangleright $}}

\bibitem{Silva:2004bv}
P.~J. Silva and O.~Oliveira, ``\textit{{Gribov copies, lattice QCD and the
  gluon propagator}},''
  \href{http://dx.doi.org/10.1016/j.nuclphysb.2004.04.020}{{\em Nucl. Phys.}
  {\bf B690} (2004)  177--198} \href{http://arxiv.org/abs/hep-lat/0403026}{{
  $\bullet$}}
\href{http://inspirehep.net/search?p=find+eprint+hep-lat/0403026}{{$
  \triangleright $}}

\bibitem{Bogolubsky:2009dc}
I.~L. Bogolubsky, E.~M. Ilgenfritz, M.~Muller-Preussker, and A.~Sternbeck,
  ``\textit{{L}attice gluodynamics computation of {L}andau gauge {G}reen's
  functions in the deep infrared},''
  \href{http://dx.doi.org/10.1016/j.physletb.2009.04.076}{{\em Phys. Lett.}
  {\bf B676} (2009)  69--73} \href{http://arxiv.org/abs/0901.0736}{{
  $\bullet$}}
\href{http://inspirehep.net/search?p=find+eprint+0901.0736}{{$ \triangleright
  $}}

\bibitem{Maas:2009ph}
A.~Maas, J.~M. Pawlowski, D.~Spielmann, A.~Sternbeck, and L.~von Smekal,
  ``\textit{{Strong-coupling study of the Gribov ambiguity in lattice Landau
  gauge}},'' \href{http://dx.doi.org/10.1140/epjc/s10052-010-1306-6}{{\em Eur.
  Phys. J.} {\bf C68} (2010)  183--195} \href{http://arxiv.org/abs/0912.4203}{{
  $\bullet$}}
\href{http://inspirehep.net/search?p=find+eprint+0912.4203}{{$ \triangleright
  $}}

\bibitem{Sternbeck:2012mf}
A.~Sternbeck and M.~Müller-Preussker, ``\textit{{Lattice evidence for the
  family of decoupling solutions of Landau gauge Yang-Mills theory}},''
  \href{http://dx.doi.org/10.1016/j.physletb.2013.08.017}{{\em Phys. Lett.}
  {\bf B726} (2013)  396--403} \href{http://arxiv.org/abs/1211.3057}{{
  $\bullet$}}
\href{http://inspirehep.net/search?p=find+eprint+1211.3057}{{$ \triangleright
  $}}

\bibitem{Kizilersu:2000qd}
A.~Kizilersu, A.~W. Schreiber, and A.~G. Williams, ``\textit{{Regularization
  independent studies of nonperturbative field theory}},''
  \href{http://dx.doi.org/10.1016/S0370-2693(01)00022-3}{{\em Phys. Lett.} {\bf
  B499} (2001)  261--269} \href{http://arxiv.org/abs/hep-th/0010161}{{
  $\bullet$}}
\href{http://inspirehep.net/search?p=find+eprint+hep-th/0010161}{{$
  \triangleright $}}

\bibitem{Kizilersu:2001pd}
A.~Kizilersu, T.~Sizer, and A.~G. Williams, ``\textit{{Regularization
  independent study of renormalized nonperturbative quenched QED}},''
  \href{http://dx.doi.org/10.1103/PhysRevD.65.085020}{{\em Phys. Rev.} {\bf
  D65} (2002)  085020} \href{http://arxiv.org/abs/hep-ph/0101188}{{ $\bullet$}}
\href{http://inspirehep.net/search?p=find+eprint+hep-ph/0101188}{{$
  \triangleright $}}

\bibitem{Bernard:1974bq}
C.~W. Bernard, ``\textit{{Feynman Rules for Gauge Theories at Finite
  Temperature}},''
\href{http://dx.doi.org/10.1103/PhysRevD.9.3312}{{\em Phys. Rev.} {\bf D9}
  (1974)  3312}

\bibitem{Itzykson:1980rh}
C.~Itzykson and J.~B. Zuber, {\em {Quantum Field Theory}}.
\newblock International Series In Pure and Applied Physics. McGraw-Hill New
  York 1980.
\newblock
\href{http://dx.doi.org/10.1063/1.2916419}{{$ \triangleright $}}

\bibitem{Roberts:1994dr}
C.~D. Roberts and A.~G. Williams, ``\textit{{D}yson-{S}chwinger equations and
  their application to hadronic physics},''
  \href{http://dx.doi.org/10.1016/0146-6410(94)90049-3}{{\em Prog. Part. Nucl.
  Phys.} {\bf 33} (1994)  477--575}
  \href{http://arxiv.org/abs/hep-ph/9403224}{{ $\bullet$}}
\href{http://inspirehep.net/search?p=find+eprint+hep-ph/9403224}{{$
  \triangleright $}}

\bibitem{Wetterich:1992yh}
C.~Wetterich, ``\textit{{E}xact evolution equation for the effective
  potential},''
\href{http://dx.doi.org/10.1016/0370-2693(93)90726-X}{{\em Phys. Lett.} {\bf
  B301} (1993)  90--94}

\bibitem{Morris:1993qb}
T.~R. Morris, ``\textit{{The Exact renormalization group and approximate
  solutions}},'' \href{http://dx.doi.org/10.1142/S0217751X94000972}{{\em Int.
  J. Mod. Phys.} {\bf A9} (1994)  2411--2450}
  \href{http://arxiv.org/abs/hep-ph/9308265}{{ $\bullet$}}
\href{http://inspirehep.net/search?p=find+eprint+hep-ph/9308265}{{$
  \triangleright $}}

\bibitem{Berges:2000ew}
J.~Berges, N.~Tetradis, and C.~Wetterich, ``\textit{{N}onperturbative
  renormalization flow in quantum field theory and statistical physics},''
  \href{http://dx.doi.org/10.1016/S0370-1573(01)00098-9}{{\em Phys. Rept.} {\bf
  363} (2002)  223--386} \href{http://arxiv.org/abs/hep-ph/0005122}{{
  $\bullet$}}
\href{http://inspirehep.net/search?p=find+eprint+hep-ph/0005122}{{$
  \triangleright $}}

\bibitem{Pawlowski:2005xe}
J.~M. Pawlowski, ``\textit{{A}spects of the functional renormalisation
  group},'' \href{http://dx.doi.org/10.1016/j.aop.2007.01.007}{{\em Annals
  Phys.} {\bf 322} (2007)  2831--2915}
  \href{http://arxiv.org/abs/hep-th/0512261}{{ $\bullet$}}
\href{http://inspirehep.net/search?p=find+eprint+hep-th/0512261}{{$
  \triangleright $}}

\bibitem{Gies:2006wv}
H.~Gies, ``\textit{{I}ntroduction to the functional {RG} and applications to
  gauge theories},'' \href{http://dx.doi.org/10.1007/978-3-642-27320-9_6}{{\em
  Lect. Notes Phys.} {\bf 852} (2012)  287--348}
  \href{http://arxiv.org/abs/hep-ph/0611146}{{ $\bullet$}}
\href{http://inspirehep.net/search?p=find+eprint+hep-ph/0611146}{{$
  \triangleright $}}

\bibitem{Schaefer:2006sr}
B.-J. Schaefer and J.~Wambach, ``\textit{{Renormalization group approach
  towards the QCD phase diagram}},''
  \href{http://dx.doi.org/10.1134/S1063779608070083}{{\em Phys. Part. Nucl.}
  {\bf 39} (2008)  1025--1032} \href{http://arxiv.org/abs/hep-ph/0611191}{{
  $\bullet$}}
\href{http://inspirehep.net/search?p=find+eprint+hep-ph/0611191}{{$
  \triangleright $}}

\bibitem{Fischer:2006vf}
C.~S. Fischer and J.~M. Pawlowski, ``\textit{{U}niqueness of infrared
  asymptotics in {L}andau gauge {Y}ang-{M}ills theory},''
  \href{http://dx.doi.org/10.1103/PhysRevD.75.025012}{{\em Phys. Rev.} {\bf
  D75} (2007)  025012} \href{http://arxiv.org/abs/hep-th/0609009}{{ $\bullet$}}
\href{http://inspirehep.net/search?p=find+eprint+hep-th/0609009}{{$
  \triangleright $}}

\bibitem{Fischer:2009tn}
C.~S. Fischer and J.~M. Pawlowski, ``\textit{{Uniqueness of infrared
  asymptotics in Landau gauge Yang-Mills theory II}},''
  \href{http://dx.doi.org/10.1103/PhysRevD.80.025023}{{\em Phys. Rev.} {\bf
  D80} (2009)  025023} \href{http://arxiv.org/abs/0903.2193}{{ $\bullet$}}
\href{http://inspirehep.net/search?p=find+eprint+0903.2193}{{$ \triangleright
  $}}

\bibitem{Reinhardt:2013iia}
H.~Reinhardt and J.~Heffner, ``\textit{{Effective potential of the confinement
  order parameter in the Hamiltonian approach}},''
  \href{http://dx.doi.org/10.1103/PhysRevD.88.045024}{{\em Phys. Rev.} {\bf
  D88} (2013)  045024} \href{http://arxiv.org/abs/1304.2980}{{ $\bullet$}}
\href{http://inspirehep.net/search?p=find+eprint+1304.2980}{{$ \triangleright
  $}}

\bibitem{Quandt:2015aaa}
M.~Quandt and H.~Reinhardt, ``\textit{{A covariant variational approach to
  Yang-Mills Theory at finite temperatures}},''
  \href{http://dx.doi.org/10.1103/PhysRevD.92.025051}{{\em Phys. Rev.} {\bf
  D92} (2015) no.~2, 025051} \href{http://arxiv.org/abs/1503.06993}{{
  $\bullet$}}
\href{http://inspirehep.net/search?p=find+eprint+1503.06993}{{$ \triangleright
  $}}

\bibitem{Heffner:2015zna}
J.~Heffner and H.~Reinhardt, ``\textit{{Finite-temperature Yang-Mills theory in
  the Hamiltonian approach in Coulomb gauge from a compactified spatial
  dimension}},'' \href{http://dx.doi.org/10.1103/PhysRevD.91.085022}{{\em Phys.
  Rev.} {\bf D91} (2015) no.~8, 085022}
  \href{http://arxiv.org/abs/1501.05858}{{ $\bullet$}}
\href{http://inspirehep.net/search?p=find+eprint+1501.05858}{{$ \triangleright
  $}}

\bibitem{Reinhardt:2016pfe}
H.~Reinhardt and P.~Vastag, ``\textit{{Chiral and deconfinement phase
  transition in the Hamiltonian approach to QCD in Coulomb gauge}},''
  \href{http://dx.doi.org/10.1103/PhysRevD.94.105005}{{\em Phys. Rev.} {\bf
  D94} (2016) no.~10, 105005} \href{http://arxiv.org/abs/1605.03740}{{
  $\bullet$}}
\href{http://inspirehep.net/search?p=find+eprint+1605.03740}{{$ \triangleright
  $}}

\bibitem{Reinhardt:2016xci}
H.~Reinhardt, ``\textit{{Hamiltonian finite-temperature quantum field theory
  from its vacuum on partially compactified space}},''
  \href{http://dx.doi.org/10.1103/PhysRevD.94.045016}{{\em Phys. Rev.} {\bf
  D94} (2016) no.~4, 045016} \href{http://arxiv.org/abs/1604.06273}{{
  $\bullet$}}
\href{http://inspirehep.net/search?p=find+eprint+1604.06273}{{$ \triangleright
  $}}

\bibitem{Quandt:2017poi}
M.~Quandt and H.~Reinhardt, ``\textit{{Covariant variational approach to
  Yang-Mills Theory: Thermodynamics}},''
  \href{http://dx.doi.org/10.1103/PhysRevD.96.054029}{{\em Phys. Rev.} {\bf
  D96} (2017) no.~5, 054029} \href{http://arxiv.org/abs/1705.05157}{{
  $\bullet$}}
\href{http://inspirehep.net/search?p=find+eprint+1705.05157}{{$ \triangleright
  $}}

\bibitem{Quandt:2018bbu}
M.~Quandt, E.~Ebadati, H.~Reinhardt, and P.~Vastag, ``\textit{{Chiral symmetry
  restoration at finite temperature within the Hamiltonian approach to QCD in
  Coulomb gauge}},'' \href{http://dx.doi.org/10.1103/PhysRevD.98.034012}{{\em
  Phys. Rev.} {\bf D98} (2018) no.~3, 034012}
  \href{http://arxiv.org/abs/1806.04493}{{ $\bullet$}}
\href{http://inspirehep.net/search?p=find+eprint+1806.04493}{{$ \triangleright
  $}}

\bibitem{Reinhardt:2017pyr}
H.~Reinhardt, G.~Burgio, D.~Campagnari, E.~Ebadati, J.~Heffner, M.~Quandt,
  P.~Vastag, and H.~Vogt, ``\textit{{Hamiltonian approach to QCD in Coulomb
  gauge - a survey of recent results}},''
  \href{http://dx.doi.org/10.1155/2018/2312498}{{\em Adv. High Energy Phys.}
  {\bf 2018} (2018)  2312498} \href{http://arxiv.org/abs/1706.02702}{{
  $\bullet$}}
\href{http://inspirehep.net/search?p=find+eprint+1706.02702}{{$ \triangleright
  $}}

\bibitem{Gribov:1977wm}
V.~N. Gribov, ``\textit{{Quantization of Nonabelian Gauge Theories}},''
\href{http://dx.doi.org/10.1016/0550-3213(78)90175-X}{{\em Nucl. Phys.} {\bf
  B139} (1978)  1}

\bibitem{Zwanziger:1989mf}
D.~Zwanziger, ``\textit{{Local and Renormalizable Action From the Gribov
  Horizon}},''
\href{http://dx.doi.org/10.1016/0550-3213(89)90122-3}{{\em Nucl. Phys.} {\bf
  B323} (1989)  513--544}

\bibitem{Zwanziger:2004np}
D.~Zwanziger, ``\textit{{Equation of state of gluon plasma from fundamental
  modular region}},''
  \href{http://dx.doi.org/10.1103/PhysRevLett.94.182301}{{\em Phys. Rev. Lett.}
  {\bf 94} (2005)  182301} \href{http://arxiv.org/abs/hep-ph/0407103}{{
  $\bullet$}}
\href{http://inspirehep.net/search?p=find+eprint+hep-ph/0407103}{{$
  \triangleright $}}

\bibitem{Reinosa:2013twa}
U.~Reinosa, J.~Serreau, M.~Tissier, and N.~Wschebor, ``\textit{{Yang-Mills
  correlators at finite temperature: A perturbative perspective}},''
  \href{http://dx.doi.org/10.1103/PhysRevD.89.105016}{{\em Phys. Rev.} {\bf
  D89} (2014) no.~10, 105016} \href{http://arxiv.org/abs/1311.6116}{{
  $\bullet$}}
\href{http://inspirehep.net/search?p=find+eprint+1311.6116}{{$ \triangleright
  $}}

\bibitem{Reinosa:2014zta}
U.~Reinosa, J.~Serreau, M.~Tissier, and N.~Wschebor, ``\textit{{Deconfinement
  transition in SU(2) Yang-Mills theory: A two-loop study}},''
  \href{http://dx.doi.org/10.1103/PhysRevD.91.045035}{{\em Phys. Rev.} {\bf
  D91} (2015)  045035} \href{http://arxiv.org/abs/1412.5672}{{ $\bullet$}}
\href{http://inspirehep.net/search?p=find+eprint+1412.5672}{{$ \triangleright
  $}}

\bibitem{Fukushima:2013xsa}
K.~Fukushima and N.~Su, ``\textit{{Stabilizing perturbative Yang-Mills
  thermodynamics with Gribov quantization}},''
  \href{http://dx.doi.org/10.1103/PhysRevD.88.076008}{{\em Phys. Rev.} {\bf
  D88} (2013)  076008} \href{http://arxiv.org/abs/1304.8004}{{ $\bullet$}}
\href{http://inspirehep.net/search?p=find+eprint+1304.8004}{{$ \triangleright
  $}}

\bibitem{Su:2014rma}
N.~Su and K.~Tywoniuk, ``\textit{{Massless Mode and Positivity Violation in Hot
  QCD}},'' \href{http://dx.doi.org/10.1103/PhysRevLett.114.161601}{{\em Phys.
  Rev. Lett.} {\bf 114} (2015) no.~16, 161601}
  \href{http://arxiv.org/abs/1409.3203}{{ $\bullet$}}
\href{http://inspirehep.net/search?p=find+eprint+1409.3203}{{$ \triangleright
  $}}

\bibitem{Florkowski:2015dmm}
W.~Florkowski, R.~Ryblewski, N.~Su, and K.~Tywoniuk, ``\textit{{Transport
  coefficients of the Gribov-Zwanziger plasma}},''
  \href{http://dx.doi.org/10.1103/PhysRevC.94.044904}{{\em Phys. Rev.} {\bf
  C94} (2016) no.~4, 044904} \href{http://arxiv.org/abs/1509.01242}{{
  $\bullet$}}
\href{http://inspirehep.net/search?p=find+eprint+1509.01242}{{$ \triangleright
  $}}

\bibitem{Reinosa:2016iml}
U.~Reinosa, J.~Serreau, M.~Tissier, and A.~Tresmontant, ``\textit{{Yang-Mills
  correlators across the deconfinement phase transition}},''
  \href{http://dx.doi.org/10.1103/PhysRevD.95.045014}{{\em Phys. Rev.} {\bf
  D95} (2017) no.~4, 045014} \href{http://arxiv.org/abs/1606.08012}{{
  $\bullet$}}
\href{http://inspirehep.net/search?p=find+eprint+1606.08012}{{$ \triangleright
  $}}

\bibitem{Maelger:2018vow}
J.~Maelger, U.~Reinosa, and J.~Serreau, ``\textit{{Universal aspects of the
  phase diagram of QCD with heavy quarks}},''
  \href{http://arxiv.org/abs/1805.10015}{{\tt 1805.10015[hep-th]}}
  \href{http://arxiv.org/abs/1805.10015}{{ $\bullet$}}
\href{http://inspirehep.net/search?p=find+eprint+1805.10015}{{$ \triangleright
  $}}

\bibitem{Greensite:2003bk}
J.~Greensite, ``\textit{{T}he {C}onfinement problem in lattice gauge theory},''
  \href{http://dx.doi.org/10.1016/S0146-6410(03)90012-3}{{\em Prog. Part. Nucl.
  Phys.} {\bf 51} (2003)  1} \href{http://arxiv.org/abs/hep-lat/0301023}{{
  $\bullet$}}
\href{http://inspirehep.net/search?p=find+eprint+hep-lat/0301023}{{$
  \triangleright $}}

\bibitem{Alkofer:2006fu}
R.~Alkofer and J.~Greensite, ``\textit{{Q}uark {C}onfinement: {T}he {H}ard
  {P}roblem of {H}adron {P}hysics},''
  \href{http://dx.doi.org/10.1088/0954-3899/34/7/S02}{{\em J. Phys.} {\bf G34}
  (2007)  S3} \href{http://arxiv.org/abs/hep-ph/0610365}{{ $\bullet$}}
\href{http://inspirehep.net/search?p=find+eprint+hep-ph/0610365}{{$
  \triangleright $}}

\bibitem{Greensite:2011zz}
J.~Greensite, ``\textit{{An introduction to the confinement problem}},''
\href{http://dx.doi.org/10.1007/978-3-642-14382-3}{{\em Lect. Notes Phys.} {\bf
  821} (2011)  1--211}

\bibitem{Greensite:2017ajx}
J.~Greensite and K.~Matsuyama, ``\textit{{Confinement criterion for gauge
  theories with matter fields}},''
  \href{http://dx.doi.org/10.1103/PhysRevD.96.094510}{{\em Phys. Rev.} {\bf
  D96} (2017) no.~9, 094510} \href{http://arxiv.org/abs/1708.08979}{{
  $\bullet$}}
\href{http://inspirehep.net/search?p=find+eprint+1708.08979}{{$ \triangleright
  $}}

\bibitem{Fischer:2009wc}
C.~S. Fischer, ``\textit{{Deconfinement phase transition and the quark
  condensate}},'' \href{http://dx.doi.org/10.1103/PhysRevLett.103.052003}{{\em
  Phys. Rev. Lett.} {\bf 103} (2009)  052003}
  \href{http://arxiv.org/abs/0904.2700}{{ $\bullet$}}
\href{http://inspirehep.net/search?p=find+eprint+0904.2700}{{$ \triangleright
  $}}

\bibitem{Fischer:2009gk}
C.~S. Fischer and J.~A. Mueller, ``\textit{{Chiral and deconfinement transition
  from Dyson-Schwinger equations}},''
  \href{http://dx.doi.org/10.1103/PhysRevD.80.074029}{{\em Phys. Rev.} {\bf
  D80} (2009)  074029} \href{http://arxiv.org/abs/0908.0007}{{ $\bullet$}}
\href{http://inspirehep.net/search?p=find+eprint+0908.0007}{{$ \triangleright
  $}}

\bibitem{Mitter:2017iye}
M.~Mitter, M.~Hopfer, B.~J. Schaefer, and R.~Alkofer, ``\textit{{Center phase
  transition from matter propagators in (scalar) QCD}},''
  \href{http://dx.doi.org/10.1016/j.physletb.2017.12.019}{{\em Phys. Lett.}
  {\bf B777} (2018)  114--120} \href{http://arxiv.org/abs/1709.00299}{{
  $\bullet$}}
\href{http://inspirehep.net/search?p=find+eprint+1709.00299}{{$ \triangleright
  $}}

\bibitem{Braun:2007bx}
J.~Braun, H.~Gies, and J.~M. Pawlowski, ``\textit{{Quark Confinement from Color
  Confinement}},'' \href{http://dx.doi.org/10.1016/j.physletb.2010.01.009}{{\em
  Phys. Lett.} {\bf B684} (2010)  262--267}
  \href{http://arxiv.org/abs/0708.2413}{{ $\bullet$}}
\href{http://inspirehep.net/search?p=find+eprint+0708.2413}{{$ \triangleright
  $}}

\bibitem{Fister:2013bh}
L.~Fister and J.~M. Pawlowski, ``\textit{{Confinement from Correlation
  Functions}},'' \href{http://dx.doi.org/10.1103/PhysRevD.88.045010}{{\em Phys.
  Rev.} {\bf D88} (2013)  045010} \href{http://arxiv.org/abs/1301.4163}{{
  $\bullet$}}
\href{http://inspirehep.net/search?p=find+eprint+1301.4163}{{$ \triangleright
  $}}

\bibitem{Fischer:2013eca}
C.~S. Fischer, L.~Fister, J.~Luecker, and J.~M. Pawlowski, ``\textit{{P}olyakov
  loop potential at finite density},''
  \href{http://dx.doi.org/10.1016/j.physletb.2014.03.057}{{\em Phys. Lett.}
  {\bf B732} (2014)  273--277} \href{http://arxiv.org/abs/1306.6022}{{
  $\bullet$}}
\href{http://inspirehep.net/search?p=find+eprint+1306.6022}{{$ \triangleright
  $}}

\bibitem{Bilgici:2008qy}
E.~Bilgici, F.~Bruckmann, C.~Gattringer, and C.~Hagen, ``\textit{{Dual quark
  condensate and dressed Polyakov loops}},''
  \href{http://dx.doi.org/10.1103/PhysRevD.77.094007}{{\em Phys. Rev.} {\bf
  D77} (2008)  094007} \href{http://arxiv.org/abs/0801.4051}{{ $\bullet$}}
\href{http://inspirehep.net/search?p=find+eprint+0801.4051}{{$ \triangleright
  $}}

\bibitem{Gattringer:2006ci}
C.~Gattringer, ``\textit{{Linking confinement to spectral properties of the
  Dirac operator}},''
  \href{http://dx.doi.org/10.1103/PhysRevLett.97.032003}{{\em Phys. Rev. Lett.}
  {\bf 97} (2006)  032003} \href{http://arxiv.org/abs/hep-lat/0605018}{{
  $\bullet$}}
\href{http://inspirehep.net/search?p=find+eprint+hep-lat/0605018}{{$
  \triangleright $}}

\bibitem{Bruckmann:2006kx}
F.~Bruckmann, C.~Gattringer, and C.~Hagen, ``\textit{{Complete spectra of the
  Dirac operator and their relation to confinement}},''
  \href{http://dx.doi.org/10.1016/j.physletb.2007.01.043}{{\em Phys. Lett.}
  {\bf B647} (2007)  56--61} \href{http://arxiv.org/abs/hep-lat/0612020}{{
  $\bullet$}}
\href{http://inspirehep.net/search?p=find+eprint+hep-lat/0612020}{{$
  \triangleright $}}

\bibitem{Synatschke:2007bz}
F.~Synatschke, A.~Wipf, and C.~Wozar, ``\textit{{Spectral sums of the
  Dirac-Wilson operator and their relation to the Polyakov loop}},''
  \href{http://dx.doi.org/10.1103/PhysRevD.75.114003}{{\em Phys. Rev.} {\bf
  D75} (2007)  114003} \href{http://arxiv.org/abs/hep-lat/0703018}{{
  $\bullet$}}
\href{http://inspirehep.net/search?p=find+eprint+hep-lat/0703018}{{$
  \triangleright $}}

\bibitem{Synatschke:2008yt}
F.~Synatschke, A.~Wipf, and K.~Langfeld, ``\textit{{Relation between chiral
  symmetry breaking and confinement in YM-theories}},''
  \href{http://dx.doi.org/10.1103/PhysRevD.77.114018}{{\em Phys. Rev.} {\bf
  D77} (2008)  114018} \href{http://arxiv.org/abs/0803.0271}{{ $\bullet$}}
\href{http://inspirehep.net/search?p=find+eprint+0803.0271}{{$ \triangleright
  $}}

\bibitem{Polyakov:1978vu}
A.~M. Polyakov, ``\textit{{Thermal Properties of Gauge Fields and Quark
  Liberation}},''
\href{http://dx.doi.org/10.1016/0370-2693(78)90737-2}{{\em Phys. Lett.} {\bf
  72B} (1978)  477--480}

\bibitem{Susskind:1979up}
L.~Susskind, ``\textit{{Lattice Models of Quark Confinement at High
  Temperature}},''
\href{http://dx.doi.org/10.1103/PhysRevD.20.2610}{{\em Phys. Rev.} {\bf D20}
  (1979)  2610--2618}

\bibitem{Fukushima:2011jc}
K.~Fukushima, ``\textit{{QCD matter in extreme environments}},''
  \href{http://dx.doi.org/10.1088/0954-3899/39/1/013101}{{\em J. Phys.} {\bf
  G39} (2012)  013101} \href{http://arxiv.org/abs/1108.2939}{{ $\bullet$}}
\href{http://inspirehep.net/search?p=find+eprint+1108.2939}{{$ \triangleright
  $}}

\bibitem{Bruckmann:2008sy}
F.~Bruckmann, C.~Hagen, E.~Bilgici, and C.~Gattringer, ``\textit{{Dual
  condensate, dressed Polyakov loops and center symmetry from Dirac
  spectra}},'' {\em PoS} {\bf LATTICE2008} (2008)  262
  \href{http://arxiv.org/abs/0810.0899}{{ $\bullet$}}
\href{http://inspirehep.net/search?p=find+eprint+0810.0899}{{$ \triangleright
  $}}

\bibitem{Fischer:2010fx}
C.~S. Fischer, A.~Maas, and J.~A. Muller, ``\textit{{Chiral and deconfinement
  transition from correlation functions: SU(2) vs. SU(3)}},''
  \href{http://dx.doi.org/10.1140/epjc/s10052-010-1343-1}{{\em Eur. Phys. J.}
  {\bf C68} (2010)  165--181} \href{http://arxiv.org/abs/1003.1960}{{
  $\bullet$}}
\href{http://inspirehep.net/search?p=find+eprint+1003.1960}{{$ \triangleright
  $}}

\bibitem{Fischer:2011mz}
C.~S. Fischer, J.~Luecker, and J.~A. Mueller, ``\textit{{Chiral and
  deconfinement phase transitions of two-flavour QCD at finite temperature and
  chemical potential}},''
  \href{http://dx.doi.org/10.1016/j.physletb.2011.07.039}{{\em Phys. Lett.}
  {\bf B702} (2011)  438--441} \href{http://arxiv.org/abs/1104.1564}{{
  $\bullet$}}
\href{http://inspirehep.net/search?p=find+eprint+1104.1564}{{$ \triangleright
  $}}

\bibitem{Fischer:2012vc}
C.~S. Fischer and J.~Luecker, ``\textit{{Propagators and phase structure of
  Nf=2 and Nf=2+1 QCD}},''
  \href{http://dx.doi.org/10.1016/j.physletb.2012.11.054}{{\em Phys. Lett.}
  {\bf B718} (2013)  1036--1043} \href{http://arxiv.org/abs/1206.5191}{{
  $\bullet$}}
\href{http://inspirehep.net/search?p=find+eprint+1206.5191}{{$ \triangleright
  $}}

\bibitem{Marhauser:2008fz}
F.~Marhauser and J.~M. Pawlowski, ``\textit{{Confinement in Polyakov Gauge}},''
  \href{http://arxiv.org/abs/0812.1144}{{\tt 0812.1144[hep-ph]}}
  \href{http://arxiv.org/abs/0812.1144}{{ $\bullet$}}
\href{http://inspirehep.net/search?p=find+eprint+0812.1144}{{$ \triangleright
  $}}

\bibitem{Braun:2010cy}
J.~Braun, A.~Eichhorn, H.~Gies, and J.~M. Pawlowski, ``\textit{{On the Nature
  of the Phase Transition in SU(N), Sp(2) and E(7) Yang-Mills theory}},''
  \href{http://dx.doi.org/10.1140/epjc/s10052-010-1485-1}{{\em Eur. Phys. J.}
  {\bf C70} (2010)  689--702} \href{http://arxiv.org/abs/1007.2619}{{
  $\bullet$}}
\href{http://inspirehep.net/search?p=find+eprint+1007.2619}{{$ \triangleright
  $}}

\bibitem{Herbst:2015ona}
T.~K. Herbst, J.~Luecker, and J.~M. Pawlowski, ``\textit{{Confinement order
  parameters and fluctuations}},'' \href{http://arxiv.org/abs/1510.03830}{{\tt
  1510.03830[hep-ph]}} \href{http://arxiv.org/abs/1510.03830}{{ $\bullet$}}
\href{http://inspirehep.net/search?p=find+eprint+1510.03830}{{$ \triangleright
  $}}

\bibitem{Reinhardt:2012qe}
H.~Reinhardt and J.~Heffner, ``\textit{{The effective potential of the
  confinement order parameter in the Hamilton approach}},''
  \href{http://dx.doi.org/10.1016/j.physletb.2012.10.084}{{\em Phys. Lett.}
  {\bf B718} (2012)  672--677} \href{http://arxiv.org/abs/1210.1742}{{
  $\bullet$}}
\href{http://inspirehep.net/search?p=find+eprint+1210.1742}{{$ \triangleright
  $}}

\bibitem{Fukushima:2012qa}
K.~Fukushima and K.~Kashiwa, ``\textit{{Polyakov loop and QCD thermodynamics
  from the gluon and ghost propagators}},''
  \href{http://dx.doi.org/10.1016/j.physletb.2013.05.037}{{\em Phys. Lett.}
  {\bf B723} (2013)  360--364} \href{http://arxiv.org/abs/1206.0685}{{
  $\bullet$}}
\href{http://inspirehep.net/search?p=find+eprint+1206.0685}{{$ \triangleright
  $}}

\bibitem{Kashiwa:2012td}
K.~Kashiwa and Y.~Maezawa, ``\textit{{Quark back reaction to deconfinement
  transition via gluon propagators}},''
  \href{http://arxiv.org/abs/1212.2184}{{\tt 1212.2184[hep-ph]}}
  \href{http://arxiv.org/abs/1212.2184}{{ $\bullet$}}
\href{http://inspirehep.net/search?p=find+eprint+1212.2184}{{$ \triangleright
  $}}

\bibitem{Osterwalder:1973dx}
K.~Osterwalder and R.~Schrader, ``\textit{{Axioms for Euclidean Green's
  Functions}},''
\href{http://dx.doi.org/10.1007/BF01645738}{{\em Commun. Math. Phys.} {\bf 31}
  (1973)  83--112}

\bibitem{Mandula:1987rh}
J.~E. Mandula and M.~Ogilvie, ``\textit{{The Gluon Is Massive: A Lattice
  Calculation of the Gluon Propagator in the Landau Gauge}},''
\href{http://dx.doi.org/10.1016/0370-2693(87)91541-3}{{\em Phys. Lett.} {\bf
  B185} (1987)  127--132}

\bibitem{Mueller:2010ah}
J.~A. Mueller, C.~S. Fischer, and D.~Nickel, ``\textit{{Quark spectral
  properties above Tc from Dyson-Schwinger equations}},''
  \href{http://dx.doi.org/10.1140/epjc/s10052-010-1499-8}{{\em Eur. Phys. J.}
  {\bf C70} (2010)  1037--1049} \href{http://arxiv.org/abs/1009.3762}{{
  $\bullet$}}
\href{http://inspirehep.net/search?p=find+eprint+1009.3762}{{$ \triangleright
  $}}

\bibitem{Gao:2014rqa}
F.~Gao, S.-X. Qin, Y.-X. Liu, C.~D. Roberts, and S.~M. Schmidt, ``\textit{{Zero
  mode in a strongly coupled quark gluon plasma}},''
  \href{http://dx.doi.org/10.1103/PhysRevD.89.076009}{{\em Phys. Rev.} {\bf
  D89} (2014) no.~7, 076009} \href{http://arxiv.org/abs/1401.2406}{{
  $\bullet$}}
\href{http://inspirehep.net/search?p=find+eprint+1401.2406}{{$ \triangleright
  $}}

\bibitem{Fischer:2017kbq}
C.~S. Fischer, J.~M. Pawlowski, A.~Rothkopf, and C.~A. Welzbacher,
  ``\textit{{Bayesian analysis of quark spectral properties from the
  Dyson-Schwinger equation}},'' \href{http://arxiv.org/abs/1705.03207}{{\tt
  1705.03207[hep-ph]}} \href{http://arxiv.org/abs/1705.03207}{{ $\bullet$}}
\href{http://inspirehep.net/search?p=find+eprint+1705.03207}{{$ \triangleright
  $}}

\bibitem{Nickel:2006mm}
D.~Nickel, ``\textit{{Extraction of Spectral Functions from Dyson-Schwinger
  Studies via the Maximum Entropy Method}},''
  \href{http://dx.doi.org/10.1016/j.aop.2006.09.002}{{\em Annals Phys.} {\bf
  322} (2007)  1949--1960} \href{http://arxiv.org/abs/hep-ph/0607224}{{
  $\bullet$}}
\href{http://inspirehep.net/search?p=find+eprint+hep-ph/0607224}{{$
  \triangleright $}}

\bibitem{Harada:2007gg}
M.~Harada, Y.~Nemoto, and S.~Yoshimoto, ``\textit{{Quasi-quark spectrum in the
  chiral symmetric phase from the Schwinger-Dyson equation}},''
  \href{http://dx.doi.org/10.1143/PTP.119.117}{{\em Prog. Theor. Phys.} {\bf
  119} (2008)  117} \href{http://arxiv.org/abs/0708.3351}{{ $\bullet$}}
\href{http://inspirehep.net/search?p=find+eprint+0708.3351}{{$ \triangleright
  $}}

\bibitem{Harada:2009zq}
M.~Harada and S.~Yoshimoto, ``\textit{{Disappearance of quasi-fermions in the
  strongly coupled plasma from the Schwinger-Dyson equation with in-medium
  gauge boson propagator}},'' \href{http://arxiv.org/abs/0903.5495}{{\tt
  0903.5495[hep-ph]}} \href{http://arxiv.org/abs/0903.5495}{{ $\bullet$}}
\href{http://inspirehep.net/search?p=find+eprint+0903.5495}{{$ \triangleright
  $}}

\bibitem{Karsch:2007wc}
F.~Karsch and M.~Kitazawa, ``\textit{{Spectral properties of quarks above T(c)
  in quenched lattice QCD}},''
  \href{http://dx.doi.org/10.1016/j.physletb.2007.10.034}{{\em Phys. Lett.}
  {\bf B658} (2007)  45--49} \href{http://arxiv.org/abs/0708.0299}{{
  $\bullet$}}
\href{http://inspirehep.net/search?p=find+eprint+0708.0299}{{$ \triangleright
  $}}

\bibitem{Karsch:2009tp}
F.~Karsch and M.~Kitazawa, ``\textit{{Quark propagator at finite temperature
  and finite momentum in quenched lattice QCD}},''
  \href{http://dx.doi.org/10.1103/PhysRevD.80.056001}{{\em Phys. Rev.} {\bf
  D80} (2009)  056001} \href{http://arxiv.org/abs/0906.3941}{{ $\bullet$}}
\href{http://inspirehep.net/search?p=find+eprint+0906.3941}{{$ \triangleright
  $}}

\bibitem{Qin:2010pc}
S.-x. Qin, L.~Chang, Y.-x. Liu, and C.~D. Roberts, ``\textit{{Quark spectral
  density and a strongly-coupled QGP}},''
  \href{http://dx.doi.org/10.1103/PhysRevD.84.014017}{{\em Phys. Rev.} {\bf
  D84} (2011)  014017} \href{http://arxiv.org/abs/1010.4231}{{ $\bullet$}}
\href{http://inspirehep.net/search?p=find+eprint+1010.4231}{{$ \triangleright
  $}}

\bibitem{Qin:2013ufa}
S.-x. Qin and D.~H. Rischke, ``\textit{{Quark Spectral Function and
  Deconfinement at Nonzero Temperature}},''
  \href{http://dx.doi.org/10.1103/PhysRevD.88.056007}{{\em Phys. Rev.} {\bf
  D88} (2013)  056007} \href{http://arxiv.org/abs/1304.6547}{{ $\bullet$}}
\href{http://inspirehep.net/search?p=find+eprint+1304.6547}{{$ \triangleright
  $}}

\bibitem{Christiansen:2014ypa}
N.~Christiansen, M.~Haas, J.~M. Pawlowski, and N.~Strodthoff,
  ``\textit{{Transport Coefficients in Yang--Mills Theory and QCD}},''
  \href{http://dx.doi.org/10.1103/PhysRevLett.115.112002}{{\em Phys. Rev.
  Lett.} {\bf 115} (2015) no.~11, 112002}
  \href{http://arxiv.org/abs/1411.7986}{{ $\bullet$}}
\href{http://inspirehep.net/search?p=find+eprint+1411.7986}{{$ \triangleright
  $}}

\bibitem{Ilgenfritz:2017kkp}
E.-M. Ilgenfritz, J.~M. Pawlowski, A.~Rothkopf, and A.~Trunin,
  ``\textit{{Finite temperature gluon spectral functions from $N_f=2+1+1$
  lattice QCD}},'' \href{http://dx.doi.org/10.1140/epjc/s10052-018-5593-7}{{\em
  Eur. Phys. J.} {\bf C78} (2018) no.~2, 127}
  \href{http://arxiv.org/abs/1701.08610}{{ $\bullet$}}
\href{http://inspirehep.net/search?p=find+eprint+1701.08610}{{$ \triangleright
  $}}

\bibitem{Braaten:1990wp}
E.~Braaten, R.~D. Pisarski, and T.-C. Yuan, ``\textit{{Production of Soft
  Dileptons in the Quark - Gluon Plasma}},''
\href{http://dx.doi.org/10.1103/PhysRevLett.64.2242}{{\em Phys. Rev. Lett.}
  {\bf 64} (1990)  2242}

\bibitem{Peshier:1999dt}
A.~Peshier and M.~H. Thoma, ``\textit{{Quark dispersion relation and dilepton
  production in the quark gluon plasma}},''
  \href{http://dx.doi.org/10.1103/PhysRevLett.84.841}{{\em Phys. Rev. Lett.}
  {\bf 84} (2000)  841--844} \href{http://arxiv.org/abs/hep-ph/9907268}{{
  $\bullet$}}
\href{http://inspirehep.net/search?p=find+eprint+hep-ph/9907268}{{$
  \triangleright $}}

\bibitem{Arnold:2002ja}
P.~B. Arnold, G.~D. Moore, and L.~G. Yaffe, ``\textit{{Photon and gluon
  emission in relativistic plasmas}},''
  \href{http://dx.doi.org/10.1088/1126-6708/2002/06/030}{{\em JHEP} {\bf 06}
  (2002)  030} \href{http://arxiv.org/abs/hep-ph/0204343}{{ $\bullet$}}
\href{http://inspirehep.net/search?p=find+eprint+hep-ph/0204343}{{$
  \triangleright $}}

\bibitem{Kim:2015poa}
T.~Kim, M.~Asakawa, and M.~Kitazawa, ``\textit{{Dilepton production spectrum
  above Tc with a lattice quark propagator}},''
  \href{http://dx.doi.org/10.1103/PhysRevD.92.114014}{{\em Phys. Rev.} {\bf
  D92} (2015) no.~11, 114014} \href{http://arxiv.org/abs/1505.07195}{{
  $\bullet$}}
\href{http://inspirehep.net/search?p=find+eprint+1505.07195}{{$ \triangleright
  $}}

\bibitem{Alkofer:2004it}
R.~Alkofer, C.~S. Fischer, and F.~J. Llanes-Estrada, ``\textit{{V}ertex
  functions and infrared fixed point in {L}andau gauge {SU}({N}) {Y}ang-{M}ills
  theory},'' \href{http://dx.doi.org/10.1016/j.physletb.2008.11.068,
  10.1016/j.physletb.2005.02.043}{{\em Phys. Lett.} {\bf B611} (2005)
  279--288} \href{http://arxiv.org/abs/hep-th/0412330}{{ $\bullet$}}
\href{http://inspirehep.net/search?p=find+eprint+hep-th/0412330}{{$
  \triangleright $}}

\bibitem{Huber:2007kc}
M.~Q. Huber, R.~Alkofer, C.~S. Fischer, and K.~Schwenzer, ``\textit{{The
  Infrared behavior of Landau gauge Yang-Mills theory in d=2, d=3 and d=4
  dimensions}},'' \href{http://dx.doi.org/10.1016/j.physletb.2007.10.073}{{\em
  Phys. Lett.} {\bf B659} (2008)  434--440}
  \href{http://arxiv.org/abs/0705.3809}{{ $\bullet$}}
\href{http://inspirehep.net/search?p=find+eprint+0705.3809}{{$ \triangleright
  $}}

\bibitem{Eichmann:2016yit}
G.~Eichmann, H.~Sanchis-Alepuz, R.~Williams, R.~Alkofer, and C.~S. Fischer,
  ``\textit{{Baryons as relativistic three-quark bound states}},''
  \href{http://dx.doi.org/10.1016/j.ppnp.2016.07.001}{{\em Prog. Part. Nucl.
  Phys.} {\bf 91} (2016)  1--100} \href{http://arxiv.org/abs/1606.09602}{{
  $\bullet$}}
\href{http://inspirehep.net/search?p=find+eprint+1606.09602}{{$ \triangleright
  $}}

\bibitem{Huber:2018ned}
M.~Q. Huber, ``\textit{{On nonperturbative Yang-Mills correlation
  functions}},'' \href{http://arxiv.org/abs/1808.05227}{{\tt
  1808.05227[hep-ph]}} \href{http://arxiv.org/abs/1808.05227}{{ $\bullet$}}
\href{http://inspirehep.net/search?p=find+eprint+1808.05227}{{$ \triangleright
  $}}

\bibitem{Maas:2011se}
A.~Maas, ``\textit{{D}escribing gauge bosons at zero and finite temperature},''
  \href{http://dx.doi.org/10.1016/j.physrep.2012.11.002}{{\em Phys. Rept.} {\bf
  524} (2013)  203--300} \href{http://arxiv.org/abs/1106.3942}{{ $\bullet$}}
\href{http://inspirehep.net/search?p=find+eprint+1106.3942}{{$ \triangleright
  $}}

\bibitem{Haas:2013hpa}
M.~Haas, L.~Fister, and J.~M. Pawlowski, ``\textit{{Gluon spectral functions
  and transport coefficients in Yang--Mills theory}},''
  \href{http://dx.doi.org/10.1103/PhysRevD.90.091501}{{\em Phys. Rev.} {\bf
  D90} (2014)  091501} \href{http://arxiv.org/abs/1308.4960}{{ $\bullet$}}
\href{http://inspirehep.net/search?p=find+eprint+1308.4960}{{$ \triangleright
  $}}

\bibitem{Aguilar:2018epe}
A.~C. Aguilar, J.~C. Cardona, M.~N. Ferreira, and J.~Papavassiliou,
  ``\textit{{Quark gap equation with non-abelian Ball-Chiu vertex}},''
  \href{http://dx.doi.org/10.1103/PhysRevD.98.014002}{{\em Phys. Rev.} {\bf
  D98} (2018) no.~1, 014002} \href{http://arxiv.org/abs/1804.04229}{{
  $\bullet$}}
\href{http://inspirehep.net/search?p=find+eprint+1804.04229}{{$ \triangleright
  $}}

\bibitem{Skullerud:2003qu}
J.~I. Skullerud, P.~O. Bowman, A.~Kizilersu, D.~B. Leinweber, and A.~G.
  Williams, ``\textit{{N}onperturbative structure of the quark gluon vertex},''
  \href{http://dx.doi.org/10.1088/1126-6708/2003/04/047}{{\em JHEP} {\bf 04}
  (2003)  047} \href{http://arxiv.org/abs/hep-ph/0303176}{{ $\bullet$}}
\href{http://inspirehep.net/search?p=find+eprint+hep-ph/0303176}{{$
  \triangleright $}}

\bibitem{Braun:2014ata}
J.~Braun, L.~Fister, J.~M. Pawlowski, and F.~Rennecke, ``\textit{{From Quarks
  and Gluons to Hadrons: Chiral Symmetry Breaking in Dynamical QCD}},''
  \href{http://arxiv.org/abs/1412.1045}{{\tt 1412.1045[hep-ph]}}
  \href{http://arxiv.org/abs/1412.1045}{{ $\bullet$}}
\href{http://inspirehep.net/search?p=find+eprint+1412.1045}{{$ \triangleright
  $}}

\bibitem{Williams:2015cvx}
R.~Williams, C.~S. Fischer, and W.~Heupel, ``\textit{{L}ight mesons in {QCD}
  and unquenching effects from the 3{PI} effective action},''
  \href{http://dx.doi.org/10.1103/PhysRevD.93.034026}{{\em Phys. Rev.} {\bf
  D93} (2016) no.~3, 034026} \href{http://arxiv.org/abs/1512.00455}{{
  $\bullet$}}
\href{http://inspirehep.net/search?p=find+eprint+1512.00455}{{$ \triangleright
  $}}

\bibitem{Huber:2016xbs}
M.~Q. Huber, ``\textit{{An exploratory study of Yang-Mills three-point
  functions at non-zero temperature}},''
  \href{http://dx.doi.org/10.1051/epjconf/201713707009}{{\em EPJ Web Conf.}
  {\bf 137} (2017)  07009} \href{http://arxiv.org/abs/1611.06136}{{ $\bullet$}}
\href{http://inspirehep.net/search?p=find+eprint+1611.06136}{{$ \triangleright
  $}}

\bibitem{Contant:2018zpi}
R.~Contant, M.~Q. Huber, C.~S. Fischer, C.~A. Welzbacher, and R.~Williams,
  ``\textit{{On the quark-gluon vertex at non-vanishing temperature}},'' in
  {\em {10th International Winter Workshop "Excited QCD" 2018 Kopaonik, Serbia,
  March 11-15, 2018}}.
\newblock 2018.
\newblock
\href{http://arxiv.org/abs/1805.05885}{{\tt 1805.05885[hep-ph]}}.
\newblock

\bibitem{Huber:2013yqa}
M.~Q. Huber and L.~von Smekal, ``\textit{{On two- and three-point functions of
  Landau gauge Yang-Mills theory}},''
  \href{http://dx.doi.org/10.22323/1.187.0364}{{\em PoS} {\bf LATTICE2013}
  (2014)  364} \href{http://arxiv.org/abs/1311.0702}{{ $\bullet$}}
\href{http://inspirehep.net/search?p=find+eprint+1311.0702}{{$ \triangleright
  $}}

\bibitem{Fischer:2007ze}
C.~S. Fischer, D.~Nickel, and J.~Wambach, ``\textit{{Hadronic unquenching
  effects in the quark propagator}},''
  \href{http://dx.doi.org/10.1103/PhysRevD.76.094009}{{\em Phys. Rev.} {\bf
  D76} (2007)  094009} \href{http://arxiv.org/abs/0705.4407}{{ $\bullet$}}
\href{http://inspirehep.net/search?p=find+eprint+0705.4407}{{$ \triangleright
  $}}

\bibitem{Marciano:1977su}
W.~J. Marciano and H.~Pagels, ``\textit{{Quantum Chromodynamics: A Review}},''
\href{http://dx.doi.org/10.1016/0370-1573(78)90208-9}{{\em Phys. Rept.} {\bf
  36} (1978)  137}

\bibitem{Maas:2005ym}
A.~Maas, ``\textit{{Gluons at finite temperature in Landau gauge Yang-Mills
  theory}},'' \href{http://dx.doi.org/10.1142/S0217732305018049}{{\em Mod.
  Phys. Lett.} {\bf A20} (2005)  1797--1811}
  \href{http://arxiv.org/abs/hep-ph/0506066}{{ $\bullet$}}
\href{http://inspirehep.net/search?p=find+eprint+hep-ph/0506066}{{$
  \triangleright $}}

\bibitem{Fister:2011uw}
L.~Fister and J.~M. Pawlowski, ``\textit{{Yang-Mills correlation functions at
  finite temperature}},'' \href{http://arxiv.org/abs/1112.5440}{{\tt
  1112.5440[hep-ph]}} \href{http://arxiv.org/abs/1112.5440}{{ $\bullet$}}
\href{http://inspirehep.net/search?p=find+eprint+1112.5440}{{$ \triangleright
  $}}

\bibitem{Cyrol:2017qkl}
A.~K. Cyrol, M.~Mitter, J.~M. Pawlowski, and N.~Strodthoff,
  ``\textit{{Nonperturbative finite-temperature Yang-Mills theory}},''
  \href{http://dx.doi.org/10.1103/PhysRevD.97.054015}{{\em Phys. Rev.} {\bf
  D97} (2018) no.~5, 054015} \href{http://arxiv.org/abs/1708.03482}{{
  $\bullet$}}
\href{http://inspirehep.net/search?p=find+eprint+1708.03482}{{$ \triangleright
  $}}

\bibitem{Fischer:2003rp}
C.~S. Fischer and R.~Alkofer, ``\textit{{N}onperturbative propagators, running
  coupling and dynamical quark mass of {L}andau gauge {QCD}},''
  \href{http://dx.doi.org/10.1103/PhysRevD.67.094020}{{\em Phys. Rev.} {\bf
  D67} (2003)  094020} \href{http://arxiv.org/abs/hep-ph/0301094}{{ $\bullet$}}
\href{http://inspirehep.net/search?p=find+eprint+hep-ph/0301094}{{$
  \triangleright $}}

\bibitem{Maas:2011ez}
A.~Maas, J.~M. Pawlowski, L.~von Smekal, and D.~Spielmann, ``\textit{{The Gluon
  propagator close to criticality}},''
  \href{http://dx.doi.org/10.1103/PhysRevD.85.034037}{{\em Phys. Rev.} {\bf
  D85} (2012)  034037} \href{http://arxiv.org/abs/1110.6340}{{ $\bullet$}}
\href{http://inspirehep.net/search?p=find+eprint+1110.6340}{{$ \triangleright
  $}}

\bibitem{Haque:2012my}
N.~Haque, M.~G. Mustafa, and M.~Strickland, ``\textit{{Two-loop hard thermal
  loop pressure at finite temperature and chemical potential}},''
  \href{http://dx.doi.org/10.1103/PhysRevD.87.105007}{{\em Phys. Rev.} {\bf
  D87} (2013) no.~10, 105007} \href{http://arxiv.org/abs/1212.1797}{{
  $\bullet$}}
\href{http://inspirehep.net/search?p=find+eprint+1212.1797}{{$ \triangleright
  $}}

\bibitem{Qin:2010nq}
S.-x. Qin, L.~Chang, H.~Chen, Y.-x. Liu, and C.~D. Roberts, ``\textit{{Phase
  diagram and critical endpoint for strongly-interacting quarks}},''
  \href{http://dx.doi.org/10.1103/PhysRevLett.106.172301}{{\em Phys. Rev.
  Lett.} {\bf 106} (2011)  172301} \href{http://arxiv.org/abs/1011.2876}{{
  $\bullet$}}
\href{http://inspirehep.net/search?p=find+eprint+1011.2876}{{$ \triangleright
  $}}

\bibitem{Jiang:2011ke}
Y.~Jiang, L.-J. Luo, and H.-S. Zong, ``\textit{{A Model study of quark number
  susceptibility at finite temperature beyond rainbow-ladder approximation}},''
  \href{http://dx.doi.org/10.1007/JHEP02(2011)066}{{\em JHEP} {\bf 02} (2011)
  066} \href{http://arxiv.org/abs/1102.1532}{{ $\bullet$}}
\href{http://inspirehep.net/search?p=find+eprint+1102.1532}{{$ \triangleright
  $}}

\bibitem{Shi:2014zpa}
C.~Shi, Y.-L. Wang, Y.~Jiang, Z.-F. Cui, and H.-S. Zong, ``\textit{{Locate QCD
  Critical End Point in a Continuum Model Study}},''
  \href{http://dx.doi.org/10.1007/JHEP07(2014)014}{{\em JHEP} {\bf 07} (2014)
  014} \href{http://arxiv.org/abs/1403.3797}{{ $\bullet$}}
\href{http://inspirehep.net/search?p=find+eprint+1403.3797}{{$ \triangleright
  $}}

\bibitem{Shi:2016koj}
C.~Shi, Y.-L. Du, S.-S. Xu, X.-J. Liu, and H.-S. Zong, ``\textit{{Continuum
  study of the QCD phase diagram through an OPE-modified gluon propagator}},''
  \href{http://dx.doi.org/10.1103/PhysRevD.93.036006}{{\em Phys. Rev.} {\bf
  D93} (2016) no.~3, 036006} \href{http://arxiv.org/abs/1602.00062}{{
  $\bullet$}}
\href{http://inspirehep.net/search?p=find+eprint+1602.00062}{{$ \triangleright
  $}}

\bibitem{Xu:2015jwa}
S.-S. Xu, Y.~Yan, Z.-F. Cui, and H.-S. Zong, ``\textit{{2+1 flavors QCD
  equation of state at zero temperature within Dyson-Schwinger equations}},''
  \href{http://dx.doi.org/10.1142/S0217751X15502176}{{\em Int. J. Mod. Phys.}
  {\bf A30} (2015) no.~36, 1550217} \href{http://arxiv.org/abs/1506.06846}{{
  $\bullet$}}
\href{http://inspirehep.net/search?p=find+eprint+1506.06846}{{$ \triangleright
  $}}

\bibitem{Gao:2016qkh}
F.~Gao and Y.-x. Liu, ``\textit{{QCD phase transitions via a refined truncation
  of Dyson-Schwinger equations}},''
  \href{http://dx.doi.org/10.1103/PhysRevD.94.076009}{{\em Phys. Rev.} {\bf
  D94} (2016) no.~7, 076009} \href{http://arxiv.org/abs/1607.01675}{{
  $\bullet$}}
\href{http://inspirehep.net/search?p=find+eprint+1607.01675}{{$ \triangleright
  $}}

\bibitem{Gao:2015kea}
F.~Gao, J.~Chen, Y.-X. Liu, S.-X. Qin, C.~D. Roberts, and S.~M. Schmidt,
  ``\textit{{Phase diagram and thermal properties of strong-interaction
  matter}},'' \href{http://dx.doi.org/10.1103/PhysRevD.93.094019}{{\em Phys.
  Rev.} {\bf D93} (2016) no.~9, 094019}
  \href{http://arxiv.org/abs/1507.00875}{{ $\bullet$}}
\href{http://inspirehep.net/search?p=find+eprint+1507.00875}{{$ \triangleright
  $}}

\bibitem{Gao:2016hks}
F.~Gao and Y.-x. Liu, ``\textit{{Interface Effect in QCD Phase Transitions via
  Dyson-Schwinger Equation Approach}},''
  \href{http://dx.doi.org/10.1103/PhysRevD.94.094030}{{\em Phys. Rev.} {\bf
  D94} (2016) no.~9, 094030} \href{http://arxiv.org/abs/1609.08038}{{
  $\bullet$}}
\href{http://inspirehep.net/search?p=find+eprint+1609.08038}{{$ \triangleright
  $}}

\bibitem{Xin:2014ela}
X.-y. Xin, S.-x. Qin, and Y.-x. Liu, ``\textit{{Quark number fluctuations at
  finite temperature and finite chemical potential via the Dyson-Schwinger
  equation approach}},''
\href{http://dx.doi.org/10.1103/PhysRevD.90.076006}{{\em Phys. Rev.} {\bf D90}
  (2014) no.~7, 076006}

\bibitem{Chang:2010hb}
L.~Chang, Y.-X. Liu, and C.~D. Roberts, ``\textit{{D}ressed-quark anomalous
  magnetic moments},''
  \href{http://dx.doi.org/10.1103/PhysRevLett.106.072001}{{\em Phys. Rev.
  Lett.} {\bf 106} (2011)  072001} \href{http://arxiv.org/abs/1009.3458}{{
  $\bullet$}}
\href{http://inspirehep.net/search?p=find+eprint+1009.3458}{{$ \triangleright
  $}}

\bibitem{Horvatic:2007wu}
D.~Horvatic, D.~Blaschke, D.~Klabucar, and A.~E. Radzhabov,
  ``\textit{{Pseudoscalar Meson Nonet at Zero and Finite Temperature}},''
  \href{http://dx.doi.org/10.1134/S1063779608070095}{{\em Phys. Part. Nucl.}
  {\bf 39} (2008)  1033--1039} \href{http://arxiv.org/abs/hep-ph/0703115}{{
  $\bullet$}}
\href{http://inspirehep.net/search?p=find+eprint+hep-ph/0703115}{{$
  \triangleright $}}

\bibitem{Horvatic:2007qs}
D.~Horvatic, D.~Klabucar, and A.~E. Radzhabov, ``\textit{{eta and eta-prime
  mesons in the Dyson-Schwinger approach at finite temperature}},''
  \href{http://dx.doi.org/10.1103/PhysRevD.76.096009}{{\em Phys. Rev.} {\bf
  D76} (2007)  096009} \href{http://arxiv.org/abs/0708.1260}{{ $\bullet$}}
\href{http://inspirehep.net/search?p=find+eprint+0708.1260}{{$ \triangleright
  $}}

\bibitem{Horvatic:2010md}
D.~Horvatic, D.~Blaschke, D.~Klabucar, and O.~Kaczmarek, ``\textit{{Width of
  the QCD transition in a Polyakov-loop DSE model}},''
  \href{http://dx.doi.org/10.1103/PhysRevD.84.016005}{{\em Phys. Rev.} {\bf
  D84} (2011)  016005} \href{http://arxiv.org/abs/1012.2113}{{ $\bullet$}}
\href{http://inspirehep.net/search?p=find+eprint+1012.2113}{{$ \triangleright
  $}}

\bibitem{Maris:2003vk}
P.~Maris and C.~D. Roberts, ``\textit{{Dyson-Schwinger equations: A Tool for
  hadron physics}},'' \href{http://dx.doi.org/10.1142/S0218301303001326}{{\em
  Int. J. Mod. Phys.} {\bf E12} (2003)  297--365}
  \href{http://arxiv.org/abs/nucl-th/0301049}{{ $\bullet$}}
\href{http://inspirehep.net/search?p=find+eprint+nucl-th/0301049}{{$
  \triangleright $}}

\bibitem{Fischer:2006ub}
C.~S. Fischer, ``\textit{{I}nfrared properties of {QCD} from
  {D}yson-{S}chwinger equations},''
  \href{http://dx.doi.org/10.1088/0954-3899/32/8/R02}{{\em J. Phys.} {\bf G32}
  (2006)  R253--R291} \href{http://arxiv.org/abs/hep-ph/0605173}{{ $\bullet$}}
\href{http://inspirehep.net/search?p=find+eprint+hep-ph/0605173}{{$
  \triangleright $}}

\bibitem{Binosi:2009qm}
D.~Binosi and J.~Papavassiliou, ``\textit{{P}inch {T}echnique: {T}heory and
  {A}pplications},''
  \href{http://dx.doi.org/10.1016/j.physrep.2009.05.001}{{\em Phys. Rept.} {\bf
  479} (2009)  1--152} \href{http://arxiv.org/abs/0909.2536}{{ $\bullet$}}
\href{http://inspirehep.net/search?p=find+eprint+0909.2536}{{$ \triangleright
  $}}

\bibitem{Holt:2010vj}
R.~J. Holt and C.~D. Roberts, ``\textit{{Distribution Functions of the Nucleon
  and Pion in the Valence Region}},''
  \href{http://dx.doi.org/10.1103/RevModPhys.82.2991}{{\em Rev. Mod. Phys.}
  {\bf 82} (2010)  2991--3044} \href{http://arxiv.org/abs/1002.4666}{{
  $\bullet$}}
\href{http://inspirehep.net/search?p=find+eprint+1002.4666}{{$ \triangleright
  $}}

\bibitem{Cloet:2013jya}
I.~C. Cloet and C.~D. Roberts, ``\textit{{E}xplanation and {P}rediction of
  {O}bservables using {C}ontinuum {S}trong {QCD}},''
  \href{http://dx.doi.org/10.1016/j.ppnp.2014.02.001}{{\em Prog. Part. Nucl.
  Phys.} {\bf 77} (2014)  1--69} \href{http://arxiv.org/abs/1310.2651}{{
  $\bullet$}}
\href{http://inspirehep.net/search?p=find+eprint+1310.2651}{{$ \triangleright
  $}}

\bibitem{Aguilar:2015bud}
A.~C. Aguilar, D.~Binosi, and J.~Papavassiliou, ``\textit{{T}he {G}luon {M}ass
  {G}eneration {M}echanism: {A} {C}oncise {P}rimer},''
  \href{http://dx.doi.org/10.1007/s11467-015-0517-6}{{\em Front. Phys. China}
  {\bf 11} (2016) no.~2, 111203} \href{http://arxiv.org/abs/1511.08361}{{
  $\bullet$}}
\href{http://inspirehep.net/search?p=find+eprint+1511.08361}{{$ \triangleright
  $}}

\bibitem{Sanchis-Alepuz:2017jjd}
H.~Sanchis-Alepuz and R.~Williams, ``\textit{{Recent developments in
  bound-state calculations using the Dyson-Schwinger and Bethe-Salpeter
  equations}},'' \href{http://dx.doi.org/10.1016/j.cpc.2018.05.020}{{\em
  Comput. Phys. Commun.} {\bf 232} (2018)  1--21}
  \href{http://arxiv.org/abs/1710.04903}{{ $\bullet$}}
\href{http://inspirehep.net/search?p=find+eprint+1710.04903}{{$ \triangleright
  $}}

\bibitem{Mandelstam:1979xd}
S.~Mandelstam, ``\textit{{Approximation Scheme for QCD}},''
\href{http://dx.doi.org/10.1103/PhysRevD.20.3223}{{\em Phys. Rev.} {\bf D20}
  (1979)  3223}

\bibitem{Cornwall:1981zr}
J.~M. Cornwall, ``\textit{{Dynamical Mass Generation in Continuum QCD}},''
\href{http://dx.doi.org/10.1103/PhysRevD.26.1453}{{\em Phys. Rev.} {\bf D26}
  (1982)  1453}

\bibitem{Brown:1988bn}
N.~Brown and M.~R. Pennington, ``\textit{{Studies of Confinement: How the Gluon
  Propagates}},''
\href{http://dx.doi.org/10.1103/PhysRevD.39.2723}{{\em Phys. Rev.} {\bf D39}
  (1989)  2723}

\bibitem{vonSmekal:1997ohs}
L.~von Smekal, R.~Alkofer, and A.~Hauck, ``\textit{{The Infrared behavior of
  gluon and ghost propagators in Landau gauge QCD}},''
  \href{http://dx.doi.org/10.1103/PhysRevLett.79.3591}{{\em Phys. Rev. Lett.}
  {\bf 79} (1997)  3591--3594} \href{http://arxiv.org/abs/hep-ph/9705242}{{
  $\bullet$}}
\href{http://inspirehep.net/search?p=find+eprint+hep-ph/9705242}{{$
  \triangleright $}}

\bibitem{vonSmekal:1997ern}
L.~von Smekal, A.~Hauck, and R.~Alkofer, ``\textit{{A} {S}olution to {C}oupled
  {D}yson {S}chwinger {E}quations for {G}luons and {G}hosts in {L}andau
  {G}auge},'' \href{http://dx.doi.org/10.1006/aphy.1998.5806,
  10.1006/aphy.1998.5864}{{\em Annals Phys.} {\bf 267} (1998)  1--60}
  \href{http://arxiv.org/abs/hep-ph/9707327}{{ $\bullet$}}
\href{http://inspirehep.net/search?p=find+eprint+hep-ph/9707327}{{$
  \triangleright $}}

\bibitem{Atkinson:1997tu}
D.~Atkinson and J.~C.~R. Bloch, ``\textit{{Running coupling in nonperturbative
  QCD. 1. Bare vertices and y-max approximation}},''
  \href{http://dx.doi.org/10.1103/PhysRevD.58.094036}{{\em Phys. Rev.} {\bf
  D58} (1998)  094036} \href{http://arxiv.org/abs/hep-ph/9712459}{{ $\bullet$}}
\href{http://inspirehep.net/search?p=find+eprint+hep-ph/9712459}{{$
  \triangleright $}}

\bibitem{Atkinson:1998zc}
D.~Atkinson and J.~C.~R. Bloch, ``\textit{{QCD in the infrared with exact
  angular integrations}},''
  \href{http://dx.doi.org/10.1142/S0217732398001121}{{\em Mod. Phys. Lett.}
  {\bf A13} (1998)  1055--1062} \href{http://arxiv.org/abs/hep-ph/9802239}{{
  $\bullet$}}
\href{http://inspirehep.net/search?p=find+eprint+hep-ph/9802239}{{$
  \triangleright $}}

\bibitem{Fischer:2002hna}
C.~S. Fischer and R.~Alkofer, ``\textit{{Infrared exponents and running
  coupling of SU(N) Yang-Mills theories}},''
  \href{http://dx.doi.org/10.1016/S0370-2693(02)01809-9}{{\em Phys. Lett.} {\bf
  B536} (2002)  177--184} \href{http://arxiv.org/abs/hep-ph/0202202}{{
  $\bullet$}}
\href{http://inspirehep.net/search?p=find+eprint+hep-ph/0202202}{{$
  \triangleright $}}

\bibitem{Zwanziger:2001kw}
D.~Zwanziger, ``\textit{{Nonperturbative Landau gauge and infrared critical
  exponents in QCD}},''
  \href{http://dx.doi.org/10.1103/PhysRevD.65.094039}{{\em Phys. Rev.} {\bf
  D65} (2002)  094039} \href{http://arxiv.org/abs/hep-th/0109224}{{ $\bullet$}}
\href{http://inspirehep.net/search?p=find+eprint+hep-th/0109224}{{$
  \triangleright $}}

\bibitem{Lerche:2002ep}
C.~Lerche and L.~von Smekal, ``\textit{{On the infrared exponent for gluon and
  ghost propagation in Landau gauge QCD}},''
  \href{http://dx.doi.org/10.1103/PhysRevD.65.125006}{{\em Phys. Rev.} {\bf
  D65} (2002)  125006} \href{http://arxiv.org/abs/hep-ph/0202194}{{ $\bullet$}}
\href{http://inspirehep.net/search?p=find+eprint+hep-ph/0202194}{{$
  \triangleright $}}

\bibitem{Pawlowski:2003hq}
J.~M. Pawlowski, D.~F. Litim, S.~Nedelko, and L.~von Smekal,
  ``\textit{{Infrared behavior and fixed points in Landau gauge QCD}},''
  \href{http://dx.doi.org/10.1103/PhysRevLett.93.152002}{{\em Phys. Rev. Lett.}
  {\bf 93} (2004)  152002} \href{http://arxiv.org/abs/hep-th/0312324}{{
  $\bullet$}}
\href{http://inspirehep.net/search?p=find+eprint+hep-th/0312324}{{$
  \triangleright $}}

\bibitem{Aguilar:2008xm}
A.~C. Aguilar, D.~Binosi, and J.~Papavassiliou, ``\textit{{G}luon and ghost
  propagators in the {L}andau gauge: {D}eriving lattice results from
  {S}chwinger-{D}yson equations},''
  \href{http://dx.doi.org/10.1103/PhysRevD.78.025010}{{\em Phys. Rev.} {\bf
  D78} (2008)  025010} \href{http://arxiv.org/abs/0802.1870}{{ $\bullet$}}
\href{http://inspirehep.net/search?p=find+eprint+0802.1870}{{$ \triangleright
  $}}

\bibitem{Boucaud:2008ky}
P.~Boucaud, J.~P. Leroy, A.~Le~Yaouanc, J.~Micheli, O.~Pene, and
  J.~Rodriguez-Quintero, ``\textit{{O}n the {IR} behaviour of the
  {L}andau-gauge ghost propagator},''
  \href{http://dx.doi.org/10.1088/1126-6708/2008/06/099}{{\em JHEP} {\bf 06}
  (2008)  099} \href{http://arxiv.org/abs/0803.2161}{{ $\bullet$}}
\href{http://inspirehep.net/search?p=find+eprint+0803.2161}{{$ \triangleright
  $}}

\bibitem{Dudal:2008sp}
D.~Dudal, J.~A. Gracey, S.~P. Sorella, N.~Vandersickel, and H.~Verschelde,
  ``\textit{{A} {R}efinement of the {G}ribov-{Z}wanziger approach in the
  {L}andau gauge: {I}nfrared propagators in harmony with the lattice
  results},'' \href{http://dx.doi.org/10.1103/PhysRevD.78.065047}{{\em Phys.
  Rev.} {\bf D78} (2008)  065047} \href{http://arxiv.org/abs/0806.4348}{{
  $\bullet$}}
\href{http://inspirehep.net/search?p=find+eprint+0806.4348}{{$ \triangleright
  $}}

\bibitem{Alkofer:2008jy}
R.~Alkofer, M.~Q. Huber, and K.~Schwenzer, ``\textit{{Infrared singularities in
  Landau gauge Yang-Mills theory}},''
  \href{http://dx.doi.org/10.1103/PhysRevD.81.105010}{{\em Phys. Rev.} {\bf
  D81} (2010)  105010} \href{http://arxiv.org/abs/0801.2762}{{ $\bullet$}}
\href{http://inspirehep.net/search?p=find+eprint+0801.2762}{{$ \triangleright
  $}}

\bibitem{Fischer:2008uz}
C.~S. Fischer, A.~Maas, and J.~M. Pawlowski, ``\textit{{O}n the infrared
  behavior of {L}andau gauge {Y}ang-{M}ills theory},''
  \href{http://dx.doi.org/10.1016/j.aop.2009.07.009}{{\em Annals Phys.} {\bf
  324} (2009)  2408--2437} \href{http://arxiv.org/abs/0810.1987}{{ $\bullet$}}
\href{http://inspirehep.net/search?p=find+eprint+0810.1987}{{$ \triangleright
  $}}

\bibitem{Huber:2012kd}
M.~Q. Huber and L.~von Smekal, ``\textit{{On the influence of three-point
  functions on the propagators of Landau gauge Yang-Mills theory}},''
  \href{http://dx.doi.org/10.1007/JHEP04(2013)149}{{\em JHEP} {\bf 04} (2013)
  149} \href{http://arxiv.org/abs/1211.6092}{{ $\bullet$}}
\href{http://inspirehep.net/search?p=find+eprint+1211.6092}{{$ \triangleright
  $}}

\bibitem{Cucchieri:2008fc}
A.~Cucchieri and T.~Mendes, ``\textit{{C}onstraints on the {IR} behavior of the
  ghost propagator in {Y}ang-{M}ills theories},''
  \href{http://dx.doi.org/10.1103/PhysRevD.78.094503}{{\em Phys. Rev.} {\bf
  D78} (2008)  094503} \href{http://arxiv.org/abs/0804.2371}{{ $\bullet$}}
\href{http://inspirehep.net/search?p=find+eprint+0804.2371}{{$ \triangleright
  $}}

\bibitem{Cucchieri:2007rg}
A.~Cucchieri and T.~Mendes, ``\textit{{Constraints on the IR behavior of the
  gluon propagator in Yang-Mills theories}},''
  \href{http://dx.doi.org/10.1103/PhysRevLett.100.241601}{{\em Phys. Rev.
  Lett.} {\bf 100} (2008)  241601} \href{http://arxiv.org/abs/0712.3517}{{
  $\bullet$}}
\href{http://inspirehep.net/search?p=find+eprint+0712.3517}{{$ \triangleright
  $}}

\bibitem{vonSmekal:2008ws}
L.~von Smekal, ``\textit{{Landau Gauge QCD: Functional Methods versus Lattice
  Simulations}},'' in {\em {13th International Conference on Selected Problems
  of Modern Theoretical Physics (SPMTP 08): Dedicated to the 100th Anniversary
  of the Birth of D.I. Blokhintsev (1908-1979) Dubna, Russia, June 23-27,
  2008}}.
\newblock 2008.
\newblock \href{http://arxiv.org/abs/0812.0654}{{\tt 0812.0654[hep-th]}}.
\newblock
\url{http://inspirehep.net/record/804216/files/arXiv:0812.0654.pdf}.
\newblock

\bibitem{Sternbeck:2008mv}
A.~Sternbeck and L.~von Smekal, ``\textit{{Infrared exponents and the
  strong-coupling limit in lattice Landau gauge}},''
  \href{http://dx.doi.org/10.1140/epjc/s10052-010-1381-8}{{\em Eur. Phys. J.}
  {\bf C68} (2010)  487--503} \href{http://arxiv.org/abs/0811.4300}{{
  $\bullet$}}
\href{http://inspirehep.net/search?p=find+eprint+0811.4300}{{$ \triangleright
  $}}

\bibitem{Cucchieri:2009zt}
A.~Cucchieri and T.~Mendes, ``\textit{{Landau-gauge propagators in Yang-Mills
  theories at beta = 0: Massive solution versus conformal scaling}},''
  \href{http://dx.doi.org/10.1103/PhysRevD.81.016005}{{\em Phys. Rev.} {\bf
  D81} (2010)  016005} \href{http://arxiv.org/abs/0904.4033}{{ $\bullet$}}
\href{http://inspirehep.net/search?p=find+eprint+0904.4033}{{$ \triangleright
  $}}

\bibitem{Maas:2009se}
A.~Maas, ``\textit{{Constructing non-perturbative gauges using correlation
  functions}},'' \href{http://dx.doi.org/10.1016/j.physletb.2010.04.052}{{\em
  Phys. Lett.} {\bf B689} (2010)  107--111}
  \href{http://arxiv.org/abs/0907.5185}{{ $\bullet$}}
\href{http://inspirehep.net/search?p=find+eprint+0907.5185}{{$ \triangleright
  $}}

\bibitem{Dudal:2014rxa}
D.~Dudal, M.~S. Guimaraes, I.~F. Justo, and S.~P. Sorella, ``\textit{{On bounds
  and boundary conditions in the continuum Landau gauge}},''
  \href{http://dx.doi.org/10.1140/epjc/s10052-015-3303-2}{{\em Eur. Phys. J.}
  {\bf C75} (2015) no.~2, 83} \href{http://arxiv.org/abs/1411.2500}{{
  $\bullet$}}
\href{http://inspirehep.net/search?p=find+eprint+1411.2500}{{$ \triangleright
  $}}

\bibitem{Cucchieri:2016qyc}
A.~Cucchieri and T.~Mendes, ``\textit{{Bloch Waves in Minimal Landau Gauge and
  the Infinite-Volume Limit of Lattice Gauge Theory}},''
  \href{http://dx.doi.org/10.1103/PhysRevLett.118.192002}{{\em Phys. Rev.
  Lett.} {\bf 118} (2017) no.~19, 192002}
  \href{http://arxiv.org/abs/1612.01279}{{ $\bullet$}}
\href{http://inspirehep.net/search?p=find+eprint+1612.01279}{{$ \triangleright
  $}}

\bibitem{Hopfer:2014zna}
M.~Hopfer, C.~S. Fischer, and R.~Alkofer, ``\textit{{Running coupling in the
  conformal window of large-Nf QCD}},''
  \href{http://dx.doi.org/10.1007/JHEP11(2014)035}{{\em JHEP} {\bf 11} (2014)
  035} \href{http://arxiv.org/abs/1405.7031}{{ $\bullet$}}
\href{http://inspirehep.net/search?p=find+eprint+1405.7031}{{$ \triangleright
  $}}

\bibitem{Aguilar:2017dco}
A.~C. Aguilar, D.~Binosi, C.~T. Figueiredo, and J.~Papavassiliou,
  ``\textit{{Evidence of ghost suppression in gluon mass scale dynamics}},''
  \href{http://dx.doi.org/10.1140/epjc/s10052-018-5679-2}{{\em Eur. Phys. J.}
  {\bf C78} (2018) no.~3, 181} \href{http://arxiv.org/abs/1712.06926}{{
  $\bullet$}}
\href{http://inspirehep.net/search?p=find+eprint+1712.06926}{{$ \triangleright
  $}}

\bibitem{Huber:2017txg}
M.~Q. Huber, ``\textit{{On non-primitively divergent vertices of Yang-Mills
  theory}},'' \href{http://dx.doi.org/10.1140/epjc/s10052-017-5310-y}{{\em Eur.
  Phys. J.} {\bf C77} (2017) no.~11, 733}
  \href{http://arxiv.org/abs/1709.05848}{{ $\bullet$}}
\href{http://inspirehep.net/search?p=find+eprint+1709.05848}{{$ \triangleright
  $}}

\bibitem{Huber:2014tva}
M.~Q. Huber and L.~von Smekal, ``\textit{{Spurious divergences in
  Dyson-Schwinger equations}},''
  \href{http://dx.doi.org/10.1007/JHEP06(2014)015}{{\em JHEP} {\bf 06} (2014)
  015} \href{http://arxiv.org/abs/1404.3642}{{ $\bullet$}}
\href{http://inspirehep.net/search?p=find+eprint+1404.3642}{{$ \triangleright
  $}}

\bibitem{Sternbeck:2005tk}
A.~Sternbeck, E.~M. Ilgenfritz, M.~Muller-Preussker, and A.~Schiller,
  ``\textit{{T}owards the infrared limit in {SU}(3) {L}andau gauge lattice
  gluodynamics},'' \href{http://dx.doi.org/10.1103/PhysRevD.72.014507}{{\em
  Phys. Rev.} {\bf D72} (2005)  014507}
  \href{http://arxiv.org/abs/hep-lat/0506007}{{ $\bullet$}}
\href{http://inspirehep.net/search?p=find+eprint+hep-lat/0506007}{{$
  \triangleright $}}

\bibitem{Sternbeck:2016}
A.~Sternbeck, {\em private communication}

\bibitem{Strauss:2012dg}
S.~Strauss, C.~S. Fischer, and C.~Kellermann, ``\textit{{A}nalytic structure of
  the {L}andau gauge gluon propagator},''
  \href{http://dx.doi.org/10.1103/PhysRevLett.109.252001}{{\em Phys. Rev.
  Lett.} {\bf 109} (2012)  252001} \href{http://arxiv.org/abs/1208.6239}{{
  $\bullet$}}
\href{http://inspirehep.net/search?p=find+eprint+1208.6239}{{$ \triangleright
  $}}

\bibitem{Boucaud:2000nd}
P.~Boucaud, A.~Le~Yaouanc, J.~P. Leroy, J.~Micheli, O.~Pene, and
  J.~Rodriguez-Quintero, ``\textit{{Consistent OPE description of gluon two
  point and three point Green function?}},''
  \href{http://dx.doi.org/10.1016/S0370-2693(00)01149-7}{{\em Phys. Lett.} {\bf
  B493} (2000)  315--324} \href{http://arxiv.org/abs/hep-ph/0008043}{{
  $\bullet$}}
\href{http://inspirehep.net/search?p=find+eprint+hep-ph/0008043}{{$
  \triangleright $}}

\bibitem{Langfeld:2001cz}
K.~Langfeld, H.~Reinhardt, and J.~Gattnar, ``\textit{{Gluon propagators and
  quark confinement}},''
  \href{http://dx.doi.org/10.1016/S0550-3213(01)00574-0}{{\em Nucl. Phys.} {\bf
  B621} (2002)  131--156} \href{http://arxiv.org/abs/hep-ph/0107141}{{
  $\bullet$}}
\href{http://inspirehep.net/search?p=find+eprint+hep-ph/0107141}{{$
  \triangleright $}}

\bibitem{Bowman:2007du}
P.~O. Bowman, U.~M. Heller, D.~B. Leinweber, M.~B. Parappilly, A.~Sternbeck,
  L.~von Smekal, A.~G. Williams, and J.-b. Zhang, ``\textit{{Scaling behavior
  and positivity violation of the gluon propagator in full QCD}},''
  \href{http://dx.doi.org/10.1103/PhysRevD.76.094505}{{\em Phys. Rev.} {\bf
  D76} (2007)  094505} \href{http://arxiv.org/abs/hep-lat/0703022}{{
  $\bullet$}}
\href{http://inspirehep.net/search?p=find+eprint+hep-lat/0703022}{{$
  \triangleright $}}

\bibitem{Cucchieri:2011ig}
A.~Cucchieri, D.~Dudal, T.~Mendes, and N.~Vandersickel, ``\textit{{M}odeling
  the {G}luon {P}ropagator in {L}andau {G}auge: {L}attice {E}stimates of {P}ole
  {M}asses and {D}imension-{T}wo {C}ondensates},''
  \href{http://dx.doi.org/10.1103/PhysRevD.85.094513}{{\em Phys. Rev.} {\bf
  D85} (2012)  094513} \href{http://arxiv.org/abs/1111.2327}{{ $\bullet$}}
\href{http://inspirehep.net/search?p=find+eprint+1111.2327}{{$ \triangleright
  $}}

\bibitem{Oliveira:2012eh}
O.~Oliveira and P.~J. Silva, ``\textit{{The lattice Landau gauge gluon
  propagator: lattice spacing and volume dependence}},''
  \href{http://dx.doi.org/10.1103/PhysRevD.86.114513}{{\em Phys. Rev.} {\bf
  D86} (2012)  114513} \href{http://arxiv.org/abs/1207.3029}{{ $\bullet$}}
\href{http://inspirehep.net/search?p=find+eprint+1207.3029}{{$ \triangleright
  $}}

\bibitem{Ayala:2012pb}
A.~Ayala, A.~Bashir, D.~Binosi, M.~Cristoforetti, and J.~Rodriguez-Quintero,
  ``\textit{{Quark flavour effects on gluon and ghost propagators}},''
  \href{http://dx.doi.org/10.1103/PhysRevD.86.074512}{{\em Phys. Rev.} {\bf
  D86} (2012)  074512} \href{http://arxiv.org/abs/1208.0795}{{ $\bullet$}}
\href{http://inspirehep.net/search?p=find+eprint+1208.0795}{{$ \triangleright
  $}}

\bibitem{Boucaud:2017ksi}
P.~Boucaud, F.~De~Soto, J.~Rodríguez-Quintero, and S.~Zafeiropoulos,
  ``\textit{{Comment on ?Lattice gluon and ghost propagators and the strong
  coupling in pure $SU(3)$ Yang-Mills theory: Finite lattice spacing and volume
  effects?}},'' \href{http://dx.doi.org/10.1103/PhysRevD.96.098501}{{\em Phys.
  Rev.} {\bf D96} (2017) no.~9, 098501}
  \href{http://arxiv.org/abs/1704.02053}{{ $\bullet$}}
\href{http://inspirehep.net/search?p=find+eprint+1704.02053}{{$ \triangleright
  $}}

\bibitem{Biddle:2018dtc}
J.~C. Biddle, W.~Kamleh, and D.~B. Leinweber, ``\textit{{Gluon propagator on a
  centre-vortex background}},'' \href{http://arxiv.org/abs/1806.04305}{{\tt
  1806.04305[hep-lat]}} \href{http://arxiv.org/abs/1806.04305}{{ $\bullet$}}
\href{http://inspirehep.net/search?p=find+eprint+1806.04305}{{$ \triangleright
  $}}

\bibitem{Boucaud:2018xup}
P.~Boucaud, F.~De~Soto, K.~Raya, J.~Rodríguez-Quintero, and S.~Zafeiropoulos,
  ``\textit{{Discretization effects on renormalized gauge-field Green?s
  functions, scale setting, and the gluon mass}},''
  \href{http://dx.doi.org/10.1103/PhysRevD.98.114515}{{\em Phys. Rev.} {\bf
  D98} (2018) no.~11, 114515} \href{http://arxiv.org/abs/1809.05776}{{
  $\bullet$}}
\href{http://inspirehep.net/search?p=find+eprint+1809.05776}{{$ \triangleright
  $}}

\bibitem{Fischer:2003zc}
C.~S. Fischer, {\em {Nonperturbative propagators, running coupling and
  dynamical mass generation in ghost - anti-ghost symmetric gauges in QCD}}.
\newblock PhD thesis, Tubingen U. 2003, arXiv:hep-ph/0304233.
\newblock \href{http://arxiv.org/abs/hep-ph/0304233}{{ $\bullet$}}
\href{http://inspirehep.net/search?p=find+eprint+hep-ph/0304233}{{$
  \triangleright $}}

\bibitem{Gattnar:2004bf}
J.~Gattnar, K.~Langfeld, and H.~Reinhardt, ``\textit{{Signals of confinement in
  Green functions of SU(2) Yang-Mills theory}},''
  \href{http://dx.doi.org/10.1103/PhysRevLett.93.061601}{{\em Phys. Rev. Lett.}
  {\bf 93} (2004)  061601} \href{http://arxiv.org/abs/hep-lat/0403011}{{
  $\bullet$}}
\href{http://inspirehep.net/search?p=find+eprint+hep-lat/0403011}{{$
  \triangleright $}}

\bibitem{Montero:1999by}
A.~Montero, ``\textit{{Study of SU(3) vortex - like configurations with a new
  maximal center gauge fixing method}},''
  \href{http://dx.doi.org/10.1016/S0370-2693(99)01113-2}{{\em Phys. Lett.} {\bf
  B467} (1999)  106--111} \href{http://arxiv.org/abs/hep-lat/9906010}{{
  $\bullet$}}
\href{http://inspirehep.net/search?p=find+eprint+hep-lat/9906010}{{$
  \triangleright $}}

\bibitem{Faber:1999sq}
M.~Faber, J.~Greensite, and S.~Olejnik, ``\textit{{First evidence for center
  dominance in SU(3) lattice gauge theory}},''
  \href{http://dx.doi.org/10.1016/S0370-2693(00)00013-7}{{\em Phys. Lett.} {\bf
  B474} (2000)  177--181} \href{http://arxiv.org/abs/hep-lat/9911006}{{
  $\bullet$}}
\href{http://inspirehep.net/search?p=find+eprint+hep-lat/9911006}{{$
  \triangleright $}}

\bibitem{Oehme:1979ai}
R.~Oehme and W.~Zimmermann, ``\textit{{Quark and Gluon Propagators in Quantum
  Chromodynamics}},''
\href{http://dx.doi.org/10.1103/PhysRevD.21.471}{{\em Phys. Rev.} {\bf D21}
  (1980)  471}

\bibitem{Nishijima:1993fq}
K.~Nishijima, ``\textit{{Confinement of quarks and gluons}},''
\href{http://dx.doi.org/10.1142/S0217751X94001539}{{\em Int. J. Mod. Phys.}
  {\bf A9} (1994)  3799--3820}

\bibitem{Oehme:1994hf}
R.~Oehme and W.-T. Xu, ``\textit{{Asymptotic limits and sum rules for gauge
  field propagators}},''
  \href{http://dx.doi.org/10.1016/0370-2693(94)91025-1}{{\em Phys. Lett.} {\bf
  B333} (1994)  172--177} \href{http://arxiv.org/abs/hep-th/9406081}{{
  $\bullet$}}
\href{http://inspirehep.net/search?p=find+eprint+hep-th/9406081}{{$
  \triangleright $}}

\bibitem{Nishijima:1995ie}
K.~Nishijima, ``\textit{{Confinement of quarks and gluons. 2}},''
\href{http://dx.doi.org/10.1142/S0217751X95001510}{{\em Int. J. Mod. Phys.}
  {\bf A10} (1995)  3155--3167}

\bibitem{Dudal:2013yva}
D.~Dudal, O.~Oliveira, and P.~J. Silva, ``\textit{{K\"allen-Lehmann
  spectroscopy for (un)physical degrees of freedom}},''
  \href{http://dx.doi.org/10.1103/PhysRevD.89.014010}{{\em Phys. Rev.} {\bf
  D89} (2014) no.~1, 014010} \href{http://arxiv.org/abs/1310.4069}{{
  $\bullet$}}
\href{http://inspirehep.net/search?p=find+eprint+1310.4069}{{$ \triangleright
  $}}

\bibitem{Cyrol:2018xeq}
A.~K. Cyrol, J.~M. Pawlowski, A.~Rothkopf, and N.~Wink,
  ``\textit{{Reconstructing the gluon}},''
  \href{http://arxiv.org/abs/1804.00945}{{\tt 1804.00945[hep-ph]}}
  \href{http://arxiv.org/abs/1804.00945}{{ $\bullet$}}
\href{http://inspirehep.net/search?p=find+eprint+1804.00945}{{$ \triangleright
  $}}

\bibitem{Lowdon:2017uqe}
P.~Lowdon, ``\textit{{Nonperturbative structure of the photon and gluon
  propagators}},'' \href{http://dx.doi.org/10.1103/PhysRevD.96.065013}{{\em
  Phys. Rev.} {\bf D96} (2017) no.~6, 065013}
  \href{http://arxiv.org/abs/1702.02954}{{ $\bullet$}}
\href{http://inspirehep.net/search?p=find+eprint+1702.02954}{{$ \triangleright
  $}}

\bibitem{Alkofer:2003jj}
R.~Alkofer, W.~Detmold, C.~S. Fischer, and P.~Maris, ``\textit{{Analytic
  properties of the Landau gauge gluon and quark propagators}},''
  \href{http://dx.doi.org/10.1103/PhysRevD.70.014014}{{\em Phys. Rev.} {\bf
  D70} (2004)  014014} \href{http://arxiv.org/abs/hep-ph/0309077}{{ $\bullet$}}
\href{http://inspirehep.net/search?p=find+eprint+hep-ph/0309077}{{$
  \triangleright $}}

\bibitem{Dudal:2010tf}
D.~Dudal, O.~Oliveira, and N.~Vandersickel, ``\textit{{Indirect lattice
  evidence for the Refined Gribov-Zwanziger formalism and the gluon condensate
  $\langle{A^2}\rangle$ in the Landau gauge}},''
  \href{http://dx.doi.org/10.1103/PhysRevD.81.074505}{{\em Phys. Rev.} {\bf
  D81} (2010)  074505} \href{http://arxiv.org/abs/1002.2374}{{ $\bullet$}}
\href{http://inspirehep.net/search?p=find+eprint+1002.2374}{{$ \triangleright
  $}}

\bibitem{Cucchieri:2016jwg}
A.~Cucchieri, D.~Dudal, T.~Mendes, and N.~Vandersickel, ``\textit{{Modeling the
  Landau-gauge ghost propagator in 2, 3, and 4 spacetime dimensions}},''
  \href{http://dx.doi.org/10.1103/PhysRevD.93.094513}{{\em Phys. Rev.} {\bf
  D93} (2016) no.~9, 094513} \href{http://arxiv.org/abs/1602.01646}{{
  $\bullet$}}
\href{http://inspirehep.net/search?p=find+eprint+1602.01646}{{$ \triangleright
  $}}

\bibitem{Bowman:2005vx}
P.~O. Bowman, U.~M. Heller, D.~B. Leinweber, M.~B. Parappilly, A.~G. Williams,
  and J.-b. Zhang, ``\textit{{Unquenched quark propagator in Landau gauge}},''
  \href{http://dx.doi.org/10.1103/PhysRevD.71.054507}{{\em Phys. Rev.} {\bf
  D71} (2005)  054507} \href{http://arxiv.org/abs/hep-lat/0501019}{{
  $\bullet$}}
\href{http://inspirehep.net/search?p=find+eprint+hep-lat/0501019}{{$
  \triangleright $}}

\bibitem{Fukuda:1976zb}
R.~Fukuda and T.~Kugo, ``\textit{{Schwinger-Dyson Equation for Massless Vector
  Theory and Absence of Fermion Pole}},''
\href{http://dx.doi.org/10.1016/0550-3213(76)90572-1}{{\em Nucl. Phys.} {\bf
  B117} (1976)  250--264}

\bibitem{Miransky:1984ef}
V.~A. Miransky, ``\textit{{Dynamics of Spontaneous Chiral Symmetry Breaking and
  Continuum Limit in Quantum Electrodynamics}},''
\href{http://dx.doi.org/10.1007/BF02724229}{{\em Nuovo Cim.} {\bf A90} (1985)
  149--170}

\bibitem{Miransky:1986ib}
V.~A. Miransky, ``\textit{{On Dynamical Chiral Symmetry Breaking }},''
\href{http://dx.doi.org/10.1016/0370-2693(85)91254-7}{{\em Phys. Lett.} {\bf
  165B} (1985)  401--404}

\bibitem{Windisch:2016iud}
A.~Windisch, ``\textit{{Analytic properties of the quark propagator from an
  effective infrared interaction model}},''
  \href{http://dx.doi.org/10.1103/PhysRevC.95.045204}{{\em Phys. Rev.} {\bf
  C95} (2017) no.~4, 045204} \href{http://arxiv.org/abs/1612.06002}{{
  $\bullet$}}
\href{http://inspirehep.net/search?p=find+eprint+1612.06002}{{$ \triangleright
  $}}

\bibitem{Fischer:2008sp}
C.~S. Fischer, D.~Nickel, and R.~Williams, ``\textit{{On Gribov's
  supercriticality picture of quark confinement}},''
  \href{http://dx.doi.org/10.1140/epjc/s10052-008-0821-1}{{\em Eur. Phys. J.}
  {\bf C60} (2009)  47--61} \href{http://arxiv.org/abs/0807.3486}{{ $\bullet$}}
\href{http://inspirehep.net/search?p=find+eprint+0807.3486}{{$ \triangleright
  $}}

\bibitem{Chang:2006bm}
L.~Chang, Y.-X. Liu, M.~S. Bhagwat, C.~D. Roberts, and S.~V. Wright,
  ``\textit{{Dynamical chiral symmetry breaking and a critical mass}},''
  \href{http://dx.doi.org/10.1103/PhysRevC.75.015201}{{\em Phys. Rev.} {\bf
  C75} (2007)  015201} \href{http://arxiv.org/abs/nucl-th/0605058}{{
  $\bullet$}}
\href{http://inspirehep.net/search?p=find+eprint+nucl-th/0605058}{{$
  \triangleright $}}

\bibitem{Williams:2006vva}
R.~Williams, C.~S. Fischer, and M.~R. Pennington, ``\textit{{Anti-q q
  condensate for light quarks beyond the chiral limit}},''
  \href{http://dx.doi.org/10.1016/j.physletb.2006.12.055}{{\em Phys. Lett.}
  {\bf B645} (2007)  167--172} \href{http://arxiv.org/abs/hep-ph/0612061}{{
  $\bullet$}}
\href{http://inspirehep.net/search?p=find+eprint+hep-ph/0612061}{{$
  \triangleright $}}

\bibitem{Bashir:2012fs}
A.~Bashir, L.~Chang, I.~C. Cloet, B.~El-Bennich, Y.-X. Liu, C.~D. Roberts, and
  P.~C. Tandy, ``\textit{{C}ollective perspective on advances in
  {D}yson-{S}chwinger {E}quation {QCD}},''
  \href{http://dx.doi.org/10.1088/0253-6102/58/1/16}{{\em Commun. Theor. Phys.}
  {\bf 58} (2012)  79--134} \href{http://arxiv.org/abs/1201.3366}{{ $\bullet$}}
\href{http://inspirehep.net/search?p=find+eprint+1201.3366}{{$ \triangleright
  $}}

\bibitem{Horn:2016rip}
T.~Horn and C.~D. Roberts, ``\textit{{The pion: an enigma within the Standard
  Model}},'' \href{http://dx.doi.org/10.1088/0954-3899/43/7/073001}{{\em J.
  Phys.} {\bf G43} (2016) no.~7, 073001}
  \href{http://arxiv.org/abs/1602.04016}{{ $\bullet$}}
\href{http://inspirehep.net/search?p=find+eprint+1602.04016}{{$ \triangleright
  $}}

\bibitem{Maris:1997hd}
P.~Maris, C.~D. Roberts, and P.~C. Tandy, ``\textit{{P}ion mass and decay
  constant},'' \href{http://dx.doi.org/10.1016/S0370-2693(97)01535-9}{{\em
  Phys. Lett.} {\bf B420} (1998)  267--273}
  \href{http://arxiv.org/abs/nucl-th/9707003}{{ $\bullet$}}
\href{http://inspirehep.net/search?p=find+eprint+nucl-th/9707003}{{$
  \triangleright $}}

\bibitem{Maris:1997tm}
P.~Maris and C.~D. Roberts, ``\textit{{Pi- and K meson Bethe-Salpeter
  amplitudes}},'' \href{http://dx.doi.org/10.1103/PhysRevC.56.3369}{{\em Phys.
  Rev.} {\bf C56} (1997)  3369--3383}
  \href{http://arxiv.org/abs/nucl-th/9708029}{{ $\bullet$}}
\href{http://inspirehep.net/search?p=find+eprint+nucl-th/9708029}{{$
  \triangleright $}}

\bibitem{Krassnigg:2010mh}
A.~Krassnigg and M.~Blank, ``\textit{{A} covariant study of tensor mesons},''
  \href{http://dx.doi.org/10.1103/PhysRevD.83.096006}{{\em Phys. Rev.} {\bf
  D83} (2011)  096006} \href{http://arxiv.org/abs/1011.6650}{{ $\bullet$}}
\href{http://inspirehep.net/search?p=find+eprint+1011.6650}{{$ \triangleright
  $}}

\bibitem{Fischer:2014xha}
C.~S. Fischer, S.~Kubrak, and R.~Williams, ``\textit{{M}ass spectra and {R}egge
  trajectories of light mesons in the {B}ethe-{S}alpeter approach},''
  \href{http://dx.doi.org/10.1140/epja/i2014-14126-6}{{\em Eur. Phys. J.} {\bf
  A50} (2014)  126} \href{http://arxiv.org/abs/1406.4370}{{ $\bullet$}}
\href{http://inspirehep.net/search?p=find+eprint+1406.4370}{{$ \triangleright
  $}}

\bibitem{Bicudo:2001jq}
P.~Bicudo, S.~Cotanch, F.~J. Llanes-Estrada, P.~Maris, E.~Ribeiro, and
  A.~Szczepaniak, ``\textit{{C}hirally symmetric quark description of
  low-energy $\pi\pi$ scattering},''
  \href{http://dx.doi.org/10.1103/PhysRevD.65.076008}{{\em Phys. Rev.} {\bf
  D65} (2002)  076008} \href{http://arxiv.org/abs/hep-ph/0112015}{{ $\bullet$}}
\href{http://inspirehep.net/search?p=find+eprint+hep-ph/0112015}{{$
  \triangleright $}}

\bibitem{Bicudo:2003fp}
P.~Bicudo, ``\textit{{Analytic proof that the quark model complies with
  partially conserved axial current theorems}},''
  \href{http://dx.doi.org/10.1103/PhysRevC.67.035201}{{\em Phys. Rev.} {\bf
  C67} (2003)  035201} \href{http://arxiv.org/abs/hep-ph/0311277}{{ $\bullet$}}
\href{http://inspirehep.net/search?p=find+eprint+hep-ph/0311277}{{$
  \triangleright $}}

\bibitem{Heupel:2012ua}
W.~Heupel, G.~Eichmann, and C.~S. Fischer, ``\textit{{Tetraquark Bound States
  in a Bethe-Salpeter Approach}},''
  \href{http://dx.doi.org/10.1016/j.physletb.2012.11.009}{{\em Phys. Lett.}
  {\bf B718} (2012)  545--549} \href{http://arxiv.org/abs/1206.5129}{{
  $\bullet$}}
\href{http://inspirehep.net/search?p=find+eprint+1206.5129}{{$ \triangleright
  $}}

\bibitem{Eichmann:2015cra}
G.~Eichmann, C.~S. Fischer, and W.~Heupel, ``\textit{{T}he light scalar mesons
  as tetraquarks},''
  \href{http://dx.doi.org/10.1016/j.physletb.2015.12.036}{{\em Phys. Lett.}
  {\bf B753} (2016)  282--287} \href{http://arxiv.org/abs/1508.07178}{{
  $\bullet$}}
\href{http://inspirehep.net/search?p=find+eprint+1508.07178}{{$ \triangleright
  $}}

\bibitem{Jaffe:1976ig}
R.~L. Jaffe, ``\textit{{Multi-Quark Hadrons. 1. The Phenomenology of (2 Quark 2
  anti-Quark) Mesons}},''
\href{http://dx.doi.org/10.1103/PhysRevD.15.267}{{\em Phys. Rev.} {\bf D15}
  (1977)  267}

\bibitem{Amsler:2004ps}
C.~Amsler and N.~A. Tornqvist, ``\textit{{Mesons beyond the naive quark
  model}},''
\href{http://dx.doi.org/10.1016/j.physrep.2003.09.003}{{\em Phys. Rept.} {\bf
  389} (2004)  61--117}

\bibitem{Giacosa:2006tf}
F.~Giacosa, ``\textit{{Mixing of scalar tetraquark and quarkonia states in a
  chiral approach}},'' \href{http://dx.doi.org/10.1103/PhysRevD.75.054007}{{\em
  Phys. Rev.} {\bf D75} (2007)  054007}
  \href{http://arxiv.org/abs/hep-ph/0611388}{{ $\bullet$}}
\href{http://inspirehep.net/search?p=find+eprint+hep-ph/0611388}{{$
  \triangleright $}}

\bibitem{Ebert:2008id}
D.~Ebert, R.~N. Faustov, and V.~O. Galkin, ``\textit{{Masses of light
  tetraquarks and scalar mesons in the relativistic quark model}},''
  \href{http://dx.doi.org/10.1140/epjc/s10052-009-0925-2}{{\em Eur. Phys. J.}
  {\bf C60} (2009)  273--278} \href{http://arxiv.org/abs/0812.2116}{{
  $\bullet$}}
\href{http://inspirehep.net/search?p=find+eprint+0812.2116}{{$ \triangleright
  $}}

\bibitem{Parganlija:2012fy}
D.~Parganlija, P.~Kovacs, G.~Wolf, F.~Giacosa, and D.~H. Rischke,
  ``\textit{{Meson vacuum phenomenology in a three-flavor linear sigma model
  with (axial-)vector mesons}},''
  \href{http://dx.doi.org/10.1103/PhysRevD.87.014011}{{\em Phys. Rev.} {\bf
  D87} (2013) no.~1, 014011} \href{http://arxiv.org/abs/1208.0585}{{
  $\bullet$}}
\href{http://inspirehep.net/search?p=find+eprint+1208.0585}{{$ \triangleright
  $}}

\bibitem{Pelaez:2015qba}
J.~R. Pelaez, ``\textit{{From controversy to precision on the sigma meson: a
  review on the status of the non-ordinary $f_0(500)$ resonance}},''
  \href{http://arxiv.org/abs/1510.00653}{{\tt 1510.00653[hep-ph]}}
  \href{http://arxiv.org/abs/1510.00653}{{ $\bullet$}}
\href{http://inspirehep.net/search?p=find+eprint+1510.00653}{{$ \triangleright
  $}}

\bibitem{Williams:2018adr}
R.~Williams, ``\textit{{Vector mesons as dynamical resonances in the
  Bethe-Salpeter framework}},'' \href{http://arxiv.org/abs/1804.11161}{{\tt
  1804.11161[hep-ph]}} \href{http://arxiv.org/abs/1804.11161}{{ $\bullet$}}
\href{http://inspirehep.net/search?p=find+eprint+1804.11161}{{$ \triangleright
  $}}

\bibitem{Oettel:1998bk}
M.~Oettel, G.~Hellstern, R.~Alkofer, and H.~Reinhardt, ``\textit{{O}ctet and
  decuplet baryons in a covariant and confining diquark - quark model},''
  \href{http://dx.doi.org/10.1103/PhysRevC.58.2459}{{\em Phys. Rev.} {\bf C58}
  (1998)  2459--2477} \href{http://arxiv.org/abs/nucl-th/9805054}{{ $\bullet$}}
\href{http://inspirehep.net/search?p=find+eprint+nucl-th/9805054}{{$
  \triangleright $}}

\bibitem{Eichmann:2016jqx}
G.~Eichmann, ``\textit{{Progress in the calculation of nucleon transition form
  factors}},''
\newblock 2016.
\newblock
\href{http://arxiv.org/abs/1602.03462}{{\tt 1602.03462[hep-ph]}}.
\newblock

\bibitem{Oettel:2000jj}
M.~Oettel, R.~Alkofer, and L.~von Smekal, ``\textit{{Nucleon properties in the
  covariant quark diquark model}},''
  \href{http://dx.doi.org/10.1007/s100500070078}{{\em Eur. Phys. J.} {\bf A8}
  (2000)  553--566} \href{http://arxiv.org/abs/nucl-th/0006082}{{ $\bullet$}}
\href{http://inspirehep.net/search?p=find+eprint+nucl-th/0006082}{{$
  \triangleright $}}

\bibitem{Segovia:2015hra}
J.~Segovia, B.~El-Bennich, E.~Rojas, I.~C. Cloet, C.~D. Roberts, S.-S. Xu, and
  H.-S. Zong, ``\textit{{Completing the picture of the Roper resonance}},''
  \href{http://dx.doi.org/10.1103/PhysRevLett.115.171801}{{\em Phys. Rev.
  Lett.} {\bf 115} (2015) no.~17, 171801}
  \href{http://arxiv.org/abs/1504.04386}{{ $\bullet$}}
\href{http://inspirehep.net/search?p=find+eprint+1504.04386}{{$ \triangleright
  $}}

\bibitem{Eichmann:2016hgl}
G.~Eichmann, C.~S. Fischer, and H.~Sanchis-Alepuz, ``\textit{{On light baryons
  and their excitations}},'' \href{http://arxiv.org/abs/1607.05748}{{\tt
  1607.05748[hep-ph]}} \href{http://arxiv.org/abs/1607.05748}{{ $\bullet$}}
\href{http://inspirehep.net/search?p=find+eprint+1607.05748}{{$ \triangleright
  $}}

\bibitem{Eichmann:2017inp}
G.~Eichmann and C.~S. Fischer, {\em in preparation}

\bibitem{Eichmann:2009qa}
G.~Eichmann, R.~Alkofer, A.~Krassnigg, and D.~Nicmorus, ``\textit{{N}ucleon
  mass from a covariant three-quark {F}addeev equation},''
  \href{http://dx.doi.org/10.1103/PhysRevLett.104.201601}{{\em Phys. Rev.
  Lett.} {\bf 104} (2010)  201601} \href{http://arxiv.org/abs/0912.2246}{{
  $\bullet$}}
\href{http://inspirehep.net/search?p=find+eprint+0912.2246}{{$ \triangleright
  $}}

\bibitem{SanchisAlepuz:2011jn}
H.~Sanchis-Alepuz, G.~Eichmann, S.~Villalba-Chavez, and R.~Alkofer,
  ``\textit{{D}elta and {O}mega masses in a three-quark covariant {F}addeev
  approach},'' \href{http://dx.doi.org/10.1103/PhysRevD.84.096003}{{\em Phys.
  Rev.} {\bf D84} (2011)  096003} \href{http://arxiv.org/abs/1109.0199}{{
  $\bullet$}}
\href{http://inspirehep.net/search?p=find+eprint+1109.0199}{{$ \triangleright
  $}}

\bibitem{Patrignani:2016xqp}
{\bf Particle Data Group} Collaboration: C.~Patrignani {\em et al.},
  ``\textit{{Review of Particle Physics}},''
\href{http://dx.doi.org/10.1088/1674-1137/40/10/100001}{{\em Chin. Phys.} {\bf
  C40} (2016) no.~10, 100001}

\bibitem{Chen:2017pse}
C.~Chen, B.~El-Bennich, C.~D. Roberts, S.~M. Schmidt, J.~Segovia, and S.~Wan,
  ``\textit{{Structure of the nucleon's low-lying excitations}},''
  \href{http://arxiv.org/abs/1711.03142}{{\tt 1711.03142[nucl-th]}}
  \href{http://arxiv.org/abs/1711.03142}{{ $\bullet$}}
\href{http://inspirehep.net/search?p=find+eprint+1711.03142}{{$ \triangleright
  $}}

\bibitem{Fischer:2008wy}
C.~S. Fischer and R.~Williams, ``\textit{{B}eyond the rainbow: {E}ffects from
  pion back-coupling},''
  \href{http://dx.doi.org/10.1103/PhysRevD.78.074006}{{\em Phys. Rev.} {\bf
  D78} (2008)  074006} \href{http://arxiv.org/abs/0808.3372}{{ $\bullet$}}
\href{http://inspirehep.net/search?p=find+eprint+0808.3372}{{$ \triangleright
  $}}

\bibitem{Fischer:2009jm}
C.~S. Fischer and R.~Williams, ``\textit{{P}robing the gluon self-interaction
  in light mesons},''
  \href{http://dx.doi.org/10.1103/PhysRevLett.103.122001}{{\em Phys. Rev.
  Lett.} {\bf 103} (2009)  122001} \href{http://arxiv.org/abs/0905.2291}{{
  $\bullet$}}
\href{http://inspirehep.net/search?p=find+eprint+0905.2291}{{$ \triangleright
  $}}

\bibitem{Sanchis-Alepuz:2014wea}
H.~Sanchis-Alepuz, C.~S. Fischer, and S.~Kubrak, ``\textit{{Pion cloud effects
  on baryon masses}},''
  \href{http://dx.doi.org/10.1016/j.physletb.2014.04.031}{{\em Phys. Lett.}
  {\bf B733} (2014)  151--157} \href{http://arxiv.org/abs/1401.3183}{{
  $\bullet$}}
\href{http://inspirehep.net/search?p=find+eprint+1401.3183}{{$ \triangleright
  $}}

\bibitem{Sanchis-Alepuz:2015qra}
H.~Sanchis-Alepuz and R.~Williams, ``\textit{{Probing the quark-gluon
  interaction with hadrons}},''
  \href{http://dx.doi.org/10.1016/j.physletb.2015.08.067}{{\em Phys. Lett.}
  {\bf B749} (2015)  592--596} \href{http://arxiv.org/abs/1504.07776}{{
  $\bullet$}}
\href{http://inspirehep.net/search?p=find+eprint+1504.07776}{{$ \triangleright
  $}}

\bibitem{Cucchieri:2007ta}
A.~Cucchieri, A.~Maas, and T.~Mendes, ``\textit{{Infrared properties of
  propagators in Landau-gauge pure Yang-Mills theory at finite temperature}},''
  \href{http://dx.doi.org/10.1103/PhysRevD.75.076003}{{\em Phys. Rev.} {\bf
  D75} (2007)  076003} \href{http://arxiv.org/abs/hep-lat/0702022}{{
  $\bullet$}}
\href{http://inspirehep.net/search?p=find+eprint+hep-lat/0702022}{{$
  \triangleright $}}

\bibitem{Bornyakov:2010nc}
V.~G. Bornyakov and V.~K. Mitrjushkin, ``\textit{{SU(2) lattice gluon
  propagators at finite temperatures in the deep infrared region and Gribov
  copy effects}},'' \href{http://dx.doi.org/10.1103/PhysRevD.84.094503}{{\em
  Phys. Rev.} {\bf D84} (2011)  094503} \href{http://arxiv.org/abs/1011.4790}{{
  $\bullet$}}
\href{http://inspirehep.net/search?p=find+eprint+1011.4790}{{$ \triangleright
  $}}

\bibitem{Cucchieri:2011di}
A.~Cucchieri and T.~Mendes, ``\textit{{Electric and magnetic Landau-gauge gluon
  propagators in finite-temperature SU(2) gauge theory}},'' {\em PoS} {\bf
  FACESQCD} (2010)  007 \href{http://arxiv.org/abs/1105.0176}{{ $\bullet$}}
\href{http://inspirehep.net/search?p=find+eprint+1105.0176}{{$ \triangleright
  $}}

\bibitem{Cucchieri:2001tw}
A.~Cucchieri, F.~Karsch, and P.~Petreczky, ``\textit{{Propagators and
  dimensional reduction of hot SU(2) gauge theory}},''
  \href{http://dx.doi.org/10.1103/PhysRevD.64.036001}{{\em Phys. Rev.} {\bf
  D64} (2001)  036001} \href{http://arxiv.org/abs/hep-lat/0103009}{{
  $\bullet$}}
\href{http://inspirehep.net/search?p=find+eprint+hep-lat/0103009}{{$
  \triangleright $}}

\bibitem{Cucchieri:2000cy}
A.~Cucchieri, F.~Karsch, and P.~Petreczky, ``\textit{{Magnetic screening in hot
  nonAbelian gauge theory}},''
  \href{http://dx.doi.org/10.1016/S0370-2693(00)01331-9}{{\em Phys. Lett.} {\bf
  B497} (2001)  80--84} \href{http://arxiv.org/abs/hep-lat/0004027}{{
  $\bullet$}}
\href{http://inspirehep.net/search?p=find+eprint+hep-lat/0004027}{{$
  \triangleright $}}

\bibitem{Aouane:2011fv}
R.~Aouane, V.~G. Bornyakov, E.~M. Ilgenfritz, V.~K. Mitrjushkin,
  M.~Muller-Preussker, and A.~Sternbeck, ``\textit{{Landau gauge gluon and
  ghost propagators at finite temperature from quenched lattice QCD}},''
  \href{http://dx.doi.org/10.1103/PhysRevD.85.034501}{{\em Phys. Rev.} {\bf
  D85} (2012)  034501} \href{http://arxiv.org/abs/1108.1735}{{ $\bullet$}}
\href{http://inspirehep.net/search?p=find+eprint+1108.1735}{{$ \triangleright
  $}}

\bibitem{Bornyakov:2011jm}
V.~G. Bornyakov and V.~K. Mitrjushkin, ``\textit{{Lattice QCD gluon propagators
  near transition temperature}},''
  \href{http://dx.doi.org/10.1142/S0217751X12500509}{{\em Int. J. Mod. Phys.}
  {\bf A27} (2012)  1250050} \href{http://arxiv.org/abs/1103.0442}{{
  $\bullet$}}
\href{http://inspirehep.net/search?p=find+eprint+1103.0442}{{$ \triangleright
  $}}

\bibitem{Silva:2013maa}
P.~J. Silva, O.~Oliveira, P.~Bicudo, and N.~Cardoso, ``\textit{{Gluon screening
  mass at finite temperature from the Landau gauge gluon propagator in lattice
  QCD}},'' \href{http://dx.doi.org/10.1103/PhysRevD.89.074503}{{\em Phys. Rev.}
  {\bf D89} (2014) no.~7, 074503} \href{http://arxiv.org/abs/1310.5629}{{
  $\bullet$}}
\href{http://inspirehep.net/search?p=find+eprint+1310.5629}{{$ \triangleright
  $}}

\bibitem{Rajagopal:1992qz}
K.~Rajagopal and F.~Wilczek, ``\textit{{Static and dynamic critical phenomena
  at a second order QCD phase transition}},''
  \href{http://dx.doi.org/10.1016/0550-3213(93)90502-G}{{\em Nucl. Phys.} {\bf
  B399} (1993)  395--425} \href{http://arxiv.org/abs/hep-ph/9210253}{{
  $\bullet$}}
\href{http://inspirehep.net/search?p=find+eprint+hep-ph/9210253}{{$
  \triangleright $}}

\bibitem{Baker:1977hp}
G.~A. Baker, Jr., B.~G. Nickel, and D.~I. Meiron, ``\textit{{Critical Indices
  from Perturbation Analysis of the Callan-Symanzik Equation}},''
\href{http://dx.doi.org/10.1103/PhysRevB.17.1365}{{\em Phys. Rev.} {\bf B17}
  (1978)  1365--1374}

\bibitem{Alkofer:1986bm}
R.~Alkofer and P.~A. Amundsen, ``\textit{{A Model for the Chiral Phase
  Transition in {QCD}}},''
\href{http://dx.doi.org/10.1016/0370-2693(87)91117-8}{{\em Phys. Lett.} {\bf
  B187} (1987)  395--400}

\bibitem{Blank:2010bz}
M.~Blank and A.~Krassnigg, ``\textit{{T}he {QCD} chiral transition temperature
  in a {D}yson-{S}chwinger-equation context},''
  \href{http://dx.doi.org/10.1103/PhysRevD.82.034006}{{\em Phys. Rev.} {\bf
  D82} (2010)  034006} \href{http://arxiv.org/abs/1004.5301}{{ $\bullet$}}
\href{http://inspirehep.net/search?p=find+eprint+1004.5301}{{$ \triangleright
  $}}

\bibitem{Fischer:2005en}
C.~S. Fischer, P.~Watson, and W.~Cassing, ``\textit{{Probing unquenching
  effects in the gluon polarisation in light mesons}},''
  \href{http://dx.doi.org/10.1103/PhysRevD.72.094025}{{\em Phys. Rev.} {\bf
  D72} (2005)  094025} \href{http://arxiv.org/abs/hep-ph/0509213}{{ $\bullet$}}
\href{http://inspirehep.net/search?p=find+eprint+hep-ph/0509213}{{$
  \triangleright $}}

\bibitem{Son:2001ff}
D.~T. Son and M.~A. Stephanov, ``\textit{{Pion propagation near the QCD chiral
  phase transition}},''
  \href{http://dx.doi.org/10.1103/PhysRevLett.88.202302}{{\em Phys. Rev. Lett.}
  {\bf 88} (2002)  202302} \href{http://arxiv.org/abs/hep-ph/0111100}{{
  $\bullet$}}
\href{http://inspirehep.net/search?p=find+eprint+hep-ph/0111100}{{$
  \triangleright $}}

\bibitem{Pisarski:1996mt}
R.~D. Pisarski and M.~Tytgat, ``\textit{{Propagation of cool pions}},''
  \href{http://dx.doi.org/10.1103/PhysRevD.54.R2989}{{\em Phys. Rev.} {\bf D54}
  (1996)  R2989--R2993} \href{http://arxiv.org/abs/hep-ph/9604404}{{
  $\bullet$}}
\href{http://inspirehep.net/search?p=find+eprint+hep-ph/9604404}{{$
  \triangleright $}}

\bibitem{Maris:2000ig}
P.~Maris, C.~D. Roberts, S.~M. Schmidt, and P.~C. Tandy, ``\textit{{T -
  dependence of pseudoscalar and scalar correlations}},''
  \href{http://dx.doi.org/10.1103/PhysRevC.63.025202}{{\em Phys. Rev.} {\bf
  C63} (2001)  025202} \href{http://arxiv.org/abs/nucl-th/0001064}{{
  $\bullet$}}
\href{http://inspirehep.net/search?p=find+eprint+nucl-th/0001064}{{$
  \triangleright $}}

\bibitem{Schaefer:2004en}
B.-J. Schaefer and J.~Wambach, ``\textit{{The Phase diagram of the quark meson
  model}},'' \href{http://dx.doi.org/10.1016/j.nuclphysa.2005.04.012}{{\em
  Nucl. Phys.} {\bf A757} (2005)  479--492}
  \href{http://arxiv.org/abs/nucl-th/0403039}{{ $\bullet$}}
\href{http://inspirehep.net/search?p=find+eprint+nucl-th/0403039}{{$
  \triangleright $}}

\bibitem{Contant:2017gtz}
R.~Contant and M.~Q. Huber, ``\textit{{Phase structure and propagators at
  nonvanishing temperature for QCD and QCD-like theories}},''
  \href{http://dx.doi.org/10.1103/PhysRevD.96.074002}{{\em Phys. Rev.} {\bf
  D96} (2017) no.~7, 074002} \href{http://arxiv.org/abs/1706.00943}{{
  $\bullet$}}
\href{http://inspirehep.net/search?p=find+eprint+1706.00943}{{$ \triangleright
  $}}

\bibitem{Aouane:2012bk}
R.~Aouane, F.~Burger, E.~M. Ilgenfritz, M.~Muller-Preussker, and A.~Sternbeck,
  ``\textit{{Landau gauge gluon and ghost propagators from lattice QCD with
  $N_f$=2 twisted mass fermions at finite temperature}},''
  \href{http://dx.doi.org/10.1103/PhysRevD.87.114502}{{\em Phys. Rev.} {\bf
  D87} (2013) no.~11, 114502} \href{http://arxiv.org/abs/1212.1102}{{
  $\bullet$}}
\href{http://inspirehep.net/search?p=find+eprint+1212.1102}{{$ \triangleright
  $}}

\bibitem{Herbst:2013ufa}
T.~K. Herbst, M.~Mitter, J.~M. Pawlowski, B.-J. Schaefer, and R.~Stiele,
  ``\textit{{Thermodynamics of QCD at vanishing density}},''
  \href{http://dx.doi.org/10.1016/j.physletb.2014.02.045}{{\em Phys. Lett.}
  {\bf B731} (2014)  248--256} \href{http://arxiv.org/abs/1308.3621}{{
  $\bullet$}}
\href{http://inspirehep.net/search?p=find+eprint+1308.3621}{{$ \triangleright
  $}}

\bibitem{Rennecke:2016tkm}
F.~Rennecke and B.-J. Schaefer, ``\textit{{Fluctuation-induced modifications of
  the phase structure in (2+1)-flavor QCD}},''
  \href{http://dx.doi.org/10.1103/PhysRevD.96.016009}{{\em Phys. Rev.} {\bf
  D96} (2017) no.~1, 016009} \href{http://arxiv.org/abs/1610.08748}{{
  $\bullet$}}
\href{http://inspirehep.net/search?p=find+eprint+1610.08748}{{$ \triangleright
  $}}

\bibitem{Fukushima:2009dx}
K.~Fukushima, ``\textit{{Isentropic thermodynamics in the PNJL model}},''
  \href{http://dx.doi.org/10.1103/PhysRevD.79.074015}{{\em Phys. Rev.} {\bf
  D79} (2009)  074015} \href{http://arxiv.org/abs/0901.0783}{{ $\bullet$}}
\href{http://inspirehep.net/search?p=find+eprint+0901.0783}{{$ \triangleright
  $}}

\bibitem{Fu:2018qsk}
W.-j. Fu, J.~M. Pawlowski, and F.~Rennecke, ``\textit{{Strangeness Neutrality
  and QCD Thermodynamics}},'' \href{http://arxiv.org/abs/1808.00410}{{\tt
  1808.00410[hep-ph]}} \href{http://arxiv.org/abs/1808.00410}{{ $\bullet$}}
\href{http://inspirehep.net/search?p=find+eprint+1808.00410}{{$ \triangleright
  $}}

\bibitem{Fu:2018swz}
W.-j. Fu, J.~M. Pawlowski, and F.~Rennecke, ``\textit{{Strangeness neutrality
  and baryon-strangeness correlations}},''
  \href{http://arxiv.org/abs/1809.01594}{{\tt 1809.01594[hep-ph]}}
  \href{http://arxiv.org/abs/1809.01594}{{ $\bullet$}}
\href{http://inspirehep.net/search?p=find+eprint+1809.01594}{{$ \triangleright
  $}}

\bibitem{Pawlowski:2010ht}
J.~M. Pawlowski, ``\textit{{The QCD phase diagram: Results and challenges}},''
  \href{http://dx.doi.org/10.1063/1.3574945}{{\em AIP Conf. Proc.} {\bf 1343}
  (2011)  75--80} \href{http://arxiv.org/abs/1012.5075}{{ $\bullet$}}
\href{http://inspirehep.net/search?p=find+eprint+1012.5075}{{$ \triangleright
  $}}

\bibitem{Pawlowski:2014aha}
J.~M. Pawlowski, ``\textit{{Equation of state and phase diagram of strongly
  interacting matter}},''
\href{http://dx.doi.org/10.1016/j.nuclphysa.2014.09.074}{{\em Nucl. Phys.} {\bf
  A931} (2014)  113--124}

\bibitem{Alba:2014eba}
P.~Alba, W.~Alberico, R.~Bellwied, M.~Bluhm, V.~Mantovani~Sarti, M.~Nahrgang,
  and C.~Ratti, ``\textit{{Freeze-out conditions from net-proton and net-charge
  fluctuations at RHIC}},''
  \href{http://dx.doi.org/10.1016/j.physletb.2014.09.052}{{\em Phys. Lett.}
  {\bf B738} (2014)  305--310} \href{http://arxiv.org/abs/1403.4903}{{
  $\bullet$}}
\href{http://inspirehep.net/search?p=find+eprint+1403.4903}{{$ \triangleright
  $}}

\bibitem{Becattini:2016xct}
F.~Becattini, J.~Steinheimer, R.~Stock, and M.~Bleicher,
  ``\textit{{Hadronization conditions in relativistic nuclear collisions and
  the QCD pseudo-critical line}},''
  \href{http://dx.doi.org/10.1016/j.physletb.2016.11.033}{{\em Phys. Lett.}
  {\bf B764} (2017)  241--246} \href{http://arxiv.org/abs/1605.09694}{{
  $\bullet$}}
\href{http://inspirehep.net/search?p=find+eprint+1605.09694}{{$ \triangleright
  $}}

\bibitem{Vovchenko:2015idt}
V.~Vovchenko, V.~V. Begun, and M.~I. Gorenstein, ``\textit{{Hadron
  multiplicities and chemical freeze-out conditions in proton-proton and
  nucleus-nucleus collisions}},''
  \href{http://dx.doi.org/10.1103/PhysRevC.93.064906}{{\em Phys. Rev.} {\bf
  C93} (2016) no.~6, 064906} \href{http://arxiv.org/abs/1512.08025}{{
  $\bullet$}}
\href{http://inspirehep.net/search?p=find+eprint+1512.08025}{{$ \triangleright
  $}}

\bibitem{Adamczyk:2017iwn}
{\bf STAR} Collaboration: L.~Adamczyk {\em et al.}, ``\textit{{Bulk Properties
  of the Medium Produced in Relativistic Heavy-Ion Collisions from the Beam
  Energy Scan Program}},''
  \href{http://dx.doi.org/10.1103/PhysRevC.96.044904}{{\em Phys. Rev.} {\bf
  C96} (2017) no.~4, 044904} \href{http://arxiv.org/abs/1701.07065}{{
  $\bullet$}}
\href{http://inspirehep.net/search?p=find+eprint+1701.07065}{{$ \triangleright
  $}}

\bibitem{Andronic:2016nof}
A.~Andronic, P.~Braun-Munzinger, K.~Redlich, and J.~Stachel, ``\textit{{Hadron
  yields, the chemical freeze-out and the QCD phase diagram}},''
  \href{http://dx.doi.org/10.1088/1742-6596/779/1/012012}{{\em J. Phys. Conf.
  Ser.} {\bf 779} (2017) no.~1, 012012}
  \href{http://arxiv.org/abs/1611.01347}{{ $\bullet$}}
\href{http://inspirehep.net/search?p=find+eprint+1611.01347}{{$ \triangleright
  $}}

\bibitem{Andronic:2017pug}
A.~Andronic, P.~Braun-Munzinger, K.~Redlich, and J.~Stachel,
  ``\textit{{Decoding the phase structure of QCD via particle production at
  high energy}},'' \href{http://arxiv.org/abs/1710.09425}{{\tt
  1710.09425[nucl-th]}} \href{http://arxiv.org/abs/1710.09425}{{ $\bullet$}}
\href{http://inspirehep.net/search?p=find+eprint+1710.09425}{{$ \triangleright
  $}}

\bibitem{Strodthoff:2011tz}
N.~Strodthoff, B.-J. Schaefer, and L.~von Smekal,
  ``\textit{{Quark-meson-diquark model for two-color QCD}},''
  \href{http://dx.doi.org/10.1103/PhysRevD.85.074007}{{\em Phys. Rev.} {\bf
  D85} (2012)  074007} \href{http://arxiv.org/abs/1112.5401}{{ $\bullet$}}
\href{http://inspirehep.net/search?p=find+eprint+1112.5401}{{$ \triangleright
  $}}

\bibitem{Strodthoff:2013cua}
N.~Strodthoff and L.~von Smekal, ``\textit{{Polyakov-Quark-Meson-Diquark Model
  for two-color QCD}},''
  \href{http://dx.doi.org/10.1016/j.physletb.2014.03.008}{{\em Phys. Lett.}
  {\bf B731} (2014)  350--357} \href{http://arxiv.org/abs/1306.2897}{{
  $\bullet$}}
\href{http://inspirehep.net/search?p=find+eprint+1306.2897}{{$ \triangleright
  $}}

\bibitem{Khan:2015puu}
N.~Khan, J.~M. Pawlowski, F.~Rennecke, and M.~M. Scherer, ``\textit{{The Phase
  Diagram of QC2D from Functional Methods}},''
  \href{http://arxiv.org/abs/1512.03673}{{\tt 1512.03673[hep-ph]}}
  \href{http://arxiv.org/abs/1512.03673}{{ $\bullet$}}
\href{http://inspirehep.net/search?p=find+eprint+1512.03673}{{$ \triangleright
  $}}

\bibitem{Mu:2012zz}
C.-f. Mu, Y.~Jiang, P.-f. Zhuang, and Y.-x. Liu, ``\textit{{Nucleons at finite
  temperature and low density in Faddeev equation approach}},''
\href{http://dx.doi.org/10.1103/PhysRevD.85.014033}{{\em Phys. Rev.} {\bf D85}
  (2012)  014033}

\bibitem{Wang:2013wk}
K.-l. Wang, Y.-x. Liu, L.~Chang, C.~D. Roberts, and S.~M. Schmidt,
  ``\textit{{Baryon and meson screening masses}},''
  \href{http://dx.doi.org/10.1103/PhysRevD.87.074038}{{\em Phys. Rev.} {\bf
  D87} (2013) no.~7, 074038} \href{http://arxiv.org/abs/1301.6762}{{
  $\bullet$}}
\href{http://inspirehep.net/search?p=find+eprint+1301.6762}{{$ \triangleright
  $}}

\bibitem{Aarts:2017rrl}
G.~Aarts, C.~Allton, D.~De~Boni, S.~Hands, B.~J\"ager, C.~Praki, and J.-I.
  Skullerud, ``\textit{{Light baryons below and above the deconfinement
  transition: medium effects and parity doubling}},''
  \href{http://dx.doi.org/10.1007/JHEP06(2017)034}{{\em JHEP} {\bf 06} (2017)
  034} \href{http://arxiv.org/abs/1703.09246}{{ $\bullet$}}
\href{http://inspirehep.net/search?p=find+eprint+1703.09246}{{$ \triangleright
  $}}

\bibitem{Borsanyi:2012cr}
S.~Borsanyi, G.~Endrodi, Z.~Fodor, S.~D. Katz, S.~Krieg, C.~Ratti, and K.~K.
  Szabo, ``\textit{{QCD equation of state at nonzero chemical potential:
  continuum results with physical quark masses at order $mu^2$}},''
  \href{http://dx.doi.org/10.1007/JHEP08(2012)053}{{\em JHEP} {\bf 08} (2012)
  053} \href{http://arxiv.org/abs/1204.6710}{{ $\bullet$}}
\href{http://inspirehep.net/search?p=find+eprint+1204.6710}{{$ \triangleright
  $}}

\bibitem{Bender:1996bm}
A.~Bender, D.~Blaschke, Y.~Kalinovsky, and C.~D. Roberts, ``\textit{{Continuum
  study of deconfinement at finite temperature}},''
  \href{http://dx.doi.org/10.1103/PhysRevLett.77.3724}{{\em Phys. Rev. Lett.}
  {\bf 77} (1996)  3724--3727} \href{http://arxiv.org/abs/nucl-th/9606006}{{
  $\bullet$}}
\href{http://inspirehep.net/search?p=find+eprint+nucl-th/9606006}{{$
  \triangleright $}}

\bibitem{Braaten:1989mz}
E.~Braaten and R.~D. Pisarski, ``\textit{{Soft Amplitudes in Hot Gauge
  Theories: A General Analysis}},''
\href{http://dx.doi.org/10.1016/0550-3213(90)90508-B}{{\em Nucl. Phys.} {\bf
  B337} (1990)  569--634}

\bibitem{Baym:1992eu}
G.~Baym, J.-P. Blaizot, and B.~Svetitsky, ``\textit{{Emergence of new
  quasiparticles in quantum electrodynamics at finite temperature}},''
\href{http://dx.doi.org/10.1103/PhysRevD.46.4043}{{\em Phys. Rev.} {\bf D46}
  (1992)  4043--4051}

\bibitem{Blaizot:1993bb}
J.-P. Blaizot and J.-Y. Ollitrault, ``\textit{{Collective fermionic excitations
  in systems with a large chemical potential}},''
  \href{http://dx.doi.org/10.1103/PhysRevD.48.1390}{{\em Phys. Rev.} {\bf D48}
  (1993)  1390--1408} \href{http://arxiv.org/abs/hep-th/9303070}{{ $\bullet$}}
\href{http://inspirehep.net/search?p=find+eprint+hep-th/9303070}{{$
  \triangleright $}}

\bibitem{Lebedev:1989ev}
V.~V. Lebedev and A.~V. Smilga, ``\textit{{Spectrum of Quark - Gluon
  Plasma}},''
\href{http://dx.doi.org/10.1016/0003-4916(90)90225-D}{{\em Annals Phys.} {\bf
  202} (1990)  229--270}

\bibitem{Kitazawa:2005mp}
M.~Kitazawa, T.~Kunihiro, and Y.~Nemoto, ``\textit{{Quark spectrum above but
  near critical temperature of chiral transition}},''
  \href{http://dx.doi.org/10.1016/j.physletb.2005.11.076}{{\em Phys. Lett.}
  {\bf B633} (2006)  269--274} \href{http://arxiv.org/abs/hep-ph/0510167}{{
  $\bullet$}}
\href{http://inspirehep.net/search?p=find+eprint+hep-ph/0510167}{{$
  \triangleright $}}

\bibitem{Hidaka:2011rz}
Y.~Hidaka, D.~Satow, and T.~Kunihiro, ``\textit{{Ultrasoft Fermionic Modes at
  High Temperature}},''
  \href{http://dx.doi.org/10.1016/j.nuclphysa.2011.12.007}{{\em Nucl. Phys.}
  {\bf A876} (2012)  93--108} \href{http://arxiv.org/abs/1111.5015}{{
  $\bullet$}}
\href{http://inspirehep.net/search?p=find+eprint+1111.5015}{{$ \triangleright
  $}}

\bibitem{Blaizot:2014hka}
J.-P. Blaizot and D.~Satow, ``\textit{{Ultrasoft fermionic excitation at finite
  chemical potential}},''
  \href{http://dx.doi.org/10.1103/PhysRevD.89.096001}{{\em Phys. Rev.} {\bf
  D89} (2014) no.~9, 096001} \href{http://arxiv.org/abs/1402.0241}{{
  $\bullet$}}
\href{http://inspirehep.net/search?p=find+eprint+1402.0241}{{$ \triangleright
  $}}

\bibitem{Kovtun:2004de}
P.~Kovtun, D.~T. Son, and A.~O. Starinets, ``\textit{{Viscosity in strongly
  interacting quantum field theories from black hole physics}},''
  \href{http://dx.doi.org/10.1103/PhysRevLett.94.111601}{{\em Phys. Rev. Lett.}
  {\bf 94} (2005)  111601} \href{http://arxiv.org/abs/hep-th/0405231}{{
  $\bullet$}}
\href{http://inspirehep.net/search?p=find+eprint+hep-th/0405231}{{$
  \triangleright $}}

\bibitem{Dubla:2018czx}
A.~Dubla, S.~Masciocchi, J.~M. Pawlowski, B.~Schenke, C.~Shen, and J.~Stachel,
  ``\textit{{Towards QCD-assisted hydrodynamics for heavy-ion collision
  phenomenology}},'' \href{http://arxiv.org/abs/1805.02985}{{\tt
  1805.02985[nucl-th]}} \href{http://arxiv.org/abs/1805.02985}{{ $\bullet$}}
\href{http://inspirehep.net/search?p=find+eprint+1805.02985}{{$ \triangleright
  $}}

\bibitem{Schafer:2000tw}
T.~Sch\"afer, ``\textit{{Quark hadron continuity in QCD with one flavor}},''
  \href{http://dx.doi.org/10.1103/PhysRevD.62.094007}{{\em Phys. Rev.} {\bf
  D62} (2000)  094007} \href{http://arxiv.org/abs/hep-ph/0006034}{{ $\bullet$}}
\href{http://inspirehep.net/search?p=find+eprint+hep-ph/0006034}{{$
  \triangleright $}}

\bibitem{Alford:2002rz}
M.~G. Alford, J.~A. Bowers, J.~M. Cheyne, and G.~A. Cowan, ``\textit{{Single
  color and single flavor color superconductivity}},''
  \href{http://dx.doi.org/10.1103/PhysRevD.67.054018}{{\em Phys. Rev.} {\bf
  D67} (2003)  054018} \href{http://arxiv.org/abs/hep-ph/0210106}{{ $\bullet$}}
\href{http://inspirehep.net/search?p=find+eprint+hep-ph/0210106}{{$
  \triangleright $}}

\bibitem{Schmitt:2004et}
A.~Schmitt, ``\textit{{The Ground state in a spin-one color superconductor}},''
  \href{http://dx.doi.org/10.1103/PhysRevD.71.054016}{{\em Phys. Rev.} {\bf
  D71} (2005)  054016} \href{http://arxiv.org/abs/nucl-th/0412033}{{
  $\bullet$}}
\href{http://inspirehep.net/search?p=find+eprint+nucl-th/0412033}{{$
  \triangleright $}}

\bibitem{Rischke:2000ra}
D.~H. Rischke, ``\textit{{Debye screening and Meissner effect in a three flavor
  color superconductor}},''
  \href{http://dx.doi.org/10.1103/PhysRevD.62.054017}{{\em Phys. Rev.} {\bf
  D62} (2000)  054017} \href{http://arxiv.org/abs/nucl-th/0003063}{{
  $\bullet$}}
\href{http://inspirehep.net/search?p=find+eprint+nucl-th/0003063}{{$
  \triangleright $}}

\bibitem{Alford:2017ale}
M.~G. Alford, K.~Pangeni, and A.~Windisch, ``\textit{{Color superconductivity
  and charge neutrality in Yukawa theory}},''
  \href{http://dx.doi.org/10.1103/PhysRevLett.120.082701}{{\em Phys. Rev.
  Lett.} {\bf 120} (2018) no.~8, 082701}
  \href{http://arxiv.org/abs/1712.02407}{{ $\bullet$}}
\href{http://inspirehep.net/search?p=find+eprint+1712.02407}{{$ \triangleright
  $}}

\bibitem{Rajagopal:2000ff}
K.~Rajagopal and F.~Wilczek, ``\textit{{Enforced electrical neutrality of the
  color flavor locked phase}},''
  \href{http://dx.doi.org/10.1103/PhysRevLett.86.3492}{{\em Phys. Rev. Lett.}
  {\bf 86} (2001)  3492--3495} \href{http://arxiv.org/abs/hep-ph/0012039}{{
  $\bullet$}}
\href{http://inspirehep.net/search?p=find+eprint+hep-ph/0012039}{{$
  \triangleright $}}

\bibitem{Ruester:2005jc}
S.~B. Ruester, V.~Werth, M.~Buballa, I.~A. Shovkovy, and D.~H. Rischke,
  ``\textit{{The Phase diagram of neutral quark matter: Self-consistent
  treatment of quark masses}},''
  \href{http://dx.doi.org/10.1103/PhysRevD.72.034004}{{\em Phys. Rev.} {\bf
  D72} (2005)  034004} \href{http://arxiv.org/abs/hep-ph/0503184}{{ $\bullet$}}
\href{http://inspirehep.net/search?p=find+eprint+hep-ph/0503184}{{$
  \triangleright $}}

\bibitem{Abuki:2005ms}
H.~Abuki and T.~Kunihiro, ``\textit{{Extensive study of phase diagram for
  charge neutral homogeneous quark matter affected by dynamical chiral
  condensation: Unified picture for thermal unpairing transitions from weak to
  strong coupling}},''
  \href{http://dx.doi.org/10.1016/j.nuclphysa.2005.12.019}{{\em Nucl. Phys.}
  {\bf A768} (2006)  118--159} \href{http://arxiv.org/abs/hep-ph/0509172}{{
  $\bullet$}}
\href{http://inspirehep.net/search?p=find+eprint+hep-ph/0509172}{{$
  \triangleright $}}

\bibitem{Das:1997}
A.~K. Das, {\em Finite Temperature Field Theory}.
\newblock Singapore: World Scientific 1997

\end{thebibliography}\endgroup

\end{document}